\documentclass{article}
\usepackage[a4paper,left=2cm,top=1.5cm,bottom=1.5cm,right=2cm]{geometry}
\usepackage[utf8]{inputenc}

\usepackage{color}
\usepackage{xcolor}
\usepackage[hyphens]{url}
\usepackage{hyperref}
\usepackage[hyphenbreaks]{breakurl}
\definecolor{linkcolour}{rgb}{0,0.2,0.6}
\hypersetup{colorlinks,breaklinks,urlcolor=linkcolour, linkcolor=linkcolour, citecolor=blue}
\usepackage[onehalfspacing]{setspace}
\usepackage{graphicx}
\usepackage{caption}
\usepackage{amsmath}
\usepackage{amssymb}
\usepackage{multirow}
\usepackage{setspace} 
\usepackage{placeins}
\usepackage{booktabs}
\usepackage{authblk}
\usepackage{afterpage}
\usepackage{float}
\usepackage{tabularx}

\usepackage{subfigure} 
\makeatletter
\renewcommand{\@thesubfigure}{\normalsize(\textbf{\alph{subfigure}})}
\makeatother

\usepackage{feynmp} % For Feynman diagrams
\RequirePackage[style=numeric,backend=biber,sorting=none]{biblatex}
\begin{document}

\newcolumntype{C}{>{\centering\arraybackslash}X}

% Useful macros
\newcommand{\ifb}{\ensuremath{\,\mathrm{fb}^{-1}}}
\newcommand{\pb}{\ensuremath{\,\mathrm{pb}}}
\newcommand{\GeV}{\ensuremath{\,\mathrm{GeV}}}
\newcommand{\TeV}{\ensuremath{\,\mathrm{TeV}}}

\newcommand{\pT}{\ensuremath{p_\mathrm{T}}}
\newcommand{\HT}{\ensuremath{H_\mathrm{T}}}

\newcommand{\pileup}{pileup}
\newcommand{\Pileup}{Pileup}
\newcommand{\hardscatter}{hard-scatter}

\newcommand{\kT}{\ensuremath{k_t}}
\newcommand{\ntikt}{nti-\kT{}}
\newcommand{\antikt}{a\ntikt{}}
\newcommand{\Antikt}{A\ntikt{}}
\newcommand{\mallR}{mall-$R$}
\newcommand{\smallR}{s\mallR{}}
\newcommand{\SmallR}{S\mallR{}}
\newcommand{\argeR}{arge-$R$}
\newcommand{\largeR}{l\argeR{}}
\newcommand{\LargeR}{L\argeR{}}
\newcommand{\opocluster}{opo-cluster}
\newcommand{\topocluster}{t\opocluster{}}
\newcommand{\topoclusters}{\topocluster{}s}
\newcommand{\Topocluster}{T\opocluster{}}
\newcommand{\Topoclusters}{\Topocluster{}s}
\newcommand{\insitu}{{in situ}}
\newcommand{\Insitu}{{In situ}}

\title{Searching for New Physics in Hadronic Final States with Run 2 Proton--Proton Collision Data at the LHC}

\date{\today}

\author{Steven~Schramm}
\affil{Département de Physique Nucléaire et Corpusculaire, Université de Genève, Genève; Switzerland}

\maketitle

\abstract{
The symmetries of the Standard Model give rise to the forces that act on particles, and the corresponding force mediators. While the Standard Model is an excellent description of particle interactions, it has known limitations; it is therefore important to search for new physics beyond the
Standard Model, potentially indicating as-of-yet unknown symmetries of nature. The ATLAS and CMS collaborations have detailed physics programmes, involving a large number of searches for new physics in hadronic final states.
As the start of Run 3 of the LHC is imminent, now is a good time to review the progress made and the status of hadronic searches during Run 2 at a centre-of-mass collision energy of $\sqrt{s}=13\,\mathrm{TeV}$.
This review provides an overview of the motivations and challenges of hadronic final states at the LHC, followed by an introduction to jet reconstruction, calibration, and tagging.
Three classes of searches for new physics in hadronic final states are discussed: di-jet searches, searches for missing transverse momentum in association with another object, and searches for hadronic di-boson resonances.
The complementarity of these different analysis strategies is discussed, emphasising the importance of a varied hadronic physics programme in the search for new physics.
}

% Introduction

\section{Introduction}
\label{sec:intro}

The search for phenomena that are not described by the Standard Model of particle physics, often referred to as the search for physics beyond the Standard Model, is of fundamental importance to modern physics. The Standard Model describes the basic constituents of matter, and their interactions, where the interactions arise from symmetries in nature.
While the Standard Model has stood up to a plethora of tests so far, it also has limitations, and it must break down at some higher energy scale at or before the Planck scale; probing higher and higher energy scales and looking for deviations from Standard Model expectations, potentially representing the existence of new fundamental symmetries, is thus one of the key methods to search for new physics.

High-energy particle interactions are statistical rather than deterministic in nature; thus, it is important to be able to gather a large amount of data when searching for new elusive phenomena.
The LHC \cite{LHC} at CERN is instrumental to this approach: it is the highest-energy particle accelerator in the world, providing proton--proton collisions at a centre-of-mass energy of 13\TeV{}, and has delivered a huge dataset of roughly 150\ifb{} to both the ATLAS \cite{ATLAS} and CMS \cite{CMS} Experiments during the Run 2 (2015--2018) data-taking period.
As the LHC is preparing to begin Run 3 in 2022, now is an important time to review the status of searches for physics beyond the Standard Model.

Searches for new physics represent a major part of both the ATLAS and CMS physics programmes, where each collaboration consists of thousands of scientific authors, pursuing a large variety of different possible types of new physics.
A single review is insufficient to accurately represent the diversity of work that has been conducted by ATLAS and CMS, and thus this review will focus on signature-driven searches for new physics in hadronic final states.
This includes both classical hadronic signatures, where the final state involves individual light quarks and/or gluons, as well as more recently studied hadronic signatures, where the final state includes collimated hadronic decays of massive particles, such as $W$/$Z$ bosons or top quarks. This review is complementary to others covering more specific types of searches for new physics: searches with third-generation quarks \cite{Symmetry:ThirdGen}, extended Higgs sectors \cite{Symmetry:HBSM}, or di-Higgs final states \cite{Symmetry:HH}.
Analyses involving hadronic final states, but which are closely related to those other topics, are thus covered in the other review of relevance.

This review begins by summarizing the motivations and challenges of hadronic searches at the LHC in Section \ref{sec:motivation}, and discussing how hadronic final states are reconstructed as different types of jets by the ATLAS and CMS experiments in Section \ref{sec:reco}.
This background is then applied to three classes of signature-based hadronic final state searches: di-jet searches in Section \ref{sec:dijet}, missing transverse momentum searches in Section \ref{sec:monoX}, and hadronic di-boson searches in Section \ref{sec:VV}.
The complementarity of these different final states is discussed in Section \ref{sec:complementarity}, before the review concludes in Section \ref{sec:outlook}.

% Motivation for hadronic searches
\section{Motivations and Challenges for Hadronic Searches at the LHC}
\label{sec:motivation}

Hadronic signatures can be more challenging than the equivalent lepton- or photon-based searches, especially at the LHC.
A clear example of this is the discovery of the Higgs boson in 2012: this historic observation of a new particle by ATLAS \cite{ATLAS:Higgs2012} and CMS \cite{CMS:Higgs2012} was significantly driven by the leptonic and photon decay modes of the Higgs boson in both experiments, with only minor contributions to the discovery sensitivity from the hadronic decay modes.
With this in mind, it is worth discussing both where hadronic final states can be powerful tools in the search for new physics at the LHC, as well as the challenges that must be overcome in such searches.

\subsection{Motivations}
\label{sec:motivation:motivations}

% Motivations:
% Production = decay --> dijet-style searches
% ISR rate --> MET+X searches
% Branching fraction --> hadronic decay modes such as VV

One of the most striking motivations for the use of hadronic final states in the search for new physics at the LHC relates to limiting the assumptions that must be made when searching for phenomena beyond the Standard Model.
In proton--proton collisions, such as those delivered by the LHC, the overwhelming majority of the high-energy collisions will occur as the result of quark--quark, quark--gluon, or gluon--gluon interactions.
If a new particle is produced at tree-level by such interactions, then it can also decay via the same couplings to quarks and/or gluons, unless there are kinematic constraints.
In this way, some types of searches in hadronic final states can avoid making any additional assumptions about the couplings the new particle may or may not have to the other Standard Model particles.
This motivation is particularly relevant to di-jet searches, which will be discussed in Section \ref{sec:dijet}.
Feynman diagrams demonstrating this motivation are provided in Figure \ref{fig:motivation:production}.

\begin{figure}[H]
\hspace{-12pt}
% \subfigure[]{
% \includegraphics[width=0.45\textwidth]{figures/motivation/production-same.pdf}
% \label{fig:motivation:production:same}
% }
% \subfigure[]{
% \includegraphics[width=0.45\textwidth]{figures/motivation/production-diff.pdf}
% \label{fig:motivation:production:diff}
% }
\subfigure[]{
 \begin{fmffile}{treeHad}
 \begin{fmfgraph*}(150,50)
 \fmfleft{qbar,q}
 \fmfright{qbaro,qo}
 
% \fmflabel{$q$}{q}
% \fmflabel{$\bar{q}$}{qbar}
 
 \fmf{vanilla,label=$\textcolor{red}{q}$,foreground=red}{q,vertexL}
 \fmf{vanilla,label=$\textcolor{red}{\bar{q}}$,label.side=right,foreground=red}{qbar,vertexL}
 \fmf{vanilla,label=$\textcolor{red}{q}$,foreground=red}{vertexR,qo}
 \fmf{vanilla,label=$\textcolor{red}{\bar{q}}$,foreground=red}{vertexR,qbaro}
 \fmfv{decor.shape=circle,decor.filled=full,
decor.size=2thick,label=$\textcolor{red}{g_{q\bar{q}X}}$,foreground=red}{vertexL}
 \fmfv{decor.shape=circle,decor.filled=full,
decor.size=2thick,label=$\textcolor{red}{g_{q\bar{q}X}}$,foreground=red}{vertexR}
 
 \fmf{zigzag,label=$X$}{vertexL,vertexR}
 \end{fmfgraph*}
 \end{fmffile}
}
\subfigure[]{
 \begin{fmffile}{treeLep}
 \begin{fmfgraph*}(150,50)
 \fmfleft{qbar,q}
 \fmfright{lbar,l}
 
% \fmflabel{$q$}{q}
% \fmflabel{$\bar{q}$}{qbar}
 
 \fmf{vanilla,label=$\textcolor{red}{q}$,foreground=red}{q,vertexL}
 \fmf{vanilla,label=$\textcolor{red}{\bar{q}}$,label.side=right,foreground=red}{qbar,vertexL}
 \fmf{vanilla,label=$\textcolor{blue}{\ell^-}$,foreground=blue}{vertexR,l}
 \fmf{vanilla,label=$\textcolor{blue}{\ell^+}$,foreground=blue}{vertexR,lbar}
 \fmfv{decor.shape=circle,decor.filled=full,
decor.size=2thick,label=$\textcolor{red}{g_{q\bar{q}X}}$,foreground=red}{vertexL}
 \fmfv{decor.shape=circle,decor.filled=full,
decor.size=2thick,label=$\textcolor{blue}{g_{\ell^-\ell^+X}}$,foreground=blue}{vertexR}
 
 \fmf{zigzag,label=$X$}{vertexL,vertexR}
 \end{fmfgraph*}
 \end{fmffile}
}
\caption{The tree-level production of a hypothetical new particle $X$, from a pair of quarks, is one possible way in which new physics can be produced at the LHC. The particle $X$ is typically assumed to be unstable, and thus promptly decays back to Standard Model particles. (\textbf{a}) These decay products could be the same as the production mechanism, namely, a pair of quarks, meaning that the search for the new particle $X$ relies on assuming the existence of only a single coupling between the Standard Model and $X$ ($g_{q\bar{q}X}$). (\textbf{b}) In contrast, the decay could be to some other pair of Standard Model particles, such as leptons, which introduces an assumption on a second coupling between the Standard Model and $X$ ($g_{q\bar{q}X}$ and $g_{\ell^-\ell^+ X}$). While these diagrams are shown for a charge-neutral $X$, they can be adjusted for other types of $X$, thereby reaching the same conclusion: the hadronic strategy requires fewer assumptions about the couplings between the Standard Model and the new particle. \label{fig:motivation:production}}
\end{figure}
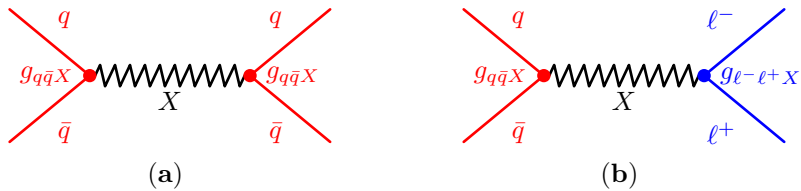

When searching for some types of particles beyond the Standard Model, it is important that the new particle is not produced in isolation, but rather it is produced together with an initial-state radiated (ISR) Standard Model particle.
This alters the momentum balance of the event, which is the entire basis of the searches described in Section \ref{sec:monoX}, and which also allows for circumventing experimental challenges faced by some of the searches discussed in Section \ref{sec:dijet}.
As the initial collision at the LHC occurs between quarks and/or gluons, the most probable source of ISR radiation is also a quark or gluon, due to the dominance of the strong coupling.
In such cases, there is therefore a statistical advantage to hadronic final states, which can lead to hadronic signatures dominating the sensitivity to new physics when the statistical sensitivity is the primary limitation.
Examples of how this can work for a possible new particle produced through quark--antiquark annihilation include the radiation of a gluon before the annihilation occurs, or one of the quarks in the annihilation originating from a gluon splitting process.
Feynman diagrams demonstrating these specific examples are provided in Figure \ref{fig:motivation:ISR}.

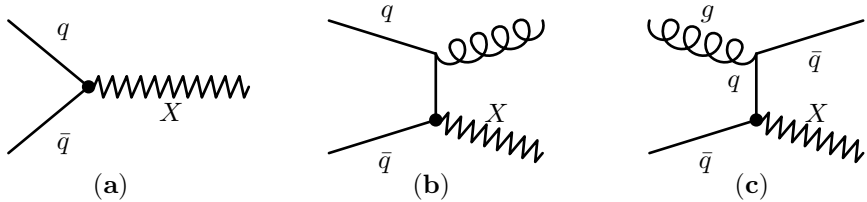
\begin{figure}[H]

% \subfigure[]{
% \includegraphics[width=0.31\textwidth]{figures/motivation/ISR-none.pdf}
% \label{fig:motivation:ISR:none}
% }
% \subfigure[]{
% \includegraphics[width=0.31\textwidth]{figures/motivation/ISR-rad.pdf}
% \label{fig:motivation:ISR:rad}
% }
% \subfigure[]{
% \includegraphics[width=0.31\textwidth]{figures/motivation/ISR-split.pdf}
% \label{fig:motivation:ISR:split}
% }
 \subfigure[]{
 \begin{fmffile}{ISRnone}
 \begin{fmfgraph*}(100,50)
 \fmfleft{qbar,q}
 \fmfright{med}
 
% \fmflabel{$q$}{q}
% \fmflabel{$\bar{q}$}{qbar}
 
 \fmf{vanilla,label=$q$,label.side=left}{q,vertex}
 \fmf{vanilla,label=$\bar{q}$}{qbar,vertex}
 \fmfdot{vertex}
 
 \fmf{zigzag,label=$X$}{vertex,med}
 \end{fmfgraph*}
 \end{fmffile}
 }
 \subfigure[]{
 \begin{fmffile}{ISRrad}
 \begin{fmfgraph*}(100,50)
 \fmfleft{qbar,q}
 \fmfright{med,g}
 
% \fmflabel{$q$}{q}
% \fmflabel{$\bar{q}$}{qbar}
 
 \fmf{vanilla,label=$q$}{q,ISRvertex}
 \fmf{vanilla}{ISRvertex,vertex}
 \fmf{vanilla,label=$\bar{q}$}{qbar,vertex}
 \fmfdot{vertex}
 
 \fmf{zigzag,label=$X$}{vertex,med}
 \fmf{gluon}{ISRvertex,g}
 \end{fmfgraph*}
 \end{fmffile}
 }
 \subfigure[]{
 \begin{fmffile}{ISRsplit}
 \begin{fmfgraph*}(100,50)
 \fmfleft{qbar,g}
 \fmfright{med,ISRq}
 
% \fmflabel{$q$}{q}
% \fmflabel{$\bar{q}$}{qbar}
 
 \fmf{gluon,label=$g$}{g,ISRvertex}
 \fmf{vanilla,label=$q$}{ISRvertex,vertex}
 \fmf{vanilla,label=$\bar{q}$}{qbar,vertex}
 \fmfdot{vertex}
 
 \fmf{zigzag,label=$X$}{vertex,med}
 \fmf{vanilla,label=$\bar{q}$}{ISRvertex,ISRq}
 \end{fmfgraph*}
 \end{fmffile}
 }
 \caption{(\textbf{a}) The original quark--antiquark annihilation process of interest, which is subsequently modified to add Initial-State Radiation (ISR) via (\textbf{b}) gluon emission off of one of the initial-state quarks resulting in an ISR gluon or (\textbf{c}) an initial gluon splitting to a pair of quarks, where one quark participates in the annihilation process and the other remains separate as an ISR quark. \label{fig:motivation:ISR}}
\end{figure}

Another motivation for hadronic signatures comes from the branching ratios of the decays of massive particles: the $W$, $Z$, and $H$ bosons, as well as the top quark, all decay primarily to hadronic final states.
As such, searches for new physics involving such massive particles can benefit from the larger branching ratios and thus increased statistical power of hadronic decays, if they can overcome the associated challenges.
This is the primary motivation for the hadronic di-boson searches described in Section \ref{sec:VV}, although it is also relevant to some of the searches in Section \ref{sec:monoX}.
Some representative branching ratios motivating the statistical potential of using hadronic decays in such cases are provided in Table \ref{tab:motivation:BRs}.
Comparing the branching fractions in this table to the aforementioned example of the Higgs boson discovery, which was primarily driven by the low-statistics $H\to{}\gamma\gamma$, $H\to{}ZZ\to{}\ell\ell\ell\ell$, and $H\to{}WW\to{}\ell\nu\ell\nu$ channels, it is clear that the statistical power of hadronic final states needs to be carefully balanced against the associated challenges.

\begin{table}[H] 
\caption{ Branching fractions for massive particles to leptonic and hadronic final states.\label{tab:motivation:BRs}}

\begin{tabularx}{\textwidth}{CCCC}
\toprule
\textbf{Particle} & \textbf{Decay} & \textbf{Fraction ($\Gamma_i/\Gamma$)} \cite{PDG} & \textbf{Ratio ($\Gamma_\mathrm{had}/\Gamma_i$)}\\
\midrule
\multirow{3}{*}{$W$ boson} & $e\nu_e + \mu\nu_\mu$ & 21.34\% & 3.159\\
 & $\tau\nu_\tau$ & 11.38\% & 5.924\\
 & Hadronic & 67.41\% & \\
\cmidrule{2-4}
\multirow{4}{*}{$Z$ boson} & $e^+e^- + \mu^+\mu^-$ & 6.7294\% & 10.389\\
 & $\tau^+\tau^-$ & 3.3696\% & 20.659\\
 & Invisible & 20.000\% & 3.4806\\
 & Hadronic & 69.911\% & \\
\cmidrule{2-4}
\multirow{5}{*}{$H$ boson} & $\gamma\gamma$ & 0.227\% & 269\\
 & $ZZ$ & 2.62\% & 23.3\\
 & $W^+W^-$ & 21.4\% & 2.85\\
 & $\tau^+\tau^-$ & 6.27\% & 9.74\\
% & $b\bar{b}$ & 58.2\% & \\
% & $c\bar{c}$ & 2.89\% & \\
 & Hadronic ($b\bar{b}+c\bar{c}$) & 61.09\% & \\
\cmidrule{2-4}
\multirow{3}{*}{top quark} & $e\nu_e b + \mu\nu_\mu b$ & 22.50\% & 2.96\\
 & $\tau\nu_\tau b$ & 10.7\% & 6.21\\
 & Hadronic ($q\bar{q}b$) & 66.5\% & \\
\bottomrule
\end{tabularx}
\end{table}

\subsection{Challenges}
\label{sec:motivation:challenges}

% Challenges:
% Enormous QCD background if pure hadronic
% Trigger thresholds due to QCD background
% Pileup at the LHC

In the search for new physics in hadronic final states at the LHC, one of the immediate challenges is the enormous background from Standard Model processes; the dominant visible scattering product from proton--proton collisions is hadronic physics.
As shown in Figure \ref{fig:motivation:xsec}, the dominant hadronic physics process (referred to as ``Jets'', to be discussed more in Section \ref{sec:reco}) has a cross-section of roughly $10^6\pb{}$, while the inclusive Higgs boson production cross-section is more than four orders of magnitude lower.
This huge difference in cross-section means that hadronic final state analyses must either measure the Standard Model background to extreme precision before searching for deviations, such as is done for the searches in Section \ref{sec:dijet}, or find a way to suppress the Standard Model background while enhancing the beyond Standard Model signal of interest, as is the case for the searches described in Sections \ref{sec:monoX} and \ref{sec:VV}.

\vspace{-6pt}
\begin{figure}[H]

\includegraphics[width=0.6666\textwidth]{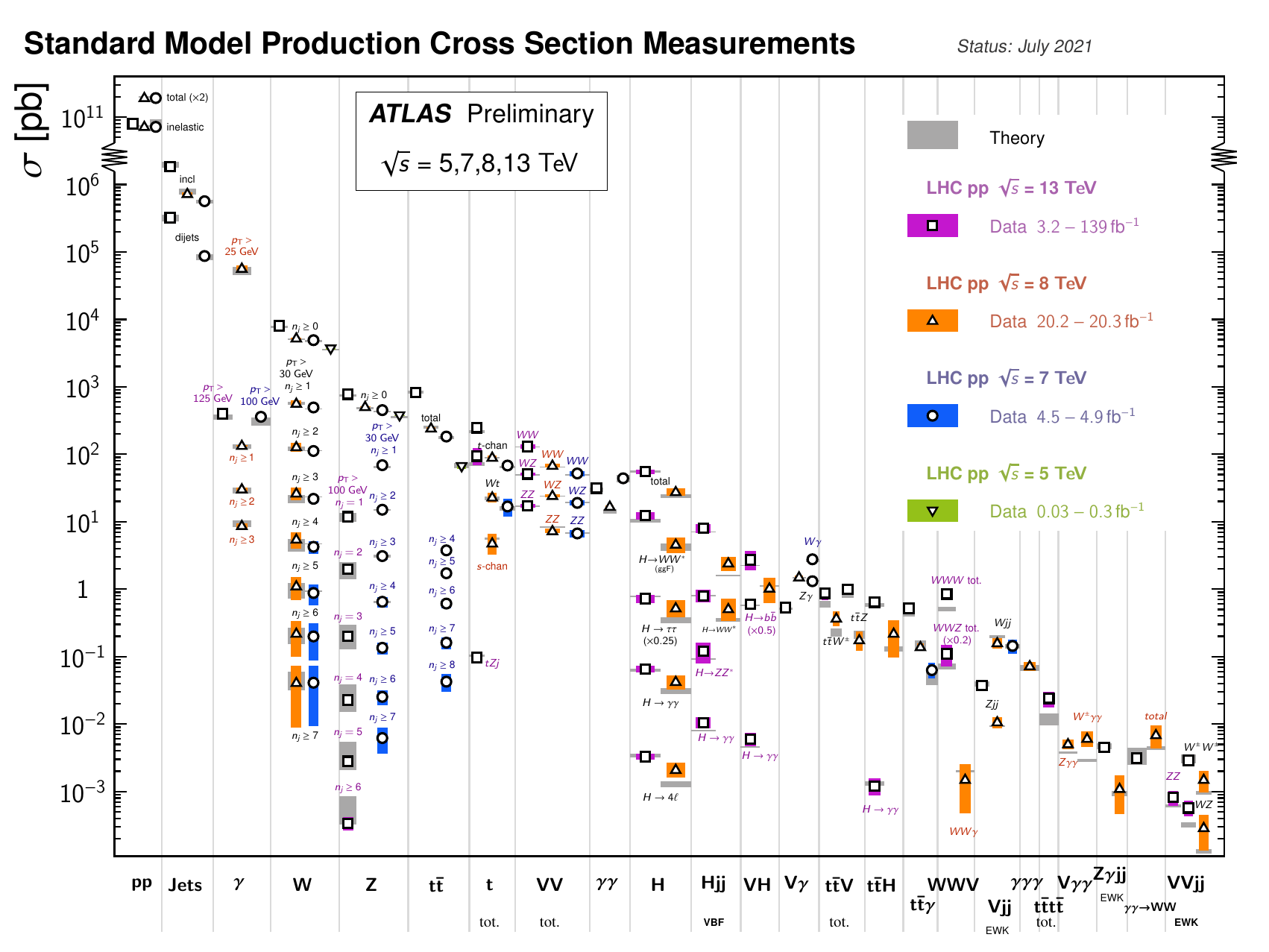}
\caption{A summary of the cross-section measurements for a variety of Standard Model processes in proton--proton (pp) collisions at the LHC, showing the dominance of hadronic processes (labelled as ``Jets'') \cite{ATLAS:SMxsec}. \label{fig:motivation:xsec}}
\end{figure}

The enormous cross-section of Standard Model hadronic physics processes also imposes significant experimental constraints on the analysis of hadronic final states.
In particular, it is not possible to record every hadronic physics process resulting from proton--proton collisions, as that would overwhelm the detector readout and data storage capabilities of the experiments.
In order to mitigate this data volume constraint, the ATLAS and CMS experiments only record very-high-energy collisions involving hadronic final states, while the lower cross-sections of leptonic Standard Model processes allow for recording all leptonic collisions down to much lower energies.
Table \ref{tab:motivation:triggers} gives example hadronic and leptonic triggers used by ATLAS and CMS during Run 2, demonstrating the order of magnitude difference in the collision energy scale that is recorded.
Searches for new physics in hadronic final states at the LHC are thus confined to the high energy regime, unless the analysis has a way to mitigate this trigger constraint, as is the case for some of the searches presented in Section \ref{sec:dijet}.

In order to maximize the potential to discover new physics at the LHC, it is important to have as large of a dataset as possible, therefore mitigating the probabilistic nature of both proton--proton collisions and the subsequent particle interactions.
One way to increase the dataset size is to collide multiple pairs of protons simultaneously, as the probability to produce a very rare interaction is thus enhanced by the number of concurrent collisions; the LHC did this during Run 2, with an average of roughly 30 simultaneous collisions.
In cases where a rare process occurs, this typically results in one rare high-energy process (usually called the \hardscatter{} collision) and several other lower-energy processes (usually called \pileup{} collisions).
Particles from these \pileup{} collisions can pollute the \hardscatter{} process, as the detector records all of the sources of energy, and it is left to reconstruction procedures to tell apart the different origins of each energy deposit in the detector.
This is a challenge for all types of analyses, but it is particularly challenging for hadronic final states as the jets used to reconstruct hadronic interactions are large and thus more susceptible to randomly overlapping contributions from \pileup{} collisions.
Furthermore, as shown in Figure \ref{fig:motivation:xsec}, hadronic processes are the most common visible by-product of proton--proton collisions at ATLAS and CMS.
It can therefore be difficult to differentiate between hadronic physics originating from the \hardscatter{} process, as opposed to the \pileup{} collisions, which is a problem that many searches for new physics in hadronic final states must address.
The searches presented in Sections \ref{sec:dijet}--\ref{sec:VV} use a variety of different techniques to mitigate the impact of \pileup{} on their respective sensitivities to new physics.

\begin{table}[H]
\caption{ Triggers used to record hadronic, leptonic, and photon physics processes during Run 2 of the LHC.
\label{tab:motivation:triggers}}
\begin{tabularx}{\textwidth}{CCC}
\toprule
\textbf{Trigger Type} & \textbf{Experiment} & \textbf{Trigger Threshold}\\
\midrule
\multirow{2}{*}{Hadronic (single-jet)}
 & ATLAS & 420\GeV{} \cite{ATLAS:triggers} \\
 & CMS & 500\GeV{} \cite{CMS:triggers} \\
\cmidrule{2-3}
% \multirow{2}{*}{\textbf{Missing transverse momentum}}
% & ATLAS & 200\GeV{} \cite{ATLAS:triggers} \\
% & CMS & 120\GeV{} \cite{CMS:triggers} \\
% \cmidrule{2-3}
\multirow{2}{*}{Electrons}
 & ATLAS & 26\GeV{} \cite{ATLAS:triggers} \\
 & CMS & 28\GeV{} \cite{CMS:triggers} \\
\cmidrule{2-3}
\multirow{2}{*}{Photons}
 & ATLAS & 140\GeV{} \cite{ATLAS:triggers} \\
 & CMS & 200\GeV{} \cite{CMS:triggers} \\
\cmidrule{2-3}
\multirow{2}{*}{Muons}
 & ATLAS & 26\GeV{} \cite{ATLAS:triggers} \\
 & CMS & 24\GeV{} \cite{CMS:triggers} \\
\bottomrule
\end{tabularx}
\end{table}

% Hadronic physics reconstruction and performance
\section{Hadronic Physics Reconstruction and Performance}
\label{sec:reco}

% Colour confinement
% Hadronic showers
% Resolved vs boosted jet interpretations

Quarks and gluons both carry a colour charge; colour confinement states that they are thus unable to exist in isolation.
Instead, they fragment and hadronise to form collimated streams of colour-neutral particles, such as pions, kaons, and other hadrons.
These collimated streams of particles are typically referred to as hadronic showers, especially when discussing their subsequent interactions with a particle detector.

While quarks and gluons cannot exist in isolation, their properties can be inferred by gathering all of the produced particles in the corresponding hadronic shower and summing the resulting set of four-vectors.
There is no single definitive way to do this, as various effects make it difficult to identify which particles or which detector energy deposits come from which originating quark and/or gluon.
Procedures that define how to group the individual four-vectors of hadronic showers are referred to as jet algorithms, and the resulting summed four-vectors are referred to as jets.
Jets are the backbone of hadronic physics analyses at the LHC, where both ATLAS and CMS predominantly use the \antikt{} jet algorithm \cite{antikt}.

The same jet algorithm can be applied to a variety of different types of input four-vectors, resulting in distinct sets of jets with different strengths and limitations.
The algorithm can also be run with different algorithmic parameters: the radius or distance parameter $R$ in particular is often manipulated to adapt the jet algorithm for different intended use cases.
Standard jets in both ATLAS and CMS, often referred to as \smallR{} jets, currently use $R=0.4$; this has been found to provide robust performance for the reconstruction of hadronic showers originating from individual light quarks (up, down, strange, charm, and bottom; also called not-top quarks) and/or gluons.

In some cases, it is advantageous to represent multiple hadronic showers as a single jet.
This is often the case for analyses involving hadronic decays of massive particles, such as $W/Z/H$ bosons or top quarks, if the parent particle has a high transverse momentum.
While the decays of such particles are back-to-back in their own reference frame, the boost to the experimental frame leads to overlapping hadronic showers, which cannot be easily disentangled.
It therefore makes sense to reconstruct the entire decay (including the subsequent parton shower and hadronisation processes) as a single jet, leading to the production of four-vectors that should correspond to the properties of the initial massive particles of interest rather than the daughter particles.
The reconstruction of such boosted jets, often referred to as \largeR{} jets, currently differs between ATLAS and CMS.
From the jet algorithm perspective, ATLAS currently uses $R=1.0$ in contrast to $R=0.8$ as used by CMS, but there are also other differences as will be discussed.

\subsection{Inputs to Jet Reconstruction}
\label{sec:reco:inputs}

% Tracks
% Calorimeter clusters
% PFlow

ATLAS and CMS are general-purpose particle physics detectors, and are built in layers; starting from the interaction point and moving radially outwards, they consist of tracking detectors, electromagnetic calorimeters, hadronic calorimeters, and muon systems \cite{ATLAS,CMS}.
Hadronic showers, as collections of particles, produce quite a complex signature in such detectors, and are primarily of relevance to the tracking detectors and both types of calorimeters.
Hadronic showers can also occasionally reach the muon systems, such as from heavy flavour decays or calorimeter punch-through; these are not the focus of this section and will thus not be discussed further here.

To first order, hadronic showers are comprised of equal fractions of the three different types of pions: $\pi^+$, $\pi^0$, and $\pi^-$.
Charged and neutral pions interact very differently with the detector.
Charged pions ($\pi^\pm$) live long enough to traverse the detector and are electrically charged; thus, they create tracks in the tracking detectors, and interact primarily through the strong force within both electromagnetic and hadronic calorimeters.
In contrast, neutral pions ($\pi^0$) have a very short lifetime, decaying quickly to pairs of photons; as photons are neutral electromagnetic particles, they do not leave tracks in the tracking detectors, and typically deposit their energy within the electromagnetic calorimeters.

\subsubsection{Calorimeter-Based Inputs}

As hadronic showers contain both neutral and charged particles, but trackers only observe charged particles, the ATLAS and CMS calorimeters play a key role in jet reconstruction.
To this end, both ATLAS and CMS have used calorimeter energy deposits as the inputs to jet reconstruction, with ATLAS building an object referred to as a topological cluster (\topocluster{}) \cite{ATLAS:topoclusters} and CMS using geometrically projected calorimeter towers \cite{CMS:Run1JETM}.
CMS has primarily moved on from such calorimeter-tower-based jet reconstruction, and thus they are not described further here.

\Topoclusters{} are reconstructed in ATLAS from topologically-adjacent calorimeter cells, using an algorithm consisting of four steps: seed-finding, expansion, boundary addition, and splitting.
Seed-finding proceeds by identifying all cells in the calorimeter which have at least four times the amount of energy as expected from noise ($|E/\sigma_E|>4$); such cells form the starting points of individual clusters.
These seeds are then expanded iteratively in all three dimensions, incorporating adjacent cells with at least two times the amount of energy expected from noise ($|E/\sigma_E|>2$).
Once this iterative expansion is finished, a final layer of all adjacent cells are added to each \topocluster{}, regardless of their energy value ($|E/\sigma_E|>0$).
After forming this initial set of \topoclusters{}, a search for multiple local maxima within each cluster is performed, after which point a given cluster may be split if multiple maxima are identified.
% a search for local energy minima within each cluster is performed, after which point a given cluster may be split if such a minimum is found.

The resulting \topoclusters{} can either be left as-is, or they can be further calibrated before being used as input to jet reconstruction.
If they are left uncalibrated, they are referred to as electromagnetic (EM) \topoclusters{}, as they do not account for calorimeter non-compensation and thus correspond to the energy scale of an electromagnetic particle (electron or photon) in the calorimeter.
They can alternatively be further calibrated using the Local Cell Weighting (LCW) procedure, which determines a probability that a given \topocluster{} corresponds to an electromagnetic or hadronic shower, and then applies a calibration weighted by that probability.
During Run 2, ATLAS used both EM and LCW \topoclusters{} as inputs to jet reconstruction in different contexts, as will be described further.

\subsubsection{Particle Flow Inputs}

While the calorimeter is instrumental in the reconstruction of hadronic showers, the tracking detector provides complementary information, which can dramatically improve jet performance in certain contexts.
Sampling calorimeters, such as are used by ATLAS and CMS, record only a fraction of the deposited shower momentum; they are therefore more sensitive to variations in shower development at low momentum, and better measure the shower properties at high momentum.
In contrast, tracking detectors rely on measuring the curvature of charged particles traversing a magnetic field in order to evaluate their momentum; the curvature is proportional to the inverse of the transverse momentum, and thus this can be done more precisely at low momentum, as at high momentum the tracks become straight.
It is therefore natural to consider combining the information from both types of detectors in order to maximally benefit across the full kinematic regime of interest; this procedure is typically referred to as particle flow.

While the ideas behind particle flow are generally similar, the actual algorithms implementing particle flow are usually experiment-specific, as the optimal balance between the tracking detector and calorimeter is detector-dependent.
ATLAS \cite{ATLAS:PFlow} and CMS \cite{CMS:PFlow} have both developed particle flow algorithms oriented around their detectors and experimental objectives, and have used these algorithms during Run 2 of the LHC.

One of the core pieces of particle flow algorithms is a procedure to match tracks to calorimeter energy deposits, thereby avoiding double-counting of the energy from any given particle.
This matching takes place by extrapolating charged-particle tracks from the tracking detector to the calorimeter, and comparing the momentum observed in the two detectors.
If they are consistent, the calorimeter signal is interpreted as corresponding to that one charged particle.
If there is instead much more energy in the calorimeter than is expected for the track in question, then the calorimeter signal is interpreted as containing both charged and neutral components, as the tracker cannot see neutral particles.
The expected energy of the track is then subtracted from the calorimeter signal, leaving two four-vectors: one corresponding to the charged-particle track, and the other corresponding to the remaining neutral calorimeter energy deposit.
If there is no track matched to a given calorimeter signal, then the energy deposit is interpreted as originating from a neutral particle.

The output of particle flow algorithms is therefore a set of four-vectors corresponding to hybrid objects: sometimes they are charged-particle tracks, other times they are the original calorimeter signals, and they can also represent subtracted calorimeter signals.
The resulting particle flow objects can then be used as the inputs to jet reconstruction, which comes with multiple advantages over calorimeter-only inputs.
Beyond improving the low-momentum measurements of charged particles, the ATLAS and CMS tracking detectors also have the spatial resolution to link a given track to a specific proton--proton collision, thereby mitigating \pileup{}-related effects for charged particles.
As hadronic showers are composed of roughly 2/3 charged particles, the improved momentum measurements and \pileup{}-mitigation effects of individual particle flow objects can translate to a sizeable improvement in the precision of the final jet four-vector.

\subsection{Standard Jet Reconstruction and Performance}
\label{sec:reco:smallR}

% q/g interpretation
% Brief JES/JER overview
% JVT, PUPPI, etc

The majority of Run 2 analyses in ATLAS and CMS focus on hadronic showers originating from not-top quarks and gluons, and thus \antikt{} jets with $R=0.4$ are the most appropriate choice.
While CMS made extensive use of particle flow inputs to jet reconstruction for all of Run 2, ATLAS started Run 2 using \topocluster{}-based jet reconstruction, and switched to particle flow inputs closer to the end of Run 2; the ATLAS analyses that will be presented thus use a mixture of the two types of jets.
After building jets from a given set of inputs, they must be calibrated to account for a variety of effects including (but not limited to) calorimeter non-compensation and differences between data and simulation.
In addition to calibrating jets, the impact of \pileup{} on jet reconstruction and performance must be suppressed; this is critical both to minimizing jet-related uncertainties and to mitigating the effect of \pileup{} contamination in searches for new physics (hadronic or otherwise).
The jets may also be further evaluated to determine their consistency with the hypothesis of originating from a given type of particle, especially in the context of bottom and charm quarks.

\subsubsection{Correcting the Jet Scale and Resolution}

Jet calibrations are designed to correct the scale and resolution of jets, where the scale represents the mean of the energy or momentum distribution and the resolution is defined as the width of the same distribution.
The jet momentum scale and resolution have a direct impact on the ability to observe new physics: taking a di-jet resonance search as an example, the scale controls the location of the invariant mass peak corresponding to the resonant mass, while the resolution impacts the width of the peak.
It is thus important to properly calibrate jets when searching for new physics in hadronic final states.

The full jet calibration chain is quite complex, although ATLAS \cite{ATLAS:SmallRcalib} and \mbox{CMS \cite{CMS:SmallRcalib,CMS:Run1JETM}} have converged on the motivations behind and procedures for the majority of the corrections within the full calibration chain.
The scale is first corrected using simulated samples, adjusting the mean of the reconstruction jet response to match the truth expectation,
thereby correcting for the calorimeter response, which varies across the ATLAS and CMS detectors.
This truth expectation is defined using truth jets, which are built by applying the \antikt{} jet algorithm with $R=0.4$ to the set of detector-stable and detector-interacting truth particles generated by a given simulated event generator.
The most significant simulation-based correction for ATLAS and CMS is shown in Figure \ref{fig:reco:smallR:MCJES}; the correction factors are expected to be different as they account for detector-specific effects.
The usage of particle flow inputs reduces the magnitude of this correction: the full track momentum is observed, in contrast to sampling calorimeters observing only a fraction of the total deposited energy.
\vspace{-9pt}
\begin{figure}[H]

\subfigure[ ATLAS]{
 \includegraphics[width=0.55\textwidth]{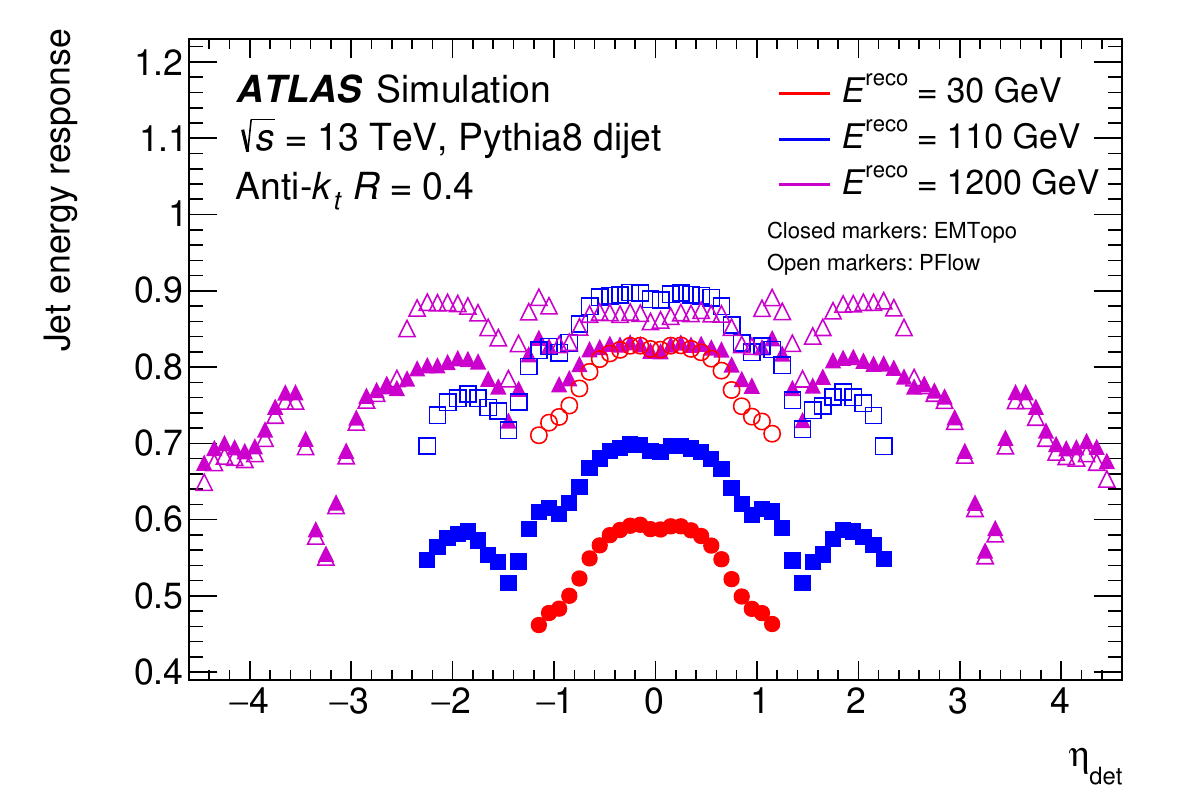}
 \label{fig:reco:smallR:MCJES:ATLAS}
}
\subfigure[ CMS]{
 \includegraphics[width=0.38\textwidth]{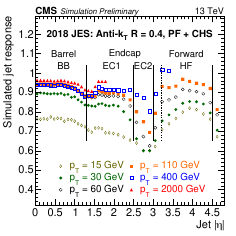}
 \label{fig:reco:smallR:MCJES:CMS}
}
\caption{The simulation-based jet response, the inverse of which is used to correct for the calorimeter response to hadronic showers, shown for (\textbf{a}) ATLAS \cite{ATLAS:SmallRcalib} and (\textbf{b}) CMS \cite{CMS:SmallRcalib}. The calibrations are expected to be different, as they correct for detector-specific effects. (\textbf{a}) The results for ATLAS are shown both using jets built from electromagnetic-scale \topoclusters{} (EMTopo, filled markers) and particle flow (PFlow, open markers) inputs; the use of PFlow dramatically reduces the magnitude of the correction factors at low energy, as the full track momentum is measured. \label{fig:reco:smallR:MCJES}}
\end{figure}

These simulation-based corrections are applied to both simulated events and data, under the assumption that the detector response is reasonably modelled in simulation.
While this is true to first order, it is important to subsequently correct the scale for differences in the jet response between data and simulation.
This is done through the derivation of a set of \insitu{} corrections, where a jet of interest is balanced against a well-defined reference object.
Reference objects include $Z$ bosons decaying to $e^+e^-/\mu^+\mu^-$, photons, the vector sum of a system of already-calibrated lower-momentum jets, or jets in different regions of the detector.
As each of these techniques has different sensitivities and covers different kinematic regimes, the individual \insitu{} correction factors are statistically combined to define the final calibration factor; this is then applied to data such that the average jet in data matches the average jet in simulation, and thus also the truth expectation.
The resulting data-to-simulation correction factors are shown in Figure \ref{fig:reco:smallR:insitu}, where a similar shift is seen for both ATLAS and CMS, suggesting that the difference between jets in data and simulated events is to first order common across the experiments.
\vspace{-9pt}

\begin{figure}[H]

\subfigure[ ATLAS]{
 \includegraphics[width=0.55\textwidth]{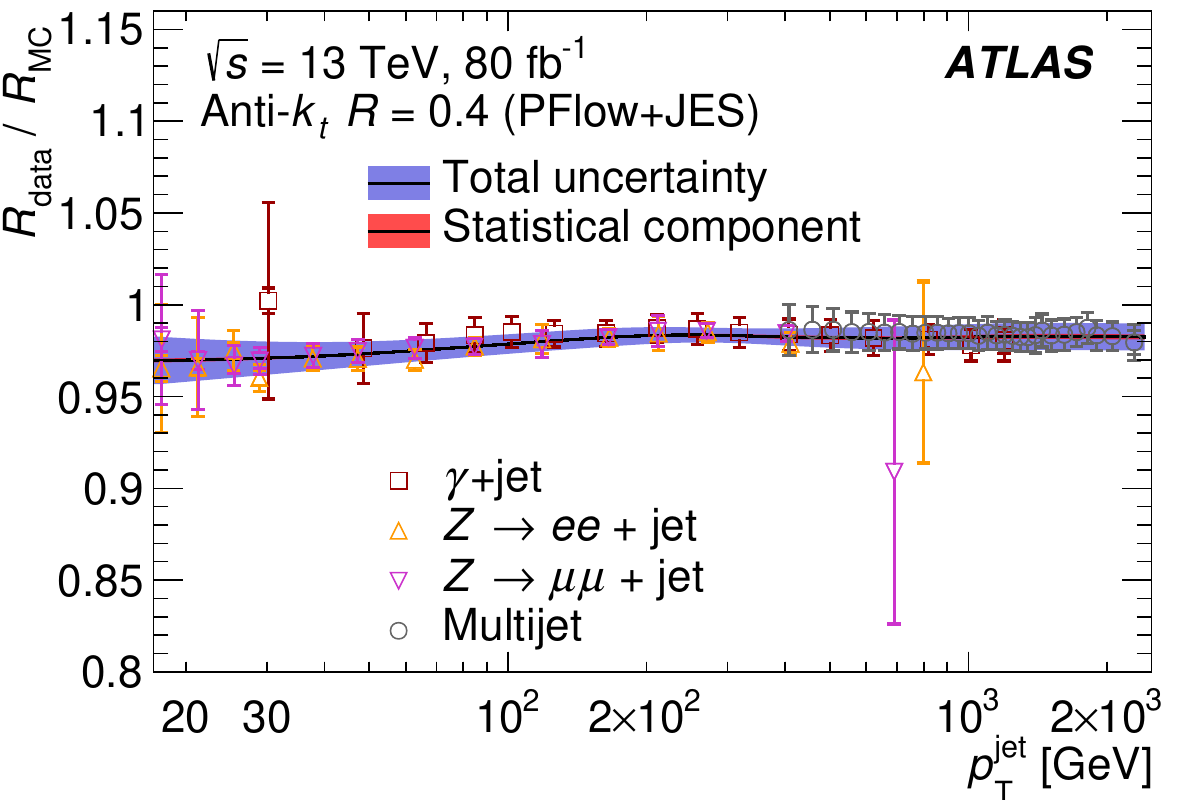}
 \label{fig:reco:smallR:insitu:ATLAS}
}
\subfigure[ CMS]{
 \includegraphics[width=0.4\textwidth]{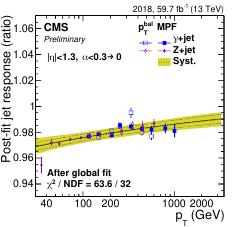}
 \label{fig:reco:smallR:insitu:CMS}
}
\caption{The data-to-simulation difference in jet response, the inverse of which is used to correct the data to match the simulation, as evaluated using a series of \insitu{} balance techniques, for (\textbf{a}) ATLAS \cite{ATLAS:SmallRcalib} and (\textbf{b}) CMS \cite{CMS:SmallRcalib}. The calibrations are very similar, suggesting that the difference between data and simulation is inherent in the generators and models used by both collaborations. (\textbf{a}) The results for ATLAS are shown using jets built from particle flow (PFlow) inputs; a similar plot is available in the same reference for jets built using electromagnetic-scale \topoclusters{}. \label{fig:reco:smallR:insitu}}
\end{figure}

Uncertainties on the scale of jets in data with respect to simulation are evaluated by combining the aforementioned measurements with other possible effects not directly evaluated \insitu{}.
These include \pileup{} uncertainties, which are dominant for the lowest momentum regime, and flavour (light-quark vs. gluon) uncertainties, which are the limiting effect for the intermediate momentum range.
The total uncertainties are quite similar for ATLAS and CMS, as shown in Figure \ref{fig:reco:smallR:unc}, peaking at roughly 5\% at very low momentum and reaching a minimum a bit below 1\% for transverse momenta above roughly 200\GeV{}.

\vspace{-9pt}
\begin{figure}[H]

\subfigure[ ATLAS]{
 \includegraphics[width=0.55\textwidth]{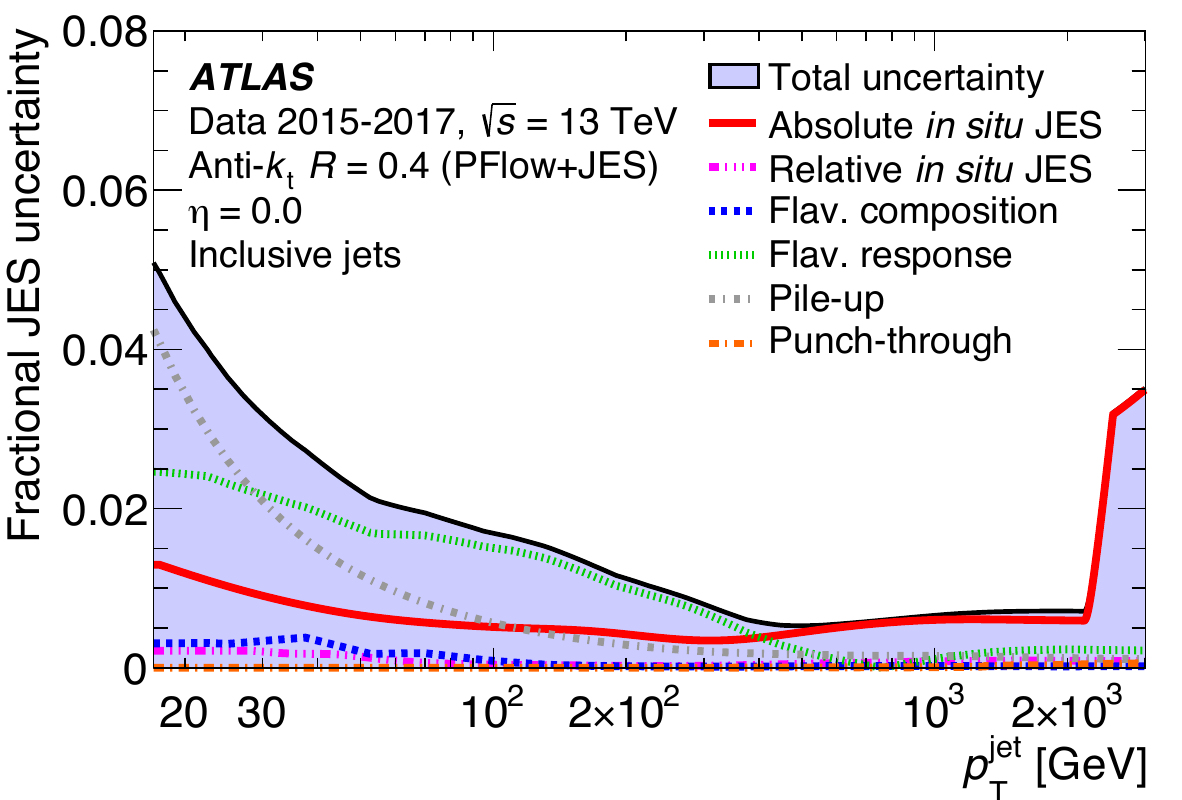}
 \label{fig:reco:smallR:unc:ATLAS}
}
\subfigure[ CMS]{
 \includegraphics[width=0.4\textwidth]{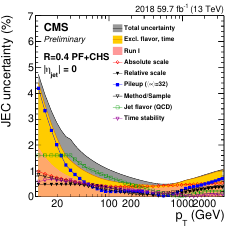}
 \label{fig:reco:smallR:unc:CMS}
}
\caption{The uncertainties on the scale of jets for (\textbf{a}) ATLAS \cite{ATLAS:SmallRcalib} and (\textbf{b}) CMS \cite{CMS:SmallRcalib}. The overall magnitude and structure of the uncertainty is reasonably similar across the experiments. (\textbf{a}) The results for ATLAS are shown using jets built from particle flow (PFlow) inputs; a similar plot is available in the same reference for jets built using electromagnetic-scale \topoclusters{}. \label{fig:reco:smallR:unc}}
\end{figure}

In addition to the scale, it is also important to quantify the jet momentum resolution.
ATLAS and CMS agree on a model to define the jet resolution \cite{ATLAS:SmallRcalib,CMS:SmallRcalib}: $\frac{N}{\pT{}} + \frac{S}{\sqrt{\pT{}}} + C$.
In this equation, $N$ represents the electronic and \pileup{} noise within the jet, $S$ represents the stochastic nature of hadronic showers in calorimeters, $C$ is a constant term defining the calorimeter's fundamental limitations, and \pT{} is the transverse momentum of the jet.
This functional form is used in fits to measurements of the jet resolution in order to extract a smooth trend, where the individual measurements are derived following \insitu{} methods, including the balance of jets in events containing exactly two jets.
ATLAS also considers the balance of randomly-defined cones as a measure of constraining the noise term, while CMS evaluates the balance of a probe jet against a reference $Z$ boson decaying to $e^+e^-/\mu^+\mu^-$.
The resulting resolution measurements are shown in Figure \ref{fig:reco:smallR:JER}, where ATLAS and CMS present their results slightly differently, but in the end they see similar trends.

\vspace{-9pt}
\begin{figure}[H]

\subfigure[ ATLAS, central $|\eta|$]{
 \includegraphics[width=0.45\textwidth]{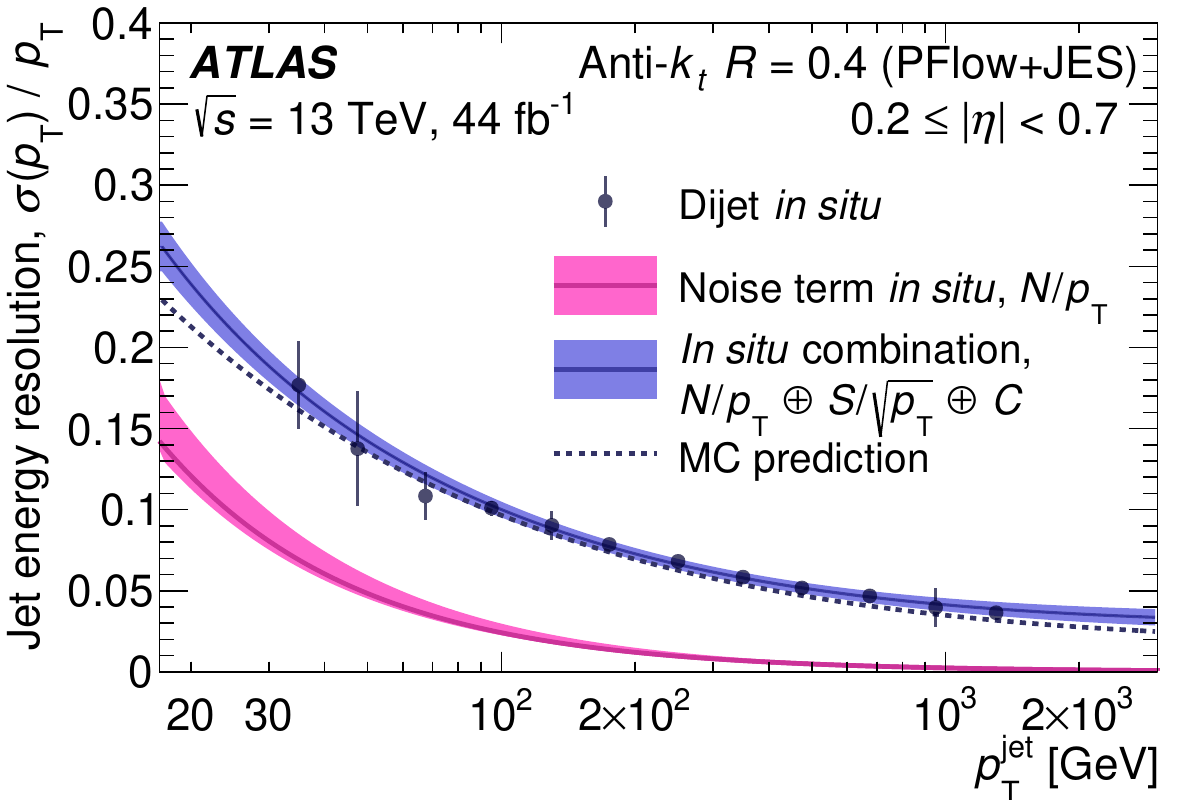}
 \label{fig:reco:smallR:JER:ATLAS}
}
\subfigure[ CMS, all $|\eta|$]{
 \includegraphics[width=0.45\textwidth]{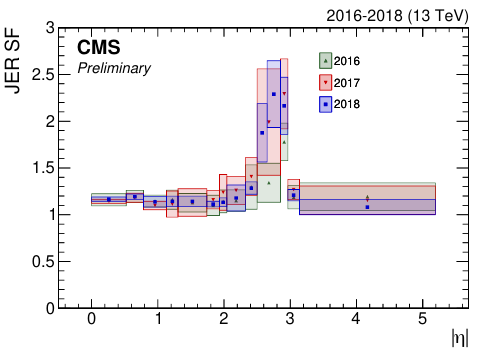}
 \label{fig:reco:smallR:JER:CMS}
}
\caption{The jet resolution for (\textbf{a}) ATLAS \cite{ATLAS:SmallRcalib} and (\textbf{b}) CMS \cite{CMS:SmallRcalib}. The agreement between data and simulation is quite good for jets in the more central parts of the detector, while differences are larger in more challenging regions of the detector. (\textbf{a}) The measurement of the resolution for ATLAS is shown in the central part of the detector, using jets built from particle flow (PFlow) inputs; a similar plot is available in the same reference for jets built using electromagnetic-scale \topoclusters{}. (\textbf{b}) The scale factor required to make the simulated resolution match the data resolution is shown across the CMS detector, highlighting the challenging EC2 region. \label{fig:reco:smallR:JER}}
\end{figure}

\subsubsection{Mitigating Pileup Effects}

\Pileup{} collisions impact jet reconstruction and performance in multiple ways, and the suppression of such effects is of key importance for analyses involving hadronic final states.
As already mentioned, \pileup{} can degrade the resolution of \hardscatter{} jets of interest, as particles from other collisions may happen to overlap with the jet of interest thereby impacting the subsequent measurement of that jet.
Particle flow already helps to mitigate such effects by linking charged energy contributions to a given vertex, a technique known as Charged Hadron Subtraction (CHS), but neutral particles from \pileup{} collisions escape such constraints.
In addition to contaminating jets from the \hardscatter{} collision, \pileup{} can also produce jets entirely separate from the \hardscatter{} collision, which must be removed from the event to properly quantify the event's properties; this is especially important when selecting events based on the number of jets in the event or the balance of the hadronic activity in the event.
As both of these selections are employed in hadronic searches for new physics, it is thus important for such searches to mitigate \pileup{}-related effects; we will now discuss some ways in which this can be done.

ATLAS and CMS have both developed \pileup{} mitigation strategies beyond CHS, and the two strategies take very different directions.
ATLAS uses a jet-based discriminant, known as the Jet Vertex Tagger (JVT), in order to reject \pileup{} jets while retaining \hardscatter{} jets with high efficiency \cite{ATLAS:JVT}.
This approach works very well for suppressing entire jets, and can be used for jets built from either electromagnetic-scale \topoclusters{} or particle flow objects as inputs, as shown in Figure \ref{fig:reco:smallR:PU}a.

While such a jet-based discriminant can efficiently reject jets originating from \pileup{} vertices, it does not help with removing neutral \pileup{} contributions within \hardscatter{} jets.
In order to improve on this, CMS has also developed an algorithm for PileUp Per Particle Identification (PUPPI) \cite{CMS:PU}.
This algorithm evaluates the consistency of each individual four-vector used in the jet reconstruction process with the hypothesis of originating from the \hardscatter{} collision, as opposed to originating from a \pileup{} collision.
Only the four-vectors that appear to originate from the \hardscatter{} collision are retained, thereby suppressing both \pileup{} contributions to \hardscatter{} jets as well as the creation of additional \pileup{}-originating jets.
The PUPPI algorithm is seen to work very well, and even outperforms the CMS jet-based \pileup{} selection algorithm, as shown in Figure \ref{fig:reco:smallR:PU}b.

\vspace{-9pt}
\begin{figure}[H]

\subfigure[ ATLAS]{
 \includegraphics[width=0.45\textwidth]{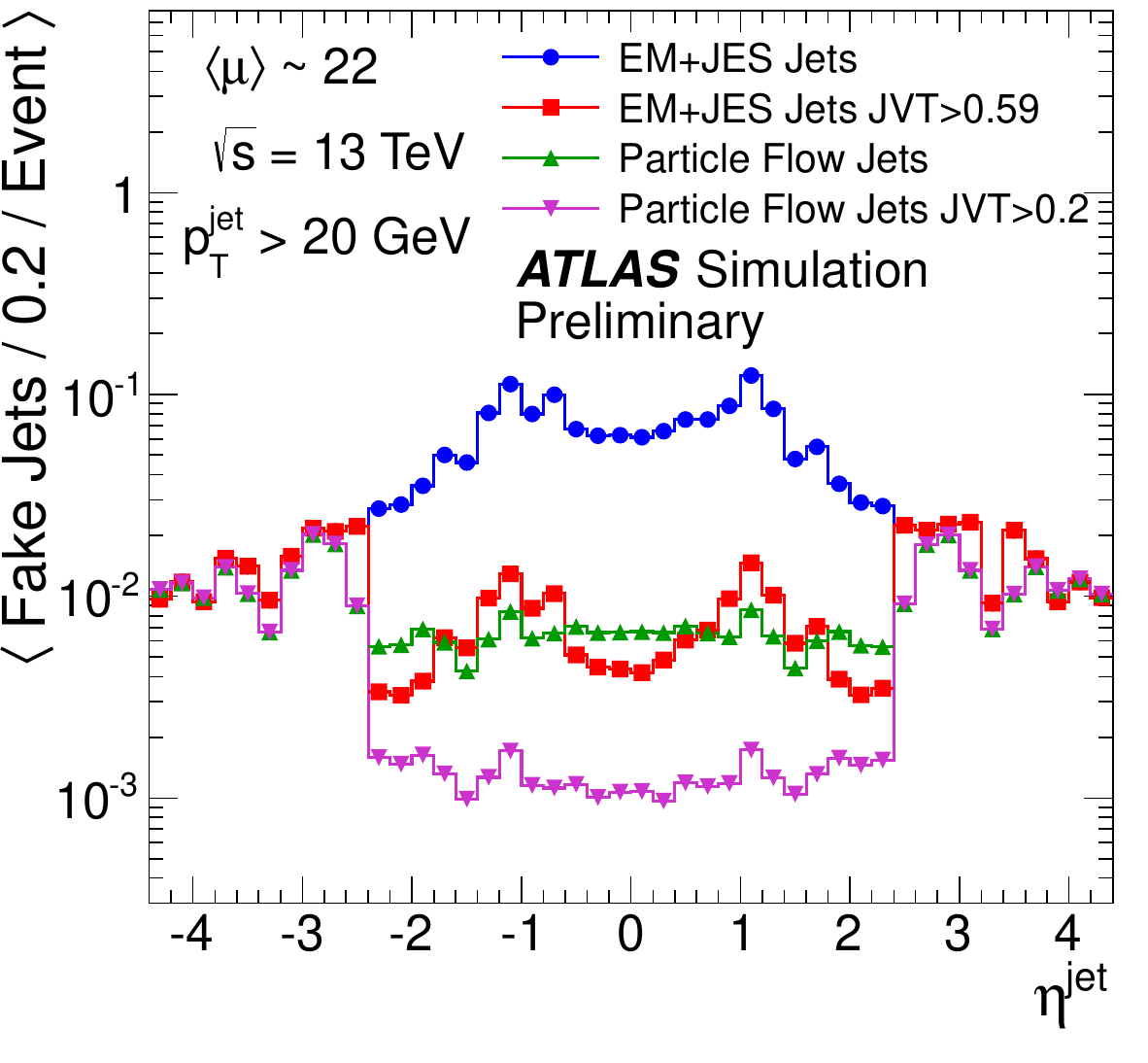}
 \label{fig:reco:smallR:PU:ATLAS}
}
\subfigure[ CMS]{
 \includegraphics[width=0.45\textwidth]{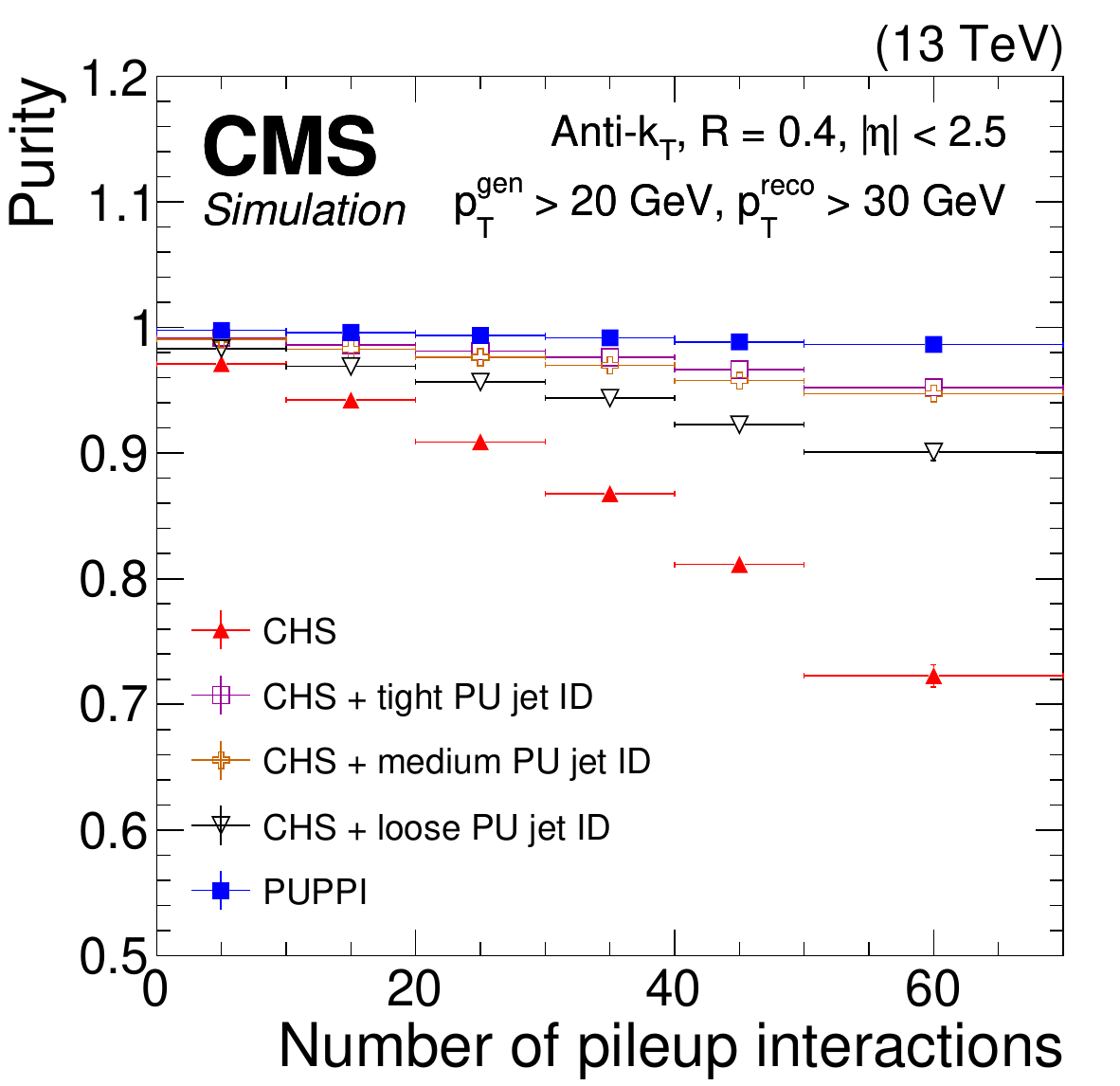}
 \label{fig:reco:smallR:PU:CMS}
}
\caption{(\textbf{a}) The number of \pileup{} (fake) jets remaining in ATLAS after the application of different reconstruction and \pileup{}-jet-tagging strategies, demonstrating the benefit of particle flow with respect to calorimeter-only (EM+JES) jet reconstruction, and the importance of applying a jet-based vertex tagger (JVT) to further suppress \pileup{} \cite{ATLAS:PU}.
(\textbf{b}) The purity of selecting only \hardscatter{} jets, for a variety of different jet reconstruction strategies: only using particle flow (charged hadron subtraction, CHS), adding a jet-based selection criterion (PU jet ID), or applying a constituent-based selection criterion during jet reconstruction (PUPPI) \cite{CMS:PU}. This shows how the performance of all techniques degrades as the number of \pileup{} collisions increases, but the constituent-based selection performs the best for the full range studied, especially as \pileup{} levels increase. \label{fig:reco:smallR:PU}}
\end{figure}

\subsubsection{Identifying Heavy Flavour Jets}

As previously mentioned, \smallR{} jets are primarily used in the context of interpreting hadronic showers from non-top quarks and gluons; so far, we have not further differentiated between the possible sources.
The heavier quarks, namely, charm and bottom, form hadrons with sufficiently long lifetimes for those hadrons to travel a non-negligible distance before decaying.
This has an important experimental implication: it is possible to observe this displacement, and thus differentiate jets involving bottom quarks, and to a lesser extent charm quarks, from those originating from lighter quarks or gluons.
This experimental capability is very useful in the context for both searches for new physics and measurements of the Standard Model.

The most straightforward experimental signature for the presence of a heavy flavour quark is the observation of a displaced vertex in the tracking detector, where displaced vertices are points that charged-particle tracks originate from, but which are spatially inconsistent with being a proton--proton collision from the crossing of the LHC proton beams.
The observation of a displaced vertex is a strong indication for the presence of a particle with a long lifetime, such as a heavy flavour hadron.
Such a displaced vertex can also come from other sources, such as tau leptons, but the ATLAS and CMS detectors provide sufficient experimental information to differentiate between displaced vertices from heavy flavour decays and those from other sources.
Displaced vertices are the most striking way in which jets involving heavy flavour can be identified, but they are not the only way: ATLAS \cite{ATLAS:btag} and CMS \cite{CMS:btag} have both designed and used a variety of increasingly complex algorithms, exploiting a variety of experimental features, in order to identify jets consistent with originating from heavy flavour decays.
This process is often referred to as flavour tagging, and a jet which passes the selection is said to be a $b$-tagged (or $c$-tagged) jet.

The flavour-tagging community has a history of developing algorithms employing modern machine learning tools in order to obtain the maximum possible flavour-tagging performance.
The exact algorithms used for flavour tagging have changed quite a bit during Run 2, and the ability to differentiate heavy-flavour jets from background jets has continued to improve.
It is not possible to discuss all of the algorithms used, rather Figure \ref{fig:reco:smallR:btag:perf} shows comparisons performed by both collaborations of the performance of their respective flavour-tagging algorithms, evaluated using partial Run 2 datasets.
Figure \ref{fig:reco:smallR:btag:perf}b in particular shows the $b$-tagging performance at CMS as evaluated with respect to backgrounds of both light-quark jets and $c$-quark jets, demonstrating that $b$-tagging is more efficient at rejecting light-quark jets than charm jets due to bottom and charm decays being more similar to each other.
Both collaborations find that the best performance is obtained using modern deep learning tools (DL1 in ATLAS, DeepCSV in CMS), although the performance of another type of machine learning classifier, boosted decision trees, is only slightly degraded in the case of $b$-tagging (MV2 in ATLAS, cMVAv2 in CMS).

\vspace{-6pt}
\begin{figure}[H]

\subfigure[ ATLAS]{
 \includegraphics[width=0.4\textwidth]{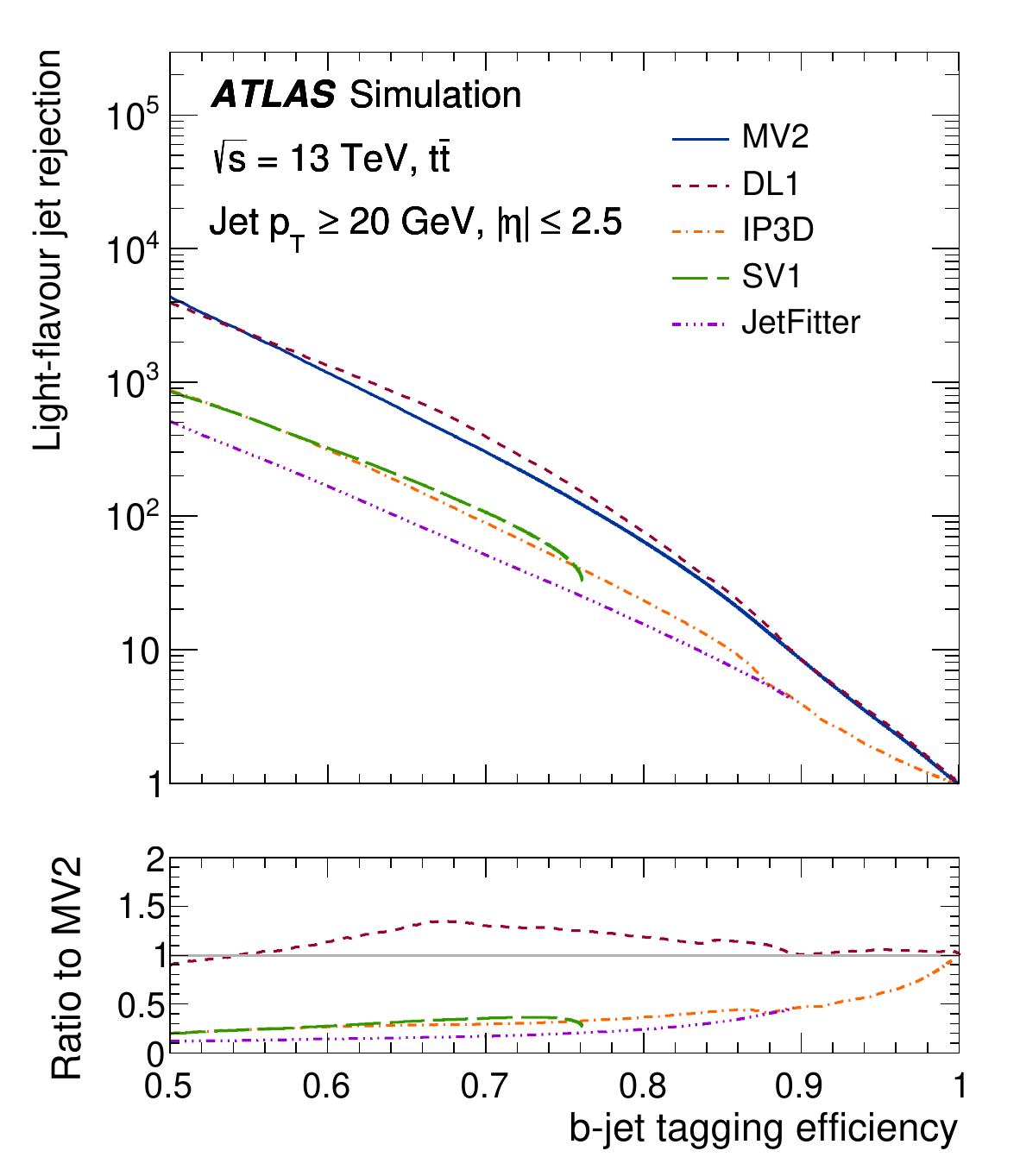}
 \label{fig:reco:smallR:btag:ATLAS:perf}
}
\subfigure[ CMS]{
 \includegraphics[width=0.5\textwidth]{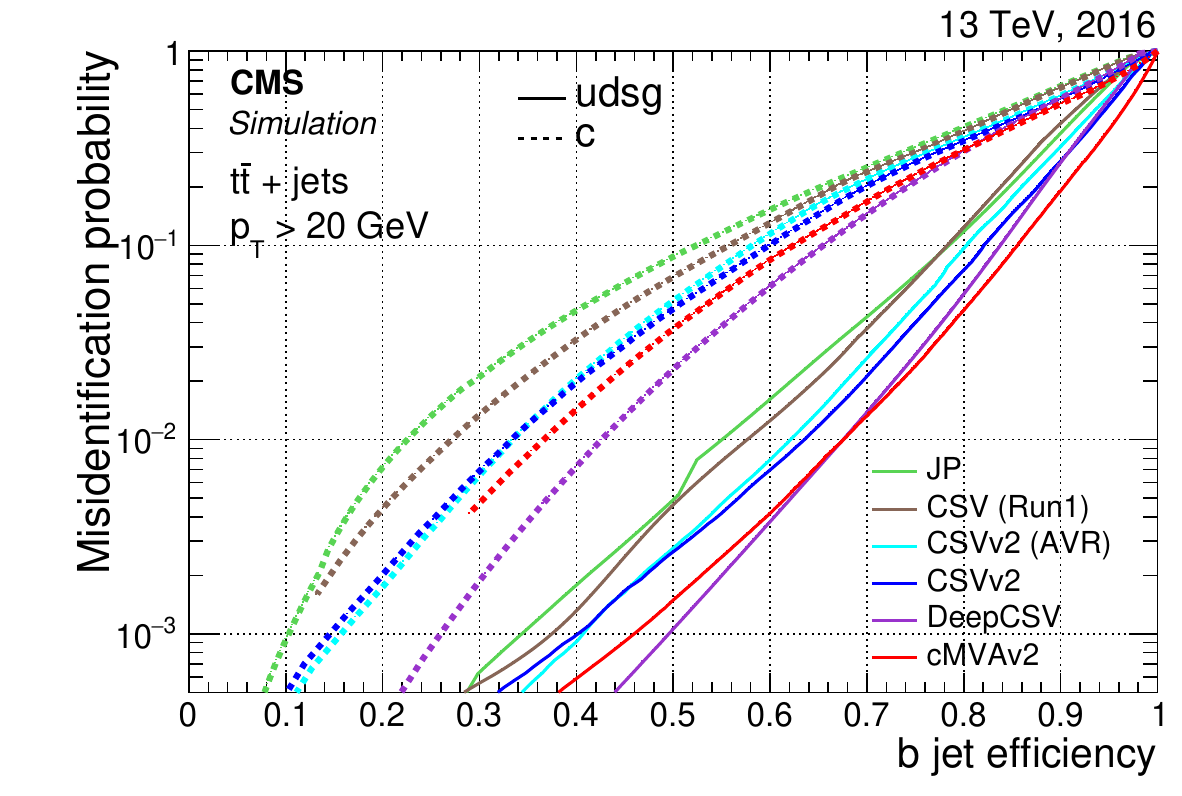}
 \label{fig:reco:smallR:btag:CMS:perf}
}
\caption{The performance of a variety of flavour-tagging algorithms, quantified in terms of the ability to identify $b$-jets with a given efficiency, and the corresponding (\textbf{a}) background rejection of light-quark and gluon jets in ATLAS \cite{ATLAS:btag} and (\textbf{b}) misidentification probability of light-quark and gluon jets (udsg) or charm jets (c) in CMS \cite{CMS:btag}. ATLAS shows results as a function of background rejection, while CMS shows results as a function of misidentification probability; these quantities are the inverse of each other. \label{fig:reco:smallR:btag:perf}}
\end{figure}

The variety of tagging algorithms developed by ATLAS and CMS are optimised based on comparisons of different simulated samples, and it is possible that the features learned by the machine learning tools are not well-modelled.
It is thus of great importance to also study the behaviour of the final taggers in both data and simulation, to extract scale factors associated with any potential differences, and to define an uncertainty on the extent to which simulation matches data.
ATLAS and CMS both have multiple methods for extracting such scale factors \cite{ATLAS:btag,CMS:btag}; representative scale factors for $b$-tagging using deep learning classifiers are shown in Figure \ref{fig:reco:smallR:btag:SF}.
While the scale factors do differ from unity, indicating that simulation does not perfectly model the data, such deviations are generally quite small and thus the tagger performance is reasonably well modelled.

\begin{figure}[H]

\subfigure[ ATLAS]{
 \includegraphics[width=0.45\textwidth]{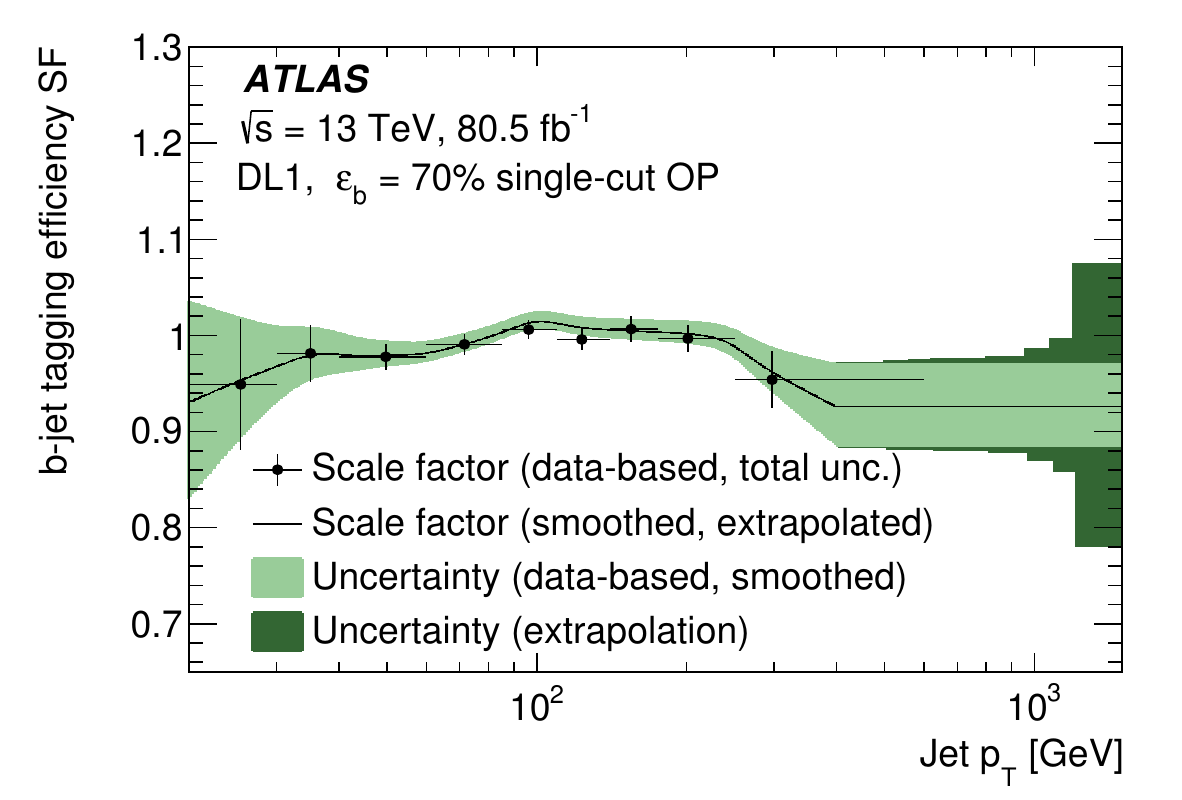}
 \label{fig:reco:smallR:btag:ATLAS:SF}
}
\subfigure[ CMS]{
 \includegraphics[width=0.45\textwidth]{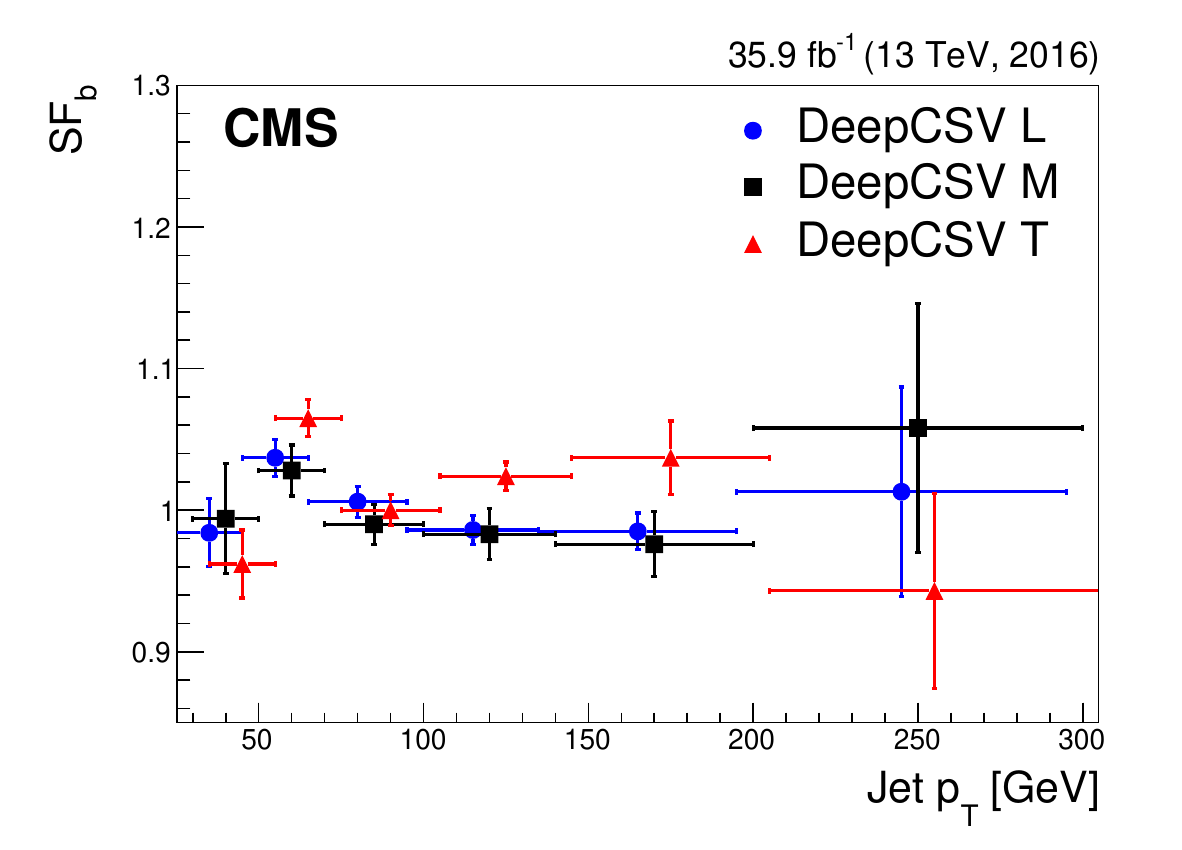}
 \label{fig:reco:smallR:btag:CMS:SF}
}
\caption{The scale factors required to correct for differences between data and simulated events for the powerful deep learning $b$-jet taggers, as extracted by (\textbf{a}) ATLAS using a partial Run 2 dataset \cite{ATLAS:btag}, and (\textbf{b}) CMS using a different partial Run 2 dataset \cite{CMS:btag}. In both cases, the extracted scale factors are close to but not equal to unity, indicating that there are some effects related to heavy flavour observables that are not fully modelled by the simulation. \label{fig:reco:smallR:btag:SF}}
\end{figure}

\subsection{Boosted Jet Reconstruction and Performance}
\label{sec:reco:largeR}

% hadronic decay interpretation
% Jet grooming
% Brief JES/JER/JMS/JMR overview
% Jet mass and substructure variables
% W/Z/H/top tagging

The very high energy scale of the LHC is sufficient to produce massive particles, such as $W/Z/H$ bosons and top quarks, with a substantial momentum in the experimental reference frame.
If this momentum is much larger than the mass of the parent particle, this implies a large Lorentz boost between the massive particle decay frame and the experimental frame; thus, the decay products are collimated in the detector.
For a two-body decay of a massive particle, the angular separation $\Delta R$ between the two daughter particles $d_1$ and $d_2$ follows the formula:
$$\Delta R_{d_1d_2} \approx \frac{1}{\sqrt{z(1-z)}}\frac{m^p}{\pT^p} \quad\implies\quad \Delta R_{d_1d_2}\gtrsim2\frac{m^p}{\pT^p}\ ,$$
where $m^p$ and $\pT^p$ are the mass and transverse momentum of the parent particle $p$, and $z$ is the momentum split between the two decay particles. Assuming an even momentum split of $z=0.5$ leads to the simplified version of the equation, which is a lower bound on the angular separation, but which is also often used as an approximation for the typical angular separation of a two-body decay.

As an example, consider the decay of a $W$ boson to a pair of light quarks.
If we require the two quarks to be fully distinguishable for an $R=0.4$ jet algorithm, then the separation between the two quarks should be at least $\Delta R=0.8$.
Using the known mass of the $W$ boson of 80.379\GeV{} \cite{PDG}, we can invert the equation to determine that this angular separation occurs when $\pT^{W}\approx200\GeV{}$.
The decays of higher \pT{} $W$ bosons will thus start to overlap for an $R=0.4$ jet algorithm, signalling the start of the boosted regime: rather than dealing with the increasingly challenge of reconstructing the two daughter particles as separate jets, it makes sense to instead reconstruct the entire massive particle decay as a single \largeR{} jet.

This transition from reconstructing massive particles using their daughter particles to reconstructing the entire decay has many implications, and is a powerful way to mitigate the challenges associated with searches for new physics in hadronic final states.
As the \largeR{} jet contains the entire decay, the jet mass should now correspond to the parent particle mass.
Additionally, the angular energy distribution of the jet should be consistent with the hypothesis of multiple regions of high energy density corresponding to the multiple daughter particles, as opposed to a single energy-dense region expected from not-top quarks and gluons.
These properties allow for the suppression of the otherwise overwhelming Standard Model hadronic physics background, supporting searches for new rare physics; such techniques were first widely discussed in the context of recovering the $b\bar{b}$ final state as a promising channel in the search for the Higgs boson \cite{JSS}.
The use of such techniques, where entire hadronic decays are reconstructed as a single jet and the internal properties of the jet are exploited, forms the basis of what is called jet substructure.

The use of jet substructure has grown dramatically since the LHC started taking data, and jet substructure techniques have become a key component of the ATLAS and CMS physics programmes, especially in the context of searches for new physics.
While reconstructing jets using the \antikt{} algorithm with a larger value of the radius parameter is a good start, more advanced strategies must be employed to really benefit from modern jet substructure techniques.
Following the reconstruction of such \largeR{} jets, they must be calibrated; the momentum must be corrected for similar reasons as for \smallR{} jets, but now the mass of the jet also has an important meaning and must be calibrated for analysis usage.
Searches for new physics using \largeR{} jets usually only want to study hadronic decays of massive particles, but \largeR{} jets by default also include the huge Standard Model background from not-top quarks and gluons; it is thus also important to suppress these large backgrounds and to quantify the amount of background remaining.

\subsubsection{Boosted Jet Reconstruction}

Just like \smallR{} jets, the reconstruction of \largeR{} jets also starts by running the \antikt{} algorithm over a given set of inputs.
However, that is not sufficient: \largeR{} jets, as the name implies, cover a large area of the detector.
This makes them particularly susceptible to both underlying event and \pileup{} effects, which must be removed in order to see the \hardscatter{} hadronic decay of interest.
This is typically done by processing all of the inputs that were grouped into a given jet by the \antikt{} algorithm, and deciding which of the inputs to keep or remove based on some criterion aimed at only retaining four-vectors consistent with a high momentum hadronic decay of a massive particle.
This procedure of further processing the inputs used in each jet, after reconstructing the original jet, is known as jet grooming; one such algorithm was used already in the aforementioned pre-LHC studies of boosted $H\to{}b\bar{b}$ decays \cite{JSS}.

There are now a large number of different jet grooming algorithms, which typically have a few different parameters that can be adjusted to optimize how aggressive the grooming procedure is.
The optimal algorithm and configuration thereof depends on many factors, and ATLAS and CMS have settled on different choices for the majority of their Run 2 searches for new physics.
In the context of a search for new physics, an ideal jet grooming algorithm will remove the underlying event and other sources of radiation while keeping the entire hadronic decay of interest, and will remove \pileup{} effects to the extent that the dependence of jet substructure variables on \pileup{} is negligible.
An example from ATLAS showing such positive results from the application of one type of grooming, known as trimming \cite{Trimming}, is shown in Figure \ref{fig:reco:largeR:groom}.
By grooming the jets, the $Z$ boson mass is recovered, background jets from not-top quarks and gluons have their masses suppressed (thereby enhancing the ability to identify $Z$ bosons), and the \pileup{} dependence is removed.
Other properly optimized grooming algorithms will provide similar benefits.
The large majority of the relevant ATLAS searches in Run 2 have used calibrated LCW \topoclusters{} as inputs, although there are a small number of analyses that have used a particle flow input type that was designed specifically for high momentum \largeR{} jet reconstruction, named Track-CaloClusters (TCCs) \cite{ATLAS:TCC,ATLAS:UFO}.
In both cases, these inputs are used to reconstruct \antikt{} $R=1.0$ jets, which are subsequently trimmed \cite{Trimming} using a \kT{} sub-jet radius of $R_\mathrm{sub}=0.2$ and a \pT{} fraction cut of $f_\mathrm{cut}=5\%$.
For jets built using \topoclusters{}, the mass of the jet is then further refined through a combination of the calorimeter jet mass with tracking information in a form of minimal particle flow, which is referred to as the combined mass \cite{ATLAS:LargeRcalib}.
The subsequent ATLAS plots will focus on jets built using calibrated \topocluster{} inputs and using the combined mass, as that was the most commonly used strategy in relevant searches.

Most of the CMS searches that will be presented have used particle flow inputs to the \antikt{} algorithm with $R=0.8$, although in some cases Cambridge--Aachen (CA) jets \cite{CA} with $R=1.5$ have been used instead, especially for searches where top quarks are at a lower momentum and thus a larger radius is needed to contain the full decay.
Most of the Run 2 results are then groomed using the Soft Drop algorithm \cite{mMDT,SoftDrop} with a soft threshold cut $z_\mathrm{cut}=0.1$ and an angular exponent $\beta=0$.
In addition to grooming, it is possible to apply further selections on the jet inputs; CMS does so by applying the same PUPPI algorithm discussed in the context of \smallR{} jets to the reconstruction of \largeR{} jets.
The subsequent CMS plots will focus on jets built using the \antikt{} algorithm with $R=0.8$ and Soft Drop grooming plus PUPPI applied, as that was the most common \largeR{} jet strategy employed by the relevant searches.

\vspace{-6pt}
\begin{figure}[H]

\subfigure[ Jet mass for signal and background]{
 \includegraphics[width=0.4\textwidth]{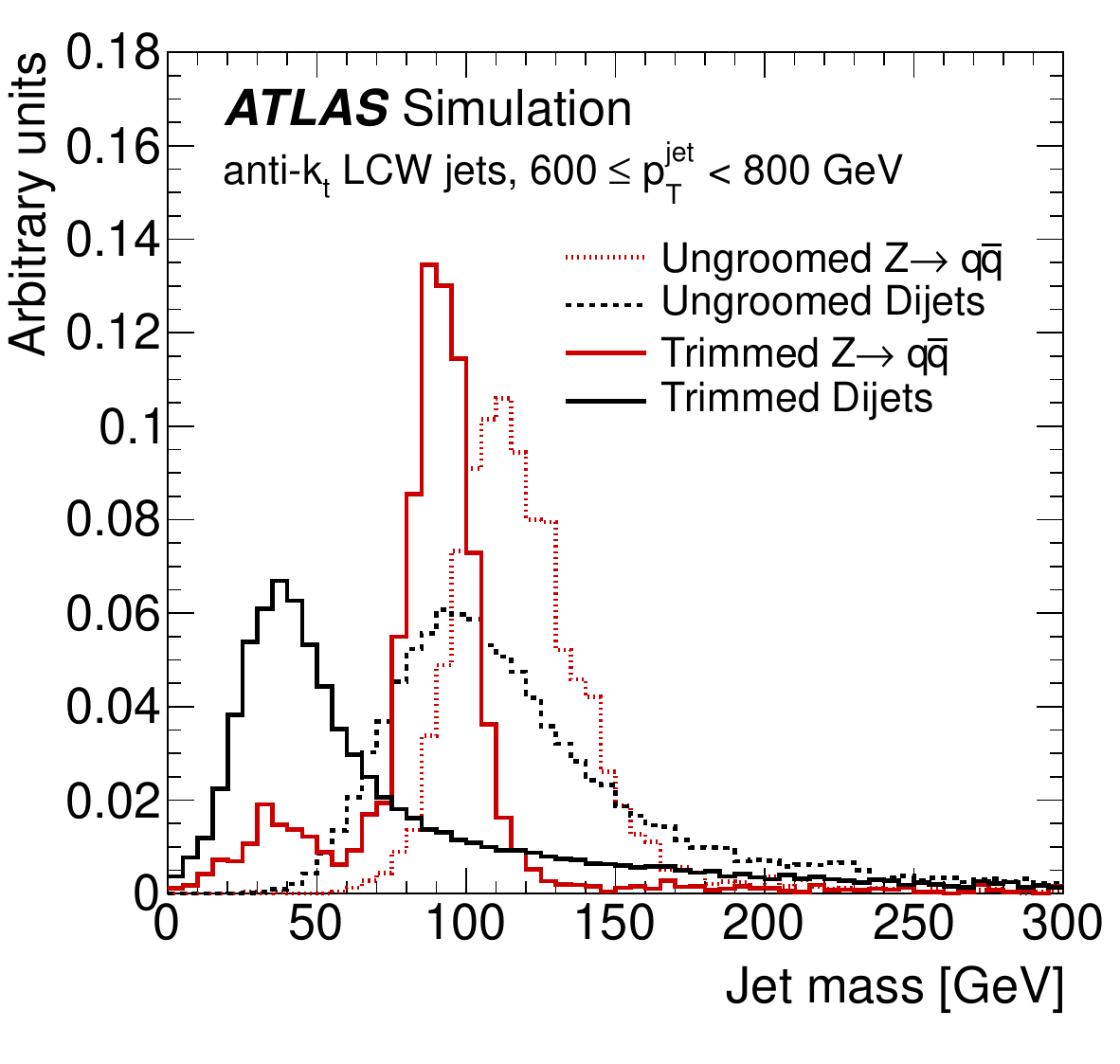}
 \label{fig:reco:largeR:groom:impact}
}
\subfigure[ Jet mass dependence on \pileup{}]{
 \includegraphics[width=0.4\textwidth]{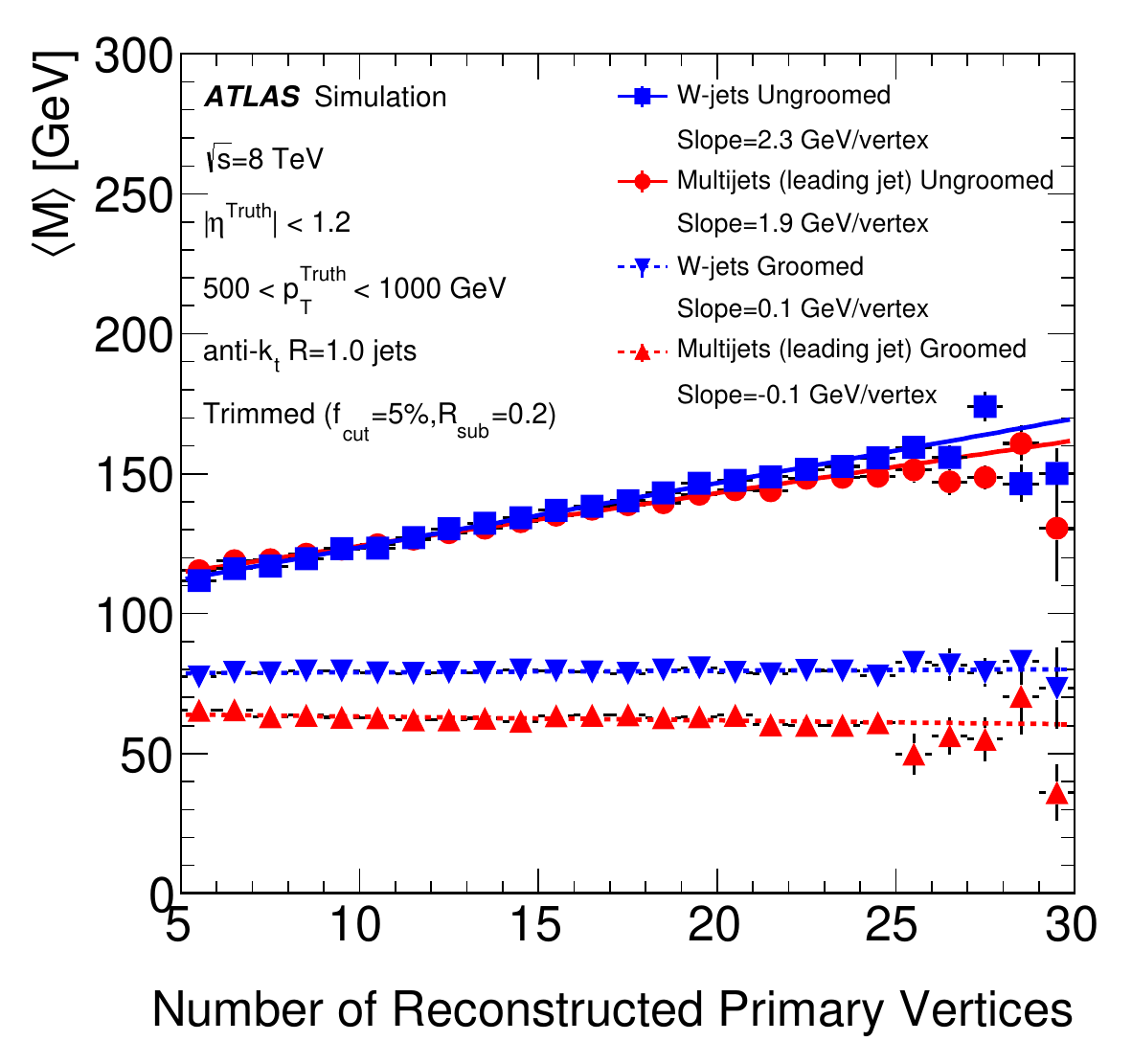}
 \label{fig:reco:largeR:groom:pileup}
}
\caption{Grooming is a key component of \largeR{} jet reconstruction, as it removes energy from the underlying event and suppresses contributions from other \pileup{} collisions. (\textbf{a}) Applying trimming, one type of grooming, is shown to recover the expected $Z$-boson mass for $Z\to{}qq$ decays, while also strongly suppressing the mass from not-top quarks and gluons (dijets) \cite{ATLAS:JSS2011}. (\textbf{b}) In addition to correcting the central value, appropriate grooming can also strongly mitigate the \pileup{} dependence of the mass of the jet, which is important to the exploitation of the high-\pileup{} LHC dataset \cite{ATLAS:WTag8TeV}. \label{fig:reco:largeR:groom}}
\end{figure}

Plots showing the mass in simulated samples, for a variety of different signal types of interest and the primary background to mitigate, are provided in Figure \ref{fig:reco:largeR:mass}.
These plots show the most commonly used mass definition and jet reconstruction strategy for both ATLAS and CMS.
The mass already provides a strong first means of differentiating between jets from hadronic decays of massive particles as opposed to background sources, although much more can be done, as will be discussed soon.
It is first important to discuss how \largeR{} jets, and especially their mass, can be calibrated.

\vspace{-3pt}

\begin{figure}[H]

\subfigure[ ATLAS]{
 \includegraphics[width=0.4\textwidth]{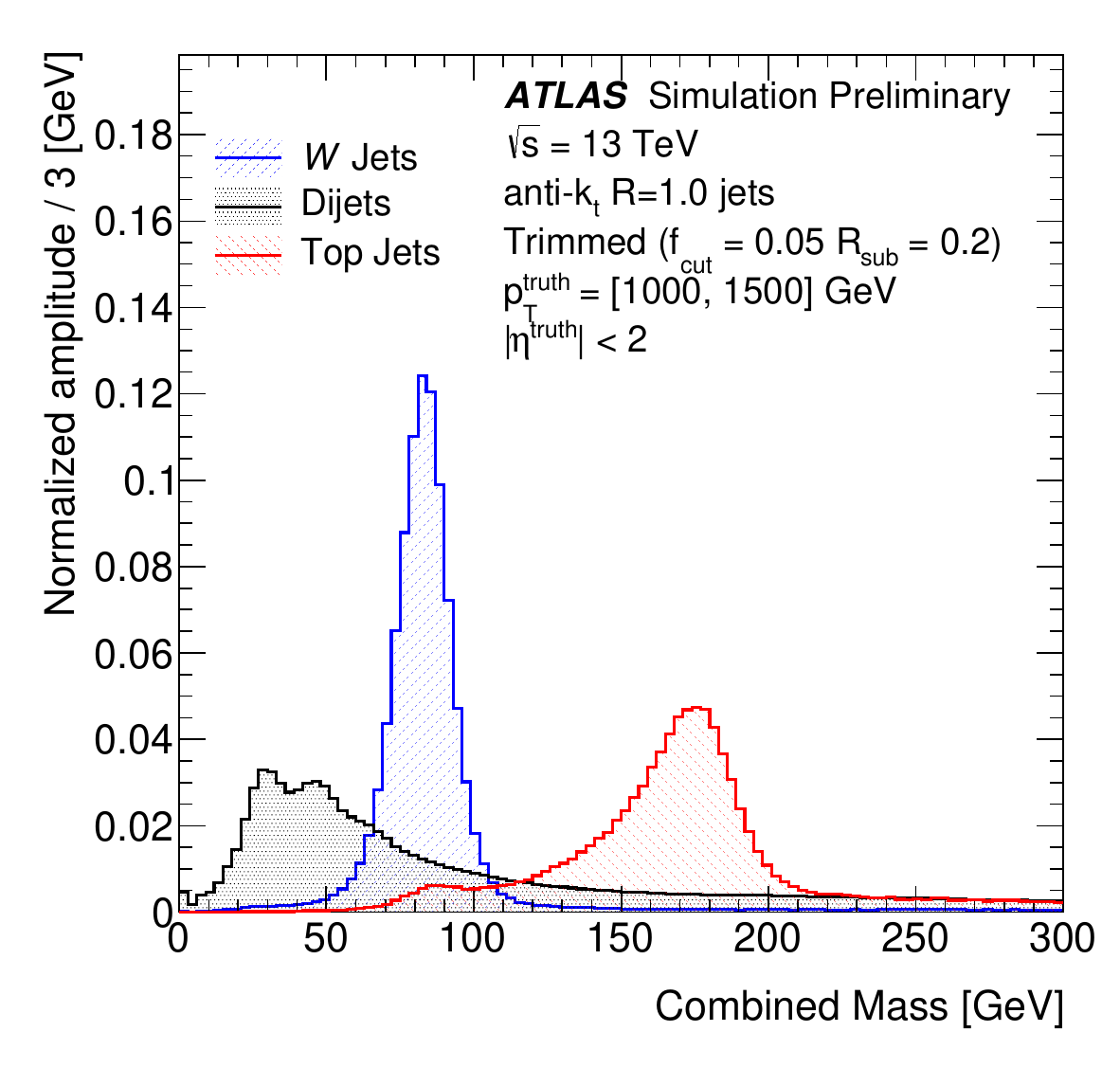}
 \label{fig:reco:largeR:mass:ATLAS}
}
\subfigure[CMS]{
 \includegraphics[width=0.4\textwidth]{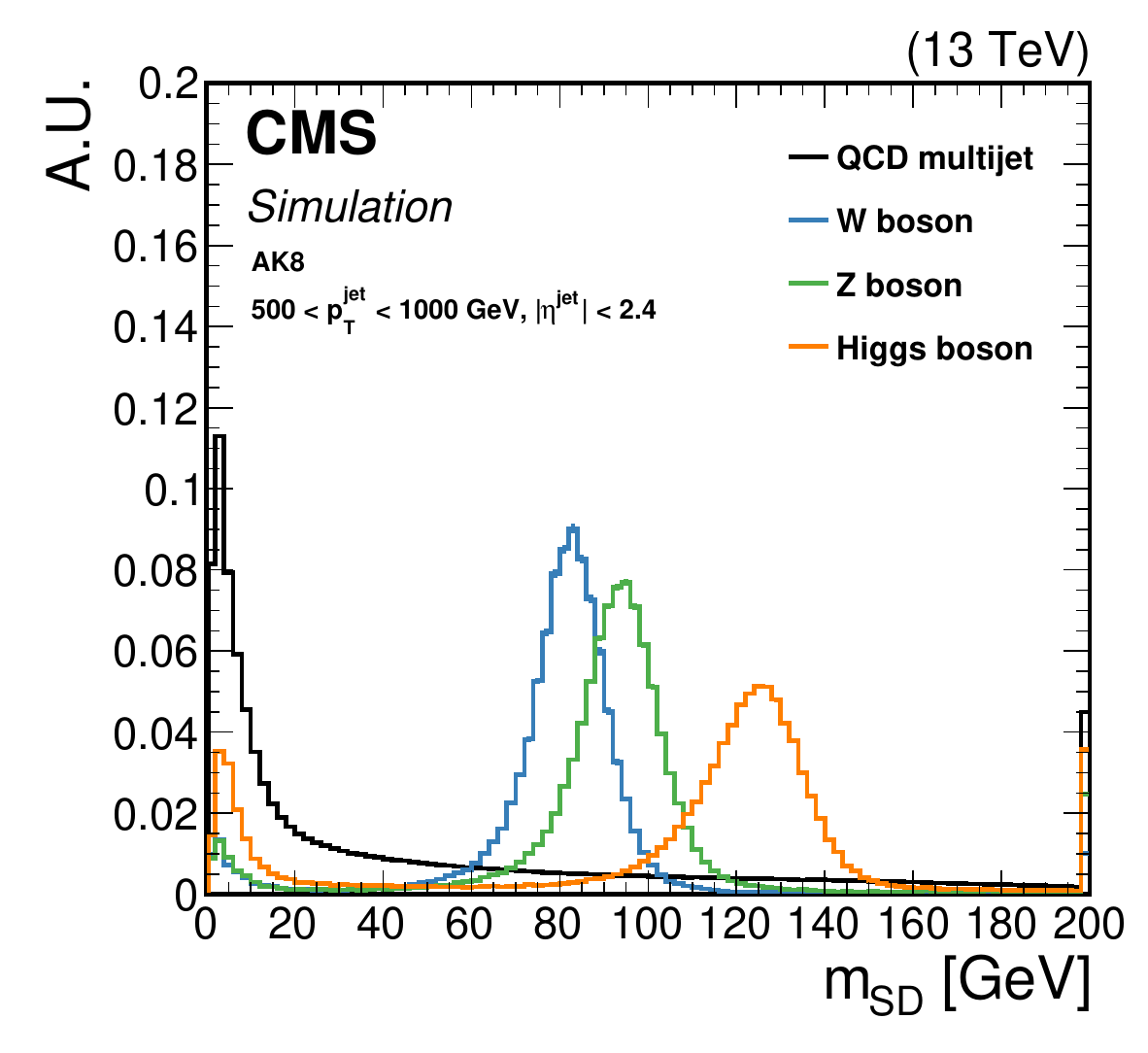}
 \label{fig:reco:largeR:mass:CMS}
}
\caption{The jet mass in simulated events for \largeR{} jets as reconstructed by (\textbf{a}) ATLAS \cite{ATLAS:WTopTagOldCONF} and (\textbf{b}) CMS \cite{CMS:MLTag}. The background of non-top quark and gluon jets (Dijets, QCD multijet) peaks at low values of mass, while the hadronic decays of various different massive particles are seen to peak at roughly the expected mass of the parent particle. \label{fig:reco:largeR:mass}}
\end{figure}

\subsubsection{Correcting the Jet Scale and Resolution}

Just like for \smallR{} jets, it is important to correct the scale and resolution of \largeR{} jets.
The momentum scale and resolution still have a direct impact on searches for new physics in hadronic final states, including being responsible for the peak position and width in hadronic resonance searches.
However, for \largeR{} jets, it is also very important to correct the mass scale and resolution: this impacts some searches directly where the \largeR{} jet is itself expected to contain a new physics resonance, but it is also key to the concept of identifying \largeR{} jets consistent with a given signal interpretation ($W/Z/H$ boson or top-quark) and rejecting jets from background sources.

The procedure for both momentum and mass calibrations starts by comparing a given reconstructed jet against its truth jet reference, but in this context the truth jet must also have the same grooming algorithm applied to remove underlying event contributions in the same way and thus represent the same type of shower in the detector.
The corrections are derived sequentially, as the momentum calibration scales the full four-vector (including the mass), and the mass is then further corrected \cite{ATLAS:LargeRcalib}.
The resulting momentum and mass calibrations, correcting for the ATLAS detector response, are provided in Figure \ref{fig:reco:largeR:MCJXS}.
In order to properly correct the response, the mass calibration is actually dependent on the jet mass; a plot corresponding to the correction for a jet with the $W$ boson mass is shown.

\vspace{-6pt}
\begin{figure}[H]

\subfigure[ ATLAS, momentum calibration]{
 \includegraphics[width=0.45\textwidth]{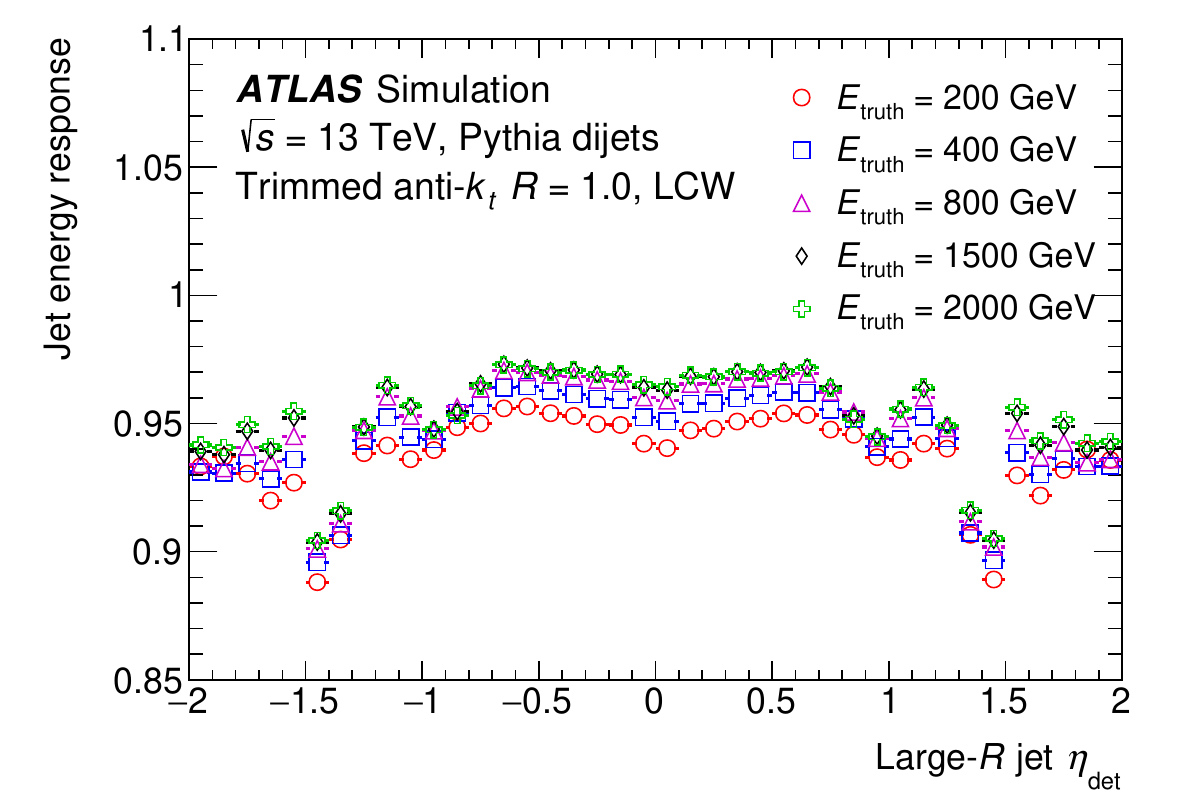}
 \label{fig:reco:largeR:MCJES}
}
\subfigure[ ATLAS, mass calibration]{
 \includegraphics[width=0.45\textwidth]{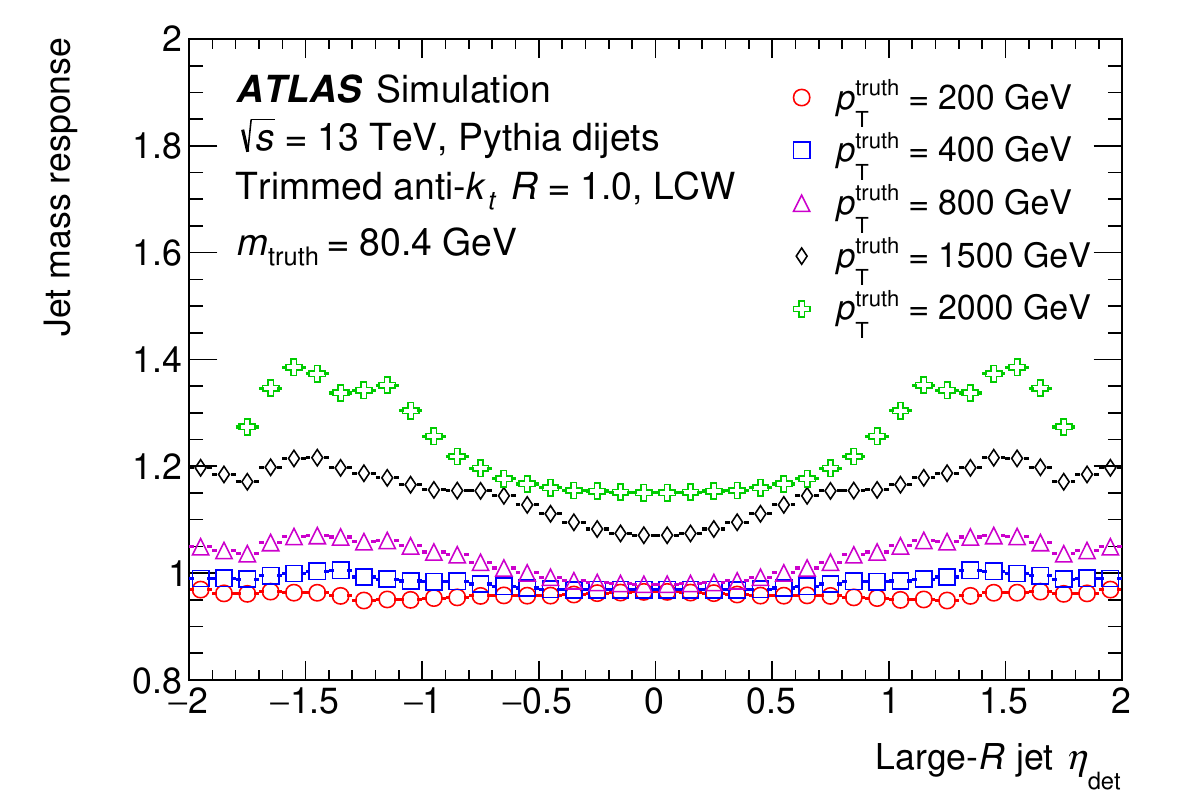}
 \label{fig:reco:largeR:MCJMS}
}
\caption{The ATLAS simulation-based jet response \cite{ATLAS:LargeRcalib}, the inverse of which is used to correct for the calorimeter response to hadronic showers, shown for (\textbf{a}) the jet momentum and (\textbf{b}) the jet mass, for a jet with a mass equal to the $W$ boson mass (other mass values are shown in the same reference). \label{fig:reco:largeR:MCJXS}}
\end{figure}

After applying a simulation-based correction, it is once again necessary to evaluate possible differences between data and simulation.
In ATLAS, this is done for the momentum scale following the same strategy as for \smallR{} jets, where the \insitu{} balance of a \largeR{} jet against a well-known reference is used \cite{ATLAS:LargeRcalib}.
The resulting momentum calibration and uncertainties, including flavour- and topology-related effects, are shown in Figure \ref{fig:reco:largeR:insituJES}.
The jet momentum resolution is also evaluated in a similar way, but with only the method using pairs of jets, as \largeR{} jets are of most relevance at higher momentum where the approach using two jets is very precise \cite{ATLAS:LargeRcalib}; plots for the \largeR{} jet momentum resolution are not shown here for brevity.

Correcting for possible differences in the jet mass between data and simulation requires a new approach with respect to \insitu{} momentum calibrations, as the jet mass is not an event-conserved quantity unlike the total transverse momentum.
It is thus important to identify a high-purity selection of signal jets, where the mass distribution should be the same, and to correct for any observed differences.
Both ATLAS and CMS do this using $t\bar{t}$ events, where the $W$ boson from one of the top quarks decays leptonically, and the $W$ boson from the other top quark decays hadronically.
These semi-leptonic $t\bar{t}$ events can be selected with high purity, and provide access to signal jets with either the $W$ boson mass or the top quark mass, depending on if the $b$-quark from the hadronically decaying top quark is inside or outside of the \largeR{} jet under study.
These high-purity selections allow for a direct extraction of the mass scale and resolution by fitting the signal component in both data and simulation, and any subsequent differences can then be corrected for.
Examples of this procedure are shown in Figure \ref{fig:reco:largeR:JMS} for top quarks in ATLAS and $W$ bosons in CMS.

\vspace{-6pt}

\begin{figure}[H]

\subfigure[ ATLAS, \insitu{} momentum calibration]{
 \includegraphics[width=0.45\textwidth]{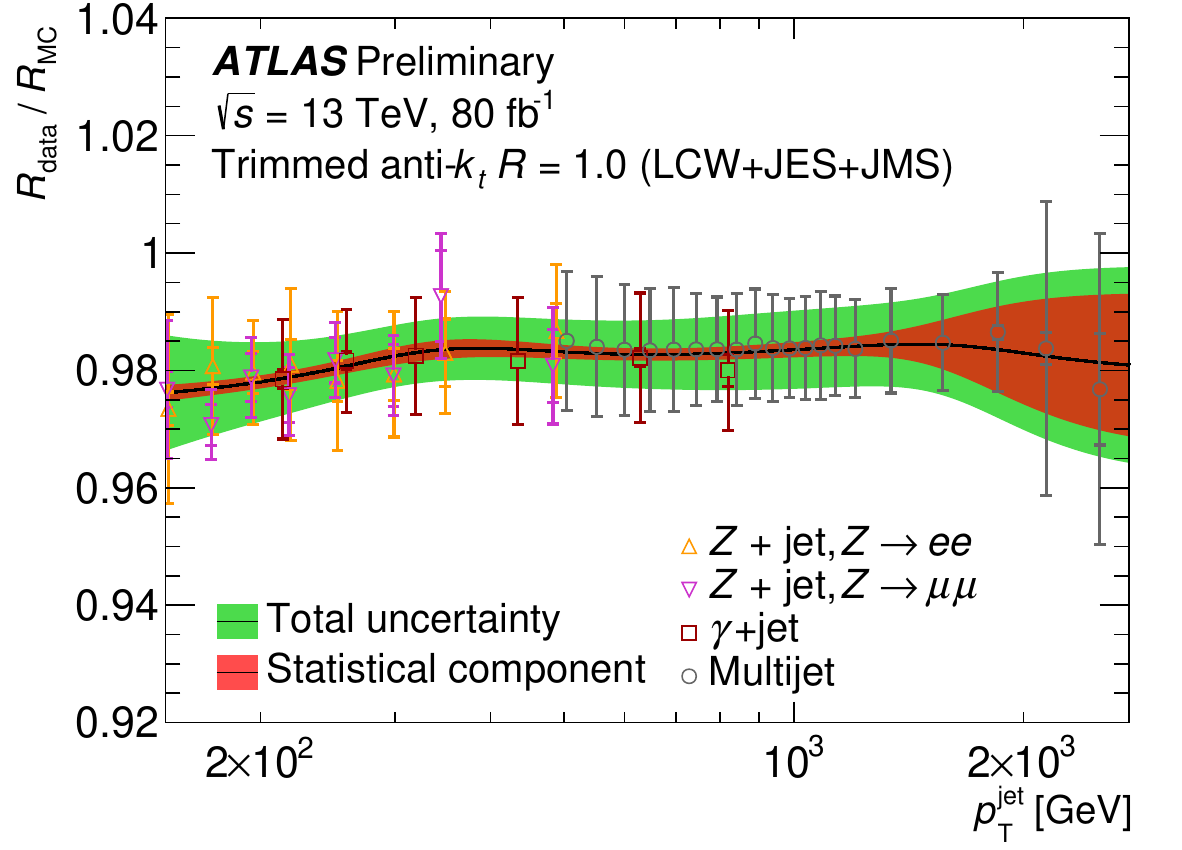}
 \label{fig:reco:largeR:insitu}
}
\subfigure[ ATLAS, momentum scale uncertainties]{
 \includegraphics[width=0.48\textwidth]{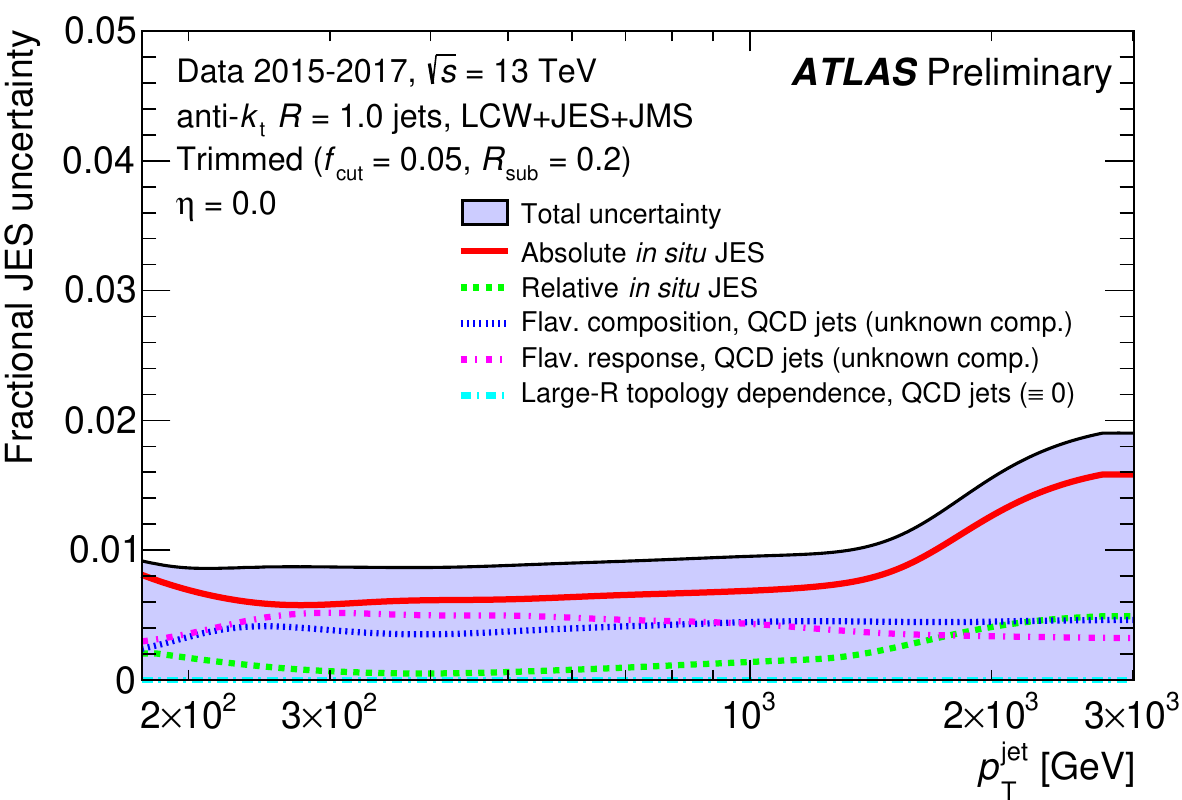}
 \label{fig:reco:largeR:unc}
}
\caption{(\textbf{a}) The ATLAS data-to-simulation difference in jet momentum response, the inverse of which is used to correct the data to match the simulation, as evaluated using a series of \insitu{} balance techniques \cite{ATLAS:LargeRcalib}. (\textbf{b}) The uncertainties on the ATLAS jet momentum scale \cite{ATLAS:LargeRcalib}. \label{fig:reco:largeR:insituJES}}
\end{figure}

\vspace{-12pt}

\begin{figure}[H]

\subfigure[ ATLAS]{
 \includegraphics[width=0.42\textwidth]{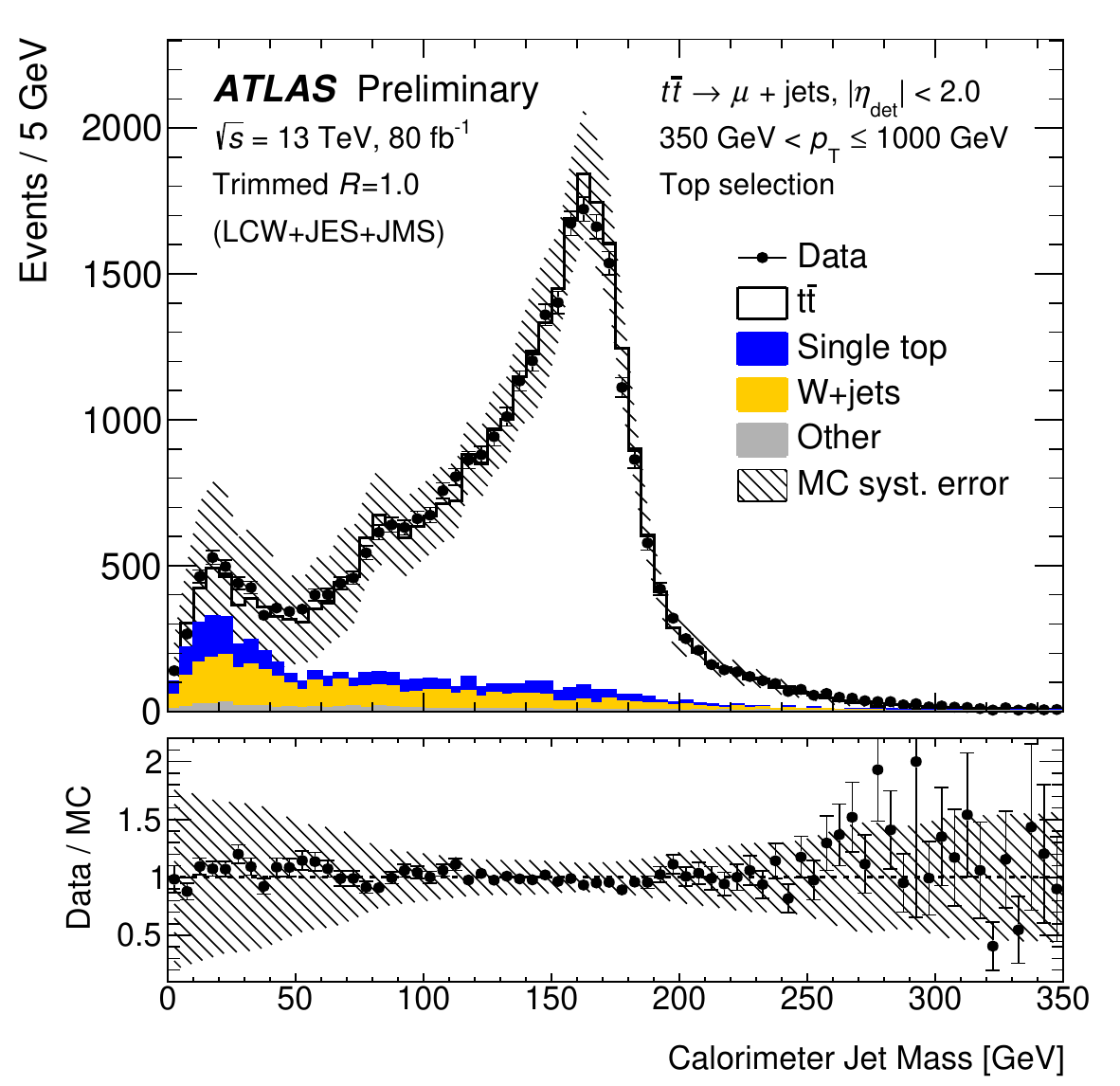}
 \label{fig:reco:largeR:JMS:ATLAS}
}
\subfigure[ CMS]{
 \includegraphics[width=0.53\textwidth]{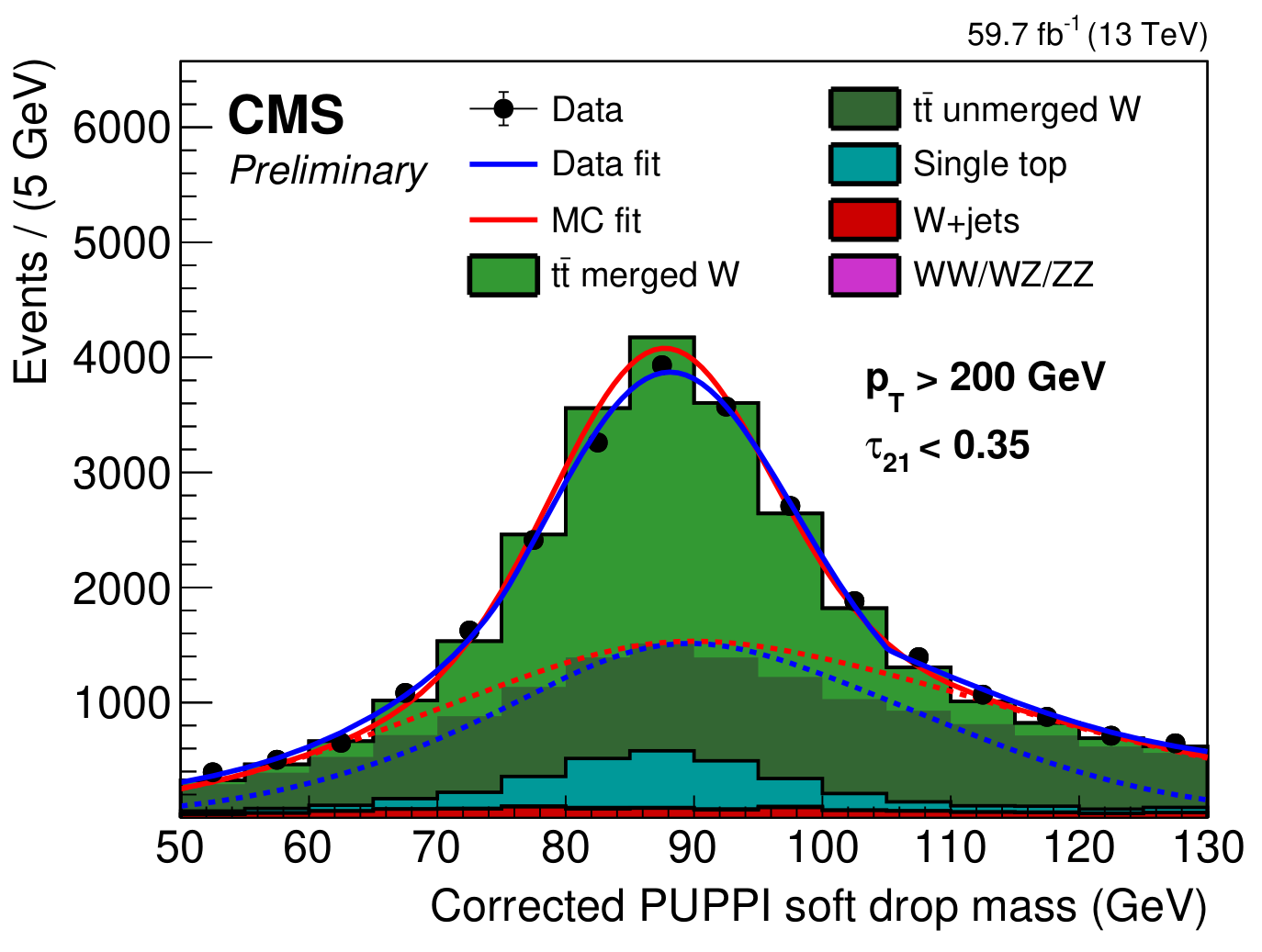}
 \label{fig:reco:largeR:JMS:CMS}
}
\caption{A comparison of the jet mass in data and simulation for \largeR{} jets as reconstructed by \linebreak (\textbf{a}) ATLAS for top-quark jets \cite{ATLAS:FFConf} and (\textbf{b}) CMS for $W$-boson jets \cite{CMS:WTopSFs}, both in semi-leptonic $t\bar{t}$ events. The high signal purity, whether the signal is a top quark or $W$ boson, allows for fitting the jet mass scale and resolution directly in both data and simulation and thereby correcting for the difference. \label{fig:reco:largeR:JMS}}
\end{figure}

\subsubsection{Identifying Boosted Hadronic Decays}
\label{sec:reco:largeR:tag}

With a well-calibrated jet mass, it is already possible to start to differentiate \largeR{} jets containing boosted hadronic decays of massive particles from \largeR{} jets containing not-top quarks and gluons.
However, the background jet distribution has a long tail as seen in Figure \ref{fig:reco:largeR:mass}, and the cross-section for such background events is dramatically higher than for signals such as $W/Z/H$ bosons and top quarks.
As such, it makes sense to design more complex algorithms to differentiate between signal and background jets; these algorithms are commonly referred to as jet taggers.

The idea of such taggers relies primarily upon the different angular energy structure of signal and background processes within \largeR{} jets.
For example, a background jet originating from a not-top quark or gluon will usually have a single region of high energy density, signal jets originating from $W/Z/H$ bosons will typically have two regions of high energy density, and signal jets originating from top quarks will typically have three regions of high energy density.
One way that this can be quantified is to impose a given number of axes on the jet, and evaluate the consistency of the jet constituent four-vectors with that number of axes.
This forms the basic idea of one of the early jet substructure variables, named $N$-subjettiness \cite{Nsubjettiness}, which is still used today.

In the nomenclature of $N$-subjettiness, a measure of the consistency of a jet with the interpretation of having one axis is $\tau_1$, two axes is $\tau_2$, and so on.
Ratios of $N$-subjettiness variables, $\tau_{XY} = \tau_X/\tau_Y$, then provide separation between jets with $X$ or more axes as opposed to $Y$ or fewer axes.
This can then be used to differentiate between signal and background jets, as shown in Figure \ref{fig:reco:largeR:JSS}, where one of the plots shows separation after having already applied a mass cut, therefore demonstrating that $N$-subjettiness ratios provide complementary information to the jet mass.
\vspace{-9pt}

\begin{figure}[H]

\subfigure[ ATLAS]{
 \includegraphics[width=0.45\textwidth]{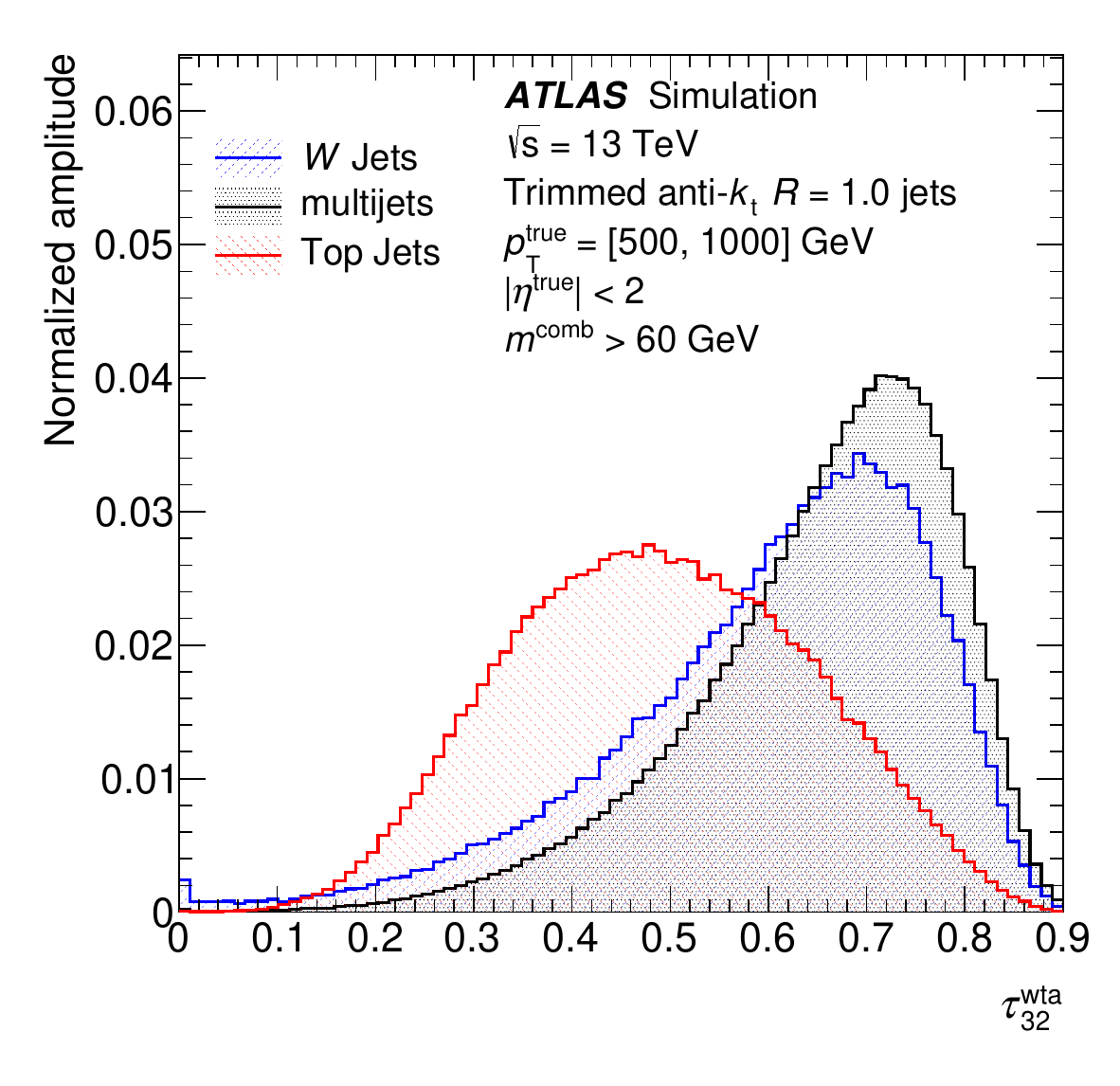}
 \label{fig:reco:largeR:JSS:ATLAS}
}
\subfigure[ CMS]{
 \includegraphics[width=0.47\textwidth]{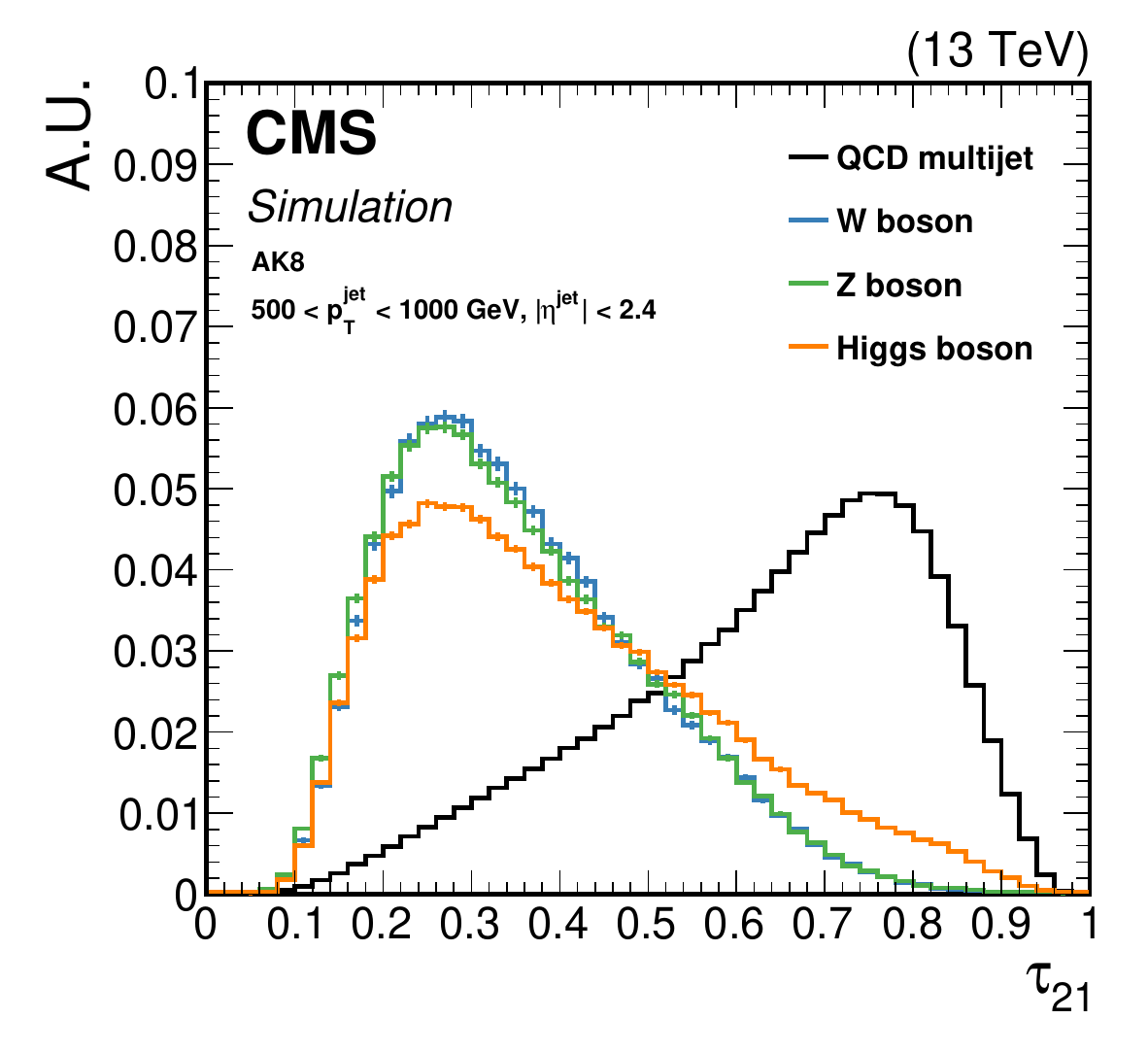}
 \label{fig:reco:largeR:JSS:CMS}
}
\caption{(\textbf{a}) The $\tau_{32} = \tau_3/\tau_2$ substructure variable in ATLAS simulated events, which is useful for identifying jets with at least three high-energy-density regions, such as top-quarks \cite{ATLAS:WTopTag}. A specific $N$-subjettiness axis definition, winner-takes-all (wta), is shown; the details of different axis definitions are beyond the scope of this review. (\textbf{b}) The $\tau_{21} = \tau_2/\tau_1$ substructure variable in CMS simulated events, which is useful for identifying jets with at least two high-energy-density regions, such as $W/Z/H$ bosons \cite{CMS:MLTag}. \label{fig:reco:largeR:JSS}}
\end{figure}

Simple cut-based jet taggers, consisting of jet mass cuts plus a single additional substructure variable, such as $\tau_{32}$ or $\tau_{21}$, have been used by many ATLAS and CMS searches in the earlier stages of Run 2.
However, more recently both ATLAS and CMS have switched to a variety of advanced jet taggers exploiting modern machine learning techniques (Boosted Decision Trees, BDTs; Deep Neural Networks, DNNs; and other modern tools).
These more advanced taggers are able to exploit non-linear correlations between the different substructure variables to further improve on few-variable cut-based taggers, or even to use the four-vectors of the individual jet constituents directly.

ATLAS and CMS have both designed and used a large variety of different jet taggers, and it would take an entire separate review to fully explore the different options that have been used during Run 2.
A few summary plots of different taggers used for the identification of jets originating from top-quarks, $W$ bosons, and $H$ bosons are shown in Figure \ref{fig:reco:largeR:tag}; $Z$ boson tagging performance is generally very similar to $W$ boson tagging.
The figures show either the background rejection (ATLAS) or background misidentification rate (CMS) for a fixed signal efficiency, where one quantity is the inverse of the other: a good tagger will have a large background rejection or, equivalently, a small misidentification rate.
Taggers using different forms of machine learning techniques are shown to provide large improvements in performance over simple cut-based taggers, as expected, and further benefits are seen from adding in jet constituent information such as is done in DeepAK8 from CMS \cite{CMS:MLTag}.

\begin{figure}[H]

\subfigure[ ATLAS, top-quark taggers]{
 \includegraphics[width=0.45\textwidth]{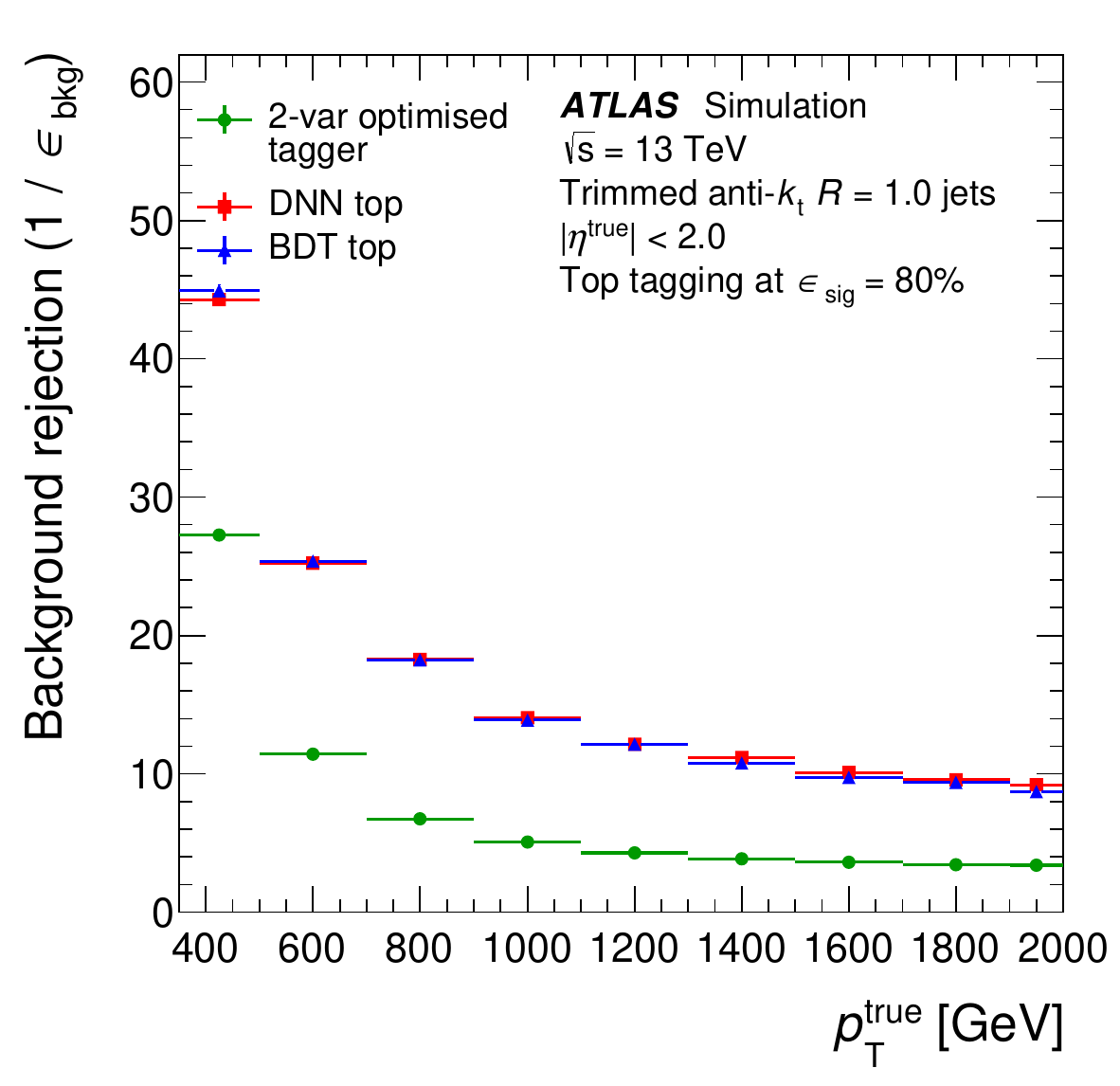}
 \label{fig:reco:largeR:tagTop:ATLAS}
}
\subfigure[ CMS, top-quark taggers]{
 \includegraphics[width=0.47\textwidth]{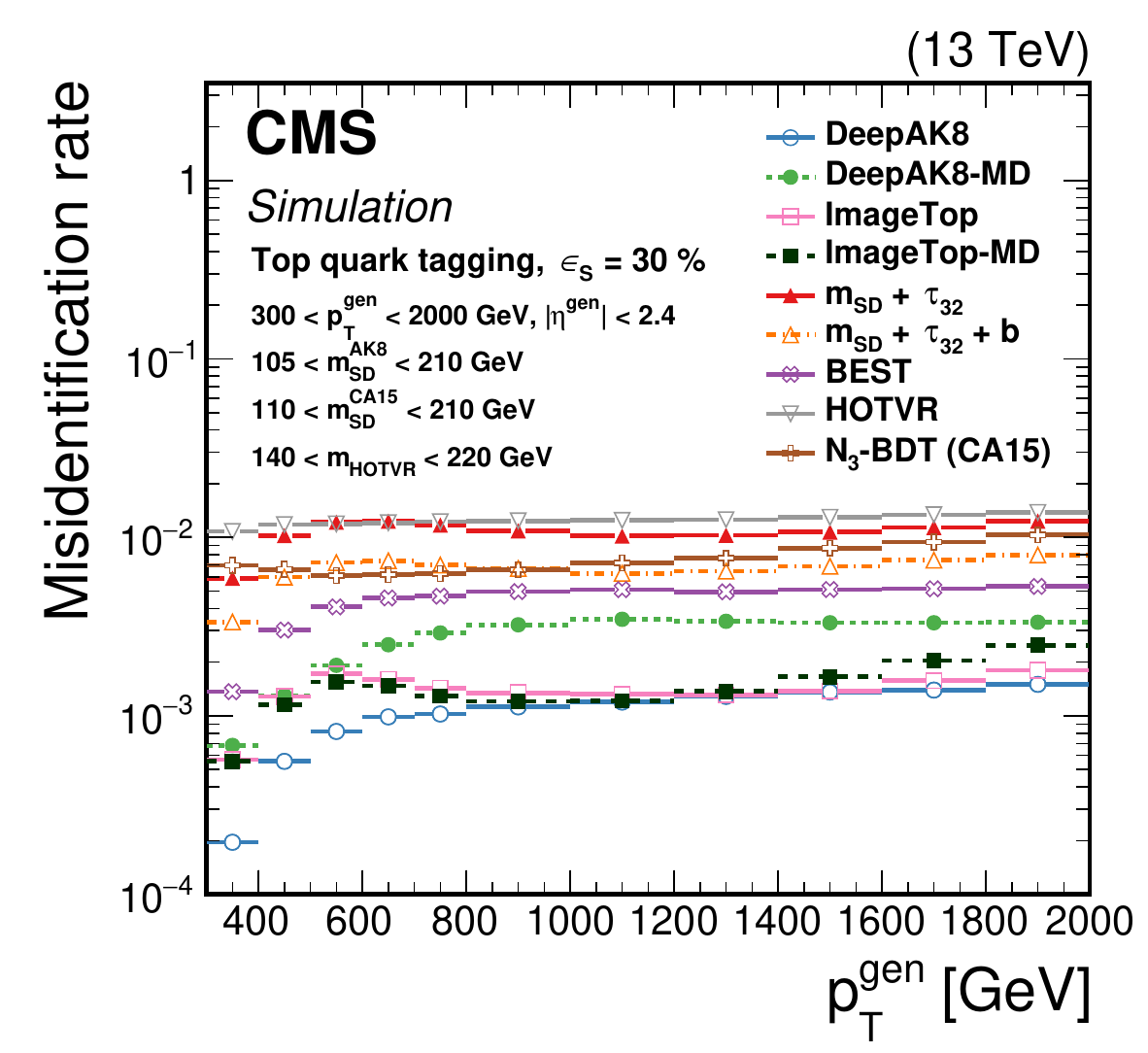}
 \label{fig:reco:largeR:tagTop:CMS}
}\\
\subfigure[ ATLAS, $W$-boson taggers]{
 \includegraphics[width=0.45\textwidth]{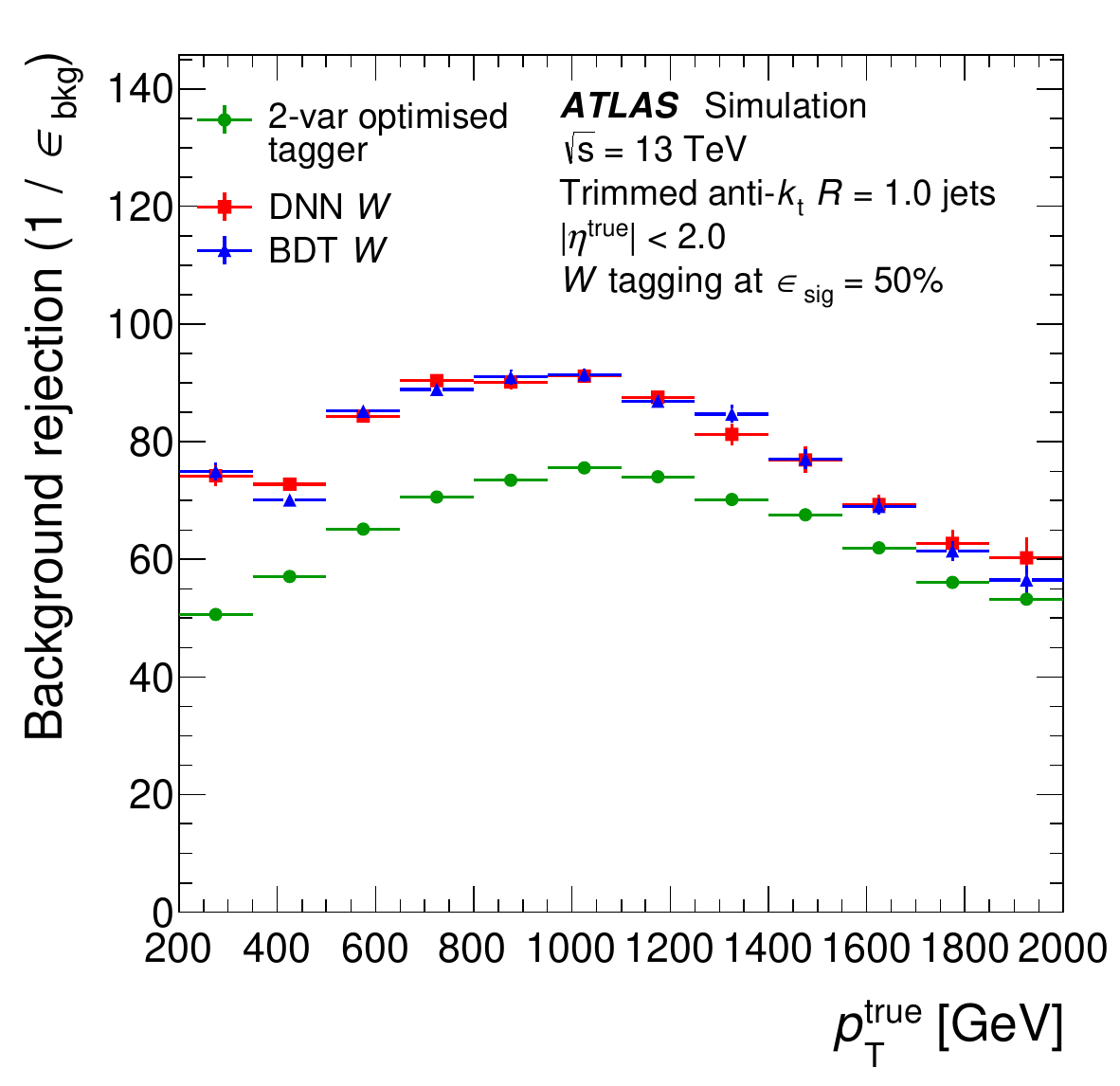}
 \label{fig:reco:largeR:tagW:ATLAS}
}
\subfigure[ CMS, $W$-boson taggers]{
 \includegraphics[width=0.47\textwidth]{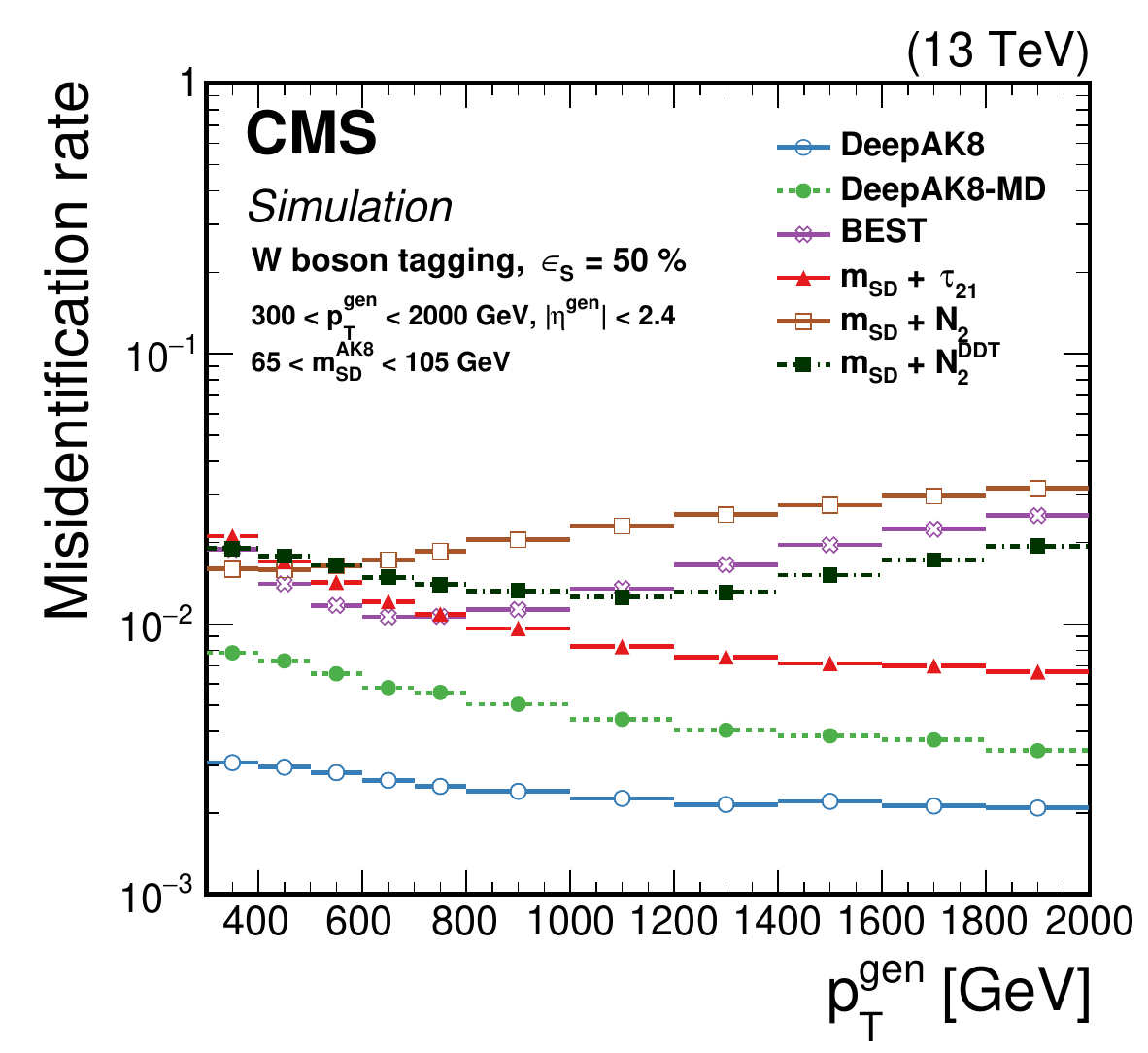}
 \label{fig:reco:largeR:tagW:CMS}
}\\
\subfigure[ ATLAS, $H$-boson taggers]{
 \includegraphics[width=0.45\textwidth]{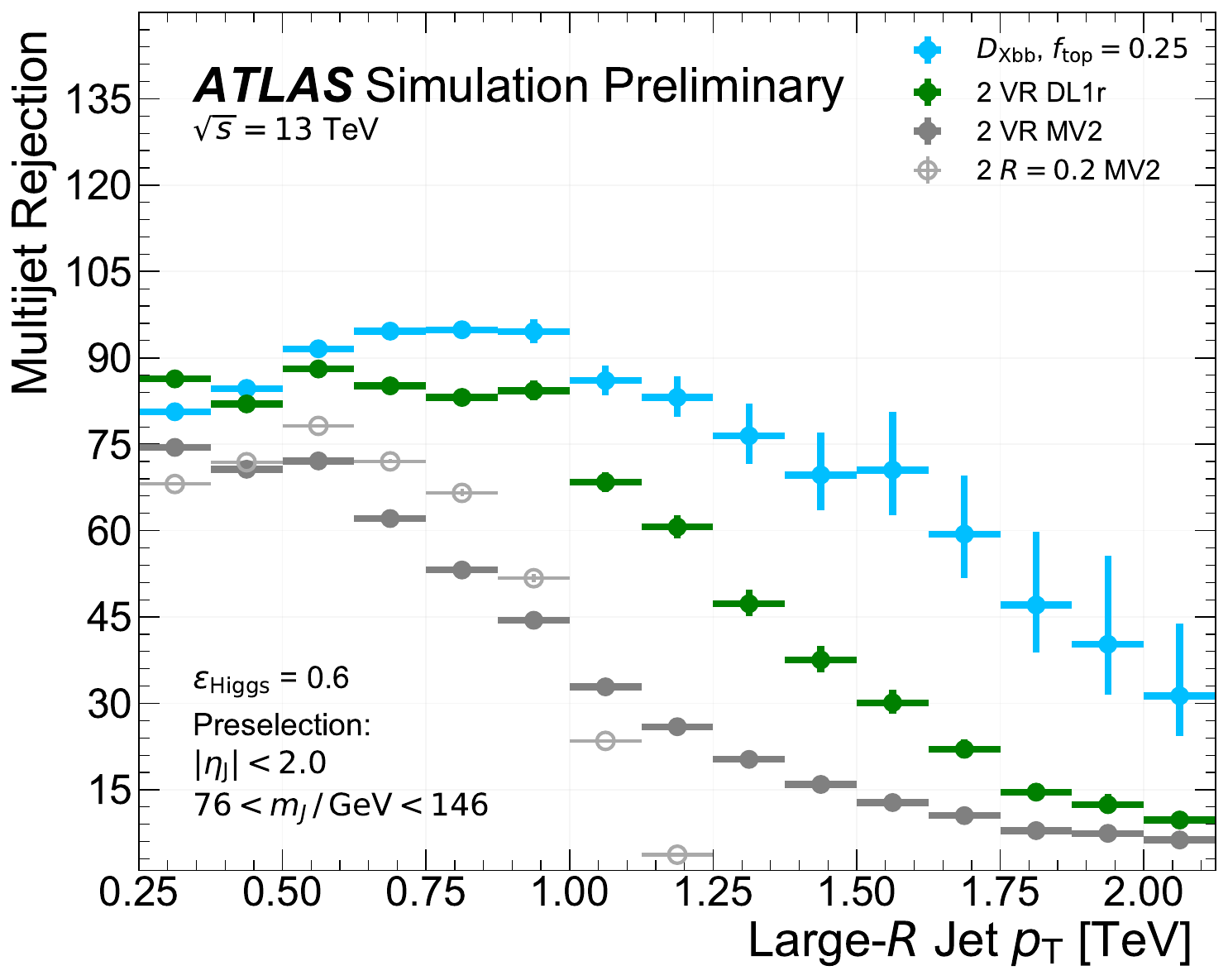}
 \label{fig:reco:largeR:tagH:ATLAS}
}
\subfigure[ CMS, $H$-boson taggers]{
 \includegraphics[width=0.47\textwidth]{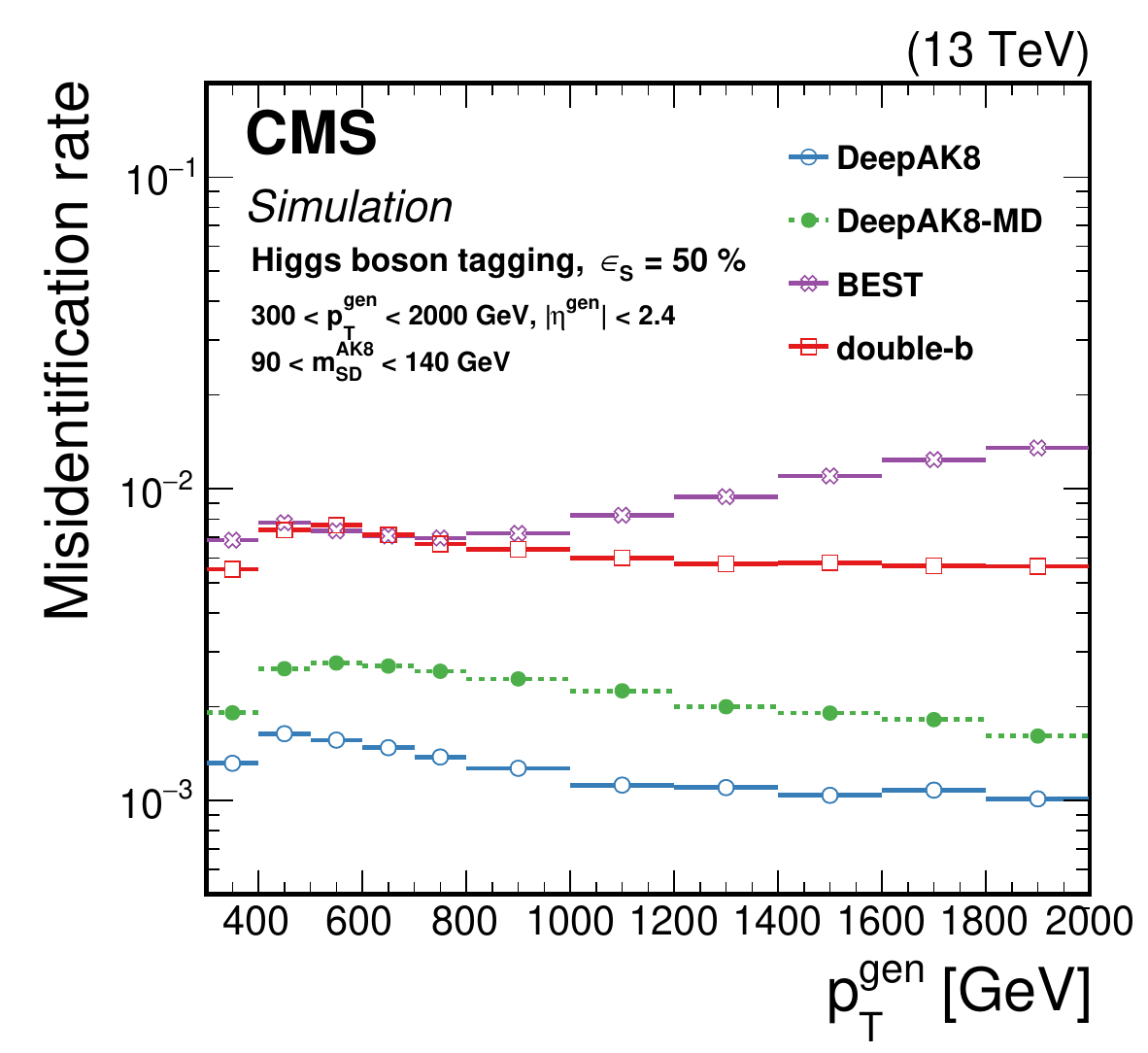}
 \label{fig:reco:largeR:tagH:CMS}
}
\caption{The performance of various \largeR{} jet taggers, in (\textbf{a},\textbf{c},\textbf{e}) ATLAS \cite{ATLAS:WTopTag,ATLAS:tagH} and (\textbf{b},\textbf{d},\textbf{f}) CMS \cite{CMS:MLTag}. Tagging performance is shown for (\textbf{a},\textbf{b}) top-quark taggers, (\textbf{c},\textbf{d}) $W$-boson taggers, and (\textbf{e},\textbf{f}) $H$-boson taggers. The performance of $Z$-boson taggers is not shown, as they are very similar to $W$-boson taggers; the equivalent plots for $Z$-boson tagging can be found in the same references. ATLAS shows results as a function of background rejection, while CMS shows results as a function of misidentification rate; these quantities are the inverse of each other. \label{fig:reco:largeR:tag}}
\end{figure}

While the ability of the tagger to reject background jets for a given signal efficiency is one very important metric, it is not always the deciding factor when choosing which tagger to use for a given search for new physics.
Taggers providing the largest background rejection often significantly sculpt the background jet mass distribution, which can be a major problem.
As an example, searches looking for resonances in the mass distribution of a single \largeR{} jet require a smooth jet mass distribution for their background estimation, and must avoid introducing artificial bumps from the use of the tagger.
In this context, the extent to which a tagger sculpts the jet mass distribution becomes a key metric.
Originally, individual substructure variables, such as $\tau_{21}$, were transformed in a way that made them independent of the jet mass; this is referred to as a Designed Decorrelated Tagger (DDT) \cite{DDT}.
However, modern advances in machine learning techniques have allowed for this to be extended to a variety of different advanced taggers, which are typically then referred to as being mass-decorrelated (MD).
ATLAS \cite{ATLAS:DDT} and CMS \cite{CMS:MLTag} have both studied the development of DDT and MD taggers, some of which can be seen in the CMS plots in Figure \ref{fig:reco:largeR:tag}; ATLAS DDT/MD results are not shown here.
The plots show that DDT techniques can actually increase the background rejection for cut-based taggers as the transformation is including more information, while MD taggers sacrifice some of their performance to avoid sculpting the mass distribution, although they remain more powerful than cut-based DDT approaches.
MD taggers can also open up new analysis opportunities, such as enabling the use of mass-related control regions, which further supports the usage of such advanced taggers.

After the development of complex jet taggers, it is necessary to derive correction factors to ensure that the taggers are selecting the same types of jets in data and simulation.
This is not a trivial statement, as the simulation may model the correlations between variables or jet constituent four-vectors differently than in data, and the taggers optimized in simulated events may then exploit these differences.
Similar to the procedure used to correct the jet mass scale and resolution, it is important to identify a high-purity selection of the object of interest and to compare the resulting data and simulated efficiencies of the taggers.
Focusing on $W$-boson and top-quark taggers, as those are of most relevance to the hadronic searches that will be discussed in this review, ATLAS and CMS both use semi-leptonic $t\bar{t}$ events for signal efficiencies; the resulting agreement between data and simulation can be seen in Figure \ref{fig:reco:largeR:SF}.
Inclusive selections of not-top quark and gluon events, or of photon plus jet events, are used to evaluate the corresponding background efficiencies for $W$-boson and top-quark taggers \cite{ATLAS:WTopTag,CMS:MLTag}.
The resulting scale factors required to correct the simulation to match the data are typically reasonably close to one, suggesting that the modelling of the quantities and correlations being exploited in the taggers is of reasonable quality.

\vspace{-8pt}
\begin{figure}[H]

\subfigure[ ATLAS, DNN top quark tagger]{
% \includegraphics[width=0.47\textwidth]{figures/reco/CMS_tag_SF_top.pdf}
% \label{fig:reco:largeR:SF:top}
 \includegraphics[width=0.45\textwidth]{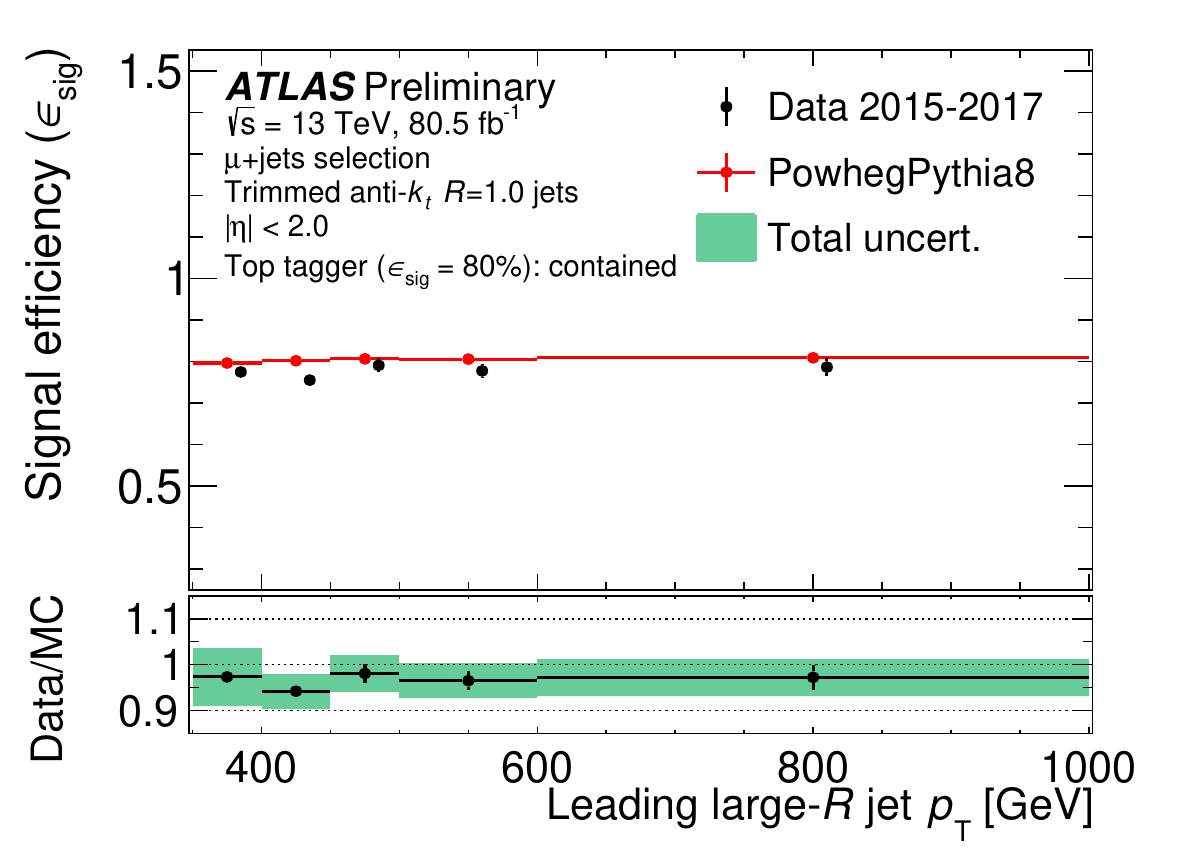}
 \label{fig:reco:largeR:SF:ATLAS}
}
\subfigure[ CMS, $W$ boson taggers]{
 \includegraphics[width=0.47\textwidth]{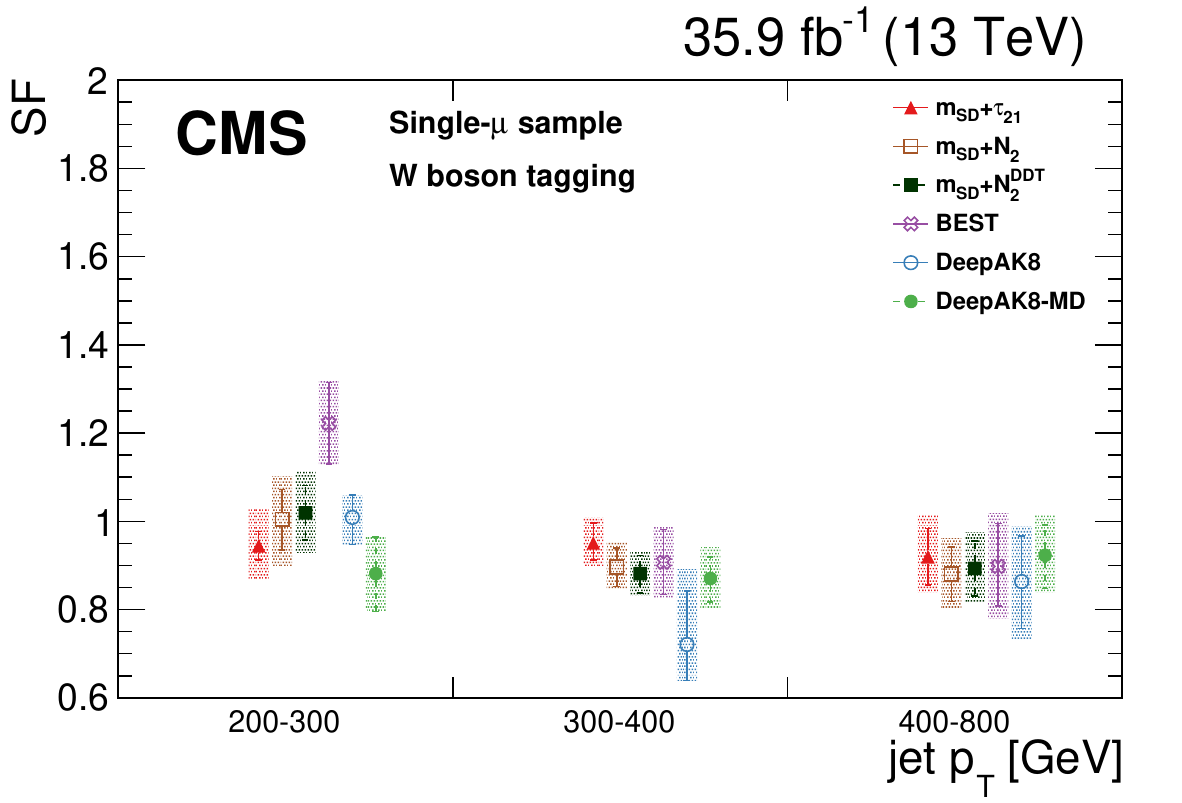}
 \label{fig:reco:largeR:SF:W}
}
\caption{Extracted scale factors to correct tagging differences between data and simulation for (\textbf{a}) a top-quark tagger in ATLAS \cite{ATLAS:WTopTagPUB} and (\textbf{b}) several $W$-boson taggers in CMS \cite{CMS:MLTag}. Scale factors are generally close to one, suggesting that the application of the taggers to data is reasonably well modelled in simulation, although small corrections are necessary. \label{fig:reco:largeR:SF}}
\end{figure}

% Dijet searches
\section{Di-Jet Searches}
\label{sec:dijet}

Di-jet searches have a long history in hadron collider experiments, and are typically among the first searches conducted upon accessing a new centre-of-mass-energy scale, as they rapidly become sensitive to new very massive physics.
Such searches are also sensitive to a wide variety of new physics models, due to their minimal assumptions about the properties of the new physics sector, as discussed in Section \ref{sec:motivation:motivations}.
This motivates a variety of different types of di-jet searches for new high-mass phenomena, as will be discussed in Section \ref{sec:dijet:classic}.

Rather than focusing solely on pushing the search for new physics to ever-higher masses, modern di-jet searches are also increasingly extending towards the lower-mass regime.
While this regime has been studied at previous colliders, new physics may still have been missed; reaching the mass scale of new physics is only sufficient to discover that new particle if the search also reaches the sensitivity required to extract the signal from the background.
In other words, new physics may be hiding in the low-mass regime if the new phenomenon is too weakly coupled to quarks and gluons to have been observed in past searches.

Given that the LHC produces an enormous amount of di-jet events, it thus provides the potential to probe low-mass hadronic physics well beyond what was possible at previous experiments, so long as you can access that data.
This is the main challenge that low-mass di-jet searches must overcome: the majority of the relevant low-mass data is not recorded by ATLAS and CMS due to trigger constraints, as discussed in Section \ref{sec:motivation:challenges}.
Low-mass di-jet searches must therefore find some way to mitigate these trigger constraints, and ATLAS and CMS have now found several complementary ways to do so, as will be discussed in Sections \ref{sec:dijet:trigger}--\ref{sec:dijet:ISRboosted}.

These different di-jet search strategies are all complementary, and must be considered as a whole to properly understand the sensitivity of the ATLAS and CMS di-jet search programmes to new physics.
During Run 2, ATLAS and CMS moved towards a harmonised means of comparing the different analyses, which are interpreted as a search for a new axial-vector $Z^\prime$ boson of a given mass $m_{Z^\prime}$ and with a given coupling of that boson to quarks $g_q$.
Tree-level Feynman diagrams for the production and subsequent decay of the searched-for $Z^\prime$ boson are shown in Figure \ref{fig:dijet:feyn}, and more details on this model as applied to dijet searches at the LHC can be found in, for example, the LPCC Dark Matter Working Group recommendation documents \cite{Theory:ZPrimeLHC1,Theory:ZPrimeLHC2,Theory:ZPrimeLHC3,Theory:ZPrimeLHC4}.
The resulting limits of the different types of searches conducted by both ATLAS and CMS are shown in Figure \ref{fig:dijet:summary}.
While these figures provide an excellent summary of the analyses conducted to the date at which the plots were last updated, it is also useful to discuss how the individual results were obtained.

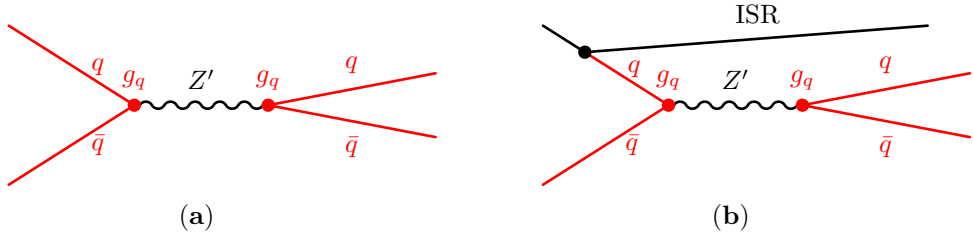
\begin{figure}[H]

\subfigure[]{
\begin{fmffile}{DijetStandard}
 \begin{fmfgraph*}(180,60)
 \fmfleft{qbar,q}
 \fmfright{balance,chibar,chi,ISR}
 
% \fmflabel{$q$}{q}
% \fmflabel{$\bar{q}$}{qbar}
 
 \fmf{vanilla,tension=2,foreground=red}{q,vertexISR}
 \fmf{vanilla,foreground=red,tension=2}{qbar,ISRphantom}
 
 \fmf{vanilla,label=$\textcolor{red}{q}$,label.side=left,label.dist=0.8,foreground=red,tension=1}{vertexISR,vertexL}
 \fmf{vanilla,label=$\textcolor{red}{\bar{q}}$,label.side=right,label.dist=0.8,foreground=red,tension=1}{ISRphantom,vertexL}
 
 \fmf{phantom,tension=0}{vertexISR,ISR}
 \fmf{phantom,tension=0}{ISRphantom,balance}
 
 \fmf{vanilla,label=$\textcolor{red}{q}$,foreground=red,tension=0.5}{vertexR,chi}
 \fmf{vanilla,label=$\textcolor{red}{\bar{q}}$,foreground=red,tension=0.5}{vertexR,chibar}
 
 \fmfv{decor.shape=circle,decor.filled=full,
decor.size=2thick,foreground=red,label=$\textcolor{red}{g_q}$,label.angle=90}{vertexL}
 \fmfv{decor.shape=circle,decor.filled=full,
decor.size=2thick,foreground=red,label=$\textcolor{red}{g_q}$,label.angle=90}{vertexR}
 
 \fmf{boson,label=$Z^\prime$,tension=1.25}{vertexL,vertexR}
 \end{fmfgraph*}
\end{fmffile}
}
\subfigure[]{
\begin{fmffile}{DijetISR}
 \begin{fmfgraph*}(180,60)
 \fmfleft{qbar,q}
 \fmfright{balance,chibar,chi,ISR}
 
% \fmflabel{$q$}{q}
% \fmflabel{$\bar{q}$}{qbar}
 
 \fmf{vanilla,tension=2}{q,vertexISR}
 \fmf{vanilla,foreground=red,tension=2}{qbar,ISRphantom}
 
 \fmf{vanilla,label=$\textcolor{red}{q}$,label.side=left,label.dist=0.8,foreground=red,tension=1}{vertexISR,vertexL}
 \fmf{vanilla,label=$\textcolor{red}{\bar{q}}$,label.side=right,label.dist=0.8,foreground=red,tension=1}{ISRphantom,vertexL}
 
 \fmf{vanilla,label=ISR,tension=0}{vertexISR,ISR}
 \fmf{phantom,tension=0}{ISRphantom,balance}
 
 \fmf{vanilla,label=$\textcolor{red}{q}$,foreground=red,tension=0.5}{vertexR,chi}
 \fmf{vanilla,label=$\textcolor{red}{\bar{q}}$,foreground=red,tension=0.5}{vertexR,chibar}

 \fmfdot{vertexISR}
 \fmfv{decor.shape=circle,decor.filled=full,
decor.size=2thick,foreground=red,label=$\textcolor{red}{g_q}$,label.angle=90}{vertexL}
 \fmfv{decor.shape=circle,decor.filled=full,
decor.size=2thick,foreground=red,label=$\textcolor{red}{g_q}$,label.angle=90}{vertexR}
 
 \fmf{boson,label=$Z^\prime$,tension=1.25}{vertexL,vertexR}
 \end{fmfgraph*}
\end{fmffile}
}
\caption{Feynman diagrams showing the s-channel production of a new axial-vector $Z^\prime$ mediator, which is produced through the annihilation of standard model quarks via a coupling $g_q$, and which decays back to standard model quarks via the same coupling. This process can be searched for either (\textbf{a}) directly, or (\textbf{b}) in association with Initial State Radiation (ISR); in the latter case, the ISR object can be used to help trigger the event or otherwise help with selecting the events of interest.\label{fig:dijet:feyn}}
\end{figure}

\begin{figure}[H]

\subfigure[ ATLAS]{
 \includegraphics[width=0.44\textwidth]{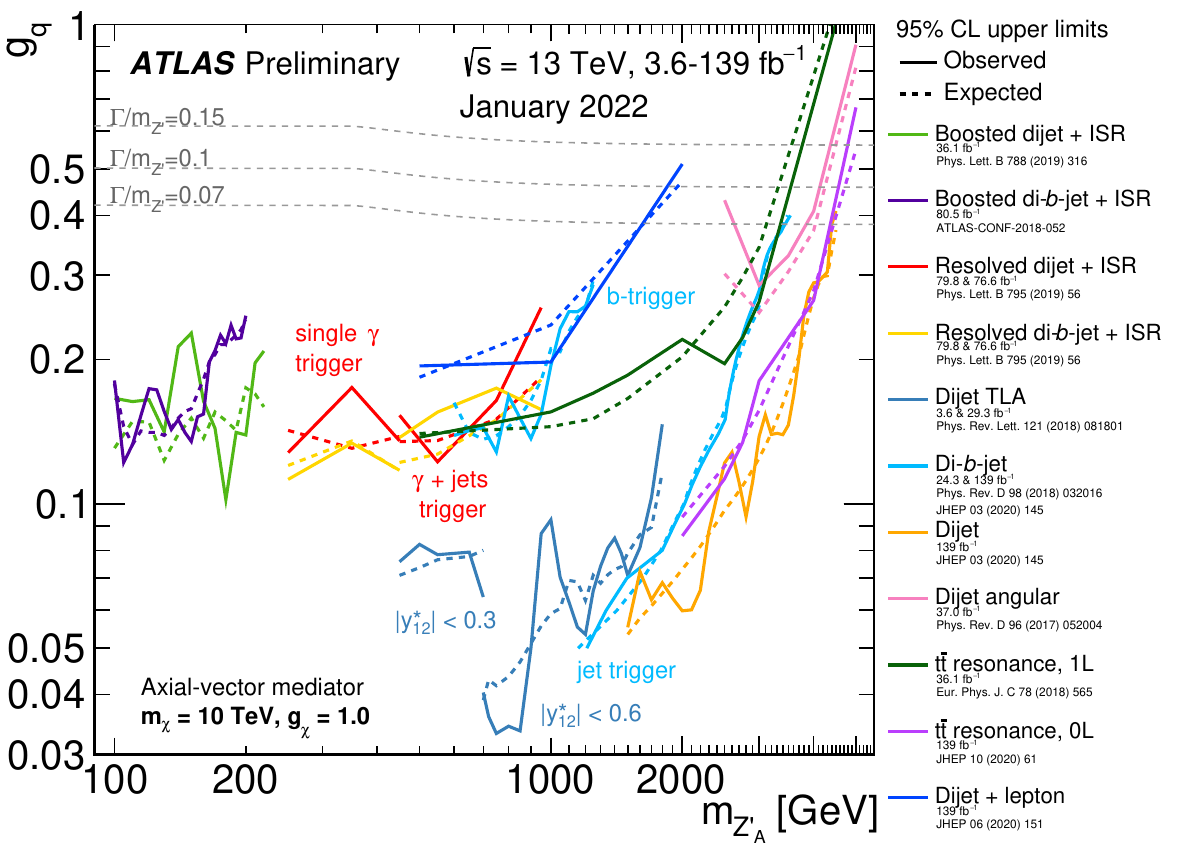}
 \label{fig:dijet:summary:ATLAS}
}
\subfigure[ CMS]{
 \includegraphics[width=0.51\textwidth]{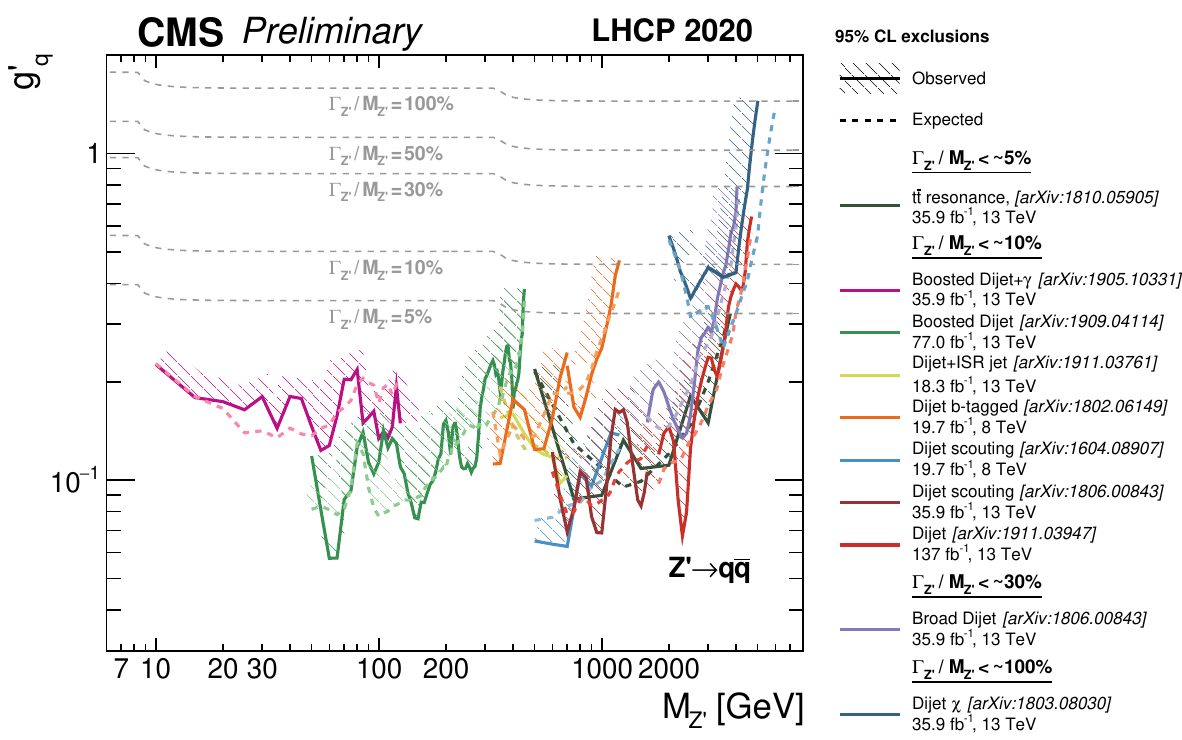}
 \label{fig:dijet:summary:CMS}
}
\caption{Summary plots showing the variety of di-jet searches and their corresponding limits on the production of a new axial-vector mediator for (\textbf{a}) ATLAS \cite{ATLAS:DMSummary} and (\textbf{b}) CMS \cite{CMS:EXOSummary}. \label{fig:dijet:summary}}
\end{figure}

\subsection{High-Mass Di-Jet Searches}
\label{sec:dijet:classic}
% High-mass resonant di-jet
% High-mass angular di-et
% High-mass resonant di-b-jet
% ttbar resonances

High-mass di-jet searches work in the regime where the corresponding jet triggers are fully efficient, and thus probe new physics with the full statistical power of the LHC.
The classic example is the di-jet resonance search, which both ATLAS and CMS have published using the full Run 2 dataset \cite{ATLAS:Dijet_HighMass,CMS:Dijet_HighMass}.
This type of search has such extensive statistical power that it includes observed events with di-jet masses of 8\TeV{}; this is the highest mass range seen by any of the searches presented in this review.
The ATLAS di-jet resonance search makes use of \smallR{} jets built from topological clusters, while the CMS search uses \smallR{} jets built from particle flow objects.
CMS furthermore uses the two leading jets in the event as seeds in the creation of ``wide jets'', whereby all other \smallR{} jets with $\Delta R < 1.1$ are added to the four-vectors of the leading jets, and the wide jets are used to define the di-jet system; this procedure reduces the impact of gluon radiation on the search.

ATLAS and CMS both fit the large and smoothly falling Standard Model background directly from data, using functions, and look for deviations from that background corresponding to new particle resonances.
CMS additionally considers another data-driven background estimation method, referred to as the ratio method, which defines signal and control regions in terms of the pseudorapidity separation of the di-jet system ($|\Delta\eta|$); these regions are then used to derive a mass-dependent transfer factor to correct the simulation to match the data expectation.
Both background estimation methods work well, and no significant deviation is observed by either ATLAS or CMS, as shown in Figure \ref{fig:dijet:highmass}a,b, respectively.
Limits are therefore set on a wide variety of different signal models, including both MC-based models and generic Gaussian signals, of various widths; examples are shown for ATLAS and CMS in Figure \ref{fig:dijet:highmass}c,d, respectively.

The di-jet resonance search provides access to the highest energy scales at the LHC, but it is possible that new resonant physics lies beyond the LHC energy scale.
In this case, it may still be possible to observe the effects of new physics through modifications to the angular structure of the highest-energy di-jet events, typically characterised using the variable $\chi = e^{|\Delta y|} \approx \frac{1+\cos\theta^*}{1-\cos\theta^*}$, where $\theta^*$ is the polar angle in the di-jet centre-of-mass frame.
Most Standard Model di-jet processes are t-channel and result in small values of $\theta^*$ (large values of $\chi$), while new physics is expected to be more isotropic, and thus may show up at smaller values of $\chi$.

ATLAS and CMS have conducted such searches using Run 2 data, but in both cases the published searches have only made use of a portion of the full dataset \cite{ATLAS:Dijet_angular,CMS:Dijet_angular}.
The angular searches are very similar to the aforementioned resonance searches, although for CMS the search uses standard \smallR{} jets rather than ``wide jets''.
For the angular search, the background is taken from simulated samples, and the resulting shape is compared to data distributions in a variety of signal regions defined in terms of di-jet mass window selections.
The resulting signal regions are shown for ATLAS in Figure \ref{fig:dijet:angular}a, while the highest mass region is shown for CMS in Figure \ref{fig:dijet:angular}b.
No significant deviations from the predicted shape are observed, and thus limits are set on the scale of new physics.

\vspace{-6pt}
\begin{figure}[H]

\subfigure[ ATLAS]{
 \includegraphics[width=0.5\textwidth]{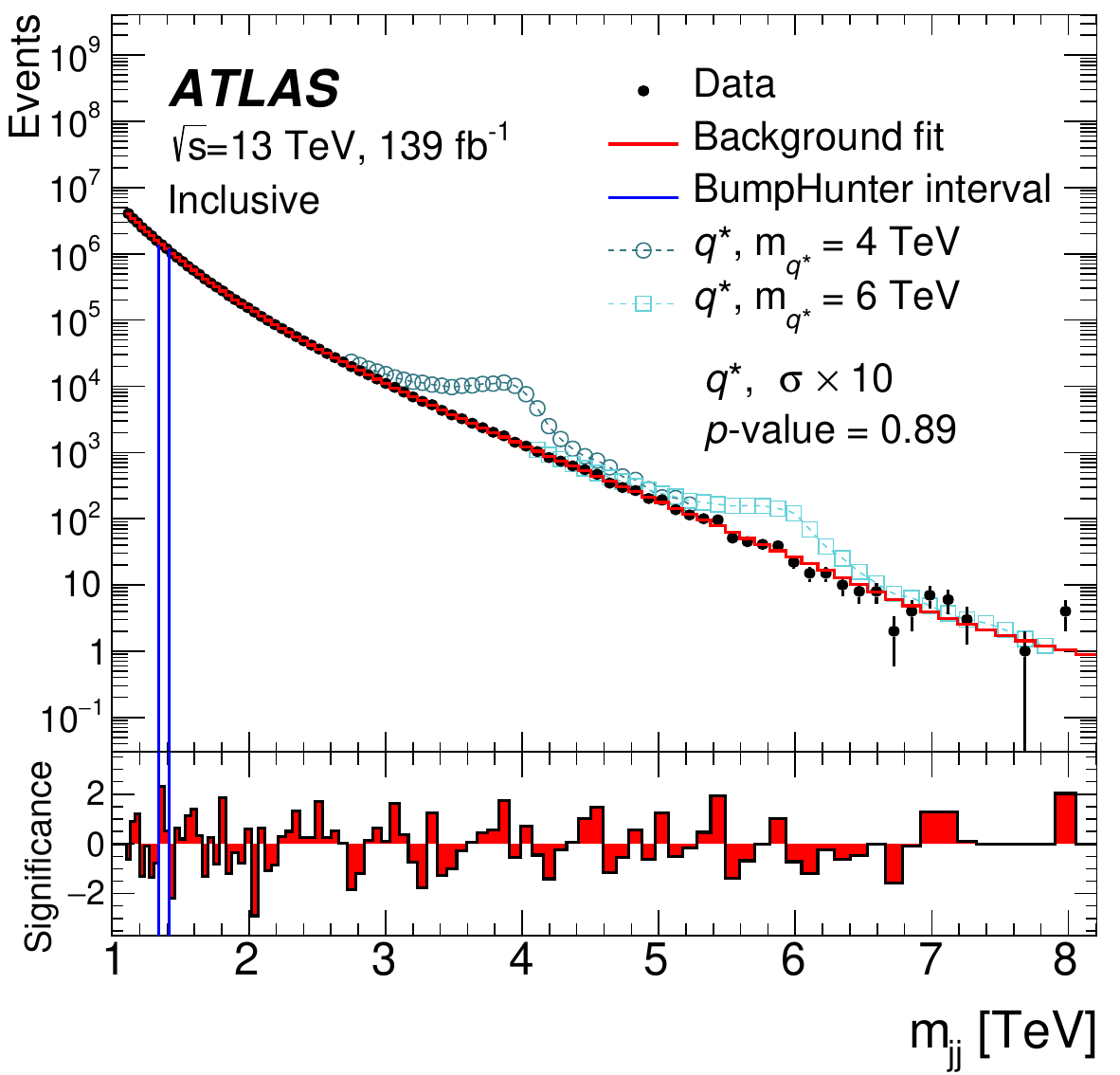}
 \label{fig:dijet:highmass:ATLAS:SR}
}
\subfigure[ CMS]{
 \includegraphics[width=0.45\textwidth]{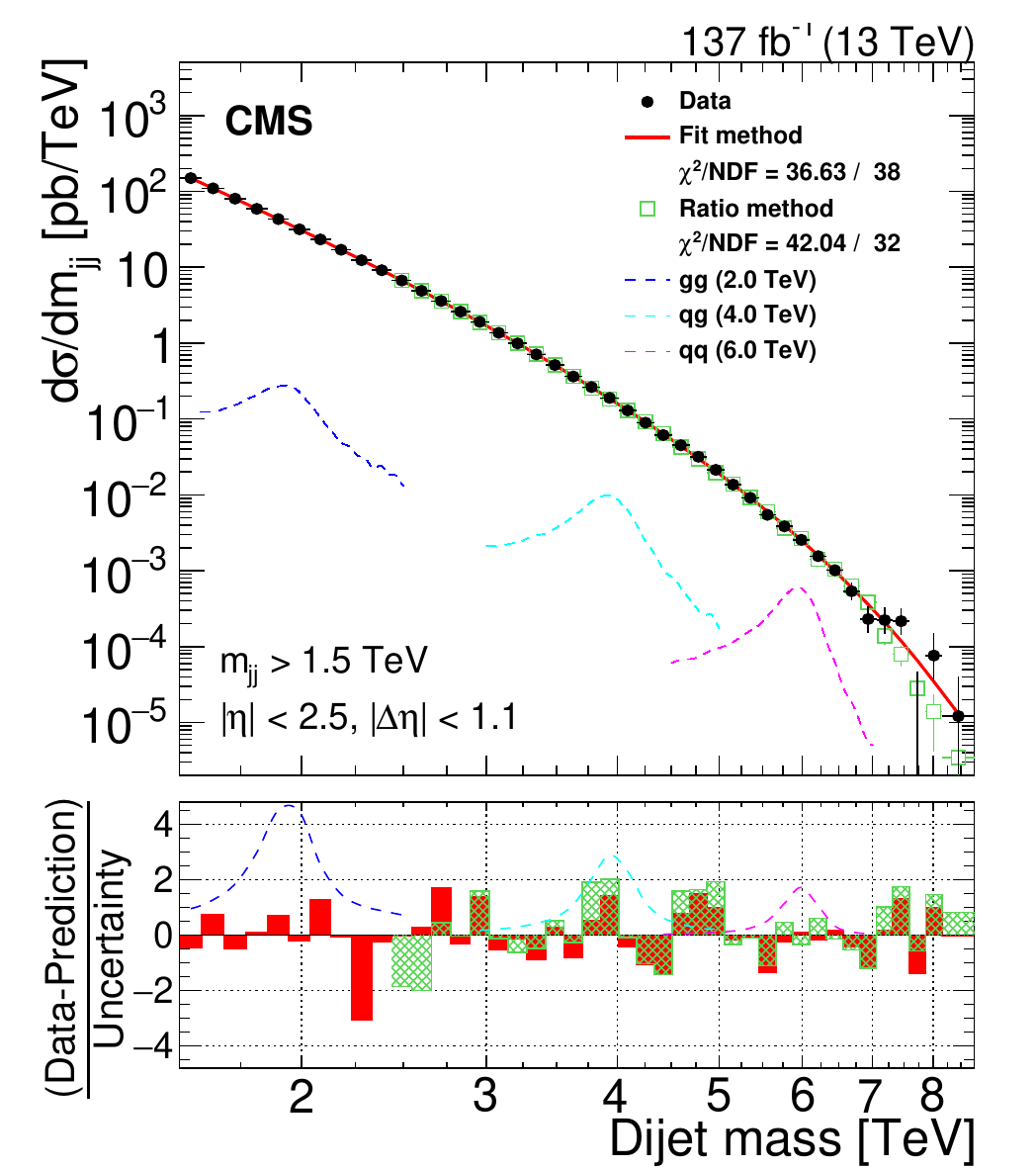}
 \label{fig:dijet:highmass:CMS:SR}
}\\
\subfigure[ ATLAS]{
 \includegraphics[width=0.48\textwidth]{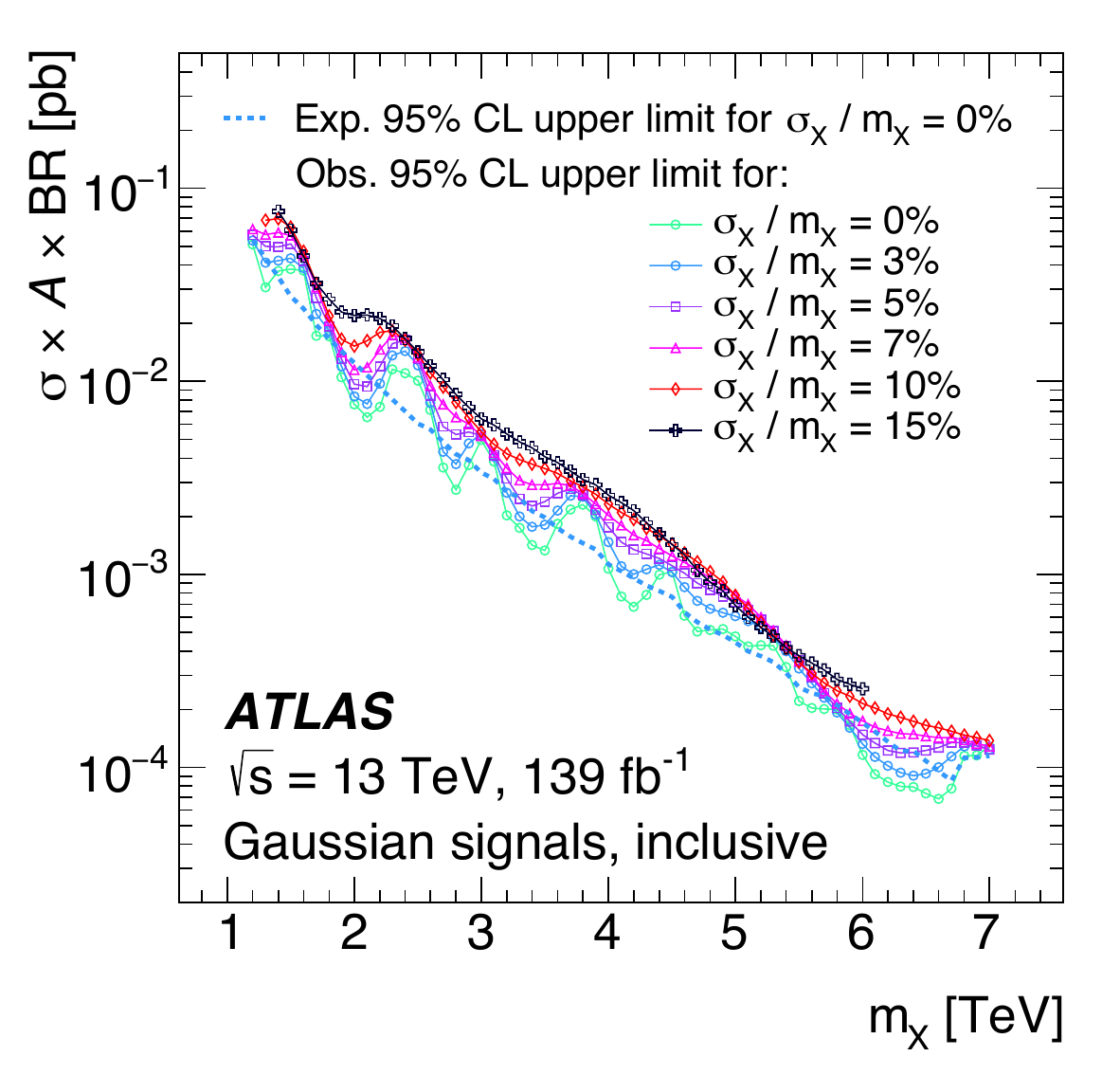}
 \label{fig:dijet:highmass:ATLAS:L}
}
\subfigure[ CMS]{
 \includegraphics[width=0.48\textwidth]{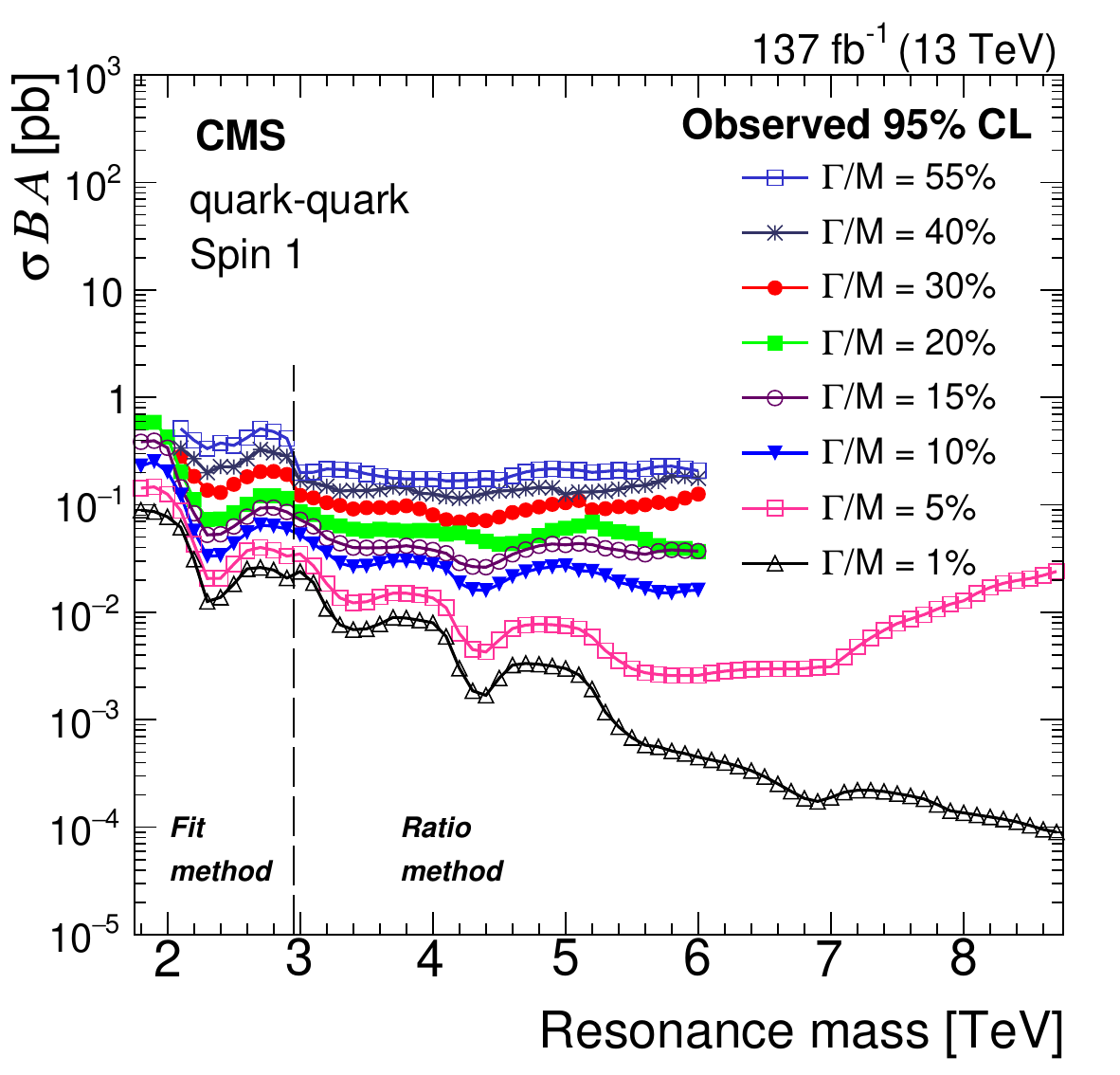}
 \label{fig:dijet:highmass:CMS:L}
}
\caption{The inclusive di-jet mass spectrum, as observed by (\textbf{a}) ATLAS \cite{ATLAS:Dijet_HighMass} and (\textbf{b}) CMS \cite{CMS:Dijet_HighMass}, using the full Run 2 dataset. No significant deviations from the background expectation are observed, and thus limits are set on a variety of signal models of interest; corresponding generic limits are shown here for results from (\textbf{c}) ATLAS and (\textbf{d}) CMS. \label{fig:dijet:highmass}}

\end{figure}

It is also possible that new physics is not uniform in its couplings to the different types of quarks.
In particular, new physics may couple preferentially to the more massive bottom and top quarks, and thus not be immediately apparent in the previously discussed inclusive di-jet searches.
Di-jet searches making use of flavour tagging algorithms can probe such possibilities by suppressing events consisting of pairs of light-quarks and/or gluons, while accepting the signal events of interest.
ATLAS has conducted both single-$b$-tagged and di-$b$-tagged di-jet searches using the full Run 2 dataset, which uses the same type of jets and the same background strategy as the inclusive search \cite{ATLAS:Dijet_HighMass}.
The resulting di-$b$-tagged di-jet mass spectrum is shown in Figure \ref{fig:dijet:highmassbb}a; no significant deviations are observed, and thus limits are set on a variety of new signal models, including Gaussian resonances as shown in Figure \ref{fig:dijet:highmassbb}b.
CMS has also conducted such searches during Run 1 \cite{CMS:Dijet_HighMassbb}, but does not yet have such a publication using Run 2 data.

\begin{figure}[H]

\subfigure[ ATLAS]{
 \includegraphics[width=0.5\textwidth]{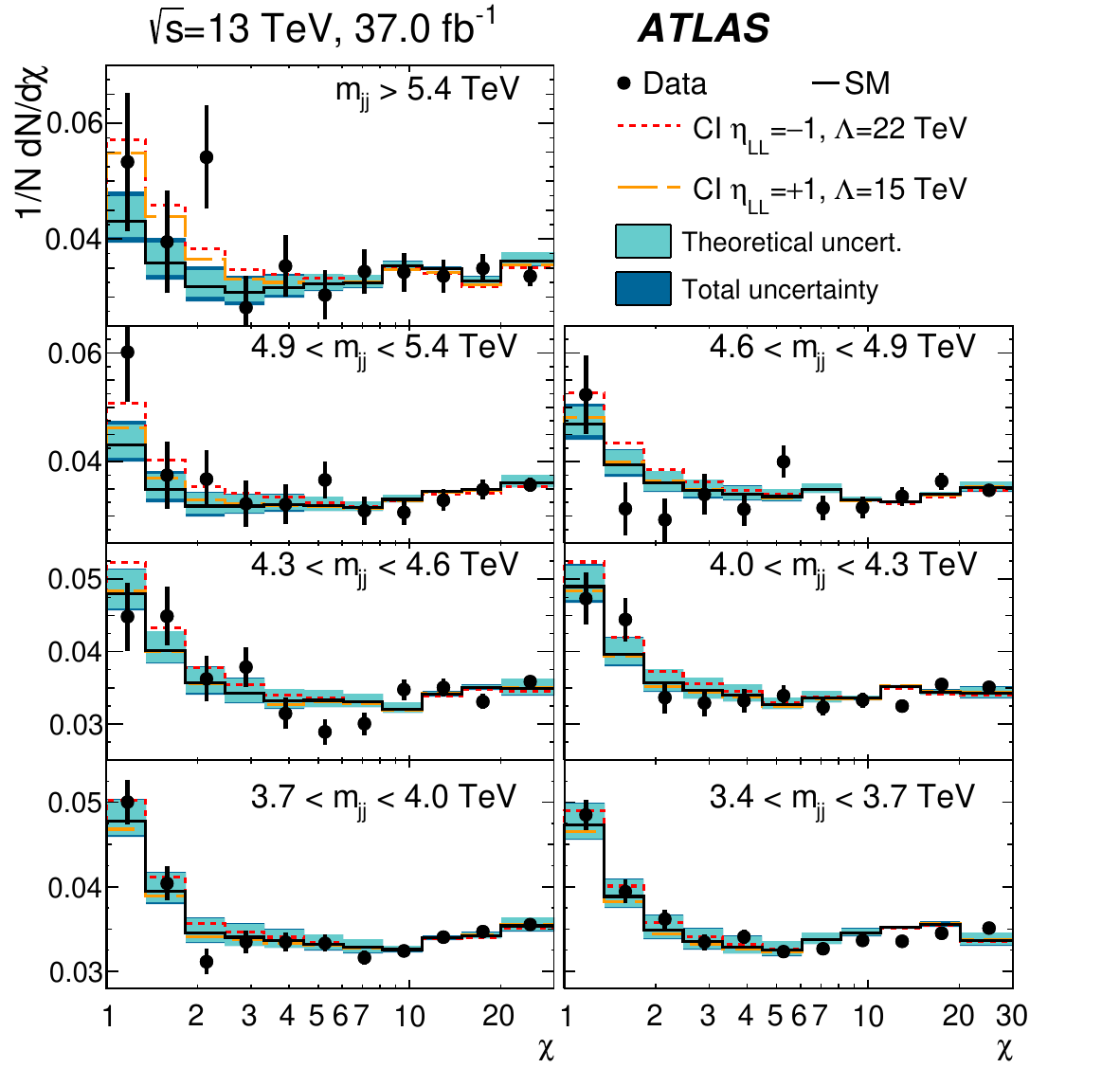}
 \label{fig:dijet:angular:ATLAS}
}
\subfigure[ CMS]{
 \includegraphics[width=0.4\textwidth]{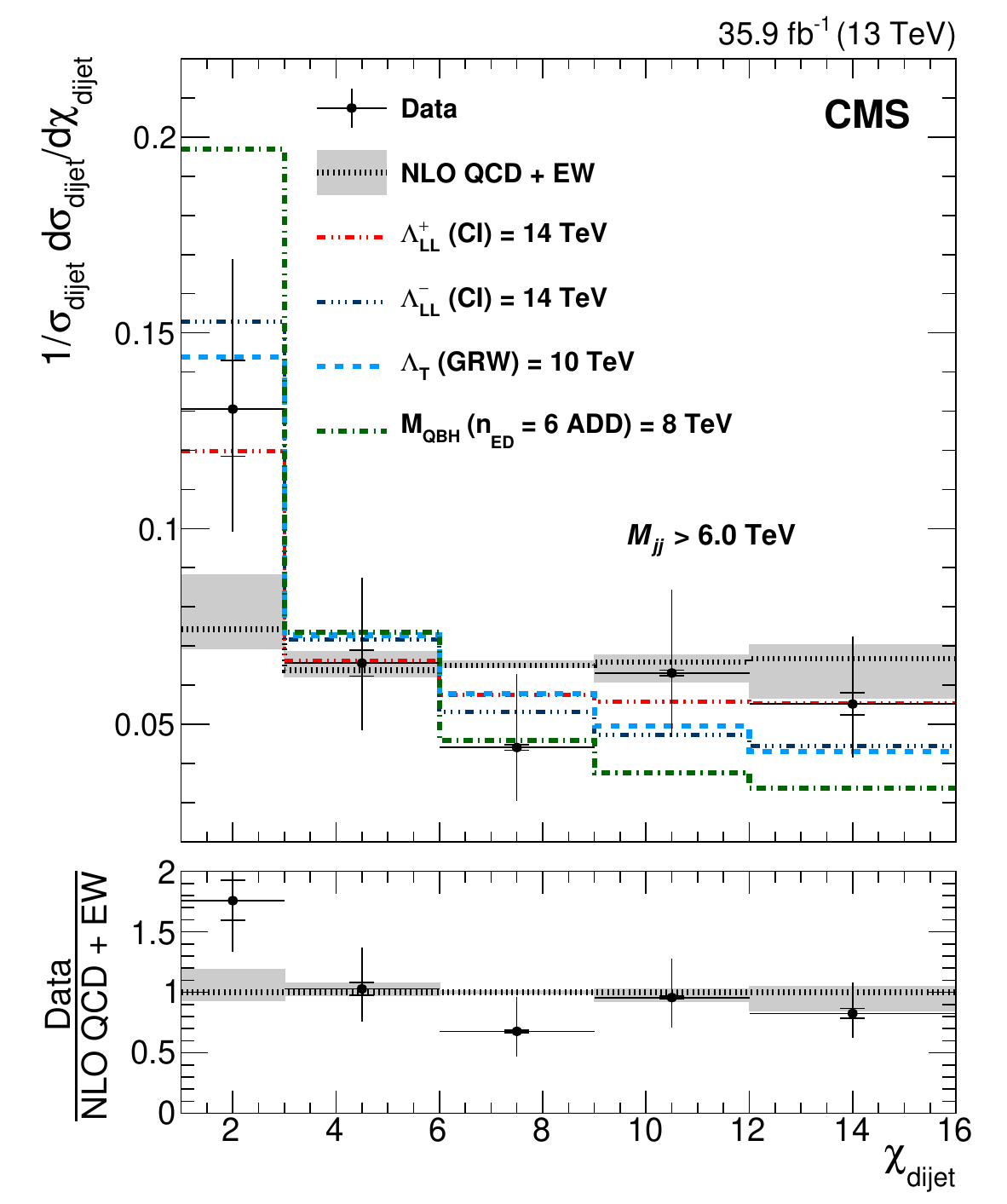}
 \label{fig:dijet:angular:CMS}
}
\caption{(\textbf{a}) The distribution of the di-jet angular variable $\chi$, as observed by (\textbf{a}) ATLAS for several different invariant mass ranges \cite{ATLAS:Dijet_angular}, and (\textbf{b}) CMS for a specific invariant mass range from 5.4\TeV{} to 6.0\TeV{} \cite{CMS:Dijet_angular}, both using a partial Run 2 dataset. No significant deviations from the background expectation are observed, and thus limits are set on models of new physics (not shown here). \label{fig:dijet:angular}}
\end{figure}

\vspace{-15pt}

\begin{figure}[H]

\subfigure[]{
 \includegraphics[width=0.45\textwidth]{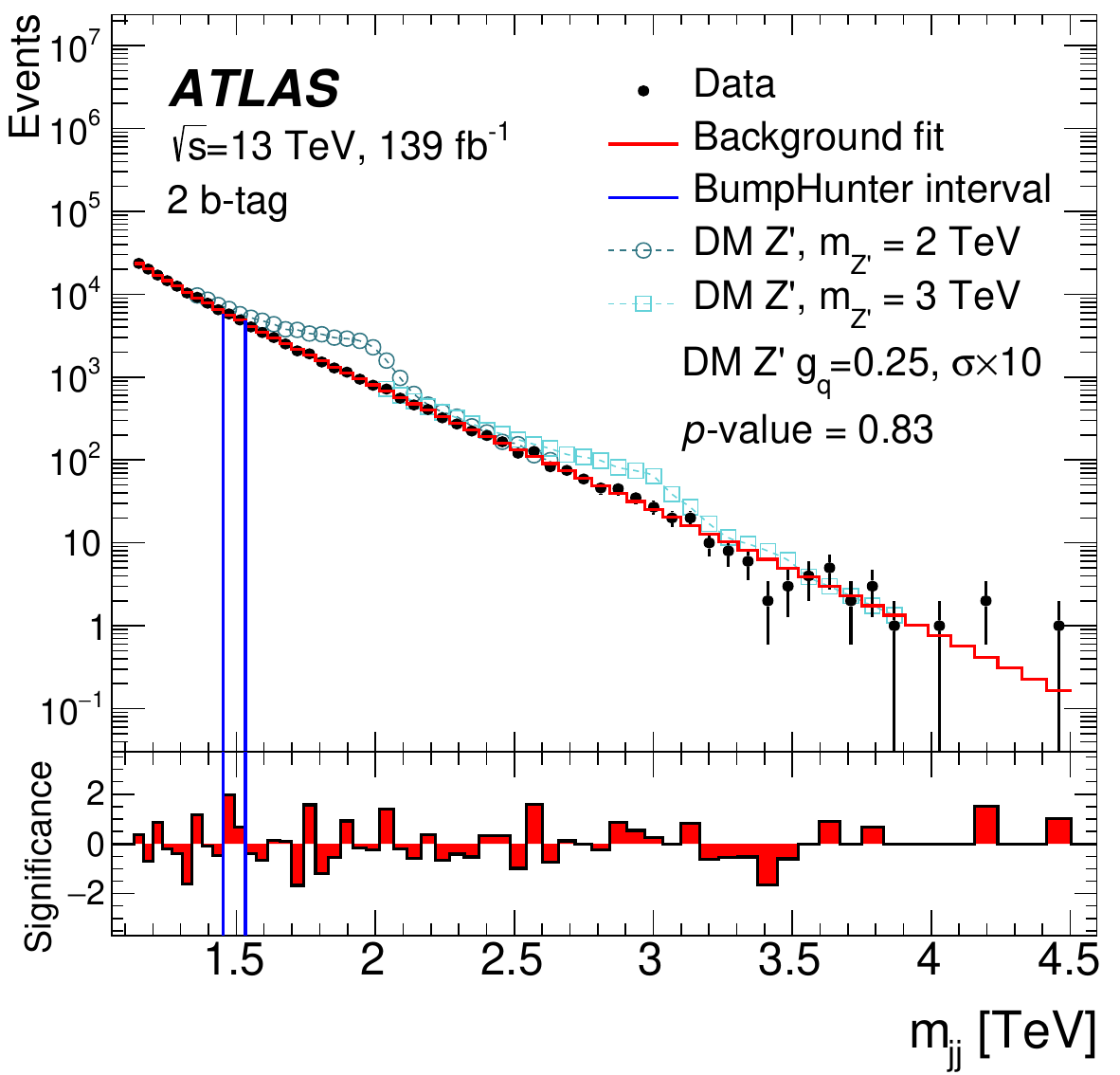}
 \label{fig:dijet:highmassbb:SR}
}
\subfigure[]{
 \includegraphics[width=0.46\textwidth]{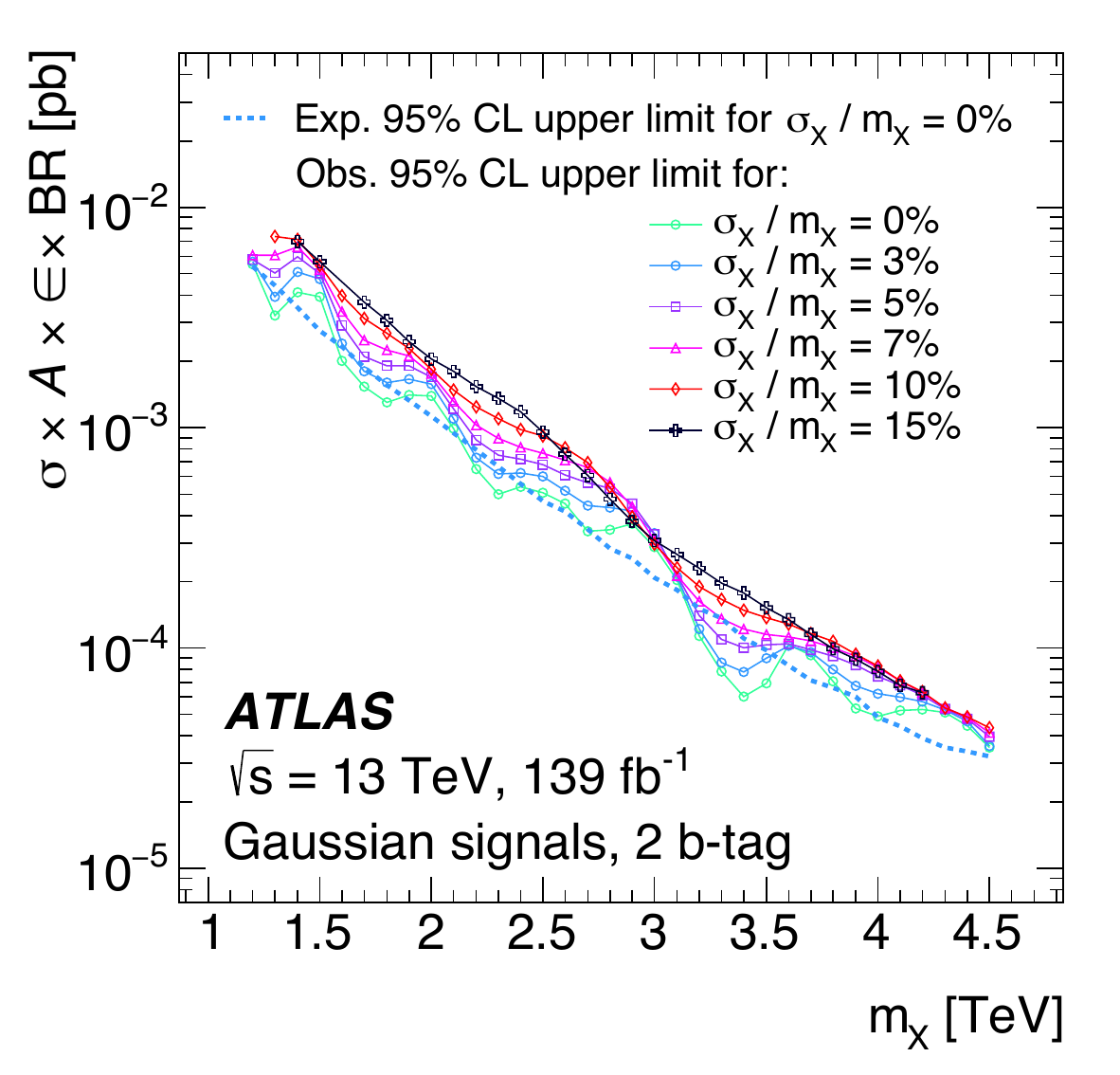}
 \label{fig:dijet:highmassbb:limits}
}
\caption{(\textbf{a}) The di-b-tagged di-jet mass spectrum, as observed by ATLAS \cite{ATLAS:Dijet_HighMass}, using the full Run 2 dataset. No significant deviations from the background expectation are observed, and thus limits are set on a variety of signal models of interest; (\textbf{b}) the corresponding Gaussian limits are shown here. \label{fig:dijet:highmassbb}}
\end{figure}

Searches for di-top-quark resonances require a bit more care, as the top quarks decay immediately, and the resulting decays produce a variety of different final states.
Focusing on the fully hadronic final state, as is the topic of this review, results in a final state with six partons: both of the two top quarks decay to a $b$ quark and a $W$ boson, and each of the $W$ bosons subsequently decays to a pair of quarks.
At the energy scale where fully hadronic di-top-quark resonance searches are conducted at the LHC, the top quarks are boosted, and thus their decays are collimated: the entire hadronic decay of a top quark is thus reconstructed as a single \largeR{} jet, forming the basis of di-top-quark resonance searches as di-\largeR{}-jet searches.

Non-top-quark and gluon jets can also be reconstructed as \largeR{} jets, forming a large background to the search for $t\bar{t}$ resonances.
Such backgrounds must be suppressed by tagging the \largeR{} jets, thereby accepting jets originating from hadronic top-quark decays and suppressing those originating from other processes.
These \largeR{} jet taggers are supplemented with $b$-tagging information to further suppress the background from light-quarks and gluons.
The events that pass these criteria are a mixture of Standard Model $t\bar{t}$ events and mis-tagged not-top-quark events, where the latter category is strongly suppressed in case of the use of a two $b$-tag requirement.

Following this set of complex taggers, the resulting mixture of background events must be evaluated in a data-driven way.
ATLAS and CMS both make use of simulated events for the $t\bar{t}$ background, which is generally well modelled.
In contrast, the surviving not-top-quark background is determined using a series of control regions enhanced in not-top-quark events, where these control regions are used to determine the corresponding selection efficiency and thus the expected contribution to the invariant mass distribution.

ATLAS has conducted the fully hadronic $t\bar{t}$ resonance search using the full Run 2 dataset \cite{ATLAS:Dijet_ttbar}, and the resulting di-$b$-tagged mass distribution is shown in Figure \ref{fig:dijet:ttbar}a.
CMS has also conducted such a search, but using a partial Run 2 dataset \cite{CMS:Dijet_ttbar}, where a similar plot is shown in Figure \ref{fig:dijet:ttbar}b.
In both cases, no significant deviations are found beyond the background expectation, and thus limits are set on the production of axial-vector $Z^\prime$ resonances as shown in Figure \ref{fig:dijet:ttbar}c,d for ATLAS and CMS, respectively.
The ATLAS limit plot is shown only for the fully hadronic $t\bar{t}$ resonance search, while the CMS limit plot combines the fully hadronic result with the semi-leptonic and fully-leptonic decay modes of the $t\bar{t}$ system.

\vspace{-6pt}
\begin{figure}[H]

\subfigure[ ATLAS]{
 \includegraphics[width=0.47\textwidth]{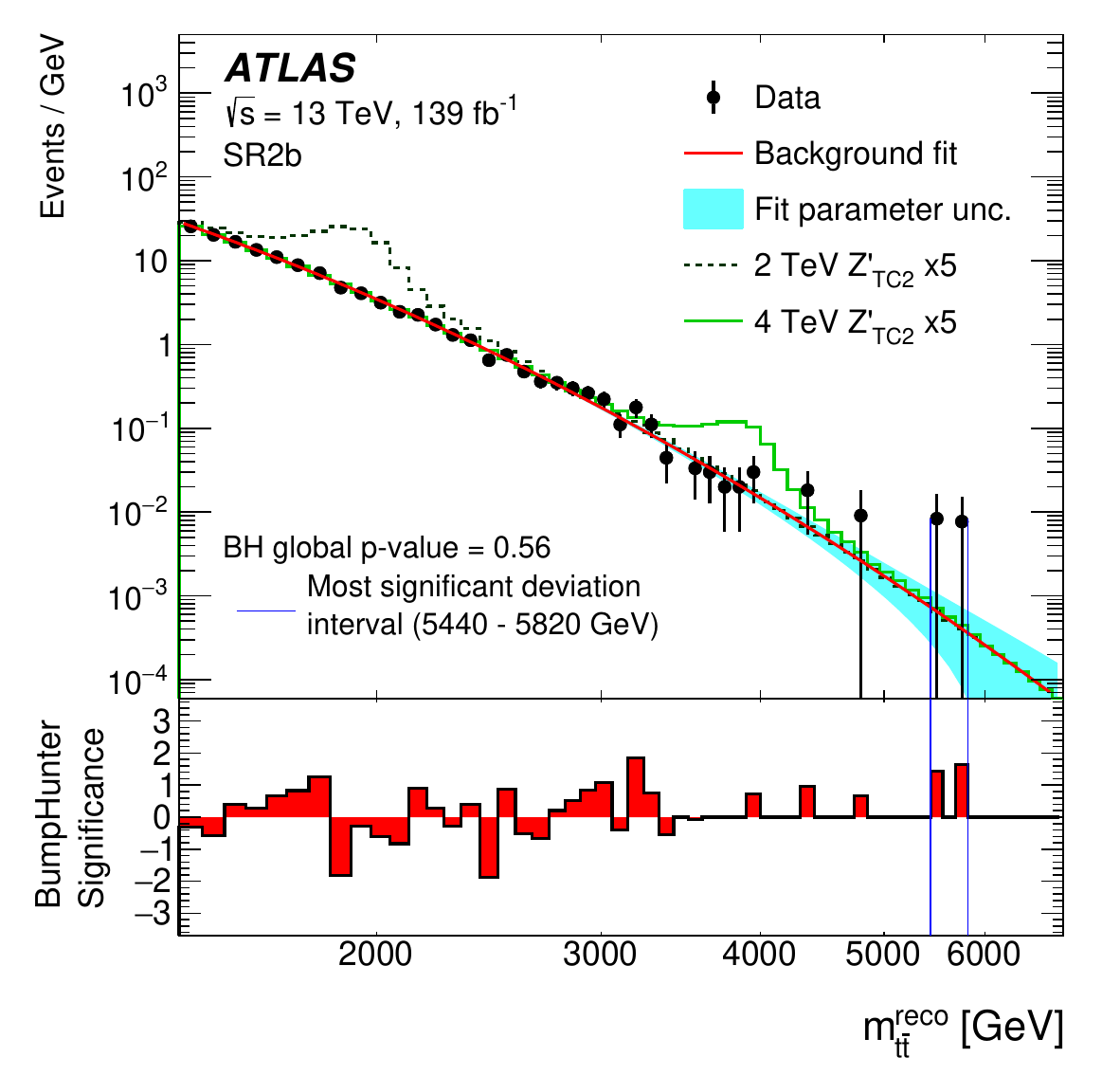}
 \label{fig:dijet:ttbar:ATLAS:SR}
}
\subfigure[ CMS]{
 \includegraphics[width=0.48\textwidth]{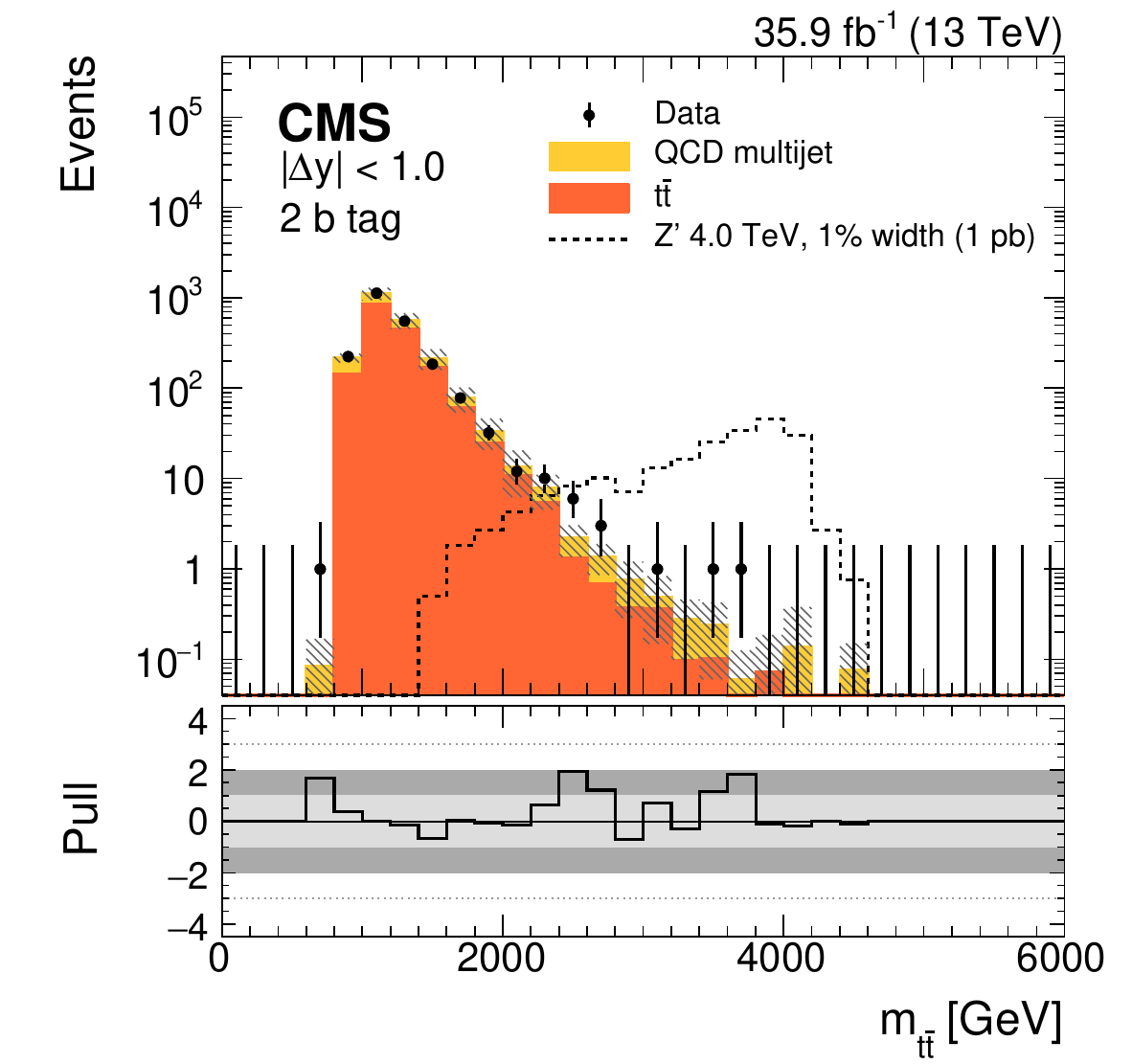}
 \label{fig:dijet:ttbar:CMS:SR}
}\\
\subfigure[ ATLAS]{
 \includegraphics[width=0.5\textwidth]{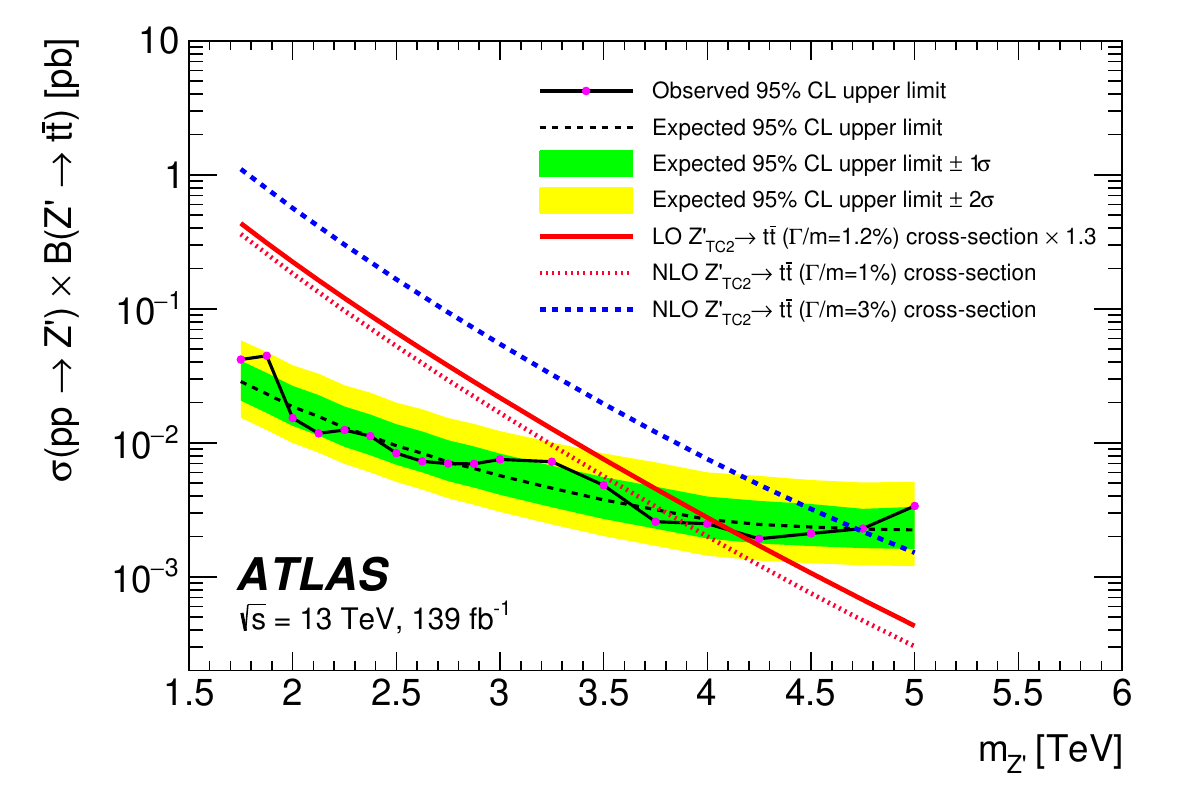}
 \label{fig:dijet:ttbar:ATLAS:L}
}
\subfigure[ CMS]{
 \includegraphics[width=0.44\textwidth]{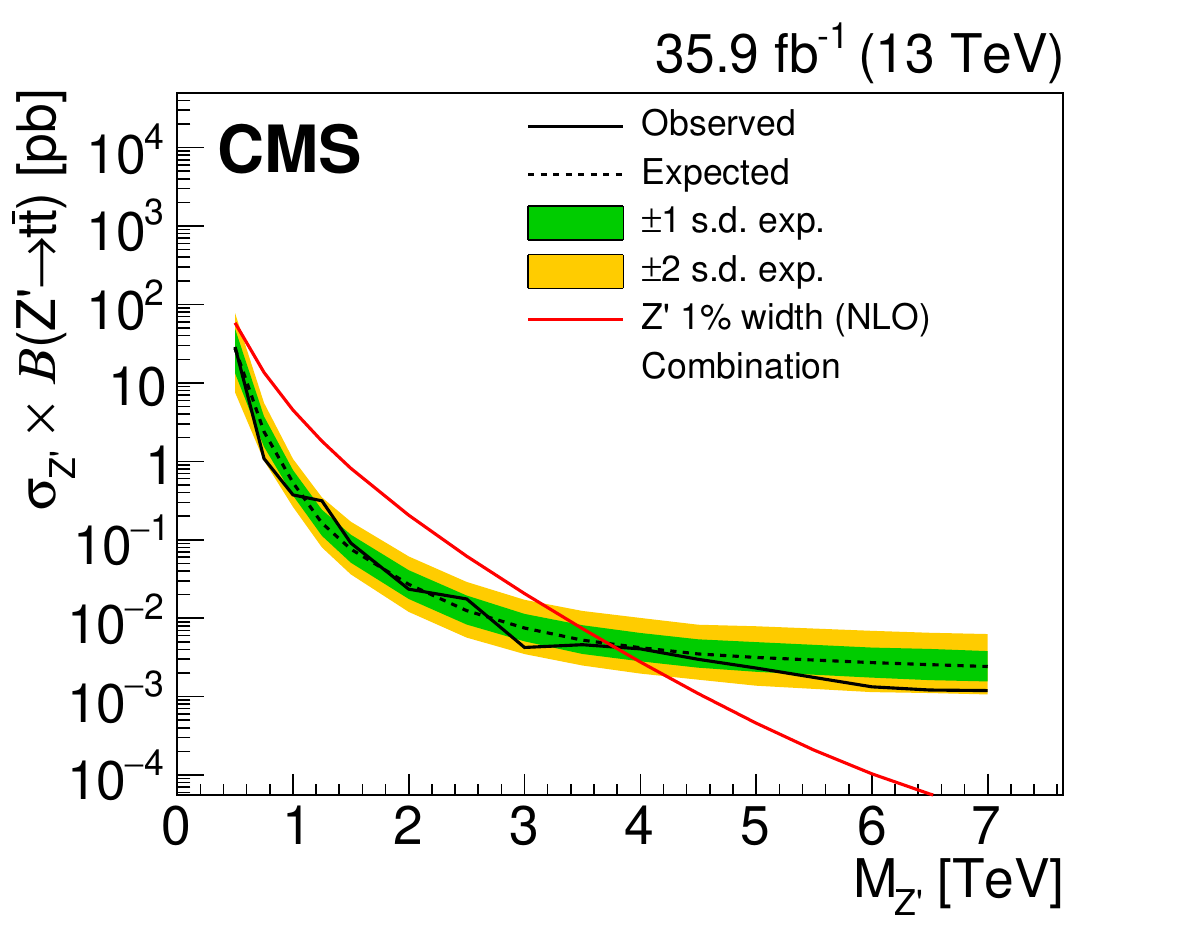}
 \label{fig:dijet:ttbar:CMS:L}
}
\caption{(\textbf{a}) The $t\bar{t}$ mass spectrum for fully-hadronic $t\bar{t}$ decays, as observed by (\textbf{a}) ATLAS using the full Run 2 dataset \cite{ATLAS:Dijet_ttbar} and (\textbf{b}) CMS using a partial Run 2 dataset \cite{CMS:Dijet_ttbar}. No significant deviations from the background expectation are observed, and thus limits are set on $Z^\prime$ models with varying widths. The corresponding limits are shown for (\textbf{c}) ATLAS using only the fully-hadronic channel, and \linebreak (\textbf{d}) CMS using a combination of different $t\bar{t}$ decay channels. \label{fig:dijet:ttbar}}
\end{figure}

\subsection{Trigger-Based Di-Jet Searches}
\label{sec:dijet:trigger}
% CMS DataScouting, ATLAS TLA

High-mass di-jet searches start at roughly 1\TeV{}, which is due to the hadronic trigger thresholds used by both ATLAS and CMS.
In order to access the lower-mass regime, it is necessary to find a way around this trigger constraint.
This could be done by using prescaled triggers, which record only a fraction of the events that would have otherwise passed the trigger selection, but such triggers are typically afforded minimal rate; the effective luminosity of such an approach is therefore too small to provide sensitivity to new physics beyond what has previously been studied.

There is, however, a way around the trigger constraint: the analysis can be performed in the trigger.
In practice, it is a bit more complex, but this idea is the foundation of the approach referred to as a Trigger-Level Analysis (TLA) by ATLAS, or Data Scouting (DS) by CMS.
Such searches exploit the fact that the trigger is bandwidth-limited, and bandwidth is the product of the event size and the trigger rate, meaning that a very small event size can enable the recording of a very-high-rate process.
As such, if all of the information needed to conduct the analysis can be calculated within the trigger system, that very small amount of information can be written out alone, while the rest of the much larger event can be discarded.
Such an approach is only useful if the precision of the information available in the trigger is sufficient for the analysis objectives.
For this reason, TLA/DS works well with jets reconstructed by the experimental software-based triggers, but not jets from the hardware-based triggers; the point at which the hardware triggers are fully efficient therefore defines a lower boundary that this approach can reasonably access.

The TLA/DS approach has allowed both ATLAS \cite{ATLAS:Dijet_TLA} and CMS \cite{CMS:Dijet_TLA} to probe much lower mass di-jet resonances with the full statistical power of the LHC dataset, albeit with some small additional uncertainties or other performance degradations related to how the trigger-reconstructed jets are typically less precisely known than those available for offline data analysis.
For example, the ATLAS di-jet TLA lacks the tracking information needed to apply part of the calibration sequence that mitigates quark-vs-gluon differences, while the CMS DS di-jet analysis uses calorimeter inputs to jet reconstruction instead of particle flow inputs.
However, these are small prices to pay for the ability to extend the search for di-jet resonances to lower masses with unprecedented statistical precision.
The resulting invariant mass spectra are fit using functional forms, similar to the high-mass di-jet resonance search, as shown in Figure \ref{fig:dijet:TLA}a for ATLAS and Figure \ref{fig:dijet:TLA}b for CMS, both of which use a partial Run 2 dataset.
No significant deviations from the background expectation are observed, and thus the limits on axial-vector $Z^\prime$ production provided by the high-mass di-jet resonance search are extended down to masses of roughly 500\GeV{}, as shown in Figures \ref{fig:dijet:TLA}c,d.

\subsection{Di-Jet Searches in Association with Other Objects}
\label{sec:dijet:ISR}
% CMS:
% - Resolved ISRjet
% - Boosted ISRjet
% - Boosted ISRphoton
%
% ATLAS:
% - Resolved+bb ISRphoton
% - Resolved ISRlepton
% - Boosted ISRjet+photon
% - Boostedbb ISRjet [CONF --> not included]

After reaching the hardware-based trigger constraints, other methods must be found to continue to probe lower-mass di-jet resonances.
One way to do this is to trigger the event based on other activity that happens to be present, but which does not relate with or otherwise impact the process of interest.
This is the case of searches for di-jet resonances in association with initial state radiation, whether that radiation is another jet, a photon, a lepton (from an ISR $W$ boson), or otherwise.
This approach can provide access to much lower di-jet masses, but it comes at the cost of the requiring both the ISR to occur and the ISR object to be energetic enough to pass the associated trigger.
ISR-based approaches thus have less statistical power than conventional approaches, but they can access regimes inaccessible to either standard or trigger-level searches, and thus they are an integral part of the search for new physics.

Once the event has been triggered, the next step is to differentiate between the di-jet system of interest and the ISR object.
This may be quite straightforward, such as is the case of a muon clearly being distinct from the di-jet system, or it may be more complex, especially when the ISR object is itself another jet.
After identifying the di-jet system, the analysis becomes more similar to one of the previously discussed di-jet searches, where a background expectation is defined (typically from a functional form fit to the data) and a search for narrow resonances over that background is performed.

\vspace{-12pt}
\begin{figure}[H]

\subfigure[ ATLAS]{
 \includegraphics[width=0.54\textwidth]{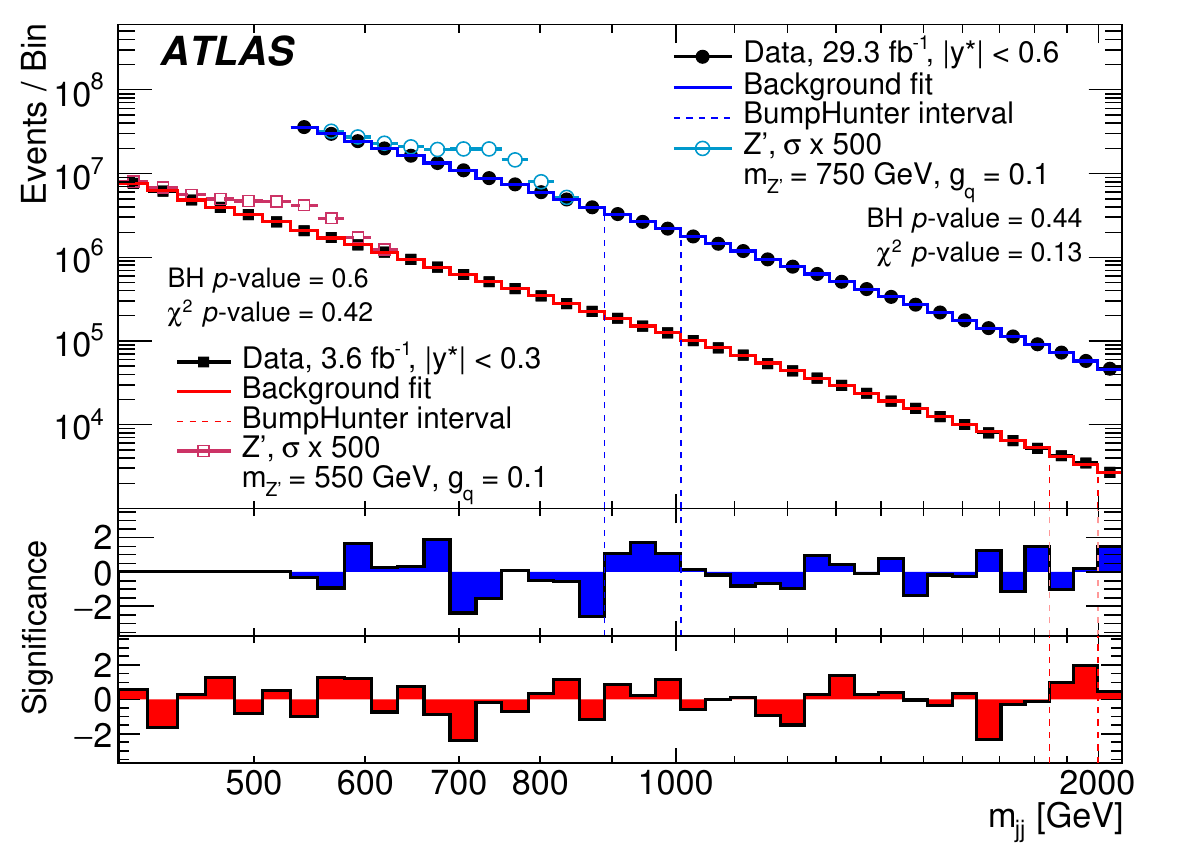}
 \label{fig:dijet:TLA:ATLAS:SR}
}
\subfigure[ CMS]{
 \includegraphics[width=0.36\textwidth]{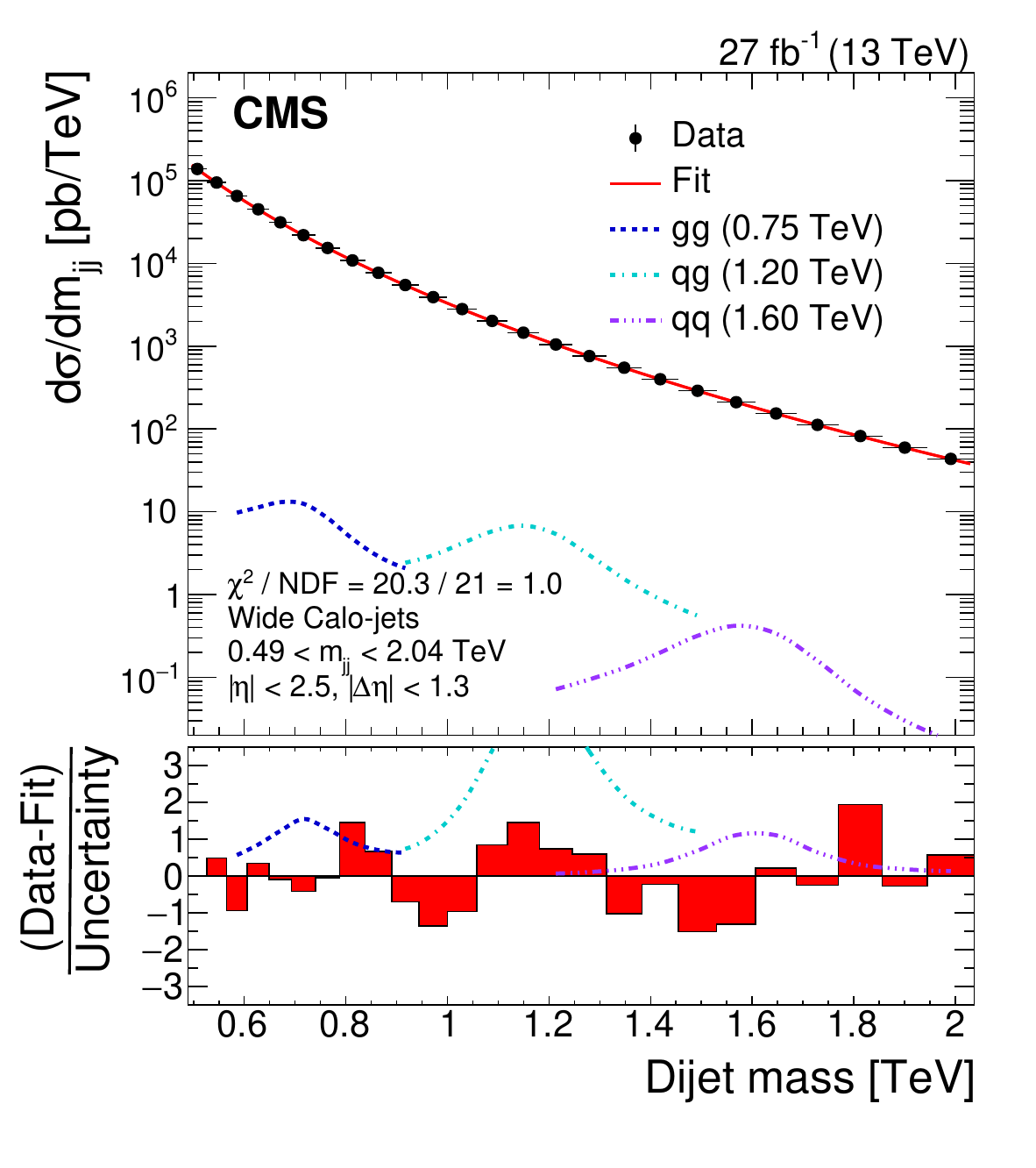}
 \label{fig:dijet:TLA:CMS:SR}
}\\
\subfigure[ ATLAS]{
 \includegraphics[width=0.53\textwidth]{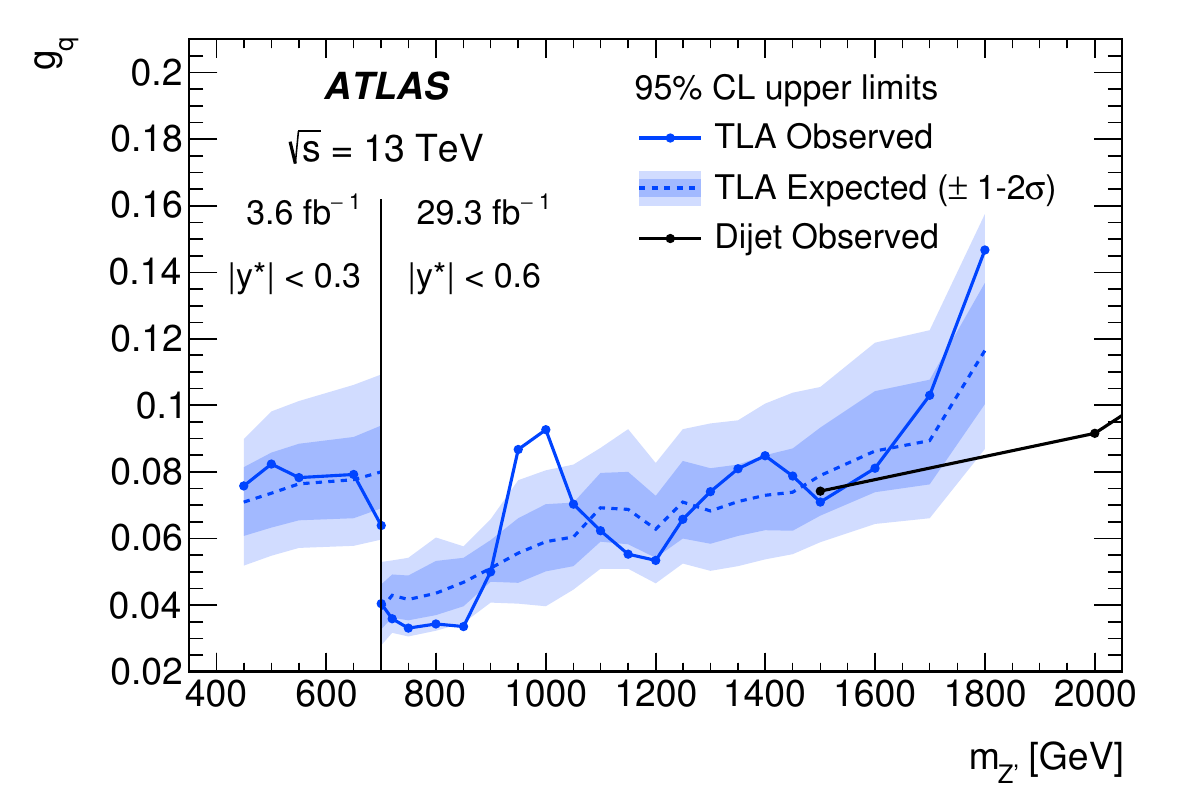}
 \label{fig:dijet:TLA:ATLAS:L}
}
\subfigure[ CMS]{
 \includegraphics[width=0.41\textwidth]{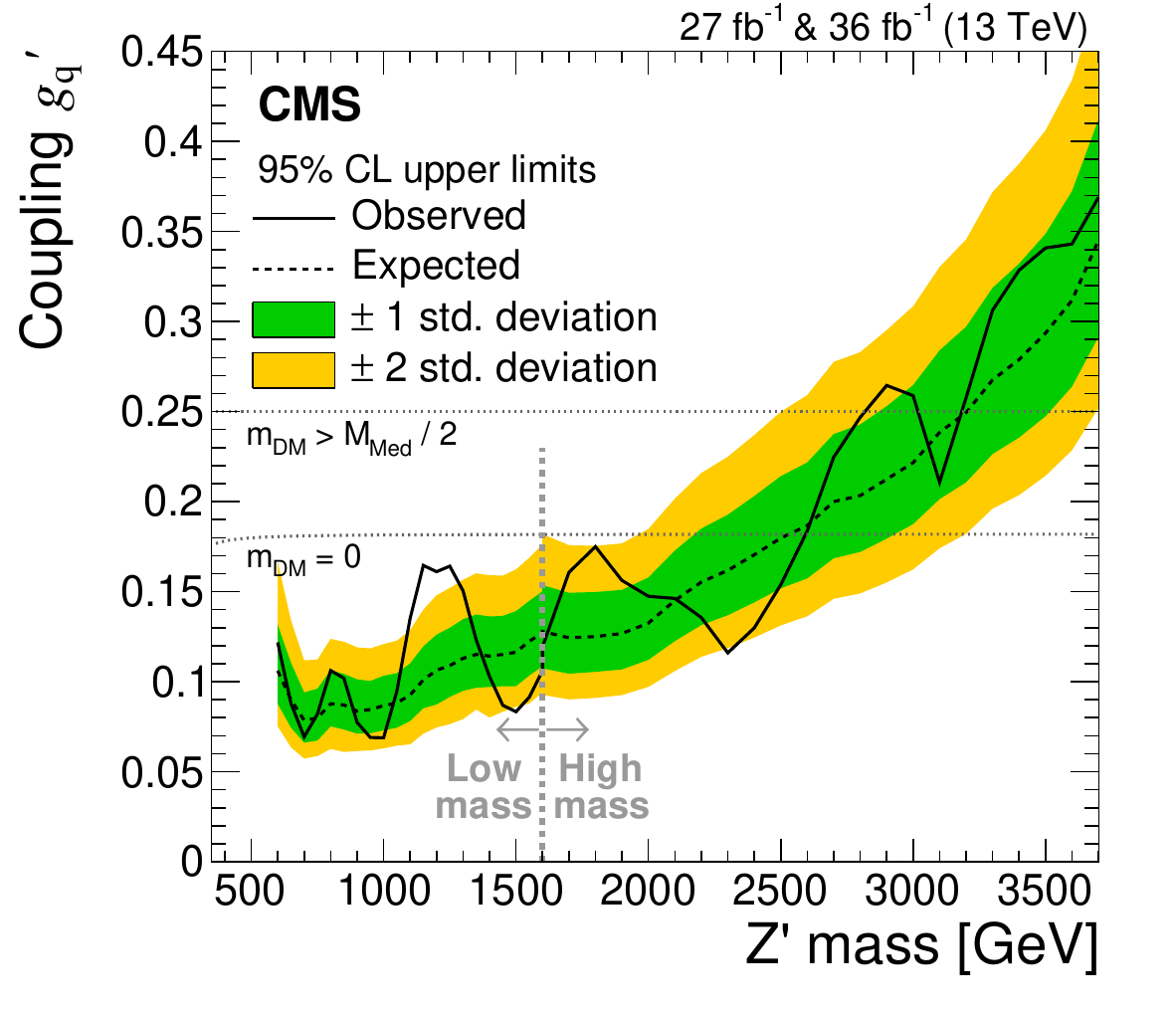}
 \label{fig:dijet:TLA:CMS:L}
}
\caption{The inclusive di-jet mass spectrum using trigger-level jets, as observed by (\textbf{a}) ATLAS \cite{ATLAS:Dijet_TLA} and (\textbf{b}) CMS \cite{CMS:Dijet_TLA}, both using a partial Run 2 dataset. No significant deviations from the background expectation are observed, and thus limits are set on a variety of signal models of interest; corresponding limits for the production of an axial-vector $Z^\prime$ are shown here for results from (\textbf{c}) ATLAS and \linebreak (\textbf{d}) CMS. \label{fig:dijet:TLA}}
\end{figure}

The strategy of searching for di-jet resonances in association with other objects is quite new at the LHC, with the first preliminary result appearing in 2016 \cite{ATLAS:FirstISRR}.
As a result, ATLAS and CMS have not yet independently considered each of the different possible types of associated objects.
However, when ATLAS and CMS searches are considered all together, they have done a good job of covering different possible associated objects.

Photons are a promising ISR object to use, as they are typically easily distinguishable from jets, and thus there is little confusion about which part of the event is the di-jet system of interest.
In addition, ATLAS and CMS photon triggers have much lower thresholds than jet triggers, as per Table \ref{tab:motivation:triggers}.
ATLAS has therefore conducted a search for \smallR{} \topocluster{}-based di-jet resonances in association with photons using a partial Run 2 dataset \cite{ATLAS:Dijet_ISRy_resolved}, which uses both a single-photon trigger for the lowest di-jet masses and a combined photon+di-jet trigger for higher di-jet masses, resulting in two separate search regions.
While both triggers record nearly the same amount of luminosity, the photon+di-jet trigger provides more statistics where it is active,
due to its use of a lower photon energy cut; this demonstrates the statistical impact of requiring an ISR object of a given energy, which is independent of the system under study.
The same search moreover considers both the inclusive di-jet spectrum, as well as the di-$b$-tagged di-jet spectrum, which are then fit using functional forms as done for traditional di-jet searches.
The resulting di-jet mass distributions for the inclusive and di-$b$-tagged selections can be found in Figure \ref{fig:dijet:ISRy_resolved}a,b, respectively.
No significant deviations are observed from the background fit expectations, and thus limits are set on the quark coupling to an axial-vector $Z^\prime$ under the assumption of a flavour-universal quark coupling, as shown in Figure \ref{fig:dijet:ISRy_resolved}c,d.
Even under this flavour-universal assumption, the di-$b$-tagged selection provides improved sensitivity to the $Z^\prime$ model under study, as the selection reduces the background contributions of the di-gluon and quark+gluon production processes, which are large at low energies.

\vspace{-6pt}
\begin{figure}[H]

\subfigure[ Inclusive]{
 \includegraphics[width=0.48\textwidth]{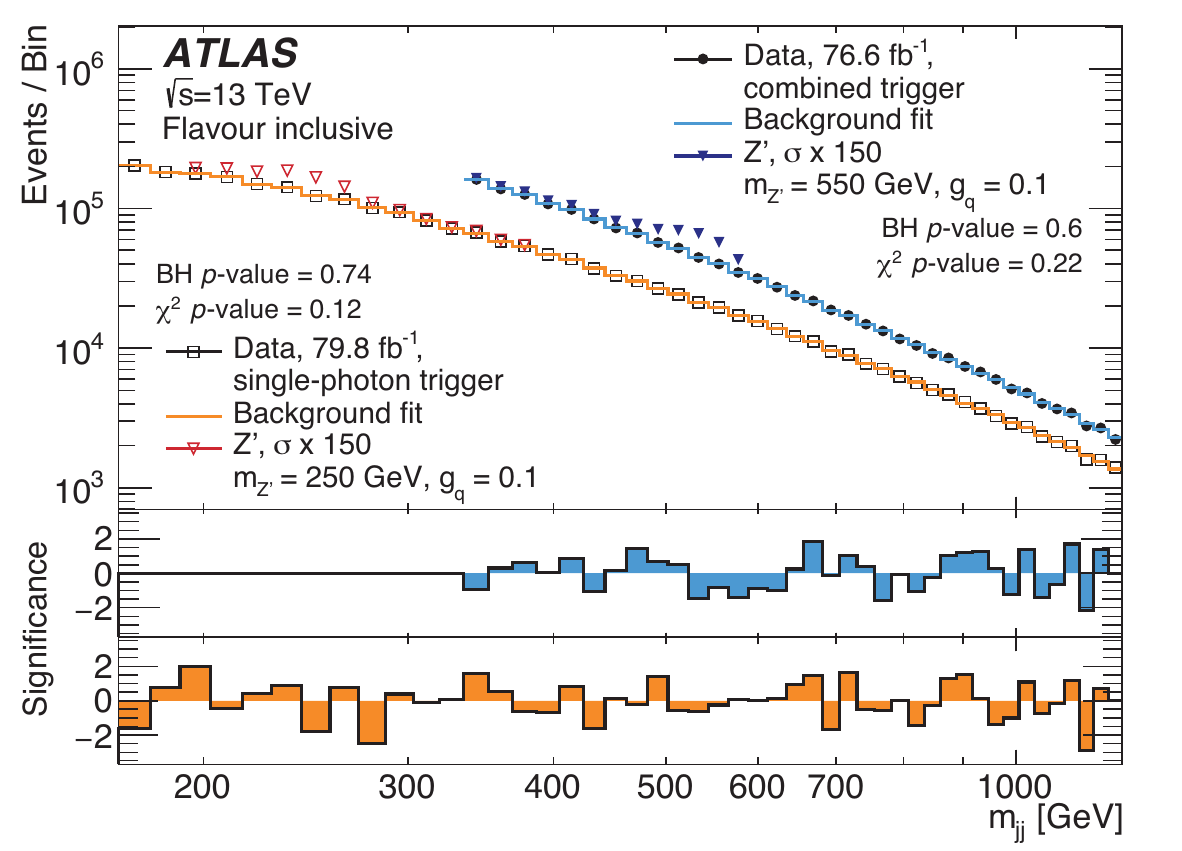}
 \label{fig:dijet:ISRyR:inc:SR}
}
\subfigure[ di-$b$-tagged]{
 \includegraphics[width=0.48\textwidth]{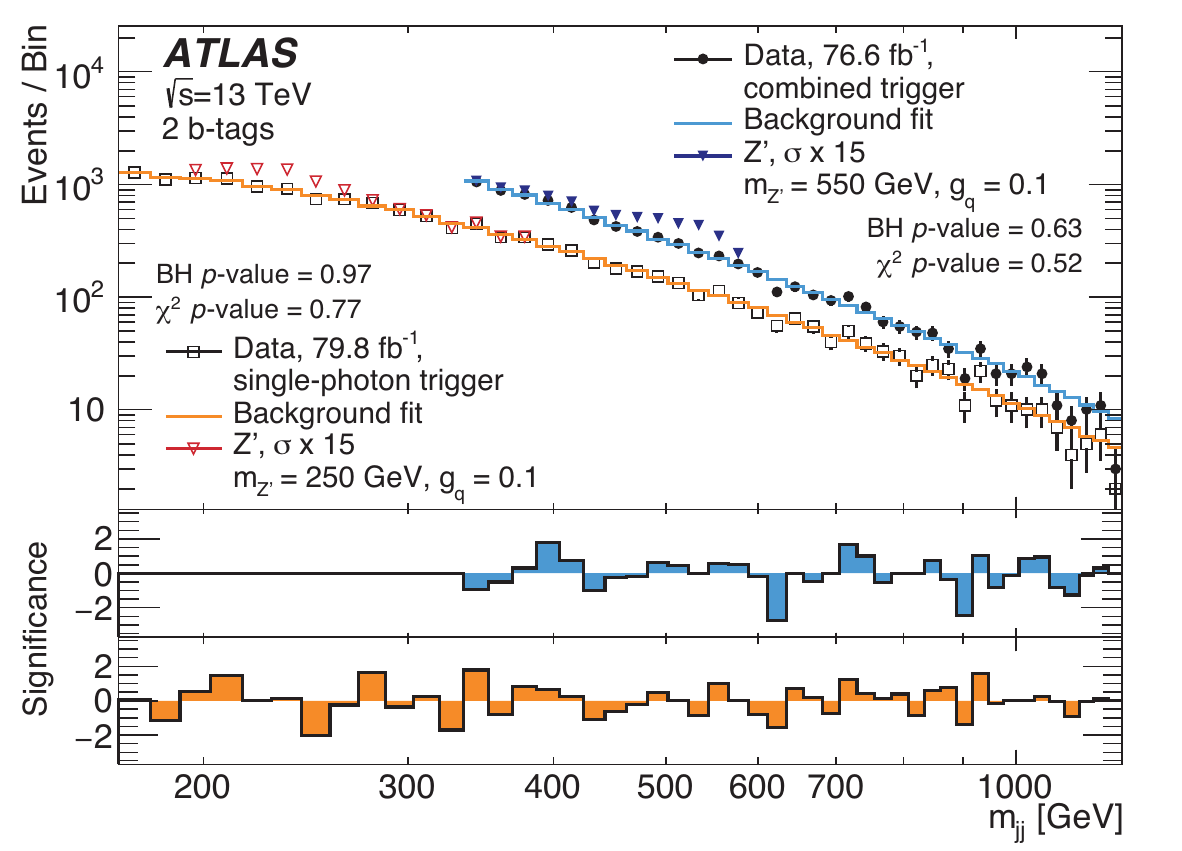}
 \label{fig:dijet:ISRyR:bb:SR}
}\\
\subfigure[ Inclusive]{
 \includegraphics[width=0.48\textwidth]{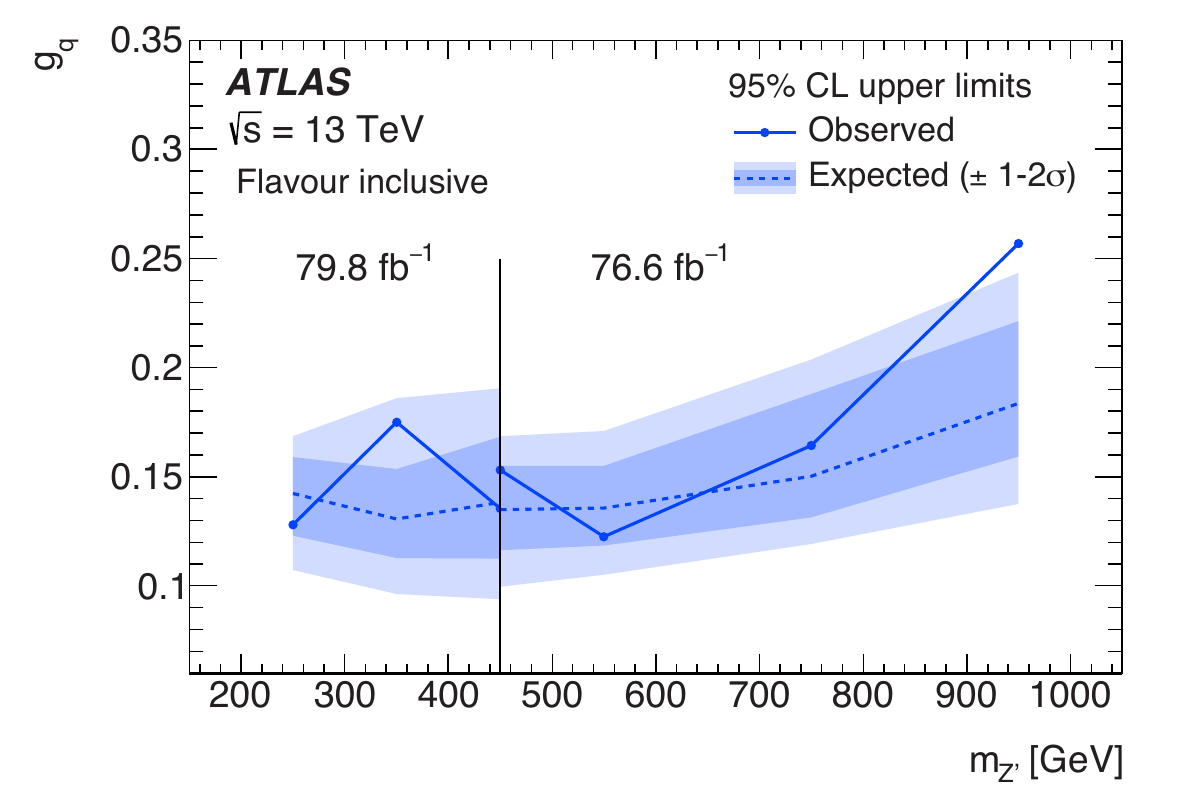}
 \label{fig:dijet:ISRyR:inc:L}
}
\subfigure[ di-$b$-tagged]{
 \includegraphics[width=0.48\textwidth]{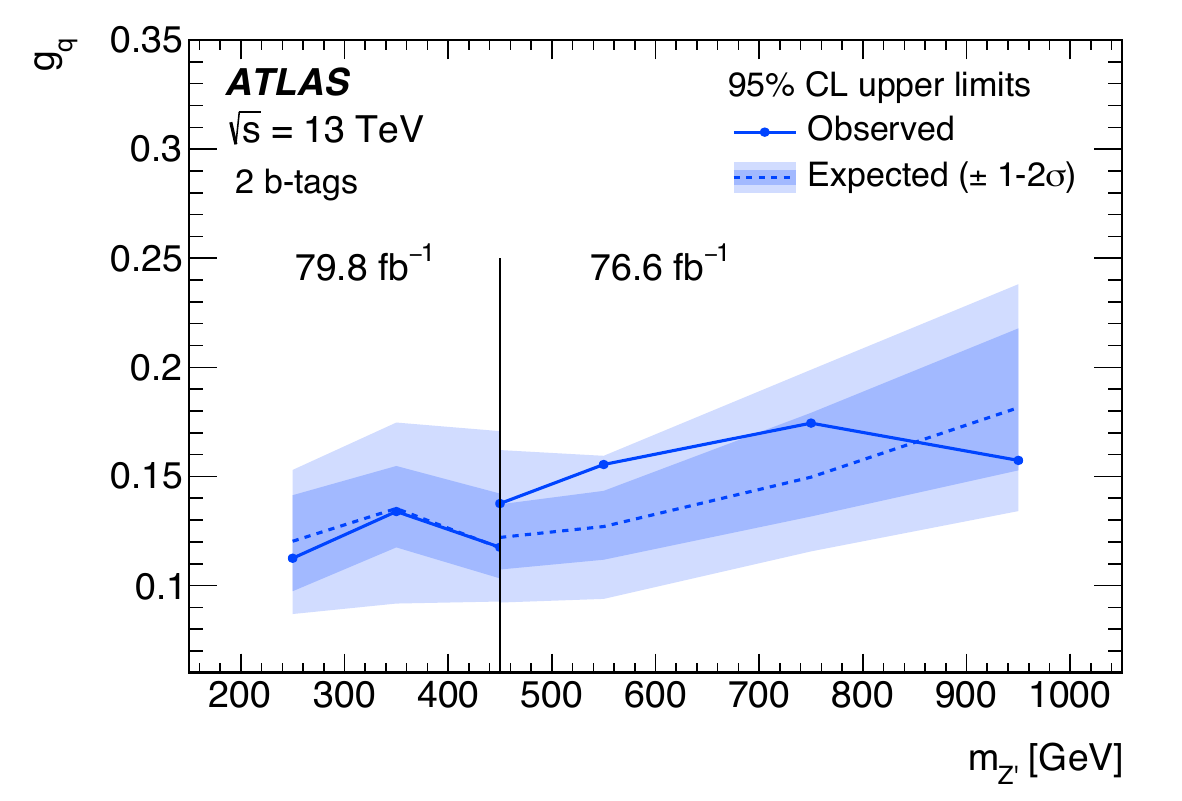}
 \label{fig:dijet:ISRyR:bb:L}
}
\caption{The inclusive di-jet mass spectrum for di-jet systems recoiling against a photon, for (\textbf{a}) flavour-inclusive and (\textbf{b}) di-$b$-tagged selections, as observed by ATLAS using a partial Run 2 dataset \cite{ATLAS:Dijet_ISRy_resolved}. No significant deviations from the background expectation are observed, and thus limits are set on the coupling of an axial-vector $Z^\prime$ to quarks, shown for both (\textbf{c}) flavour-inclusive and\linebreak (\textbf{d}) di-$b$-tagged selections. \label{fig:dijet:ISRy_resolved}}
\end{figure}

The possibility of di-jet resonances in association with an extra jet, from either quark or gluon radiation, has been studied by CMS using a partial Run 2 dataset \cite{CMS:Dijet_ISRj_resolved}.
Using jets as the associated object introduces multiple challenges, such as being subject to the same trigger constraints as before without the introduction of dedicated three-jet triggers.
The CMS result therefore developed and used a DS stream optimised for multi-jet events, with cuts on \HT{} at both the hardware- and software-levels, where \HT{} is the scalar sum of the \pT{} of the jets in the event.
This stream was then used to build three ``wide jets'', similar to what has been previously described, from the calorimeter-based \smallR{} jets available in the trigger system.
An additional complication of using a jet as the associated object is that there is no a priori way to define which two of the three jets correspond to the di-jet system of interest, and which is the additional jet.
The analysis investigated multiple ways to select the jets constituting the di-jet system, but settled on using the two jets with the highest \pT{}; it was found that this was correct more often than not, especially for the higher-mass resonances considered.
After identifying the di-jet system, the di-jet mass spectrum is fit with a functional form, defining the background estimation.
No significant deviation from the background estimation is observed, as seen in Figure \ref{fig:dijet:ISRj_resolved}a; thus, limits are set on the coupling of quarks to an axial-vector $Z^\prime$ model as shown in Figure \ref{fig:dijet:ISRj_resolved}b.

\vspace{-14pt}
\begin{figure}[H]

\subfigure[]{
 \includegraphics[width=0.44\textwidth]{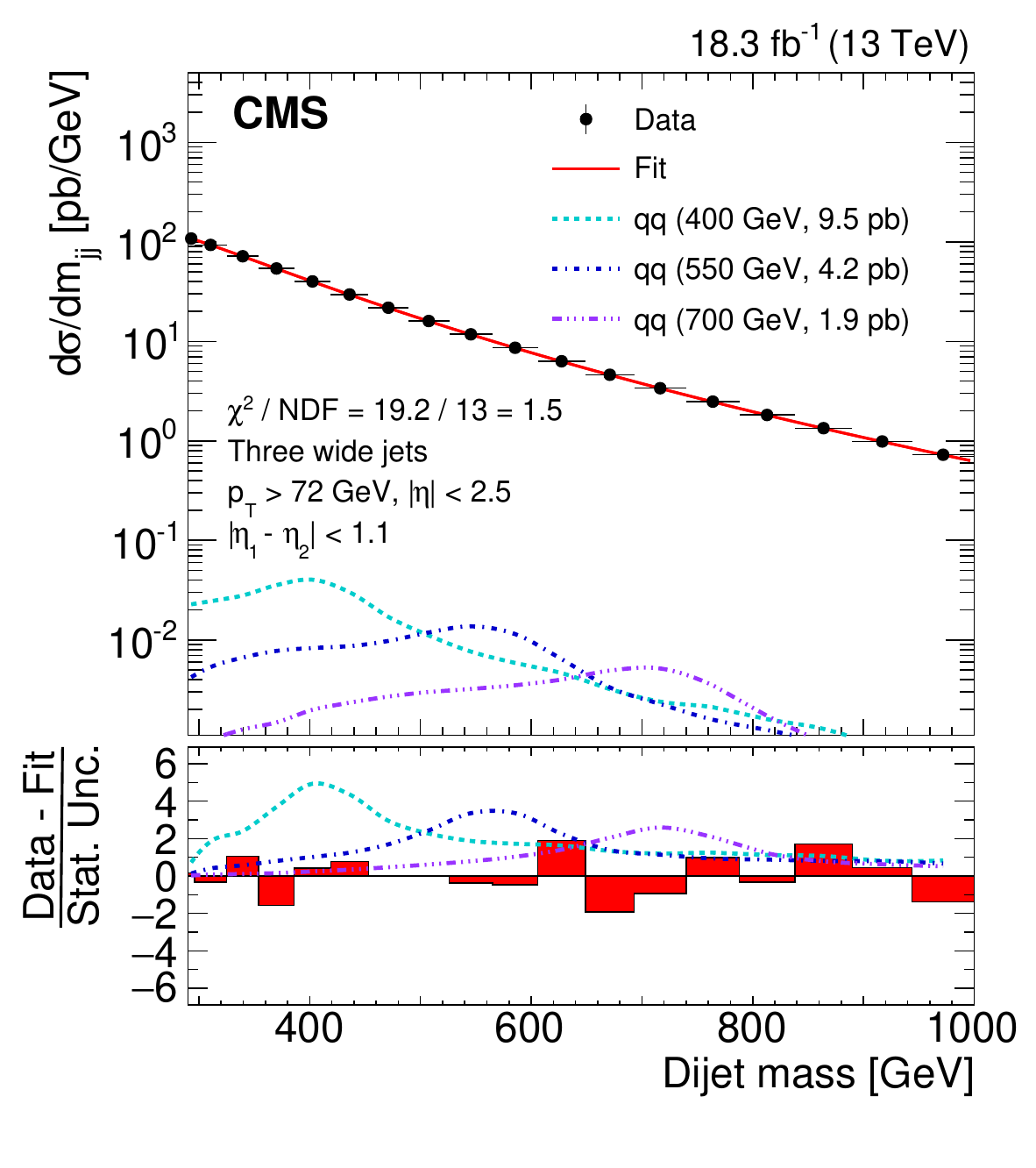}
 \label{fig:dijet:ISRjR:SR}
}
\subfigure[]{
 \includegraphics[width=0.52\textwidth]{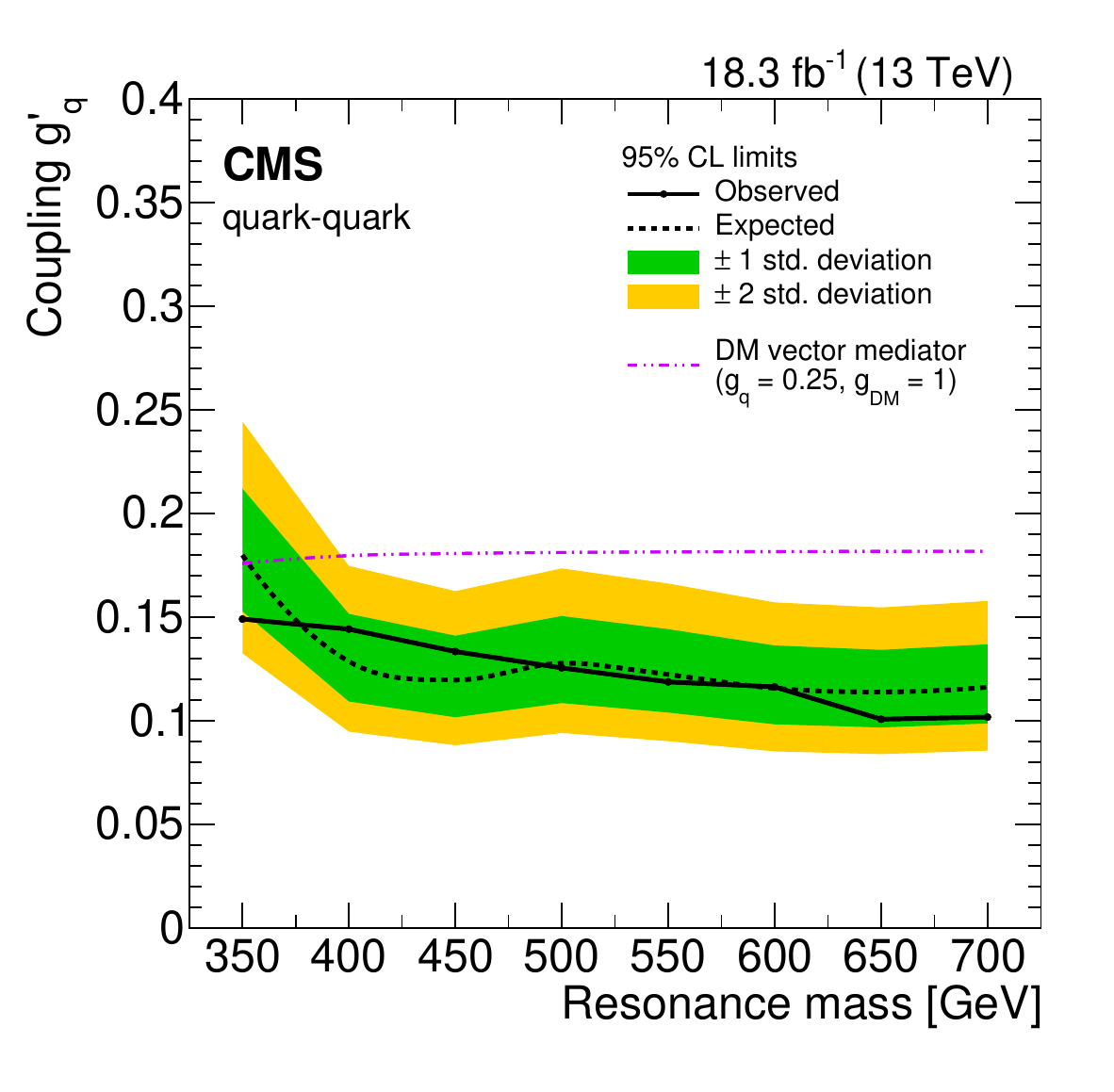}
 \label{fig:dijet:ISRjR:L}
}
\caption{(\textbf{a}) The inclusive di-jet mass spectrum for di-jet systems recoiling against a jet, as observed by CMS using a partial Run 2 dataset \cite{CMS:Dijet_ISRj_resolved}. (\textbf{b}) No significant deviations from the background expectation are observed, and thus limits are set on the coupling of an axial-vector $Z^\prime$ to quarks. \label{fig:dijet:ISRj_resolved}}
\end{figure}

A third type of associated object that can be used is a lepton (such as an electron or muon), with the interpretation of it originating from initial state $W$ boson radiation.
ATLAS has conducted such a search using the full Run 2 dataset \cite{ATLAS:Dijet_ISRl_resolved}, where the di-jet resonance in association with a lepton is one of the interpretations considered.
As listed in Table \ref{tab:motivation:triggers}, lepton triggers in ATLAS and CMS have very low thresholds, which can help to overcome the rarity of $W$ boson radiation with respect to photon or jet radiation.
Leptons are also very easy to differentiate from jets, and thus it is easy to define the di-jet system and fit the data using a functional form to obtain a background estimation.
The resulting di-jet mass spectrum is shown in Figure \ref{fig:dijet:ISRl_resolved}a; no significant deviations from the prediction are observed, and thus limits are set on the production of an axial-vector $Z^\prime$ in Figure \ref{fig:dijet:ISRl_resolved}b.

\vspace{-9pt}

\begin{figure}[H]

\subfigure[]{
 \includegraphics[width=0.55\textwidth]{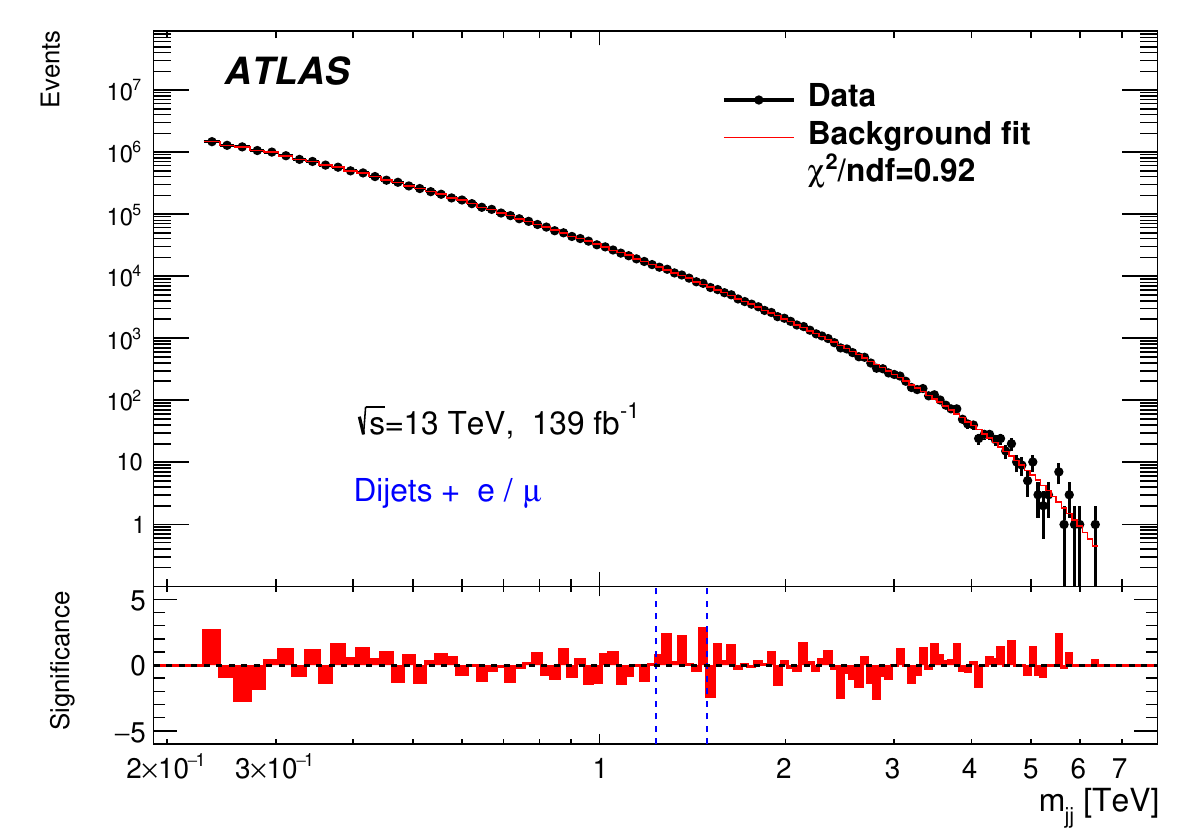}
 \label{fig:dijet:ISRlR:SR}
}
\subfigure[]{
 \includegraphics[width=0.41\textwidth]{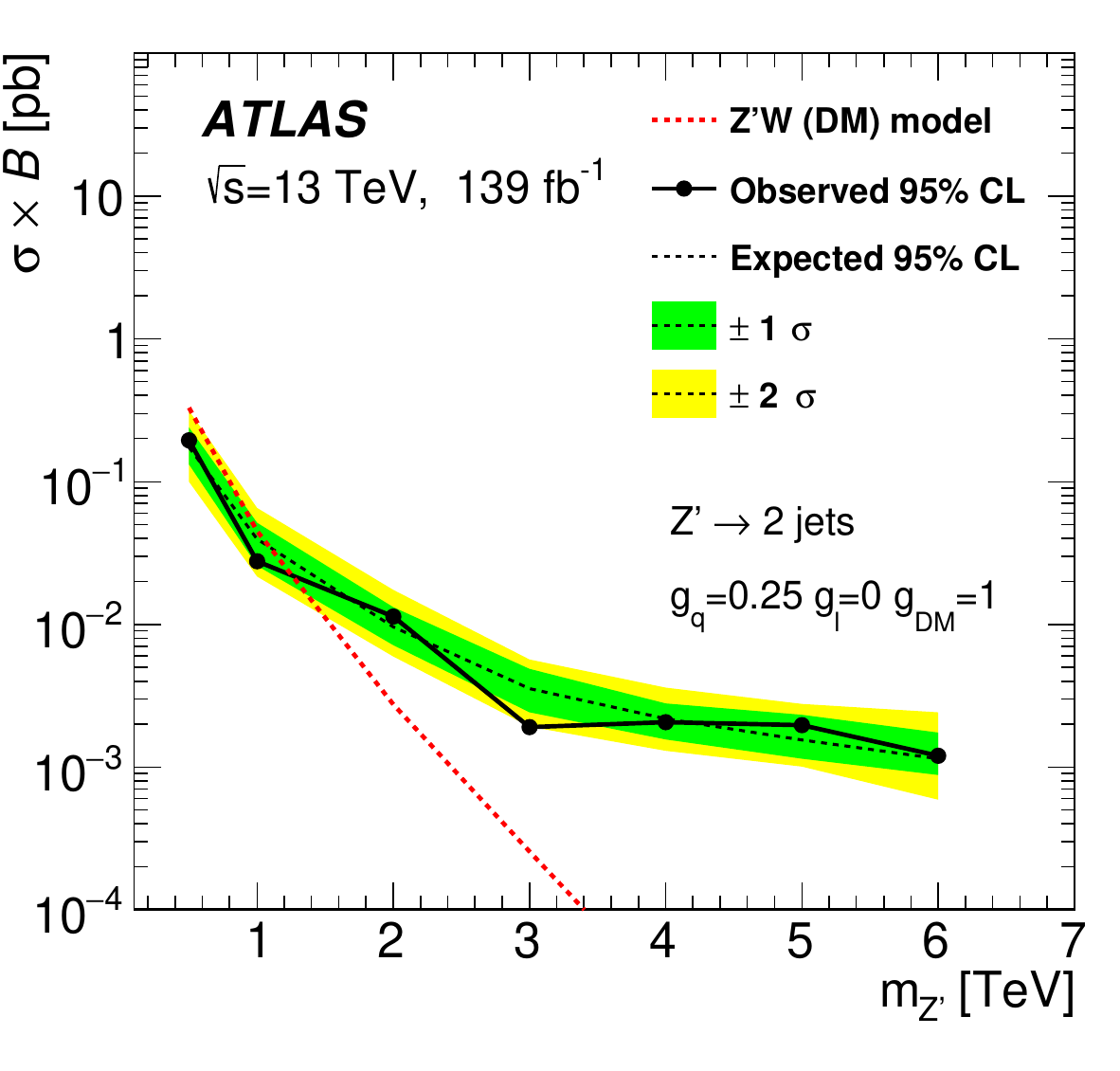}
 \label{fig:dijet:ISRlR:L}
}
\caption{(\textbf{a}) The inclusive di-jet mass spectrum for di-jet systems recoiling against an electron or muon, as observed by ATLAS using the full Run 2 dataset \cite{ATLAS:Dijet_ISRl_resolved}. (\textbf{b}) No significant deviations from the background expectation are observed, and thus limits are set on the production of an axial-vector $Z^\prime$. \label{fig:dijet:ISRl_resolved}}
\end{figure}

\subsection{Boosted Di-Jet Searches in Association with Other Objects}
\label{sec:dijet:ISRboosted}

The previous analyses have focused on di-jet masses at the level of a few hundred \!\GeV{}, which means that there is still the possibility that new physics is hiding at even lower mass scales.
Accessing lower mass scales with such techniques is challenging, as the di-jet system eventually becomes so low in mass that the decay becomes collimated and the jets overlap.
At this point, it is still possible to search for di-jet resonances, but a new technique is needed: the di-jet system must now be represented as a single \largeR{} jet, and the mass of that single jet represents the possible resonance of a new particle.
This approach is referred to as the boosted di-jet topology, in contrast with the resolved topology that we have been discussing so far.

The first LHC search for a boosted di-jet system in association with another object happened at almost the same time as the first resolved search, also occurring in 2016 \cite{CMS:FirstISRB}.
Any non-top quark or gluon can be reconstructed as a \largeR{} jet, which complicates such searches, as the Standard Model di-jet production process is a background to boosted jet + ISR jet searches, while the Standard Model photon+jet process is a background to boosted jet + ISR photon searches.
Advanced jet substructure techniques are the solution to this problem, as jet substructure variables can be used to reject jets from non-top quarks and gluons while accepting jets consistent with the interpretation of containing a di-jet system.
This rejection of the backgrounds must be done carefully, as jet substructure techniques are generally correlated with the jet mass; cutting on the substructure variable can therefore bias the observable of interest to the search for new physics.
The solution to this challenge lies in specifically designing a substructure variable to be uncorrelated with respect to the jet mass, and thus cuts on the variable can reject the background without biasing the search.
This idea was first proposed via Designing Decorrelated Taggers (DDT) \cite{DDT} in the context of such searches, but has since grown to include additional analytic and machine learning strategies for designing selections to reject the Standard Model background without shaping the mass distribution, which are now used in a variety of applications.

ATLAS has conducted a search for such boosted di-jet resonances in association with both photons and jets \cite{ATLAS:Dijet_ISR_boosted}, where the DDT technique is used to suppress the Standard Model photon+jet and di-jet backgrounds, respectively.
The background is evaluated by inverting the DDT cut and calculating transfer factors from the control to signal region, which are then subsequently smoothed using a Gaussian process regression.
The resulting background estimation strategy is validated by applying it to the $W/Z$ boson mass peak and confirming that the extracted significance of the mass peak matches the Standard Model expectation.
With this validation done, the analysis then proceeds to evaluate the consistency of data with the background estimation in both the photon and jet channels, which are shown in Figure \ref{fig:dijet:ISR_boosted_ATLAS}a,b, respectively.
No significant deviation is observed beyond the Standard Model background, and thus limits are set on the coupling of quarks to an axial-vector $Z^\prime$ mediator as shown in Figure \ref{fig:dijet:ISR_boosted_ATLAS}c, where the photon and jet channels are combined in the limit-setting procedure.

CMS has conducted a search for boosted di-jet systems in association with a photon \cite{CMS:Dijet_ISRy_boosted}, using a partial Run 2 dataset, which holds the current record in accessing the low-mass regime.
The analysis makes use of the DDT procedure to suppress the Standard Model photon+jet background, although the remaining background is still comprised primarily of events which survive this cut.
Resonant backgrounds from $W/Z$+photons and $t\bar{t}$ are taken from simulated samples, while the photon+jet background is estimated by defining a transfer factor from a control region in which the DDT cut has been inverted.
The resulting background estimation is compared to data in Figure \ref{fig:dijet:ISRy_boosted_CMS}a, and no significant deviations are observed; thus, limits are set on the quark coupling to an axial-vector $Z^\prime$ as shown in Figure \ref{fig:dijet:ISRy_boosted_CMS}b.
These limits are exceptional in that they probe all the way down to $Z^\prime$ masses of 10\GeV{}, which is the lowest mass scale probed by any current di-jet resonance search at the LHC.

\begin{figure}[H]
\centering
\subfigure[ ISR photon]{
 \includegraphics[width=0.42\textwidth]{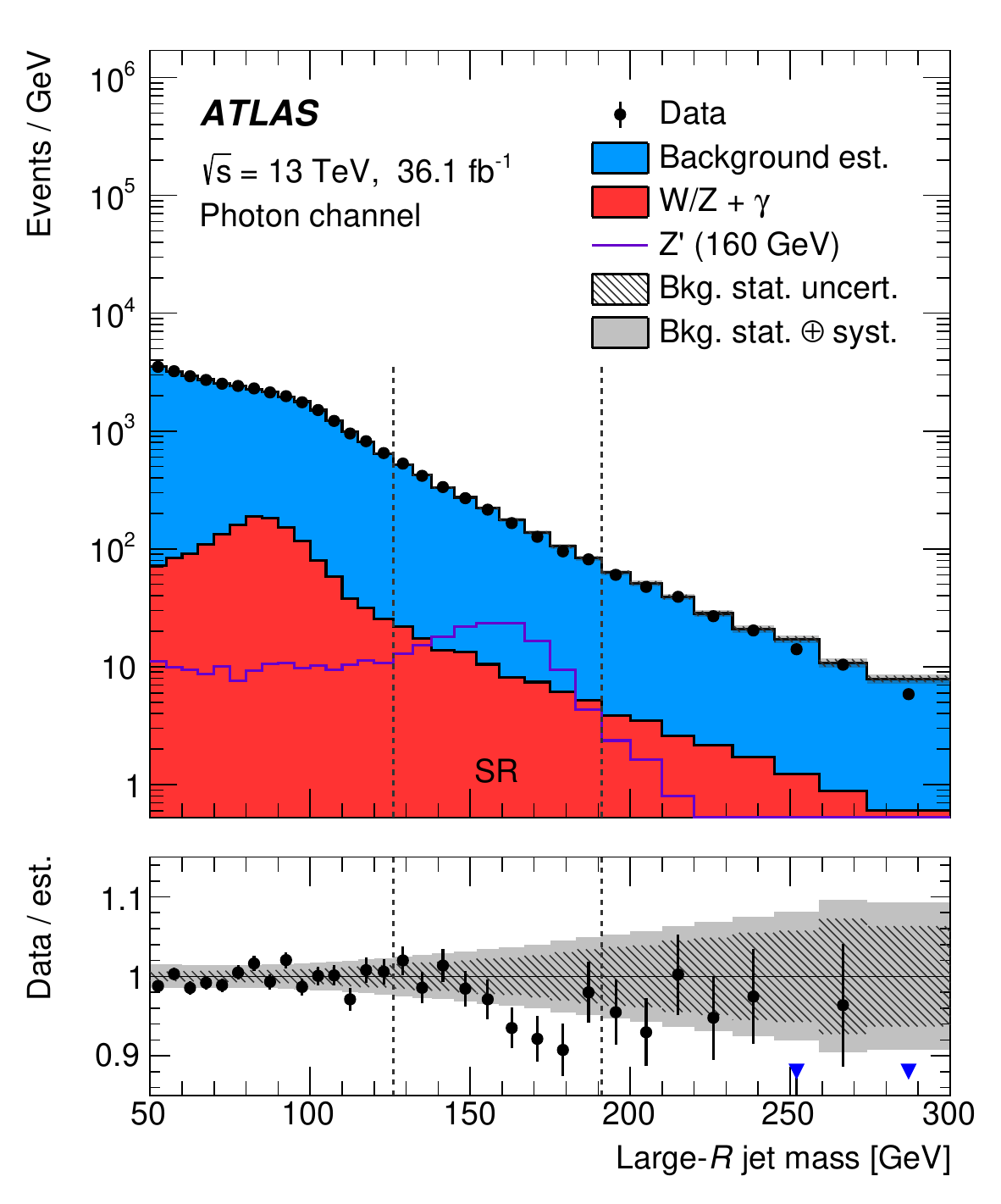}
 \label{fig:dijet:ISRyB:ATLAS:SR}
}
\subfigure[ ISR jet]{
 \includegraphics[width=0.42\textwidth]{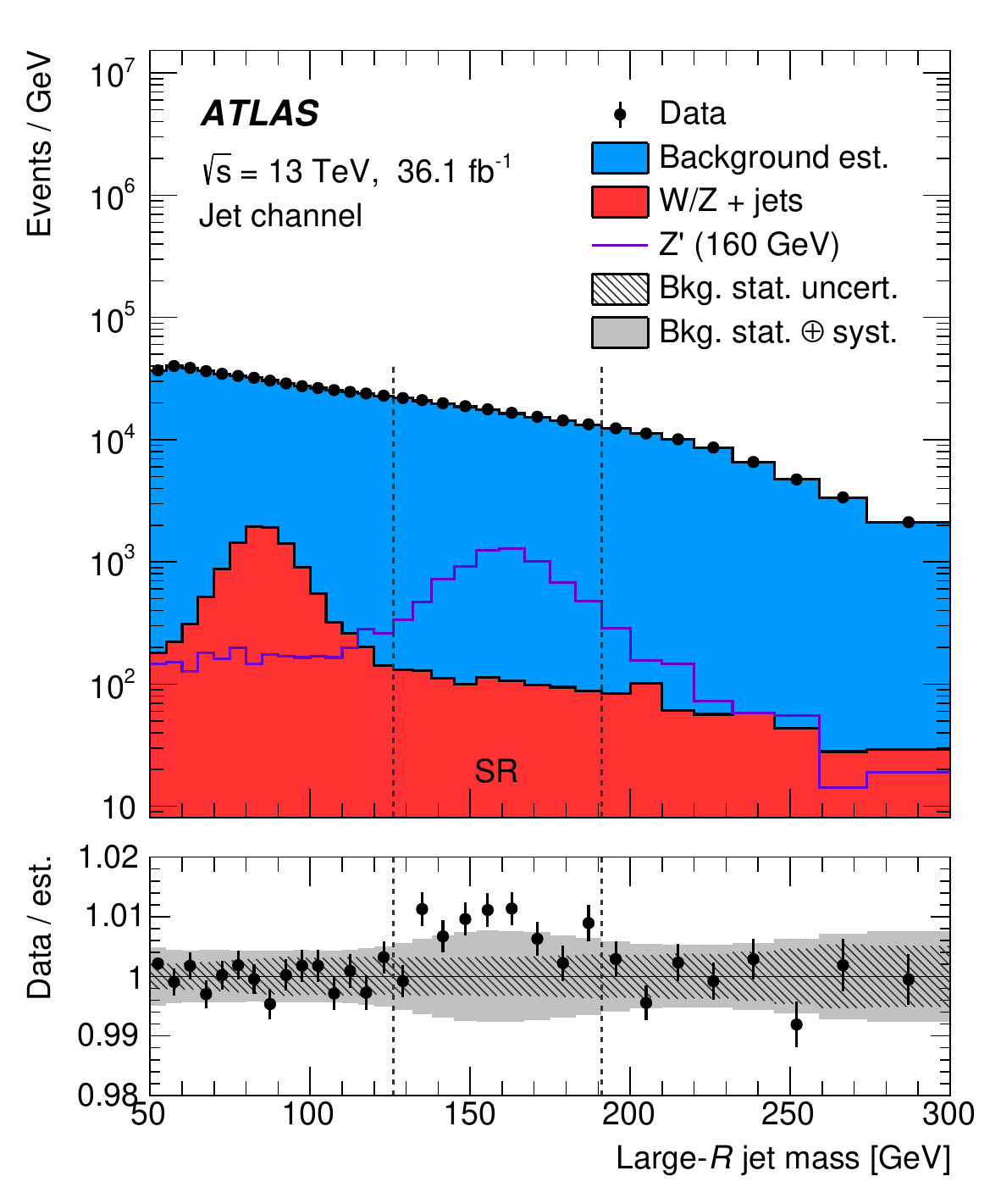}
 \label{fig:dijet:ISRjB:ATLAS:SR}
}\\

\subfigure[]{

 \includegraphics[width=0.42\textwidth]{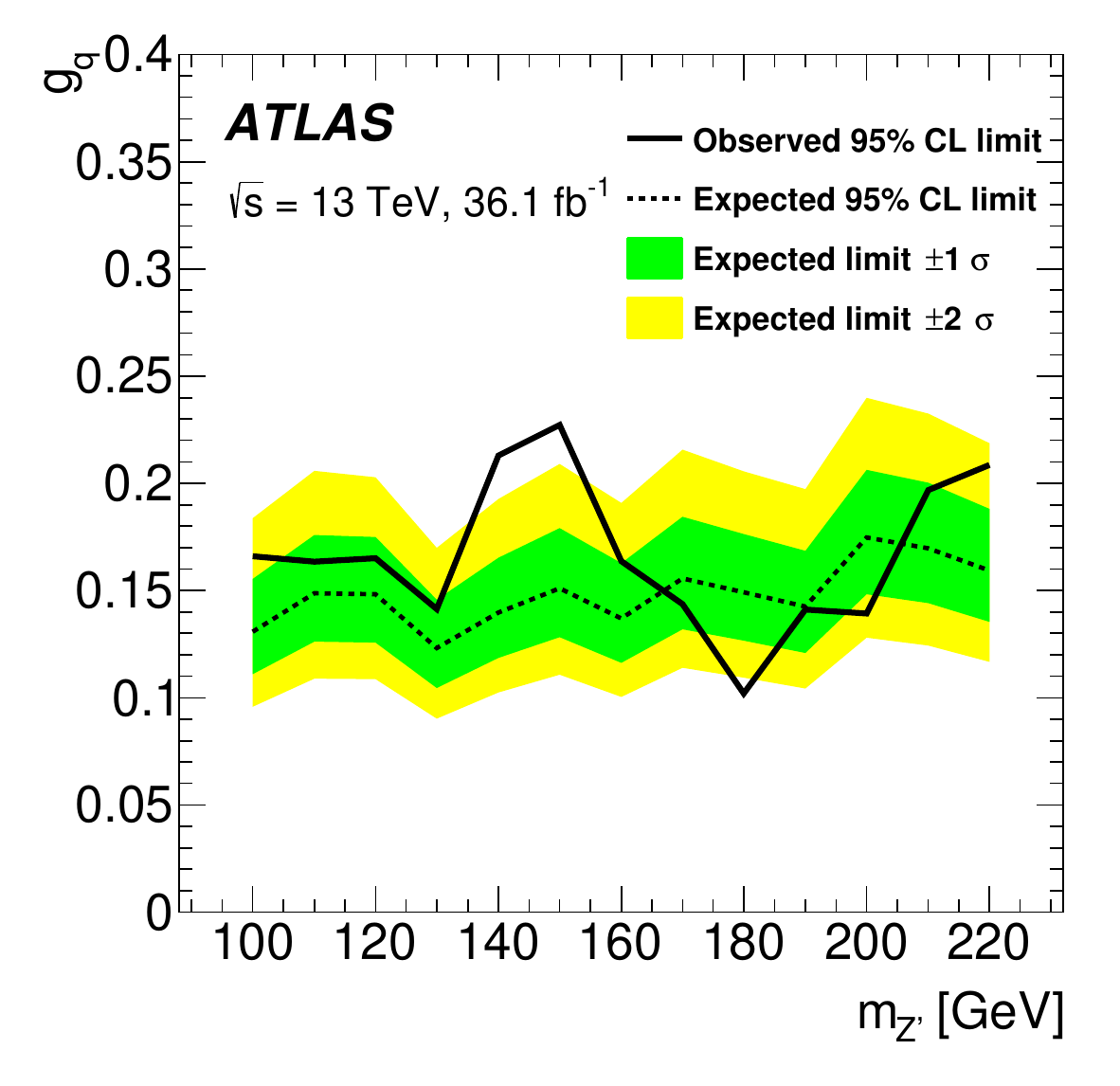}
 \label{fig:dijet:ISRB:ATLAS:L}
}
\caption{The inclusive jet mass spectrum for boosted di-jet systems recoiling against (\textbf{a}) a photon and (\textbf{b}) a jet, as observed by ATLAS using a partial Run 2 dataset \cite{ATLAS:Dijet_ISR_boosted}. (\textbf{c}) No significant deviations from the background expectation are observed, and thus limits are set by both the photon and jet channels, which are combined to provide a single limit on the coupling of an axial-vector $Z^\prime$ to quarks. The largest excess in (\textbf{c}) corresponds to a local significance of $2.4\sigma$ and a global significance of $1.2\sigma$ \label{fig:dijet:ISR_boosted_ATLAS}}
\end{figure}

\vspace{-14pt}

\begin{figure}[H]

\subfigure[]{
 \includegraphics[width=0.43\textwidth]{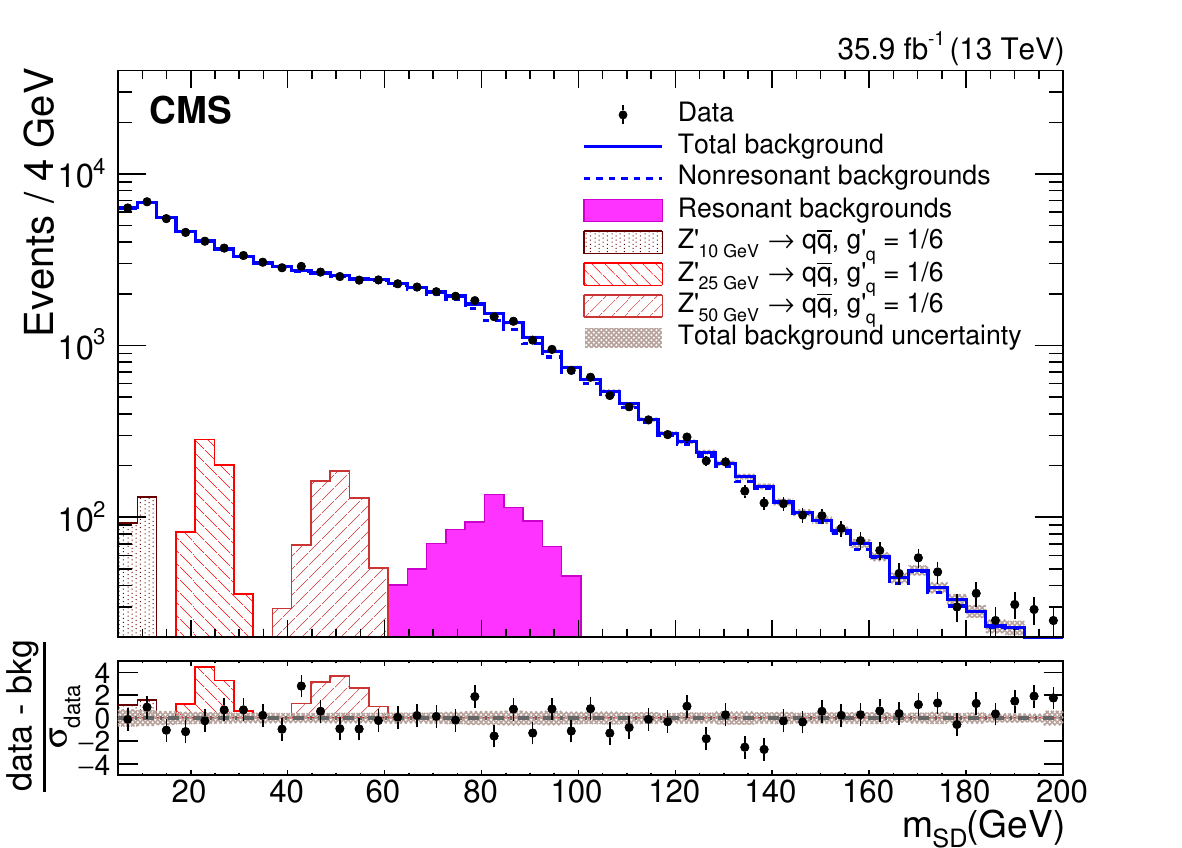}
 \label{fig:dijet:ISRyB:CMS:SR}
}
\subfigure[]{
 \includegraphics[width=0.43\textwidth]{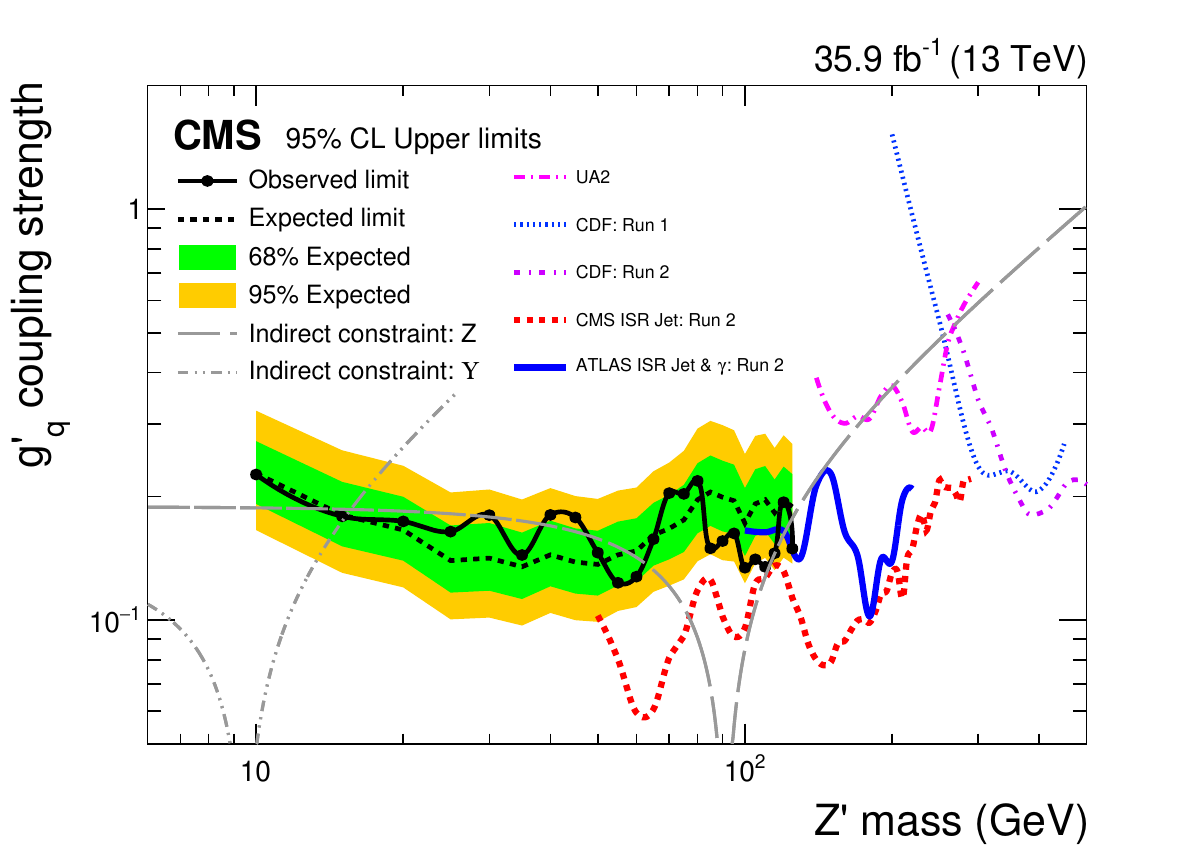}
 \label{fig:dijet:ISRyB:CMS:L}
}
\caption{(\textbf{a}) The inclusive jet mass spectrum for boosted di-jet systems recoiling against a \mbox{photon \cite{CMS:Dijet_ISRy_boosted}}, as observed by CMS using a partial Run 2 dataset. (\textbf{b}) No significant deviation from the background expectation is observed, and thus limits are set on the coupling of an axial-vector $Z^\prime$ to quarks. \label{fig:dijet:ISRy_boosted_CMS}}
\end{figure}

The search for boosted di-jet systems in association with another jet has also been studied by CMS \cite{CMS:Dijet_ISRj_boosted}, again using a partial Run 2 dataset.
The analysis is split into two separate regions corresponding to different mass regimes: the lower mass regime uses the standard CMS \largeR{} definition of \antikt{} $R=0.8$ jets, while the higher mass regime expands to using the Cambridge--Aachen algorithm with $R=1.5$; in this way, the $R=1.5$ jet can contain a more massive di-jet resonance than for $R=0.8$ at the same di-jet system \pT{}.
The DDT procedure is again used to define a substructure variable that can reject the Standard Model background from not-top quarks and gluons while retaining possible di-jet resonances, without sculpting the \largeR{} jet mass distribution.
This decorrelation is performed separately for the two different jet definitions, and is done differently than in the previous analyses: the decorrelation is defined to reject 95\% of Standard Model background jets from di-jet processes in all regions of the 2D parameter space considered.
Due to this choice, it is known that only 5\% of the di-jet background is accepted in all regions studied, up to possible simulation limitations in modelling the parameter space of interest.
The background estimation procedure thus focuses on evaluating potential differences in the DDT modelling, and is done by simultaneously fitting a function to the events passing and failing the DDT cut.
This procedure defines the background estimate for the dominant background process, while the smaller backgrounds from $W/Z$+jet, $t\bar{t}$, and single-top are taken from simulation.
The resulting background estimates are derived for different jet \pT{} bins; one such bin is shown for \antikt{} $R=0.8$ jets in Figure \ref{fig:dijet:ISRj_boosted_CMS}a and Cambridge--Aachen $R=1.5$ jets in Figure \ref{fig:dijet:ISRj_boosted_CMS}b.
No significant deviations are observed from the respective background expectations, and thus limits are set on the quark coupling to an axial-vector $Z^\prime$, as shown in Figure \ref{fig:dijet:ISRj_boosted_CMS}c.
The resulting limits in the lower-mass regime are further combined with a previous analysis, as shown in Figure \ref{fig:dijet:ISRj_boosted_CMS}d; the previous analysis was done similarly and using a distinct partial Run 2 dataset, detailed in Ref. \cite{CMS:Dijet_ISRj_boosted_old}.
This previous result saw a potential excess at a resonance mass a bit above 100\GeV{}, but the result discussed here does not confirm it, and thus the combined significance is reduced with respect to Ref. \cite{CMS:Dijet_ISRj_boosted_old}.
\vspace{-6pt}

\begin{figure}[H]

\subfigure[ \Antikt{} $R=0.8$ jets]{
 \includegraphics[width=0.34\textwidth]{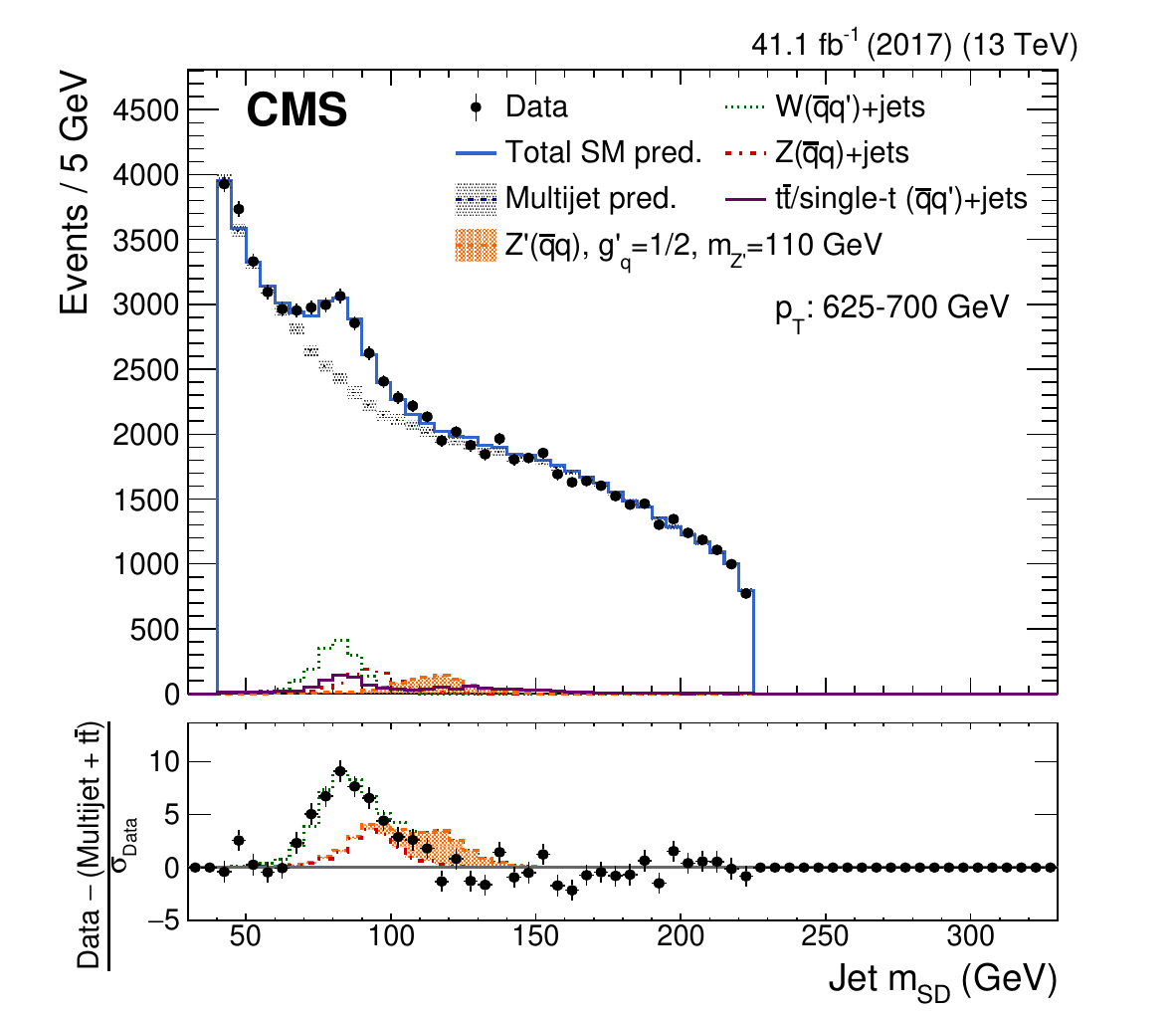}
 \label{fig:dijet:ISRjB:CMS:SR}
}
\subfigure[ Cambridge--Aachen $R=1.5$ jets]{
 \includegraphics[width=0.34\textwidth]{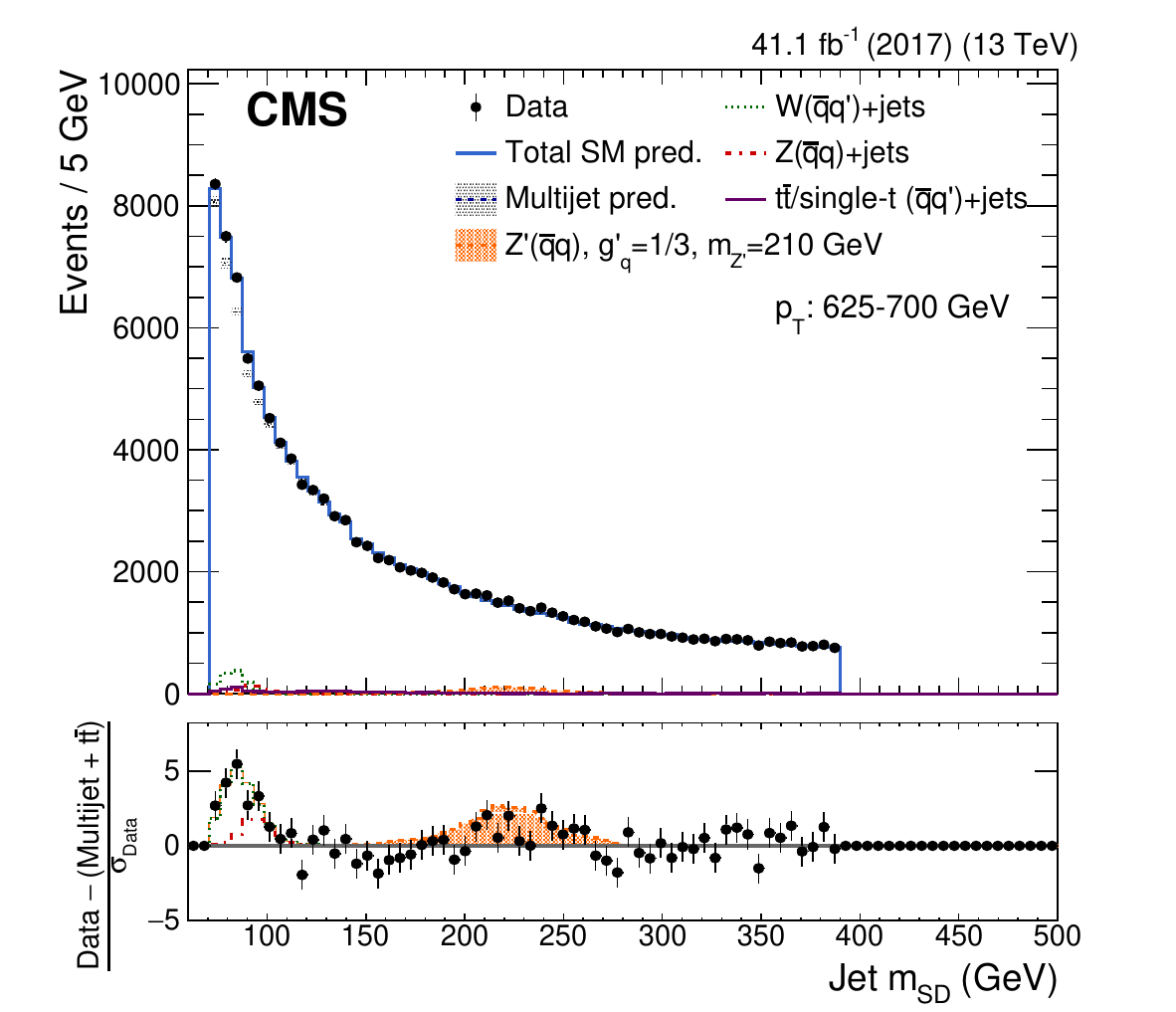}
 \label{fig:dijet:ISRjB:CMS:SRCA}
}\\
\subfigure[]{
 \includegraphics[width=0.34\textwidth]{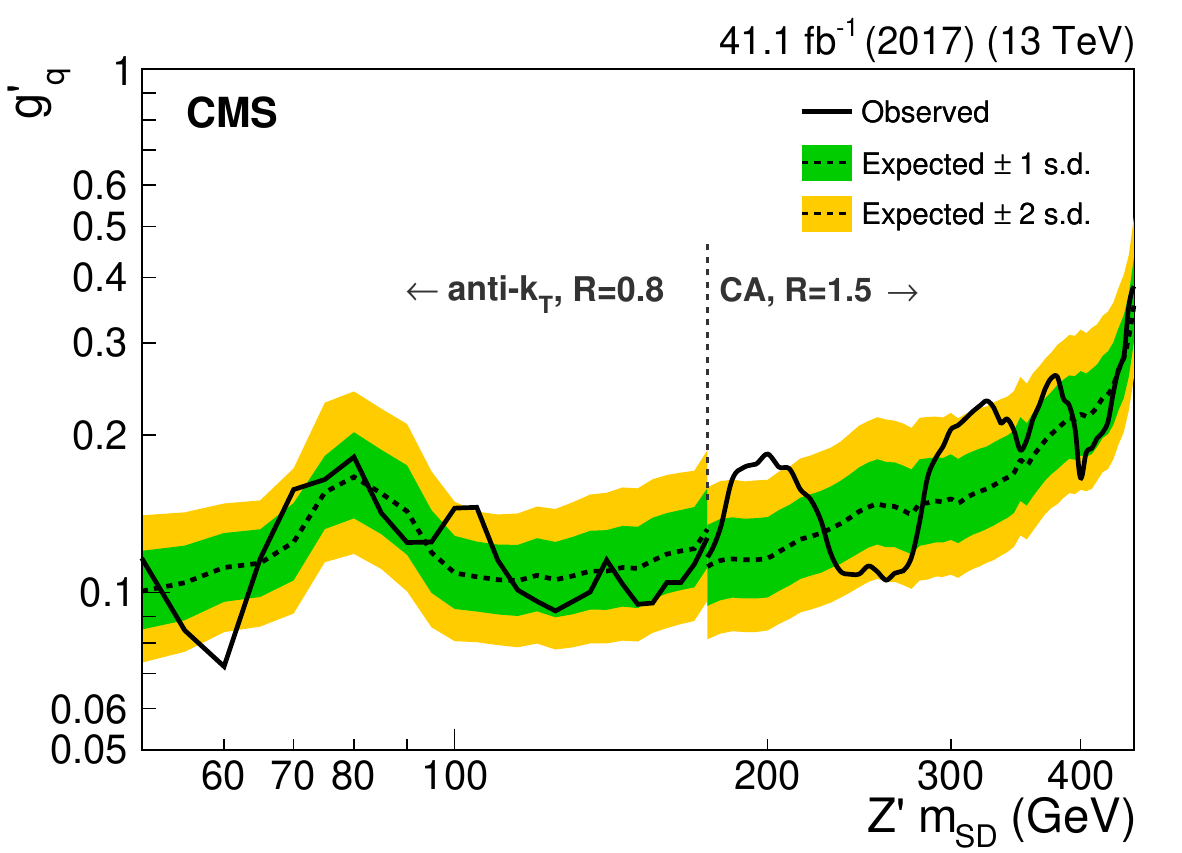}
 \label{fig:dijet:ISRjB:CMS:L}
}
\subfigure[]{
 \includegraphics[width=0.34\textwidth]{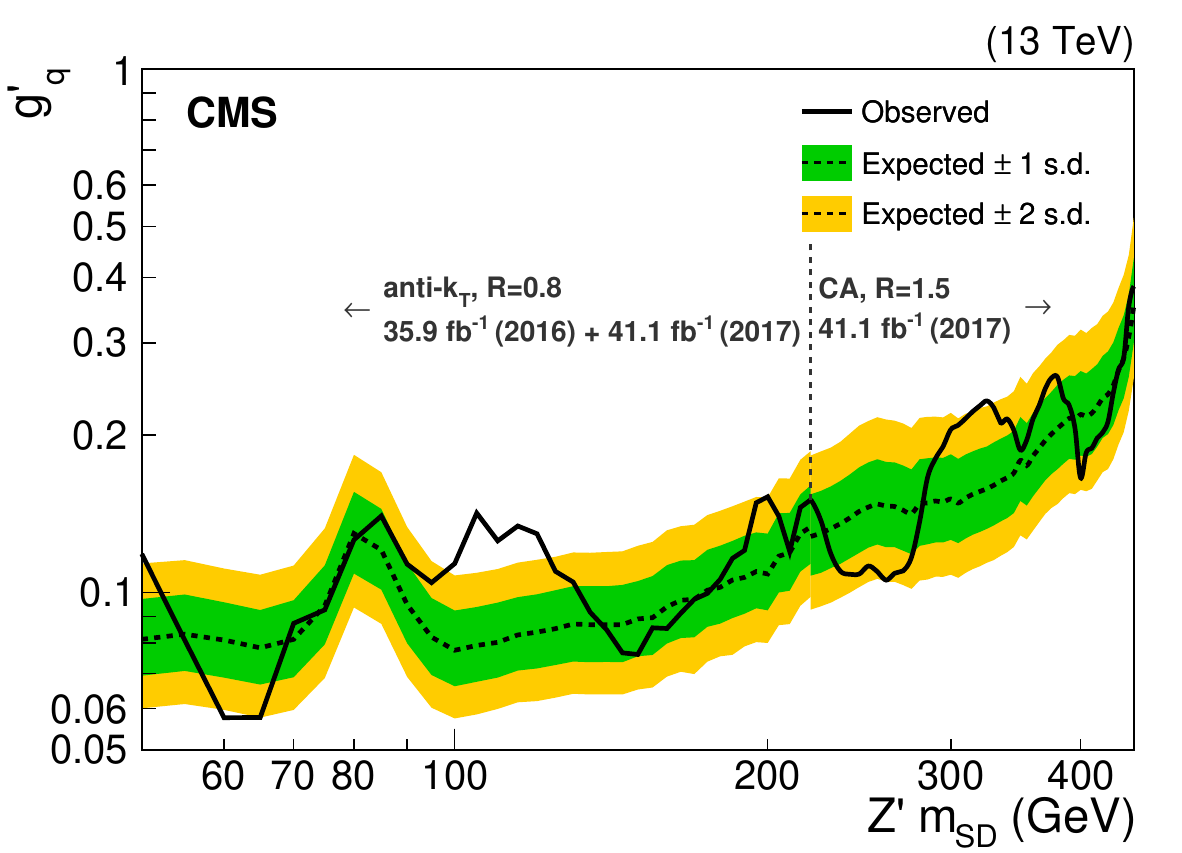}
 \label{fig:dijet:ISRjB:CMS:Lcomb}
}
\caption{The inclusive di-jet mass spectrum for boosted di-jet systems recoiling against a jet, as observed by CMS using a partial Run 2 dataset \cite{CMS:Dijet_ISRj_boosted}, for (\textbf{a}) \antikt{} $R=0.8$ jets and (\textbf{b}) Cambridge--Aachen $R=1.5$ jets. (\textbf{c}) No significant deviation from the background expectation is observed, and thus limits are set on the coupling of an axial-vector $Z^\prime$ to quarks. \linebreak (\textbf{d}) The low-mass range of the jet channel is further combined with a separate partial Run 2 dataset from Ref. \cite{CMS:Dijet_ISRj_boosted_old}; the largest excess corresponds to a $2.2\sigma$ local significance after combination, and is only present in the previous result, not the new dataset shown in (\textbf{a},\textbf{b},\textbf{c}). \label{fig:dijet:ISRj_boosted_CMS}}
\end{figure}

% MET+X searches
\section{Missing Transverse Momentum Plus X Searches}
\label{sec:monoX}

% Define MET
% Give diagrams for ISR DM
% Notion of complementarity with other DM searches (detector stable, sensitivity)

Searches for invisible or otherwise very weakly interacting particles are challenging at ATLAS and CMS, as they escape the detector without leaving any visible energy signature to indicate their presence.
The Standard Model already includes one such type of particle, the neutrino, the production of which forms an irreducible background to any search for other detector-invisible particles.
There is, however, a candidate for an invisible particle beyond the Standard Model and which is of great interest to the particle physics community: particulate dark matter.
If dark matter has a particle origin, then it happens that a weakly-interacting massive particle (WIMP) with a mass at the weak scale would naturally produce the observed relic abundance of dark matter in the universe \cite{WIMP}; this is known as the ``WIMP Miracle''.
As the LHC is particularly sensitive to particles at the weak scale, there is both strong interest in and motivation for searches for such dark matter candidates.
There is therefore a large physics programme at the LHC oriented around the search for dark matter, including common LHC recommendations on how to interpret the results of such searches \cite{Theory:ZPrimeLHC1,Theory:ZPrimeLHC2,Theory:ZPrimeLHC3,Theory:ZPrimeLHC4}.

One of the original approaches to such searches at the LHC, and one which is still of great relevance, is the search for the pair-production of dark matter particles $\chi$ through the decay of a new s-channel mediator, such as a $Z^\prime$ boson; more details on such $Z^\prime$ models as used at the LHC can be found in, for example, the above-mentioned LPCC Dark Matter Working Group recommendation documents.
However, if the collision is entirely described by the process $q\bar{q}\to{}Z^\prime\to{}\chi\bar{\chi}$, then the events will be invisible to the ATLAS and CMS detectors, as the final state only involves detector-invisible particles.
Searches therefore must add an additional experimental constraint, in the form of requiring that ISR accompanies the production of the $Z^\prime$, as shown in Figure \ref{fig:monoX:feyn}.
This ISR requirement does not add any additional assumption about the new physics production or decay couplings, as the radiation occurs independently of the new physics of interest, and thus does not bias the search to specific models.
The presence of ISR is rather an experimental consideration, which comes with the price of a reduced cross-section, but which is required to observe such events.
The addition of ISR adds a visible component to the collision by-products, which the $Z^\prime$ and thus the dark matter particles must recoil against.
Thanks to the conservation of transverse momentum in LHC collisions, this imbalance between visible activity in one part of the detector and nothing in the opposite part of the detector can be quantified; the imbalance is referred to as missing transverse momentum, and large values of missing transverse momentum imply the presence of invisible particles.

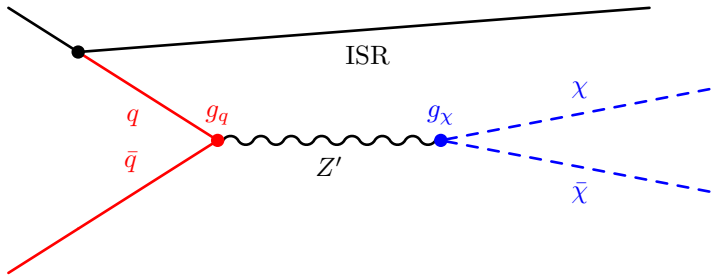
\begin{figure}[H]
\hspace{-20pt}
\begin{fmffile}{MonoISR}
 \begin{fmfgraph*}(300,100)
 \fmfleft{qbar,q}
 \fmfright{balance,chibar,chi,ISR}
 
% \fmflabel{$q$}{q}
% \fmflabel{$\bar{q}$}{qbar}
 
 \fmf{vanilla,tension=2}{q,vertexISR}
 \fmf{vanilla,foreground=red,tension=2}{qbar,ISRphantom}
 
 \fmf{vanilla,label=$\textcolor{red}{q}$,label.side=right,foreground=red,tension=1}{vertexISR,vertexL}
 \fmf{vanilla,label=$\textcolor{red}{\bar{q}}$,label.side=left,foreground=red,tension=1}{ISRphantom,vertexL}
 
 \fmf{vanilla,label=ISR,tension=0}{vertexISR,ISR}
 \fmf{phantom,tension=0}{ISRphantom,balance}
 
 \fmf{dashes,label=$\textcolor{blue}{\chi}$,foreground=blue,tension=0.5}{vertexR,chi}
 \fmf{dashes,label=$\textcolor{blue}{\bar{\chi}}$,foreground=blue,tension=0.5}{vertexR,chibar}

 \fmfdot{vertexISR}
 \fmfv{decor.shape=circle,decor.filled=full,
decor.size=2thick,foreground=red,label=$\textcolor{red}{g_q}$,label.angle=90}{vertexL}
 \fmfv{decor.shape=circle,decor.filled=full,
decor.size=2thick,foreground=blue,label=$\textcolor{blue}{g_\chi}$,label.angle=90}{vertexR}
 
 \fmf{boson,label=$Z^\prime$,tension=1.25}{vertexL,vertexR}
 \end{fmfgraph*}
\end{fmffile}
\caption{A Feynman diagram showing the s-channel production of a new axial-vector $Z^\prime$ mediator, which is produced through the annihilation of standard model quarks via a coupling $g_q$, and which decays to dark matter particles $\chi$ through a coupling $g_\chi$. Such a process would be invisible in the ATLAS and CMS detectors, as the dark matter particles would escape the detector unobserved. The presence of Initial State Radiation (ISR) is therefore required, creating a balance between the $Z^\prime$ and the ISR object, and thus providing a means of inferring the presence of invisible particles through a visible momentum imbalance.\label{fig:monoX:feyn}}
\end{figure}

Searches for missing transverse momentum balancing some other visible object, usually assumed to be from ISR, are thus a prominent means of searching for the production of dark matter at the LHC.
There are many such searches, and they will not be covered in detail here, as that could be the subject of an entire separate review.
Instead, this review will focus on analyses related to hadronic final states, which happen to have the leading sensitivity to a variety of different types of possible mediators between the Standard Model and postulated dark sectors.

\subsection{Missing Transverse Momentum Plus Jet Searches}
\label{sec:monoX:jets}

Quarks and gluons, or collectively jets, are the most common source of ISR at the LHC.
From a statistical perspective, this means that searches for invisible particles using missing transverse momentum in association with ISR jets should lead the sensitivity.
This is indeed generally true, and the resulting search is often referred to as a mono-jet search due to the presence of only a single visible jet in the detector.
This name has stuck, even though modern iterations of the mono-jet search allow for more than one jet to be present, so long as there is at least one high-energy jet in association with large missing transverse momentum.

ATLAS \cite{ATLAS:MonoX_jet} and CMS \cite{CMS:MonoX_jet} have both published mono-jet analyses using the full Run 2 dataset.
The analyses are generally quite similar in concept, with a focus on precisely evaluating the expected contribution of the irreducible Standard Model background of $Z(\to{}\nu\nu)$ + jets from a variety of control regions.
ATLAS and CMS both use \mbox{$Z(\to\ell\ell)$ + jets} and $W(\to\ell\nu)$ + jets control regions, for $\ell=\{e,\mu\}$, while ATLAS defines an additional $t\bar{t}$ + single-top control region, and CMS benefits from an additional $\gamma$+jets control region.
The combination of all of these control regions allows for a very precise determination of the dominant and irreducible $Z(\to\nu\nu)$ + jets process in the signal region, as well as the secondary contributions from other $Z$+jets and $W$+jets processes; these high-precision estimations of the signal region contributions are key to the final sensitivity of the analysis.
Both analyses also have dedicated control regions to estimate the contributions of other processes to the signal region, such as those from multi-jet backgrounds.
This use of many dedicated control regions to estimate the relevant background processes in the signal region results in a very sensitive analysis, which is predominantly limited by systematic uncertainties, both experimental (object reconstruction and scale) and theoretical (in the process of extrapolating from control regions to signal regions).

An example of the $W(\to{}e\nu)$ + jets control region is shown in Figure \ref{fig:monoX:jet:CRSR}a for ATLAS and Figure \ref{fig:monoX:jet:CRSR}c for CMS, while the signal region expectations are shown in Figure \ref{fig:monoX:jet:CRSR}b,d, respectively.
A small deviation is seen in one bin of the ATLAS signal region, but there is a related fluctuation in the $W(\to{}e\nu)$+jets control region; thus, it is possible that the effect is correlated with a feature in the control region.

Another similar but distinct hadronic final state considers the possibility of a hadronically decaying $W$ or $Z$ boson.
The resulting analysis, often referred to as the hadronic mono-$V$ search, follows a very similar background estimation strategy to the mono-jet search.
In ATLAS, the search has been conducted using a partial Run 2 dataset \cite{ATLAS:MonoX_Vqq}, and with the aforementioned $Z(\to\ell\ell)$ + jets and $W(\to\ell\nu)$ + jets control regions.
The CMS hadronic mono-$V$ search was conducted together with the mono-jet search, and thus includes all of the different control regions discussed previously, and uses the full Run 2 dataset \cite{CMS:MonoX_jet}.
While the background estimation procedure is similar, the analysis definition is quite different.
Both ATLAS and CMS use \largeR{} jets to represent the hadronically decaying $W/Z$ boson, and employ jet taggers along the lines of those discussed in Section \ref{sec:reco:largeR:tag} to reject backgrounds from non-top-quark or gluon jets while selecting those consistent with hadronic $W$ or $Z$ boson decays.
Both ATLAS and CMS additionally have both low-purity (LP) and high-purity (HP) selections; the signal regions for the high-purity selections are shown in Figure \ref{fig:monoX:Vqq} for both ATLAS and CMS.
ATLAS additionally considers single- and double-$b$-tagged selections, which further enhance the purity with which $W$ and $Z$ boson decays can be retained.
ATLAS further studies resolved categories, where the two quarks from the $W$ or $Z$ decay are not sufficiently collimated to be adequately represented by a single \largeR{} jet; pairs of \smallR{} jets are thus used instead, and the invariant mass of that pair of jets is required to be consistent with the interpretation of them originating from the decay of a $W$ or $Z$ boson.

\begin{figure}[H]

\subfigure[ ATLAS, control region]{
 \includegraphics[width=0.43\textwidth]{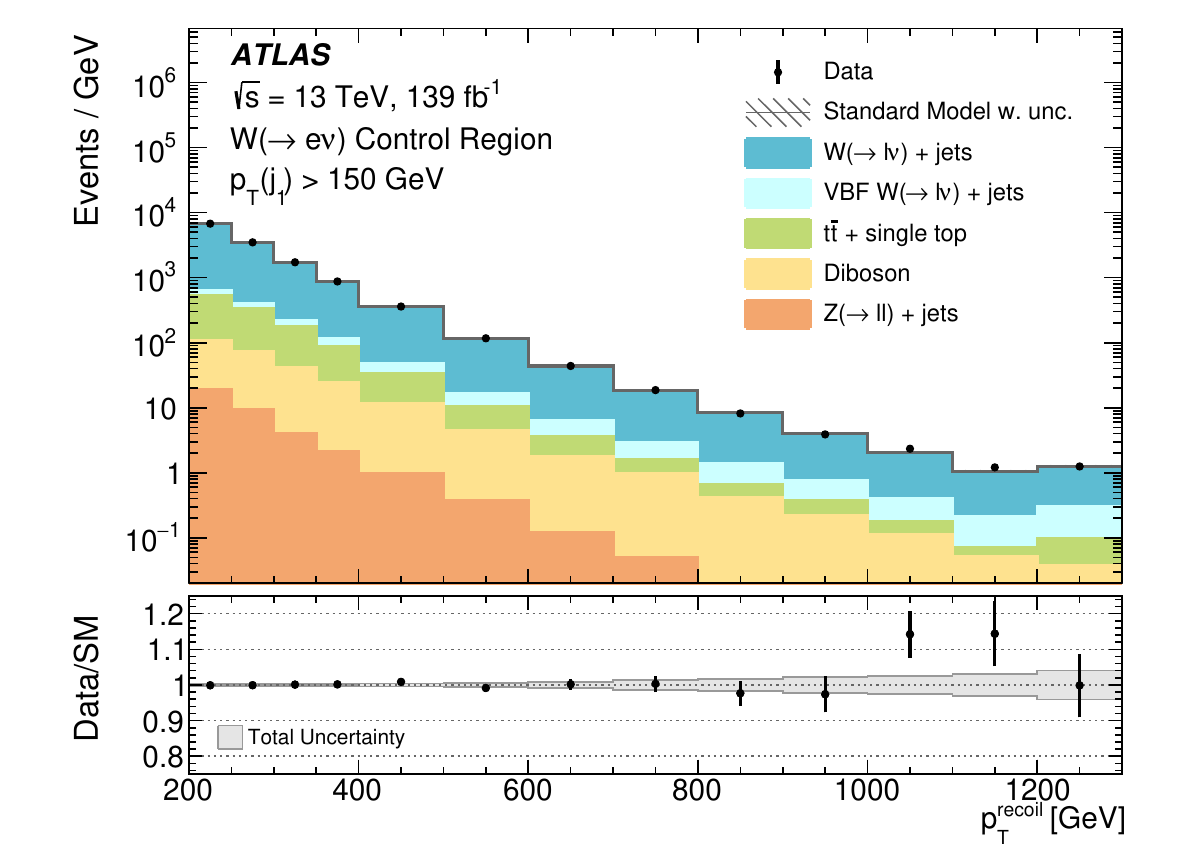}
 \label{fig:monoX:jet:ATLAS:CR}
}
\subfigure[ ATLAS, signal region]{
 \includegraphics[width=0.43\textwidth]{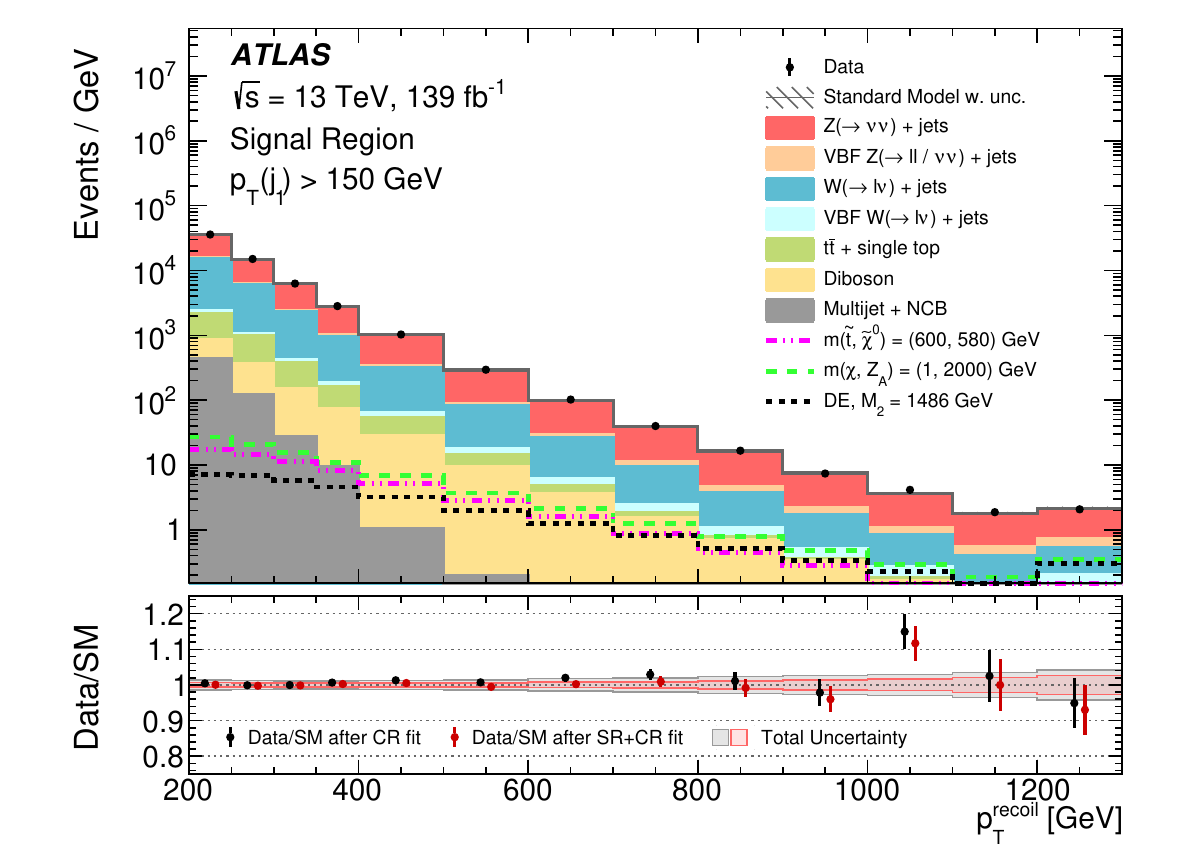}
 \label{fig:monoX:jet:ATLAS:SR}
}\\
\subfigure[ CMS, control region]{
 \includegraphics[width=0.4\textwidth]{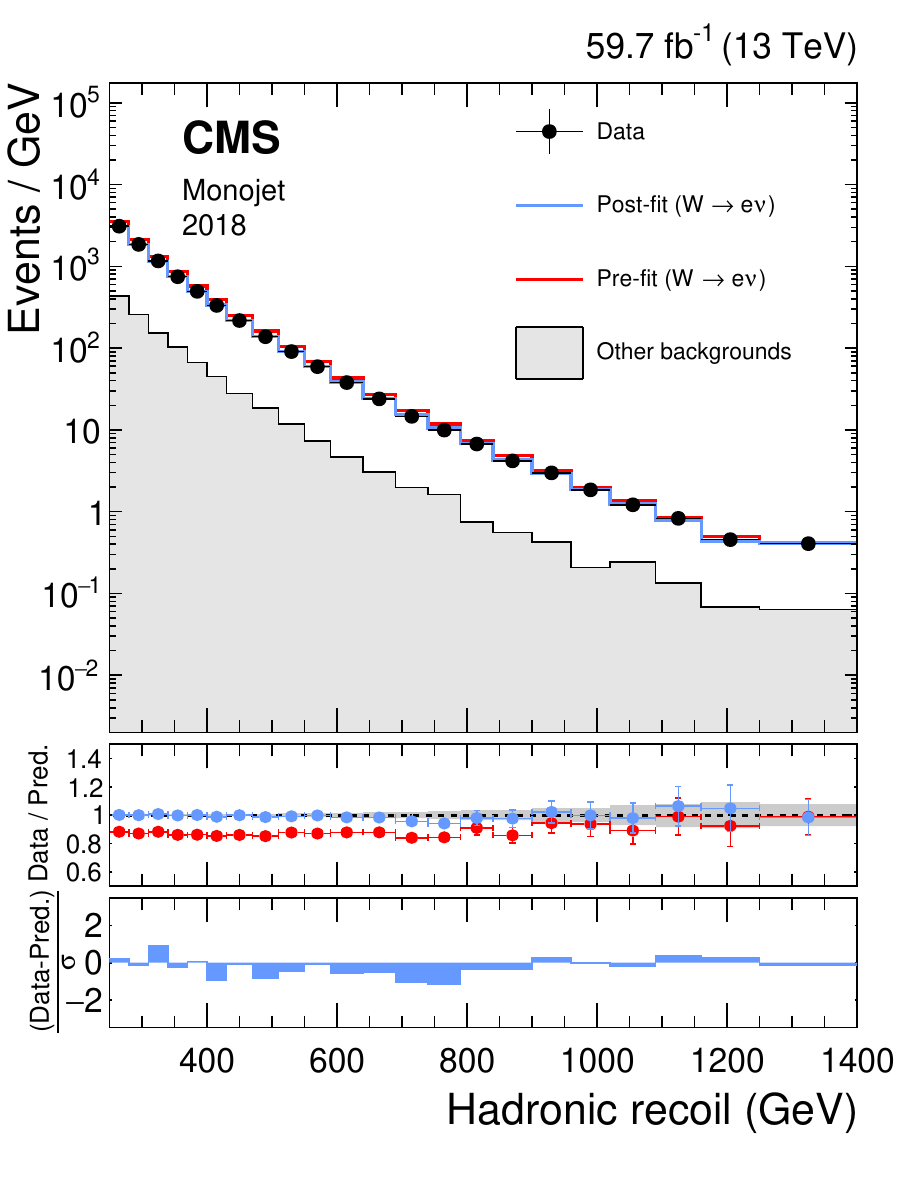}
 \label{fig:monoX:jet:CMS:CR}
}\hspace{0.5cm}
\subfigure[ CMS, signal region]{
 \includegraphics[width=0.4\textwidth]{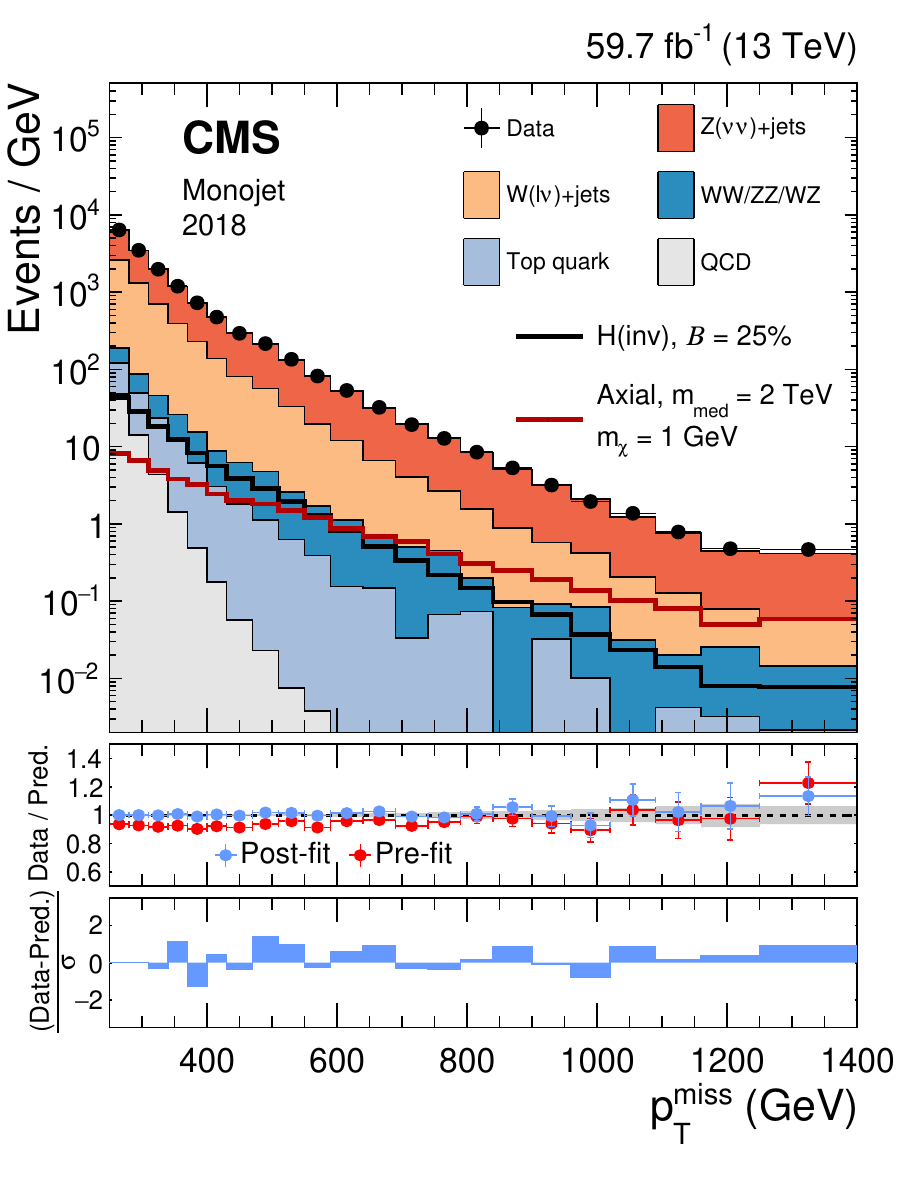}
 \label{fig:monoX:jet:CMS:SR}
}
\caption{(\textbf{a}) The $W\to{}e\nu$ control region and (\textbf{b}) signal region of the ATLAS monojet search, shown for the full Run 2 dataset \cite{ATLAS:MonoX_jet}. (\textbf{c}) The $W\to{}e\nu$ control region and (\textbf{d}) signal region of the CMS monojet search, shown for the 2018 dataset as part of a full Run 2 dataset result \cite{CMS:MonoX_jet}. \label{fig:monoX:jet:CRSR}}
\end{figure}

\vspace{-15pt}

\begin{figure}[H]

\subfigure[ ATLAS]{
 \includegraphics[width=0.5\textwidth]{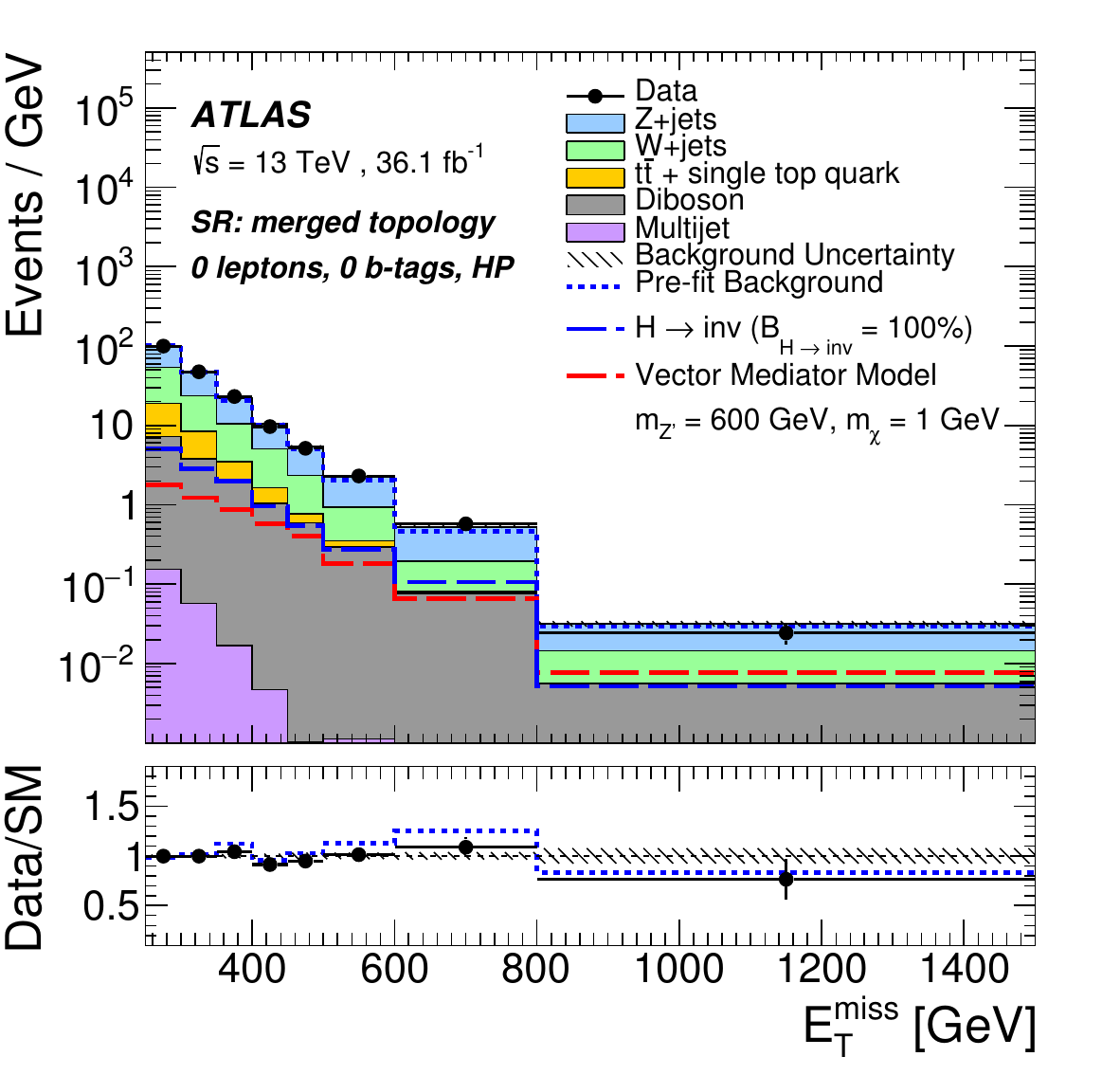}
 \label{fig:monoX:Vqq:ATLAS:SR2}
}
\subfigure[ CMS]{
 \includegraphics[width=0.38\textwidth]{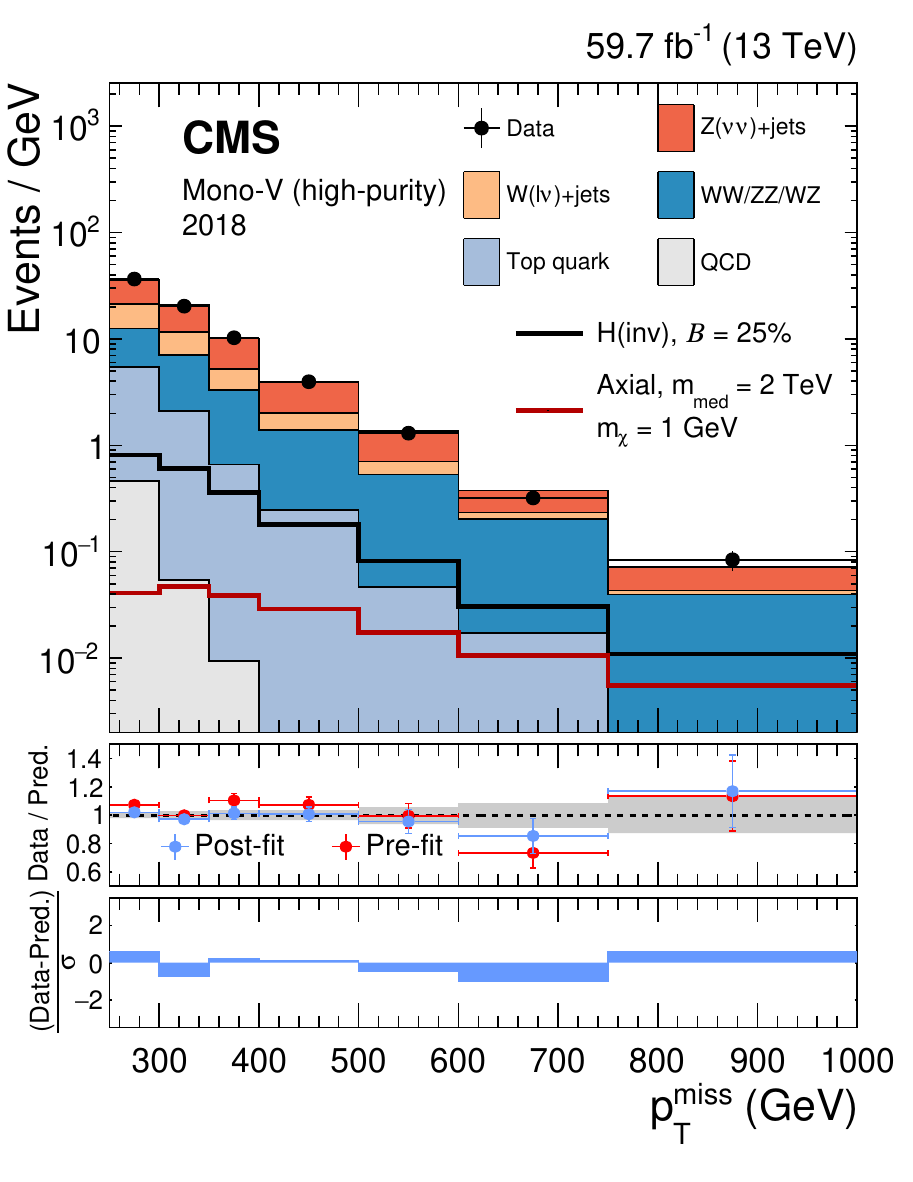}
 \label{fig:monoX:Vqq:CMS:SR2}
}
\caption{(\textbf{a}) The high-purity zero-$b$-tag signal region for the ATLAS hadronic mono-$V$ search, using a partial Run 2 dataset \cite{ATLAS:MonoX_Vqq}. (\textbf{b}) The high-purity signal region for the CMS hadronic mono-$V$ search, shown for the 2018 dataset as a part of a full Run 2 dataset result \cite{CMS:MonoX_jet}. \label{fig:monoX:Vqq}}
\end{figure}

No significant deviations from the background prediction are observed in any of the mono-jet or hadronic mono-$V$ searches, and thus limits are set on the production of various different mediators that couple the Standard Model to the dark sector.
Two key benchmark models studied by both ATLAS and CMS are the production of axial-vector mediators, and the production of pseudo-scalar mediators.
Limits on the production of both of these types of processes, as a function of the mediator mass and the dark matter mass, are shown for a given choice of the coupling between quarks and the mediator ($g_q=0.25$ for axial-vector, $g_q=1.0$ for pseudo-scalar) and the coupling between the mediator and dark matter ($g_\chi=1.0$), in Figure \ref{fig:monoX:jet:L}.
These limits are all shown for the full Run 2 dataset, and are for the mono-jet analysis signal region for ATLAS \cite{ATLAS:MonoX_jet}, while the CMS results include contributions from both the mono-jet and hadronic mono-$V$ signal regions \cite{CMS:MonoX_jet}.
% However, comparing the ATLAS mono-jet limits to the hadronic mono-$V$ limits for these specific models, the contribution of the mono-$V$ channel would be small.

\vspace{-8pt}
\begin{figure}[H]

\subfigure[ ATLAS, axial-vector mediator]{
 \includegraphics[width=0.45\textwidth]{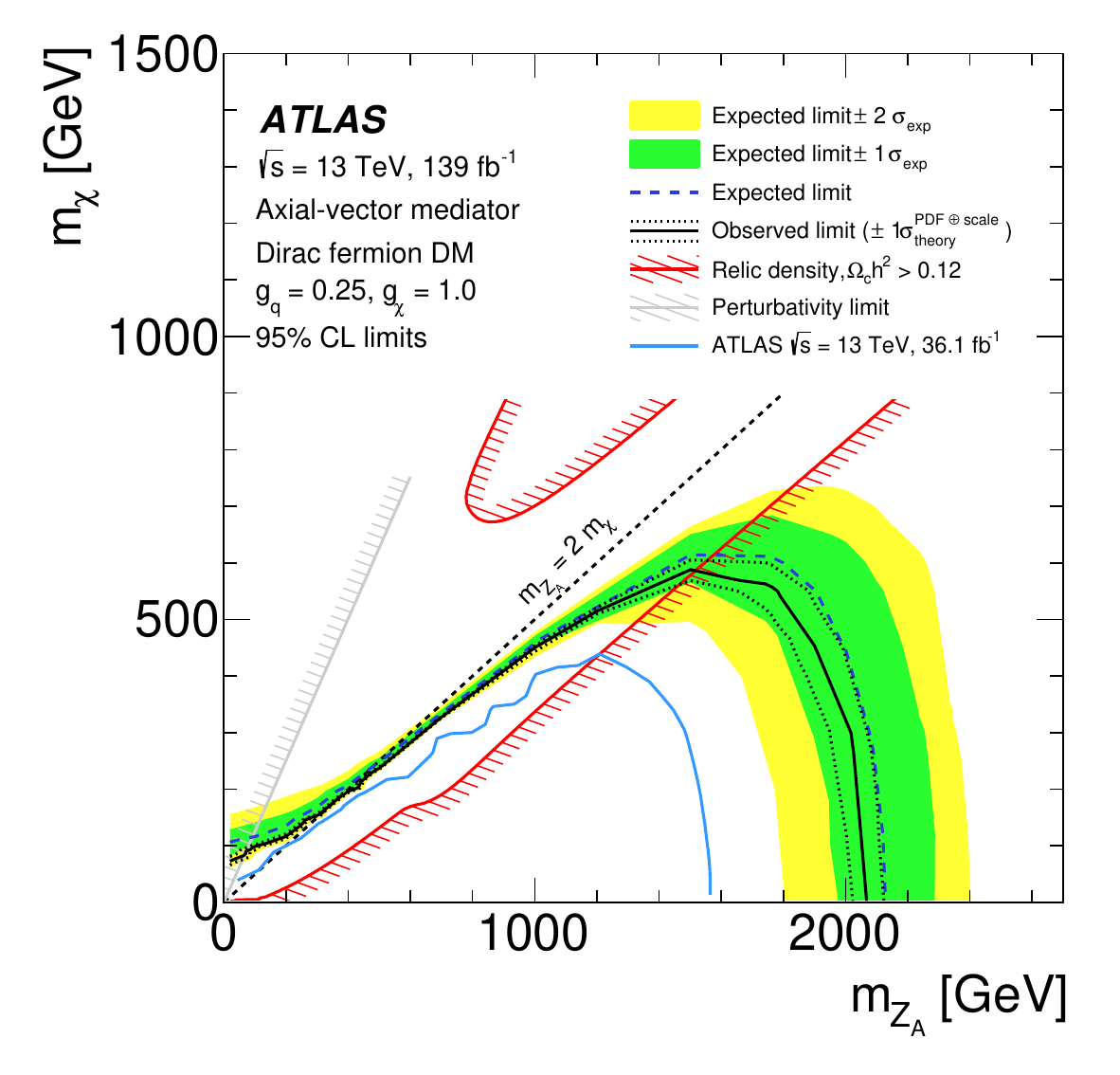}
 \label{fig:monoX:jet:ATLAS:LV}
}\hspace{0.5cm}
\subfigure[ ATLAS, pseudo-scalar mediator]{
 \includegraphics[width=0.45\textwidth]{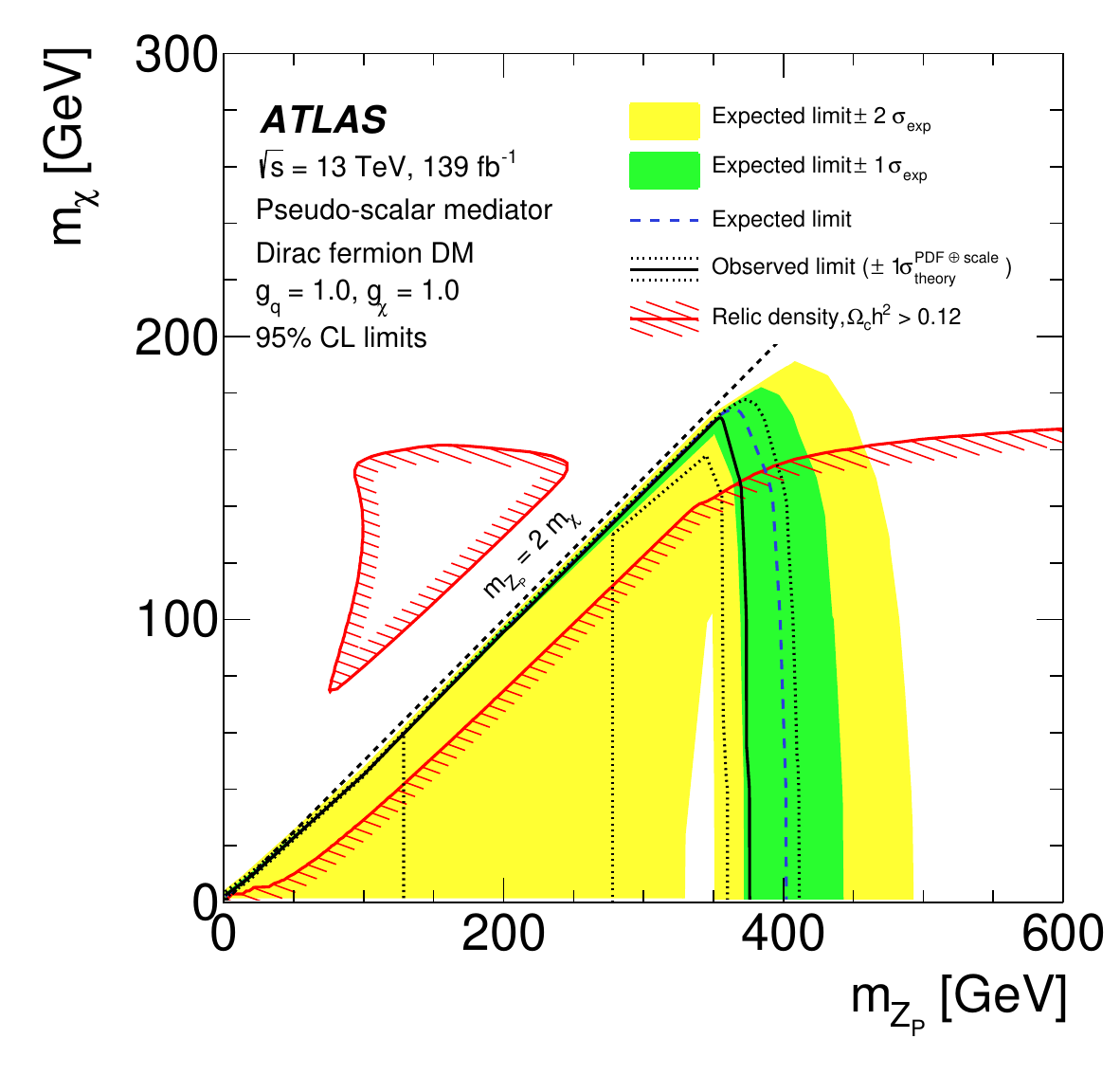}
 \label{fig:monoX:jet:ATLAS:LS}
}\\
\subfigure[ CMS, axial-vector mediator]{
 \includegraphics[width=0.48\textwidth]{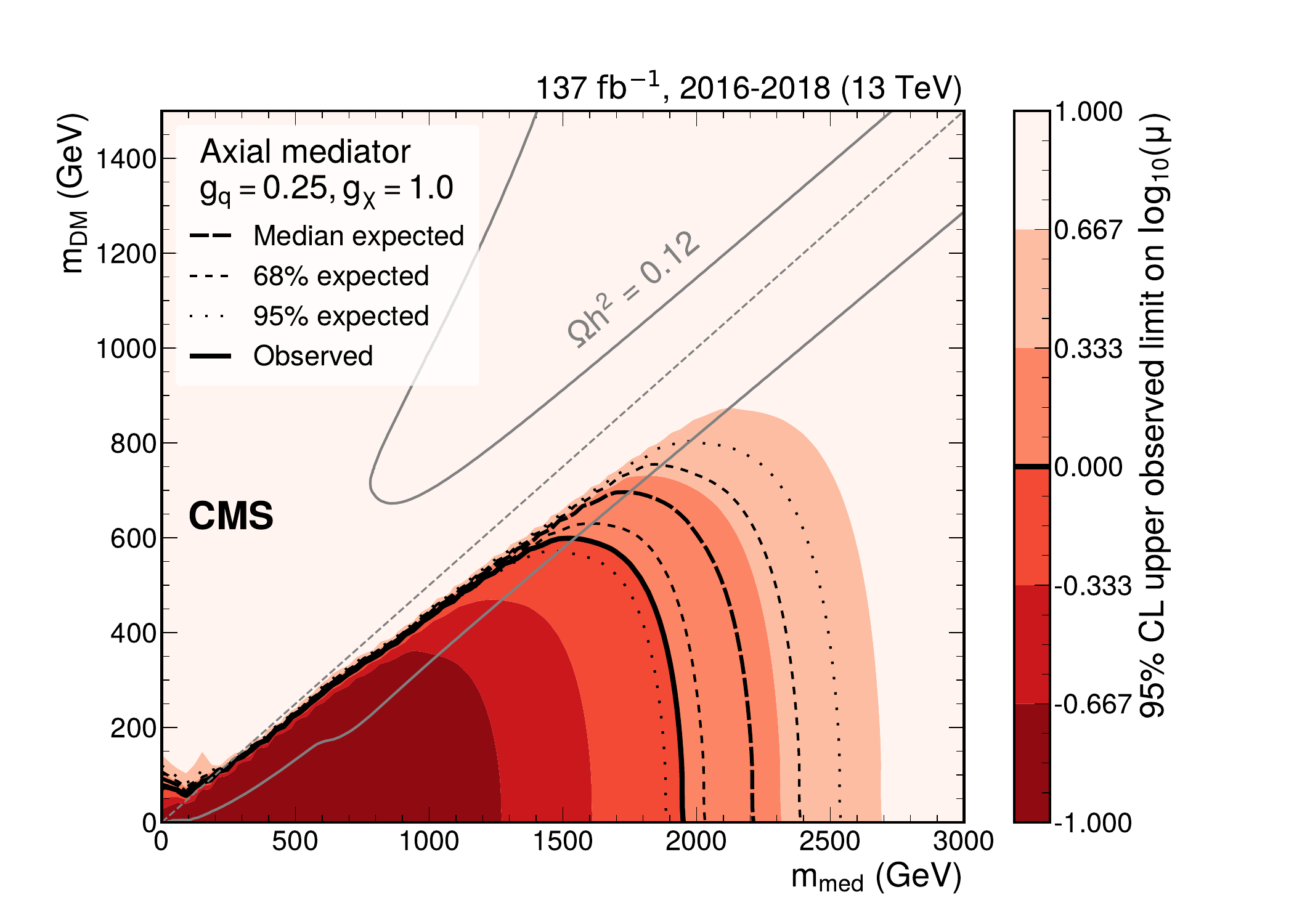}
 \label{fig:monoX:jet:CMS:LV}
}
\subfigure[ CMS, pseudo-scalar mediator]{
 \includegraphics[width=0.48\textwidth]{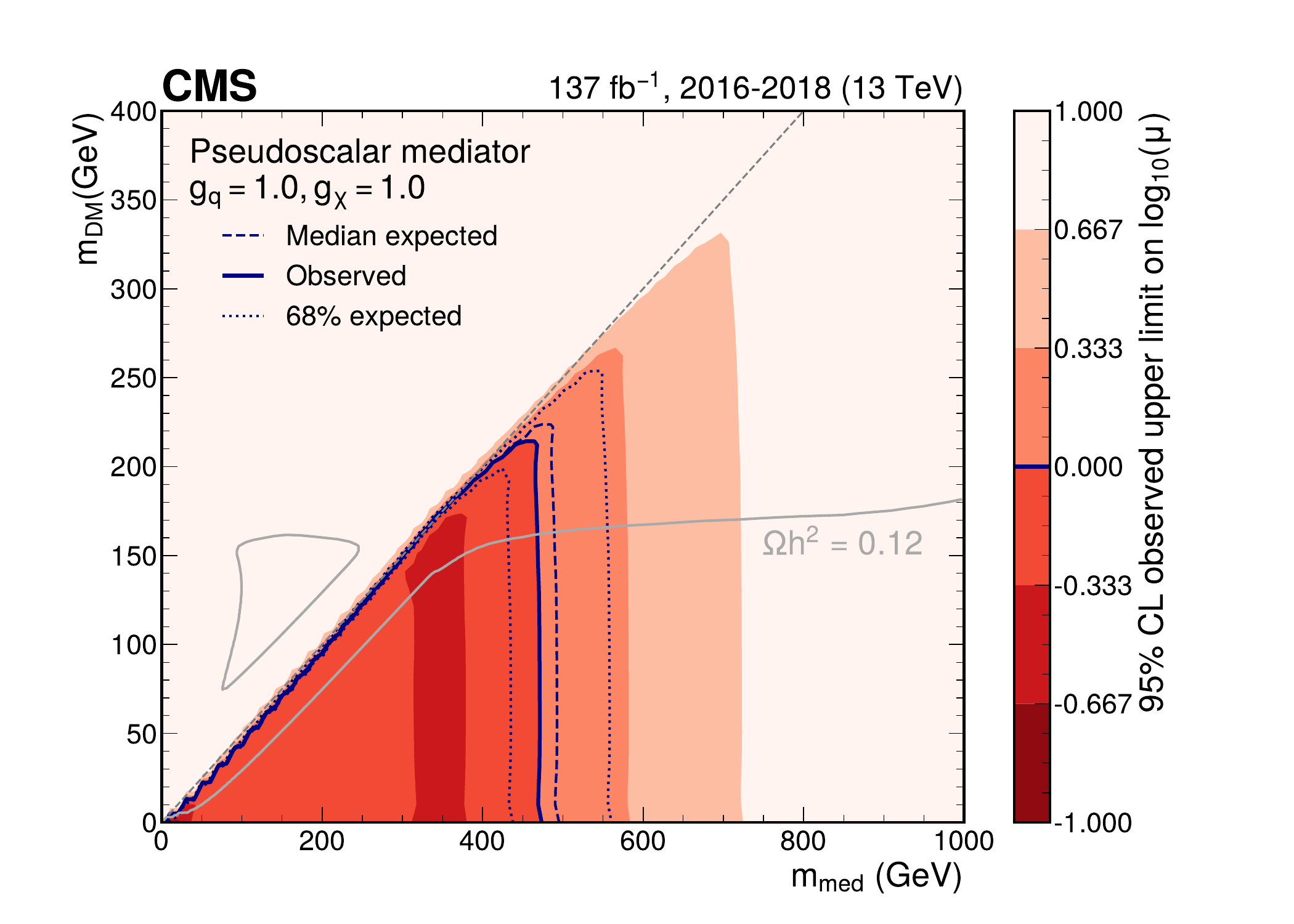}
 \label{fig:monoX:jet:CMS:LS}
}
\caption{The limits set by the ATLAS mono-jet search on the production of new (\textbf{a}) axial-vector mediators of mass $m_{Z_{A}}$ and (\textbf{b}) pseudo-scalar mediators of mass $m_{Z_{p}}$, which in turn decay to pairs of dark matter particles of a given mass $m_\chi$ \cite{ATLAS:MonoX_jet}. The limits set by the CMS mono-jet and hadronic mono-$V$ search on the production of new (\textbf{c}) axial-vector mediators and (\textbf{d}) pseudo-scalar mediators, both of mass $m_\mathrm{med}$, which in turn decay to pairs of dark matter particles of a given mass $m_\mathrm{DM}$ \cite{CMS:MonoX_jet}. \label{fig:monoX:jet:L}}
\end{figure}

\subsection{Other Missing Transverse Momentum Searches}
\label{sec:monoX:other}
% So many variants, mention as overview not in detail
% Give examples of relevance to scalar/pseudoscalar for heavy flavour

While the mono-jet final state is generally the most sensitive to the previously presented dark matter models, the second most abundant source of ISR at ATLAS and CMS is photons, not $W$ and $Z$ bosons.
The missing transverse momentum plus ISR photon search is thus also an important part of the search programme, and along the same lines as the mono-jet and hadronic mono-$V$ searches, it is often referred to as the mono-photon analysis.
ATLAS has published a mono-photon analysis using the full Run 2 dataset \cite{ATLAS:MonoX_photon}, while the corresponding CMS analysis currently uses a partial Run 2 dataset \cite{CMS:MonoX_photon}.
Similar control regions to the mono-jet search are used to estimate the $Z(\to\nu\nu)+\gamma$ background and other $W/Z+\gamma$ backgrounds, just with the associated jet replaced by an associated photon.
The CMS search is further divided into ``vertical'' and ``horizontal'' signal regions, which are defined in such a way that the contribution from background beam halo events can be determined.
The resulting signal regions for ATLAS and CMS are shown in Figure \ref{fig:monoX:photon}a,b, respectively.
No significant deviations from the background prediction are observed, and thus both analyses proceed to set limits on the production of axial-vector mediators coupling the Standard Model to dark matter, which are correspondingly shown in Figure \ref{fig:monoX:photon}c,d.
Comparing these limits with those shown in Figure \ref{fig:monoX:jet:L}, it is clear that the mono-jet analysis is more sensitive to the models shown here.

\vspace{-8pt}

\begin{figure}[H]

\subfigure[]{
 \includegraphics[width=0.51\textwidth]{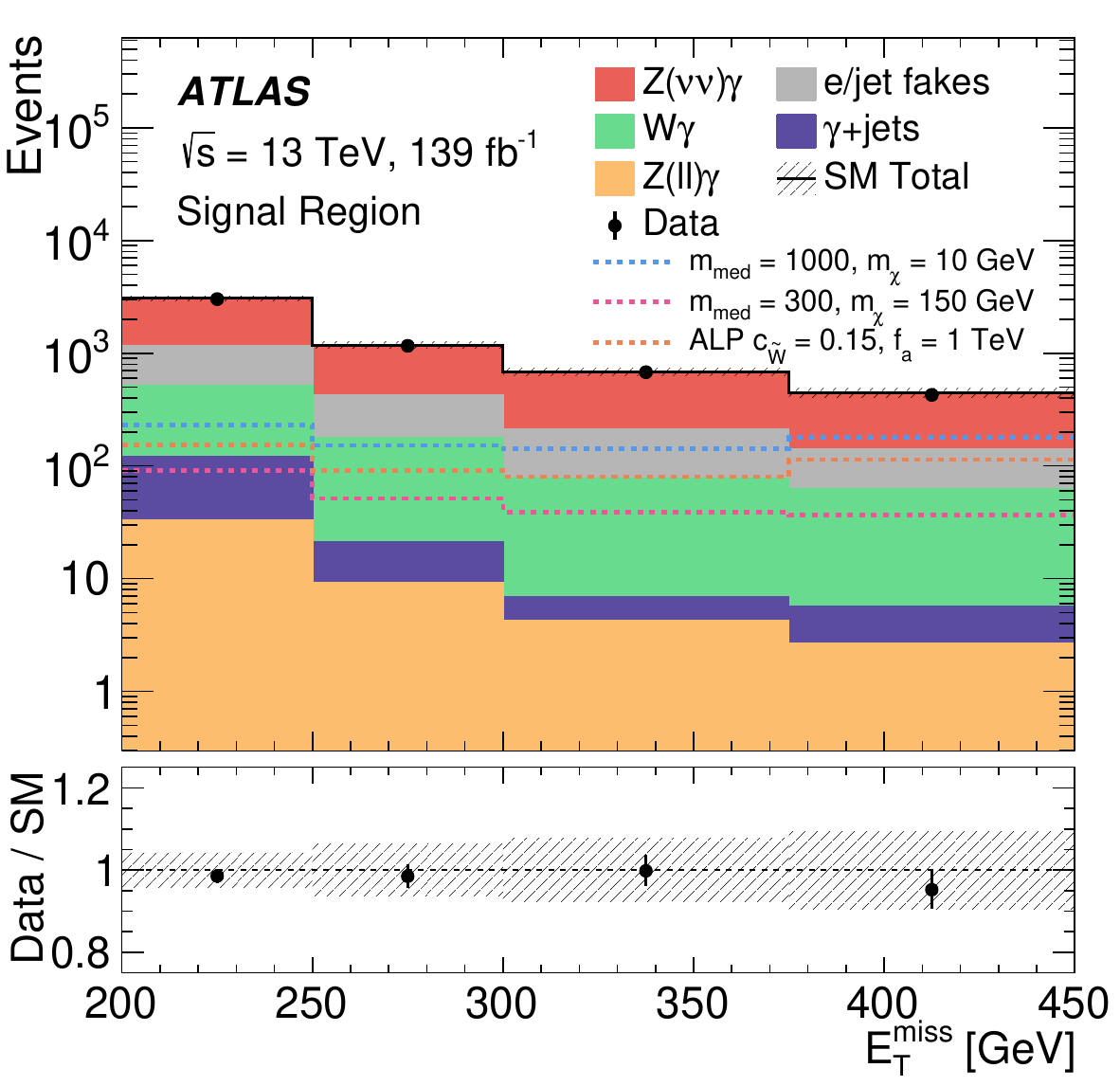}
 \label{fig:monoX:photon:ATLAS:SR}
}
\subfigure[]{
 \includegraphics[width=0.45\textwidth]{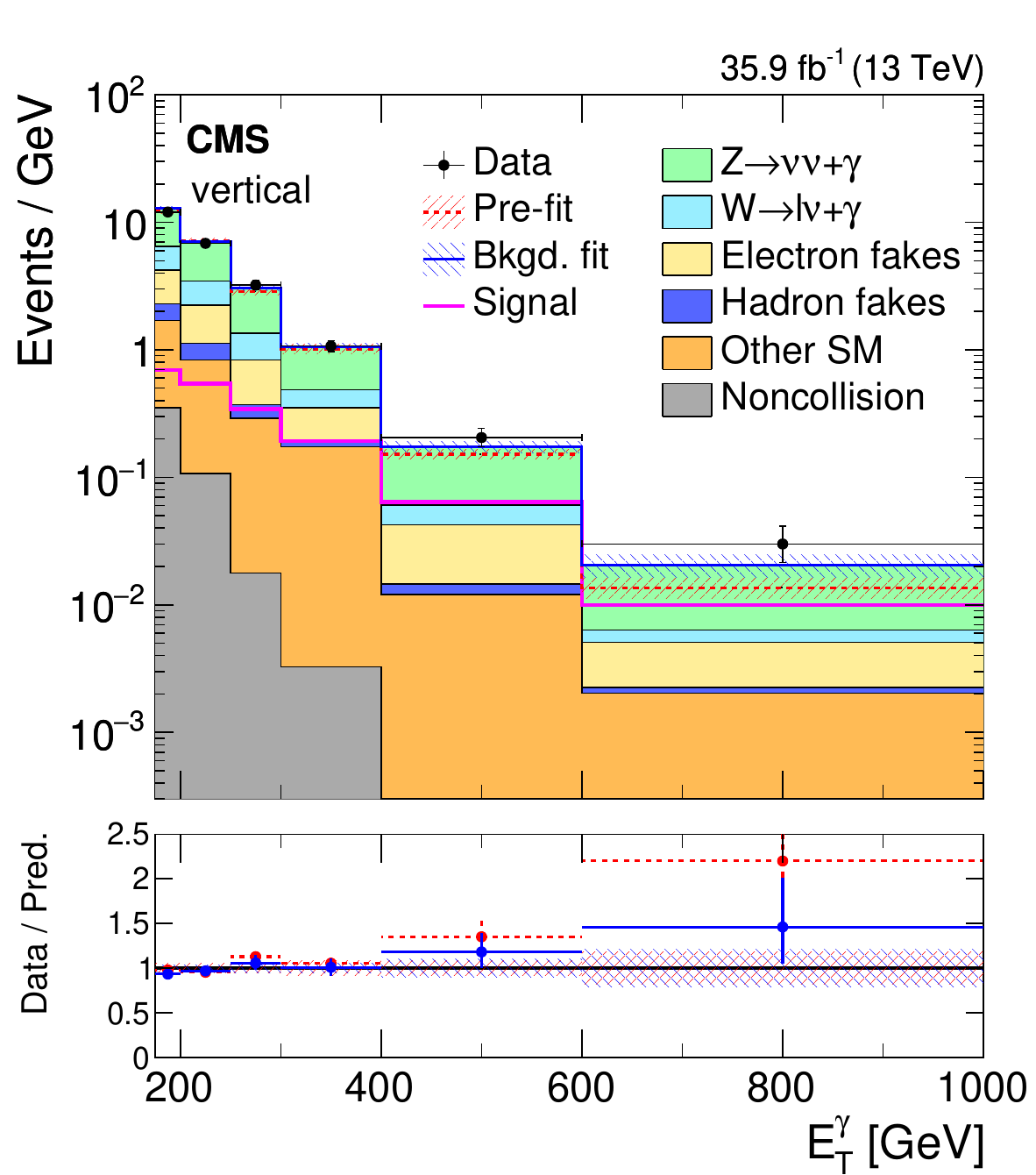}
 \label{fig:monoX:photon:CMS:SR}
}\vspace{-9pt} \\
\subfigure[]{
 \includegraphics[width=0.45\textwidth]{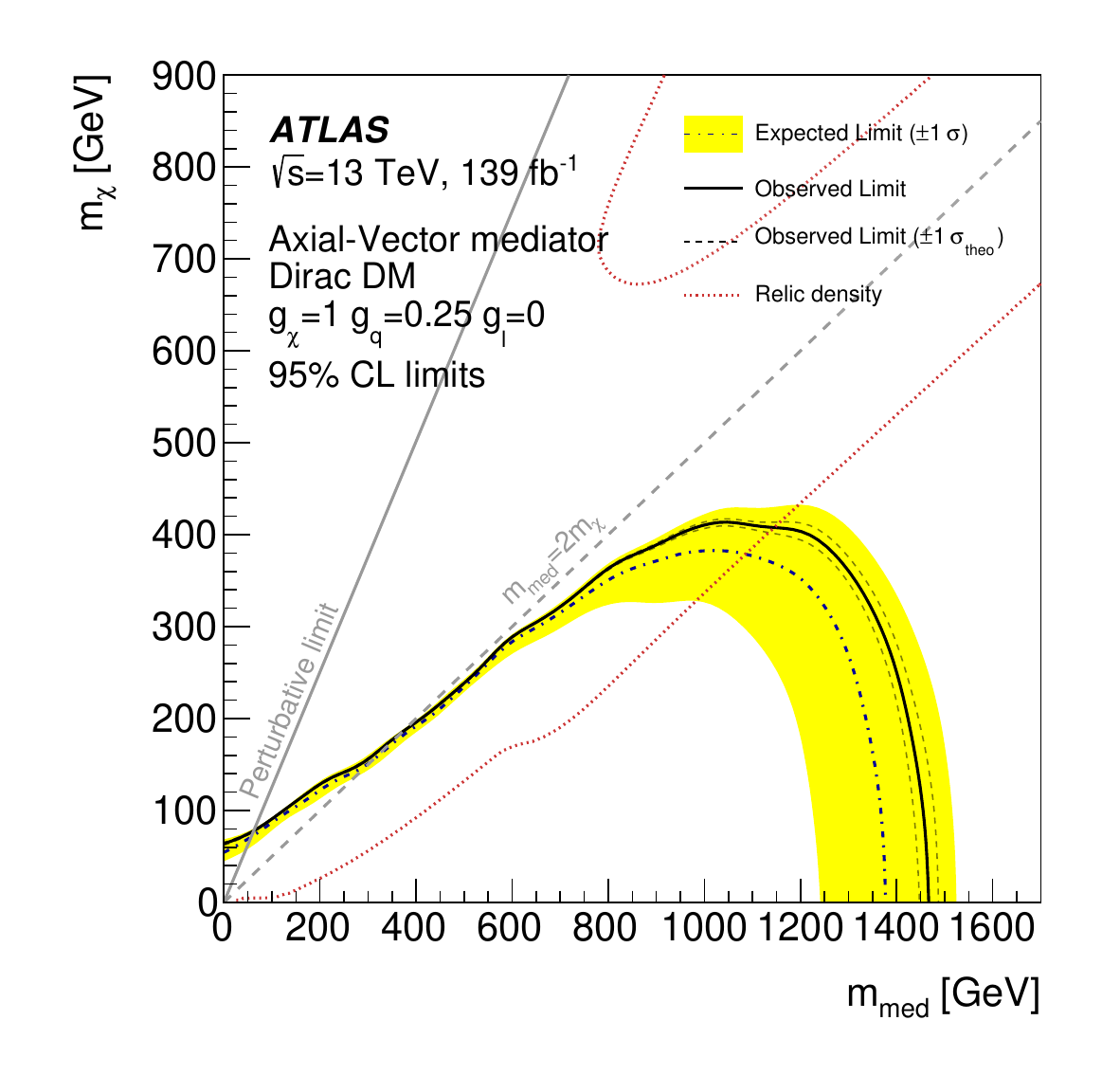}
 \label{fig:monoX:photon:ATLAS:LV}
}
\subfigure[]{
 \includegraphics[width=0.51\textwidth]{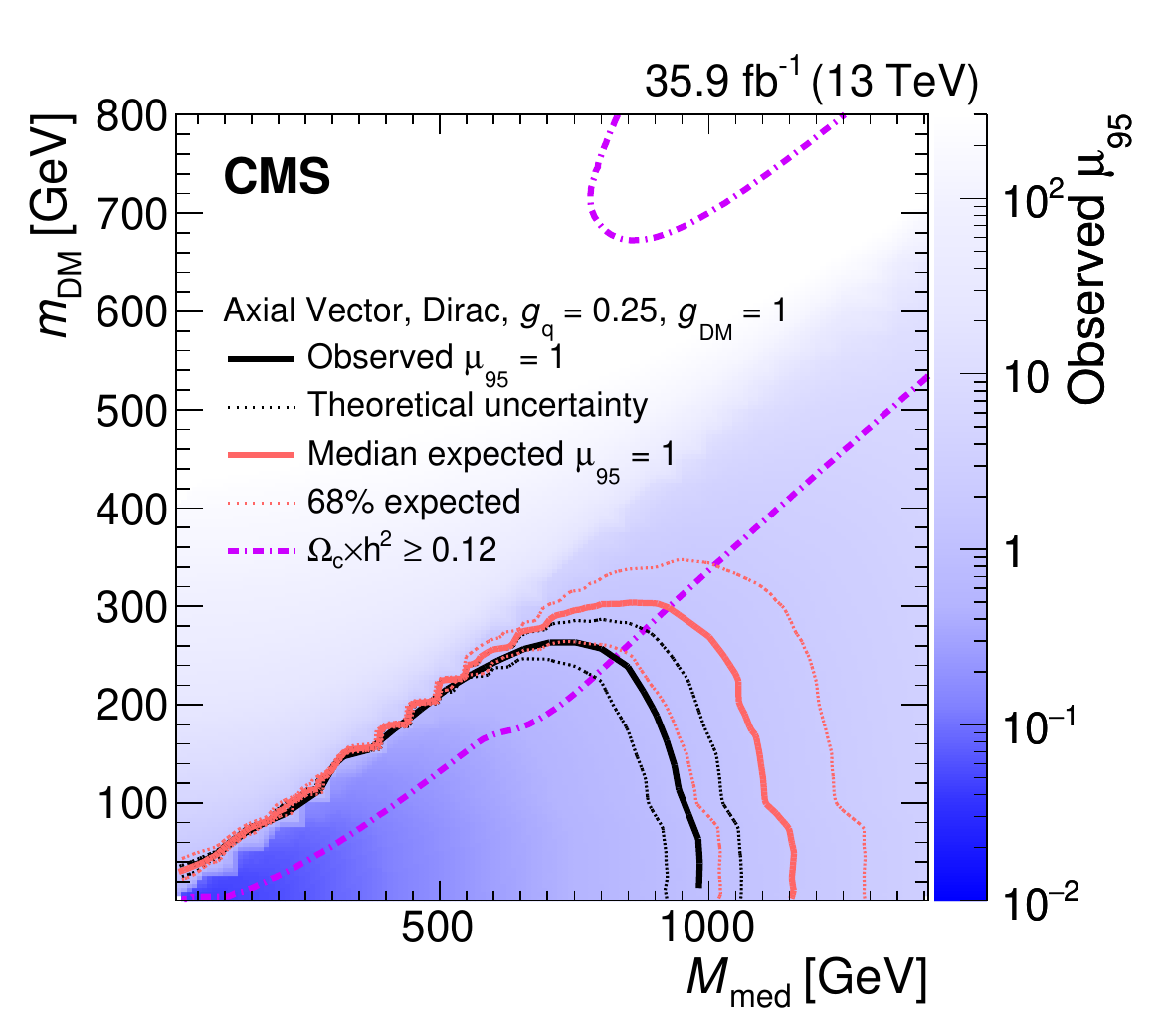}
 \label{fig:monoX:photon:CMS:LV}
}
\caption{(\textbf{a}) The signal region of the mono-photon analysis, as observed by ATLAS using the full Run 2 dataset \cite{ATLAS:MonoX_photon}. (\textbf{b}) The vertical signal region of the mono-photon analysis, as observed by CMS using a partial Run 2 dataset \cite{CMS:MonoX_photon}. No significant deviation from the background expectation is observed in either analysis, and thus both set corresponding limits on the production of new axial-vector mediators of mass $m_\mathrm{med}$, which in turn decay to pairs of dark matter particles of a given mass $m_\chi$ or $m_\mathrm{DM}$, shown for (\textbf{c}) ATLAS and (\textbf{d}) CMS.\label{fig:monoX:photon}}
\end{figure}

In addition to the mono-(jet/$V$/$\gamma$) searches presented so far, there are a vast number of other searches for missing transverse momentum in association with other objects that can be interpreted in the context of mediators connecting the Standard Model to dark matter.
In particular, the sensitivity to scalar mediators with Yukawa couplings is enhanced by the presence of massive objects, such as top or bottom quarks.
Such signatures include large amounts of missing transverse momentum in addition to single-top quarks, $t\bar{t}$ pairs, single-bottom quarks, $b\bar{b}$ pairs, or top+bottom.
While many of these signatures include hadronic final states, they are not discussed in detail here; these types of signatures are of great relevance to other types of models, and are discussed in a parallel review \cite{Symmetry:ThirdGen}.

In order to briefly demonstrate the relevance of such other final states in the search for scalar and pseudo-scalar mediators decaying to dark matter, plots showing the sensitivity of a variety of different signatures are shown for both ATLAS \cite{ATLAS:DMSummary} and CMS \cite{CMS:EXOSummary} in Figure \ref{fig:monoX:comb}.
The mono-jet process may have the largest cross-section, but the quark--antiquark annihilation usually occurs between low-mass quarks.
As such, much more rare processes including the production of pairs of top quarks in association with the scalar or pseudo-scalar mediator can actually have better sensitivity, as the coupling to the top quark mass compensates for the lower cross-section.
These other signatures are therefore an important part of the ATLAS and CMS physics programme in the context of the search for the pair-production of dark matter at the LHC.
\vspace{-4pt}

\begin{figure}[H]

\subfigure[ ATLAS, scalar mediator]{
 \includegraphics[width=0.48\textwidth]{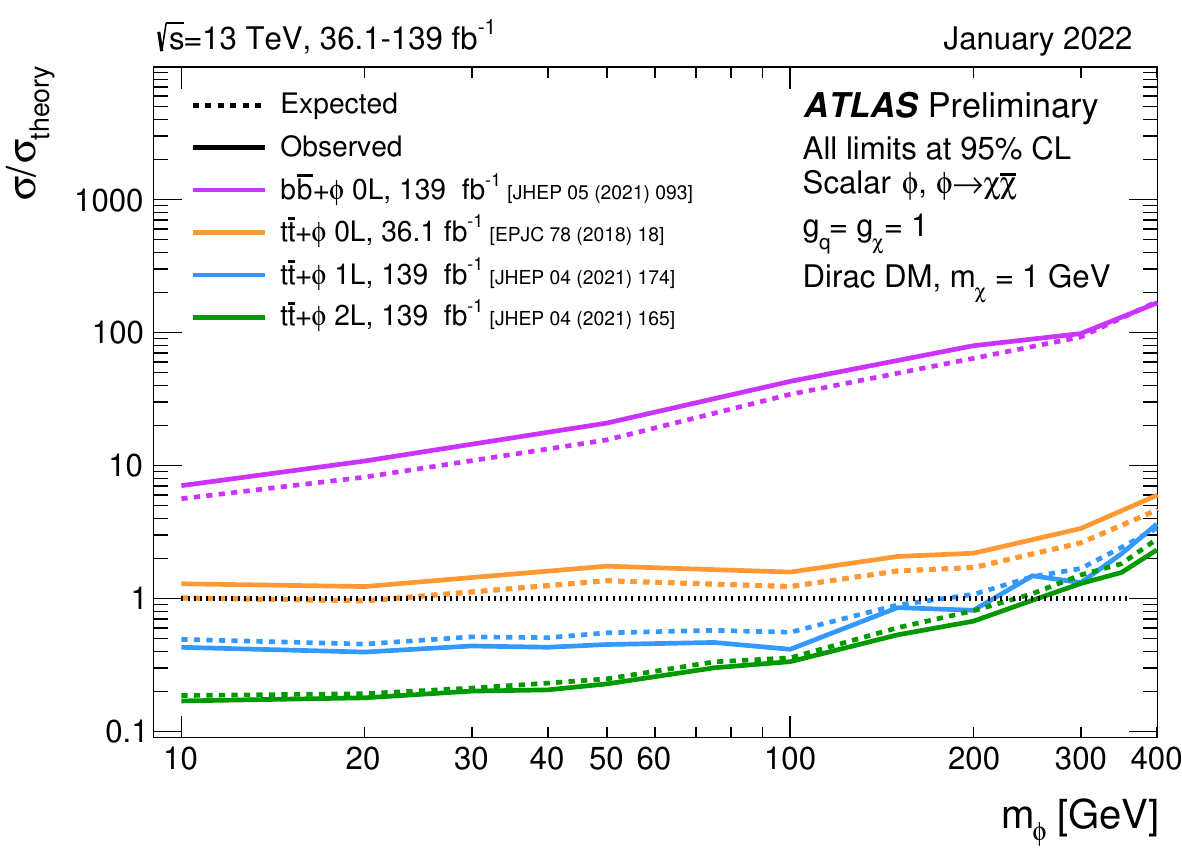}
 \label{fig:monoX:comb:ATLAS:scalar}
}
\subfigure[ ATLAS, pseudo-scalar mediator]{
 \includegraphics[width=0.48\textwidth]{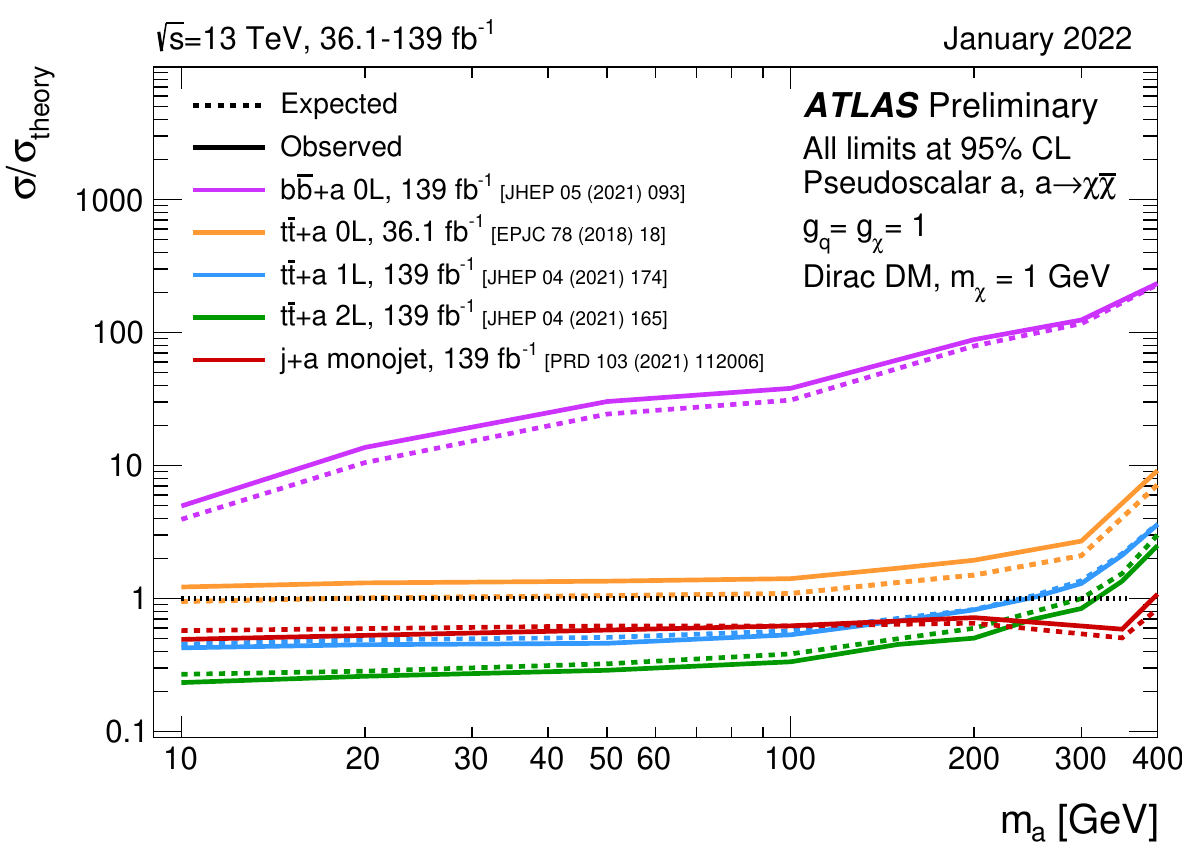}
 \label{fig:monoX:comb:ATLAS:pseudoscalar}
}\\
\subfigure[ CMS, scalar mediator]{
 \includegraphics[width=0.48\textwidth]{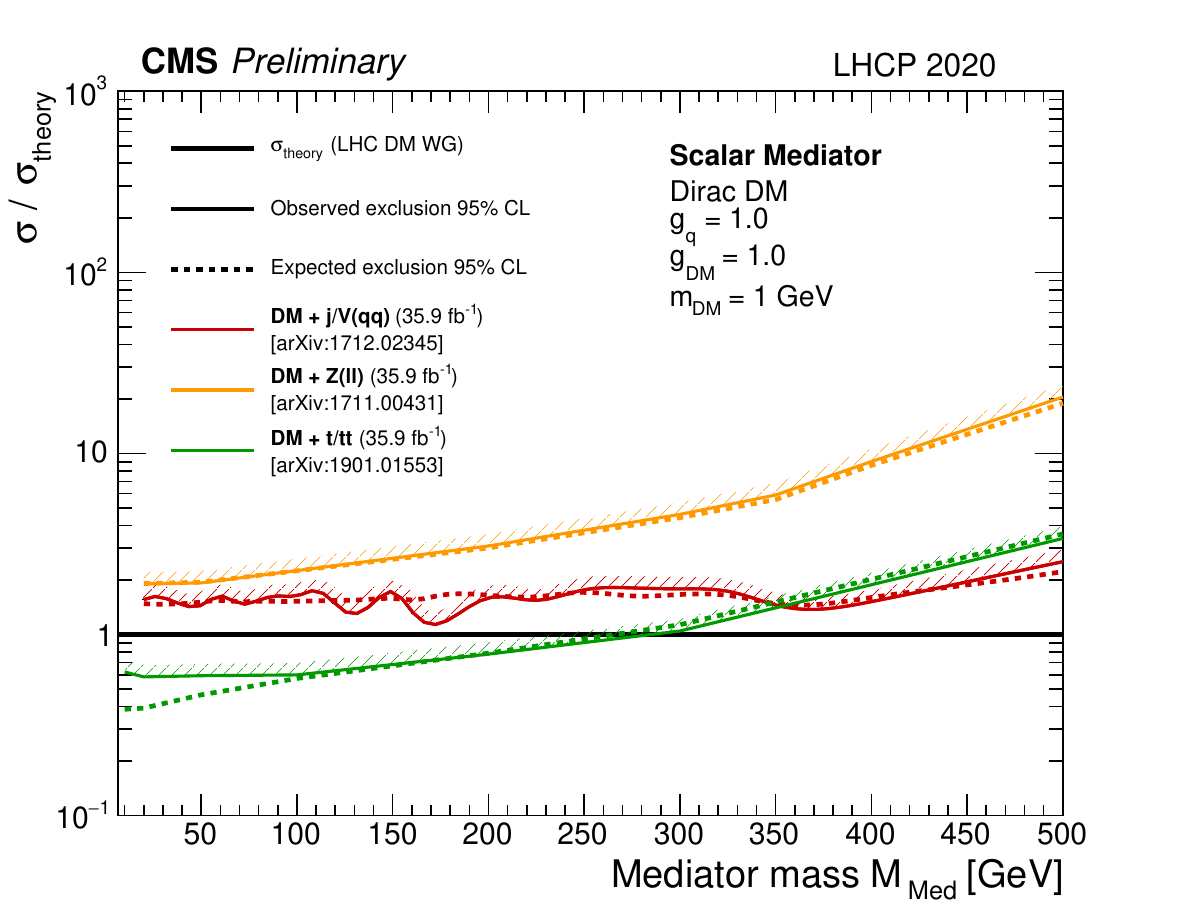}
 \label{fig:monoX:comb:CMS:scalar}
}
\subfigure[ CMS, pseudo-scalar mediator]{
 \includegraphics[width=0.48\textwidth]{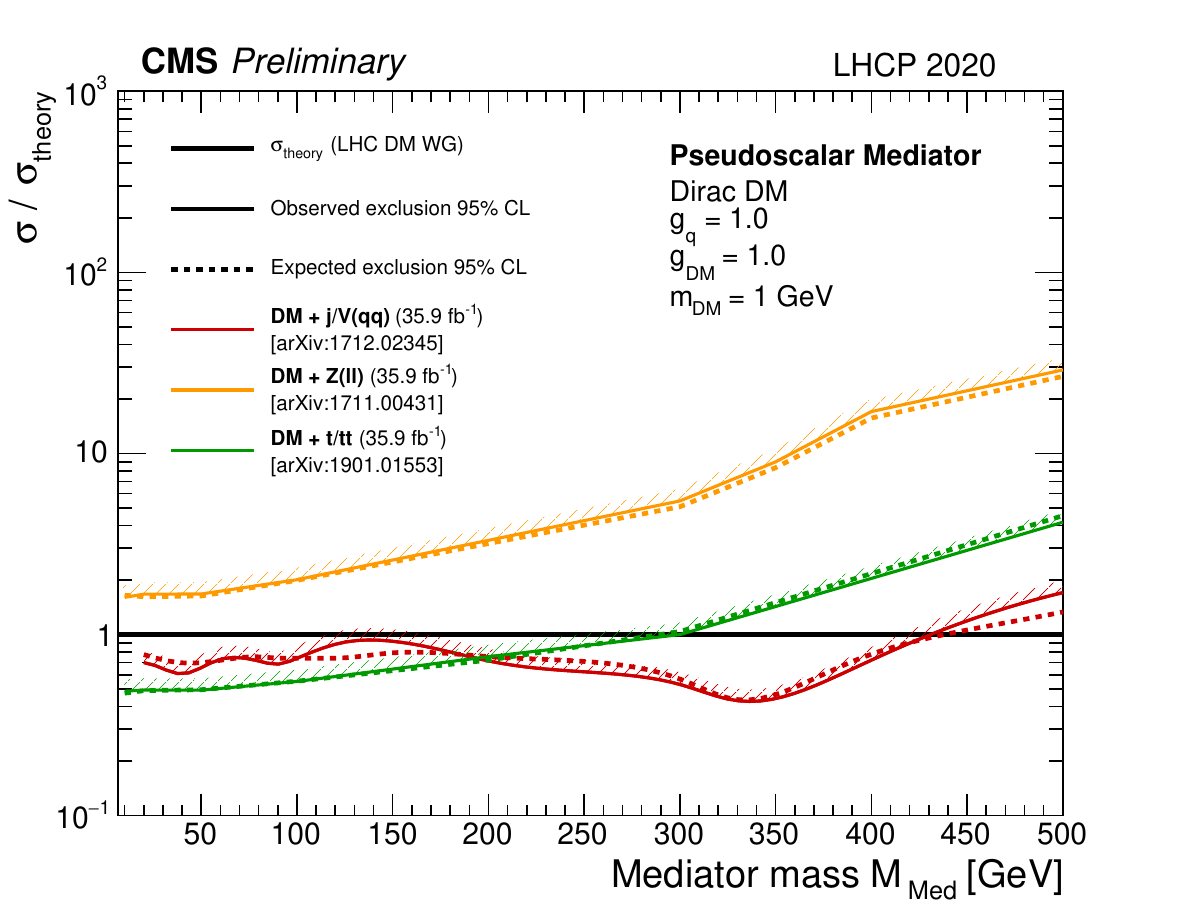}
 \label{fig:monoX:comb:CMS:pseudoscalar}
}
\caption{A comparison of the sensitivity of various different searches for (\textbf{a},\textbf{c}) scalar or (\textbf{b},\textbf{d}) pseudo-scalar mediators coupling the Standard Model to dark matter, as studied by (\textbf{a},\textbf{b}) ATLAS \cite{ATLAS:DMSummary} and (\textbf{c},\textbf{d}) CMS \cite{CMS:EXOSummary}. Scalar and pseudo-scalar mediators are sensitive to the mass of the radiated particle, and thus processes with smaller cross-sections but more massive particles can still have comparable or better sensitivity than the low-mass but highest-cross-section mono-jet process. \label{fig:monoX:comb}}
\end{figure}

% Di-boson searches
\section{Hadronic Di-Boson Searches}
\label{sec:VV}

Searches for new particles decaying to pairs of electroweak bosons are sensitive to a wide variety of models of new physics.
Pairs of bosons can be the by-product of new mediators of spin 0, spin 1, or spin 2, thus probing many interesting possibilities.
The link with spin 0 mediators means that such di-boson final states are also of great interest to searches for Higgs couplings that do not match Standard Model expectations, or to new scalar particles; most di-boson searches have thus been covered in a parallel review to this one \cite{Symmetry:HBSM}.
This review will therefore focus on searches for di-boson production in the fully hadronic final state, and where neither boson is a Higgs boson, the Feynman diagram for which is shown in Figure \ref{fig:VV:feyn}.
Results of such searches are typically interpreted in terms of benchmark models, including spin 0 Radions \cite{Theory:Radion1}, spin 1 Heavy Vector Triplets (HVTs) or $W^\prime$ and $Z^\prime$ bosons \cite{Theory:HVT1,Theory:HVT2,Theory:HVT3}, and spin 2 bulk RS gravitons \cite{Theory:Graviton1,Theory:Graviton2,Theory:Graviton3,Theory:Graviton4,Theory:Graviton5,Theory:Graviton6,Theory:Graviton7,Theory:Graviton8,Theory:Graviton9,Theory:Graviton10}.

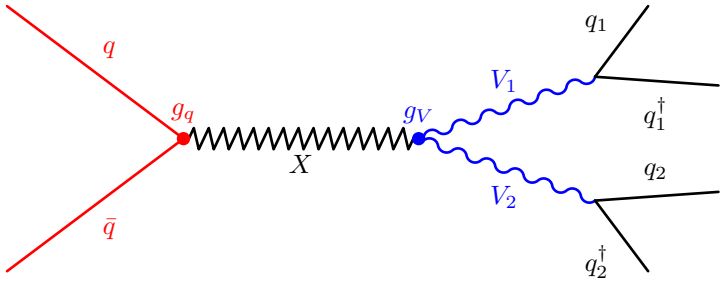
\begin{figure}[H]
\hspace{-25pt}
\begin{fmffile}{VVres}
 \begin{fmfgraph*}(300,100)
 \fmfleft{qbar,q}
 \fmfright{qbar2,q2,qbar1,q1}
 
 \fmf{vanilla,foreground=red,label=$\textcolor{red}{q}$}{q,vertexL}
 \fmf{vanilla,foreground=red,label=$\textcolor{red}{\bar{q}}$}{qbar,vertexL}
 
 \fmf{zigzag,tension=1.5,label=$X$}{vertexL,vertexR}
 
 \fmf{boson,foreground=blue,label=$\textcolor{blue}{V_1}$,label.side=left}{vertexR,boson1}
 \fmf{boson,foreground=blue,label=$\textcolor{blue}{V_2}$}{vertexR,boson2}
 
 \fmf{vanilla,label=$q_1$,label.side=left}{boson1,q1}
 \fmf{vanilla,label=$q^\dagger_1$,label.side=right}{boson1,qbar1}
 \fmf{vanilla,label=$q_2$,label.side=left}{boson2,q2}
 \fmf{vanilla,label=$q^\dagger_2$}{boson2,qbar2}
 
 \fmfv{decor.shape=circle,decor.filled=full,
decor.size=2thick,foreground=red,label=$\textcolor{red}{g_q}$,label.angle=90}{vertexL}
 \fmfv{decor.shape=circle,decor.filled=full,
decor.size=2thick,foreground=blue,label=$\textcolor{blue}{g_V}$,label.angle=90}{vertexR}
 \end{fmfgraph*}
\end{fmffile}
\caption{Feynman diagrams showing the s-channel production of a generic new mediator $X$, which is produced through the annihilation of standard model quarks via a coupling $g_q$, and which decays to pairs of standard model electroweak bosons ($WW$, $WZ$, or $ZZ$) via a coupling $g_V$. The standard model electroweak bosons further decay to pairs of quarks.\label{fig:VV:feyn}}
\end{figure}

\subsection{Searches with Standard Model Bosons}
\label{sec:VV:ewk}
% VVJJ
% WVLJ
% ZVLJ
% ZVnJ

% ATLAS VVJJ \cite{ATLAS:VV_JJ} and CMS VVJJ \cite{CMS:VV_JJ}
% 
% CMS VVNJ \cite{CMS:VV_NJ}
% 
% ATLAS VVLJ \cite{ATLAS:VV_LJ} and CMS VVLJ \cite{CMS:VV_LJ}
% 
% All but VVJJ and CWOLA covered by Arnaud's review \cite{Symmetry:HBSM}.

Searches for di-boson production in the fully hadronic final state are primarily motivated by the large branching ratios of the Standard Model bosons to pairs of quarks, as discussed in Section \ref{sec:motivation:motivations} and Table \ref{tab:motivation:BRs}.
The hadronic final state can thus have a larger statistical power than fully leptonic or semi-leptonic final states, and may therefore be the first final state to observe new physics, especially in the highest-accessible-energy regime.
This means that fully hadronic searches are primarily of interest where the resulting $W$ and $Z$ bosons are produced at very high energy, and thus their subsequent decays to pairs of quarks are highly collimated.
These collimated decays are then reconstructed as pairs of \largeR{} jets, where each of the $W$ and/or $Z$ bosons form one such jet, using the techniques discussed in Section \ref{sec:reco:largeR}.

Fully hadronic di-boson searches thus take the form of searches for di-\largeR{}-jet events; many techniques thus overlap with the di-jet searches discussed in Section \ref{sec:dijet}.
However, in the case of di-boson searches, the enormous Standard Model multi-jet and $W/Z$+jet backgrounds must be overcome in order to be sensitive to rare new di-boson physics.
The key to achieving this requirement is the development of powerful taggers, which can differentiate between \largeR{} jets consistent with originating from $W$ or $Z$ bosons against those originating from non-top-quarks or gluons, as discussed in Section \ref{sec:reco:largeR:tag}.
These taggers suppress the Standard Model backgrounds by several orders of magnitude, leaving behind a small but not insignificant fraction of multi-jet and $W/Z$+jet events.
% This poses a further challenge, as the corresponding simulated samples are currently not reliable to 

The search for fully hadronic di-boson resonances starts from this baseline.
The background estimate must be derived in a data-driven way, as the simulated samples cannot be trusted to properly represent the tiny fractions of background events that survive the jet taggers, yet these surviving background events remain a sizeable contribution with respect to the expected number of signal events.
The strategy of searching for resonances therefore allows for the smoothly-falling background to be estimated using functional forms, with the signal hypothesis confined to a narrow region of the spectrum.
This is the approach followed by both ATLAS \cite{ATLAS:VV_JJ} and CMS \cite{CMS:VV_JJ,CMS:VV_JJ_prelim} in their latest searches for fully hadronic di-boson resonances, where the published ATLAS result and the preliminary CMS result both use the full Run 2 dataset.

Both searches cut very tightly on their jet taggers in order to suppress the Standard Model background, leaving only a small fraction of background events behind.
However, in order to obtain the required background suppression, they must also discard a large number of potential signal events.
The fraction of signal events which survive these strong selection criteria are shown in Figure \ref{fig:VV:JJ:SigEff}; ATLAS only considers one category where the final signal acceptance is at the level of 5\%, while CMS considers five separate categories, which all-together bring the signal acceptance to the level of 20\%.
The five CMS categories come from the tagger targets ($V=W/Z$, or $H$) and requirements on the tagger (HP = high purity, or LP = low purity), where the combinations considered are $VV$-HPHP, $VV$-HPLP, $VH$-HPHP, $VH$-HPLP, and $VH$-LPHP.
HP denotes the use of tighter selections on the boson candidate(s), while LP indicates looser selections, thus increasing the background contamination in order to retain additional signal events.
The $VV$-dedicated regions contribute roughly 10\% of the 20\% signal acceptance for $VV$ final states; the remaining 10\% of retained $VV$ signal events comes from the $VH$ categories, due to the possibility to confuse $H$ bosons and $W$/$Z$ bosons, and thus the $VH$ categories are important even for $VV$ interpretations as presented in this review.

\vspace{-12pt}

\begin{figure}[H]

\subfigure[ ATLAS]{
 \includegraphics[width=0.54\textwidth]{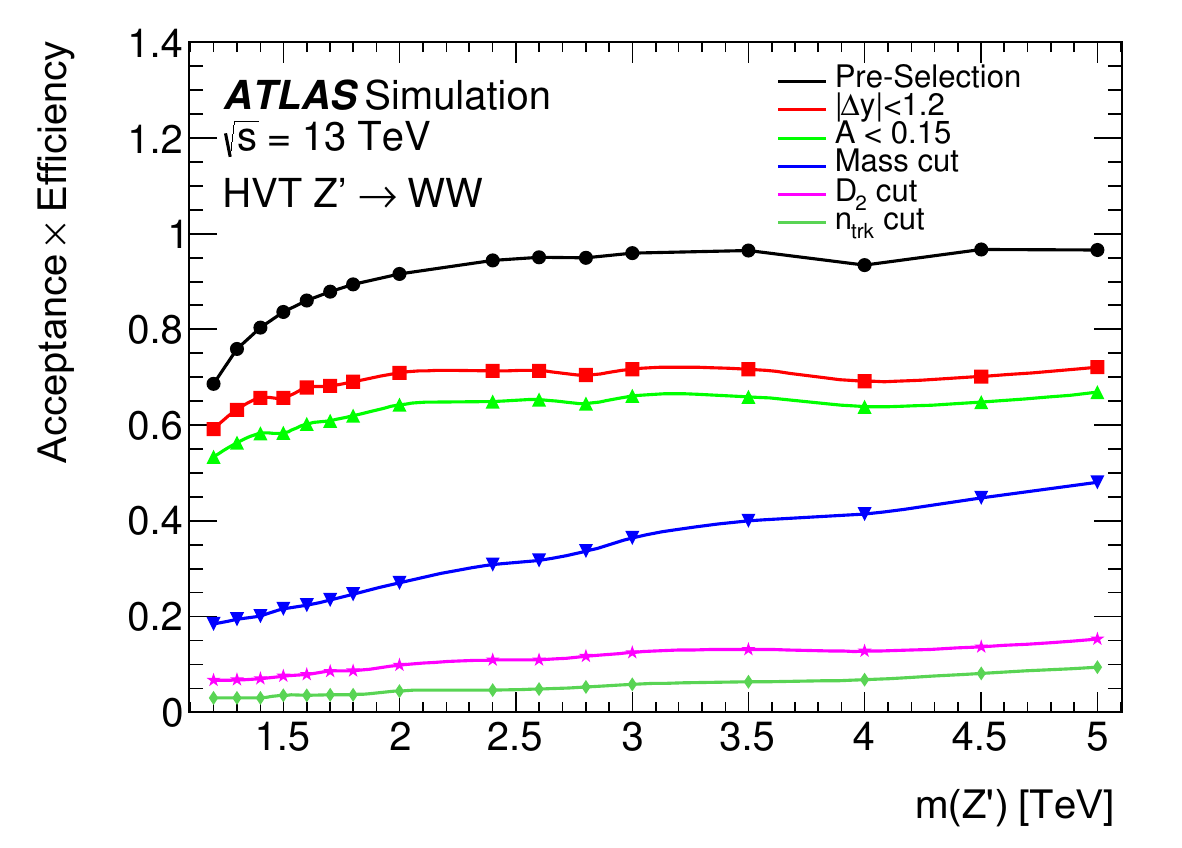}
 \label{fig:VV:JJ:ATLAS:SigEff}
}
\subfigure[ CMS]{
 \includegraphics[width=0.42\textwidth]{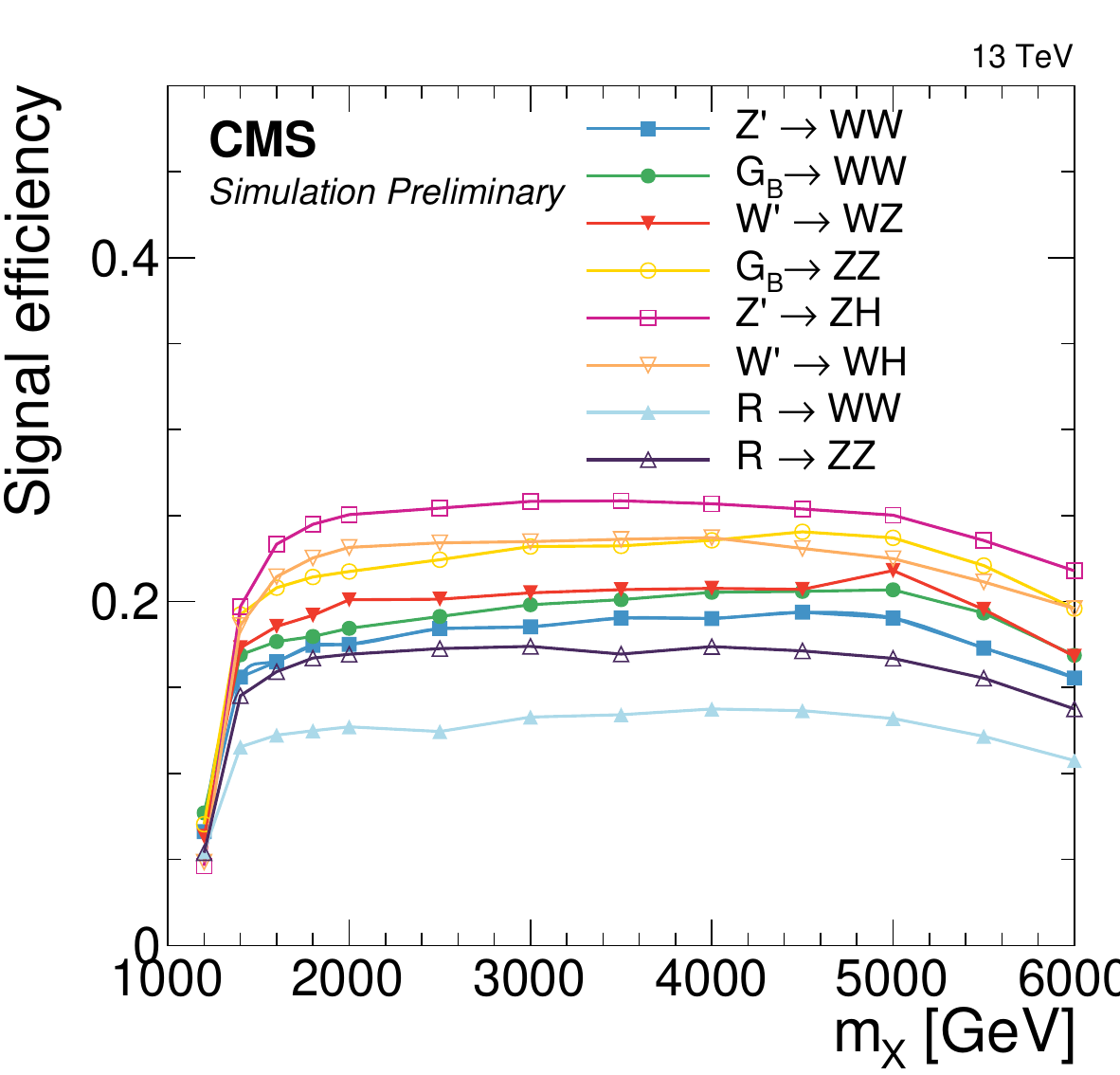}
 \label{fig:VV:JJ:CMS:SigEff}
}
\caption{(\textbf{a}) The fraction of signal events, from the $Z^\prime\to{}WW$ interpretation, accepted by each stage of the analysis selection for the ATLAS search for fully hadronic di-boson resonances \cite{ATLAS:VV_JJ}. (\textbf{b}) The fraction of signal events, from a variety of different signal interpretations, accepted by any of the five signal-enriched categories used by the CMS search for fully hadronic di-boson resonances \cite{CMS:VV_JJ_prelim}. \label{fig:VV:JJ:SigEff}}
\end{figure}

As the jet taggers are so crucial to the analysis sensitivity, it is also important to understand the extent to which the tagger performance differs between data and simulated events.
ATLAS and CMS thus evaluate the performance of their taggers in control regions, although the techniques used by the two collaborations differ.
ATLAS makes use of a dedicated $W/Z$+jets ($V$+jets) control region, where one jet is required to pass the tagger requirements other than the jet mass, and the other is required to fail one of the tagger selections.
The resulting distribution, shown in Figure \ref{fig:VV:JJ:CR}a, is still dominated by multi-jet events; however, there is a clear $W/Z$+jets peak on top of the smooth multi-jet distribution, which can therefore be fit in both data and simulation in order to extract the required tagger efficiency scale factors.
This approach works, but it is sensitive to the ability to extract the $W/Z$+jet peak from the larger multi-jet background, which limits the precision of the method.
In contrast, CMS uses a dedicated semi-leptonic $t\bar{t}$ control region, as shown in Figure \ref{fig:VV:JJ:CR}b, where the selected events are dominated by $W$ bosons.
This approach may have a higher purity of the object of interest, but does not include $Z$ bosons, and must be performed at lower energy than the analysis regime of interest as otherwise the selection becomes dominated by top quarks instead of $W$ bosons.
The two techniques used by ATLAS and CMS thus both come with their own benefits and limitations with regards to their ability to accurately measure the tagger efficiency in the kinematic regime of interest to fully hadronic di-boson resonance searches.

\begin{figure}[H]

\subfigure[ ATLAS, $W/Z$+jets]{
 \includegraphics[width=0.54\textwidth]{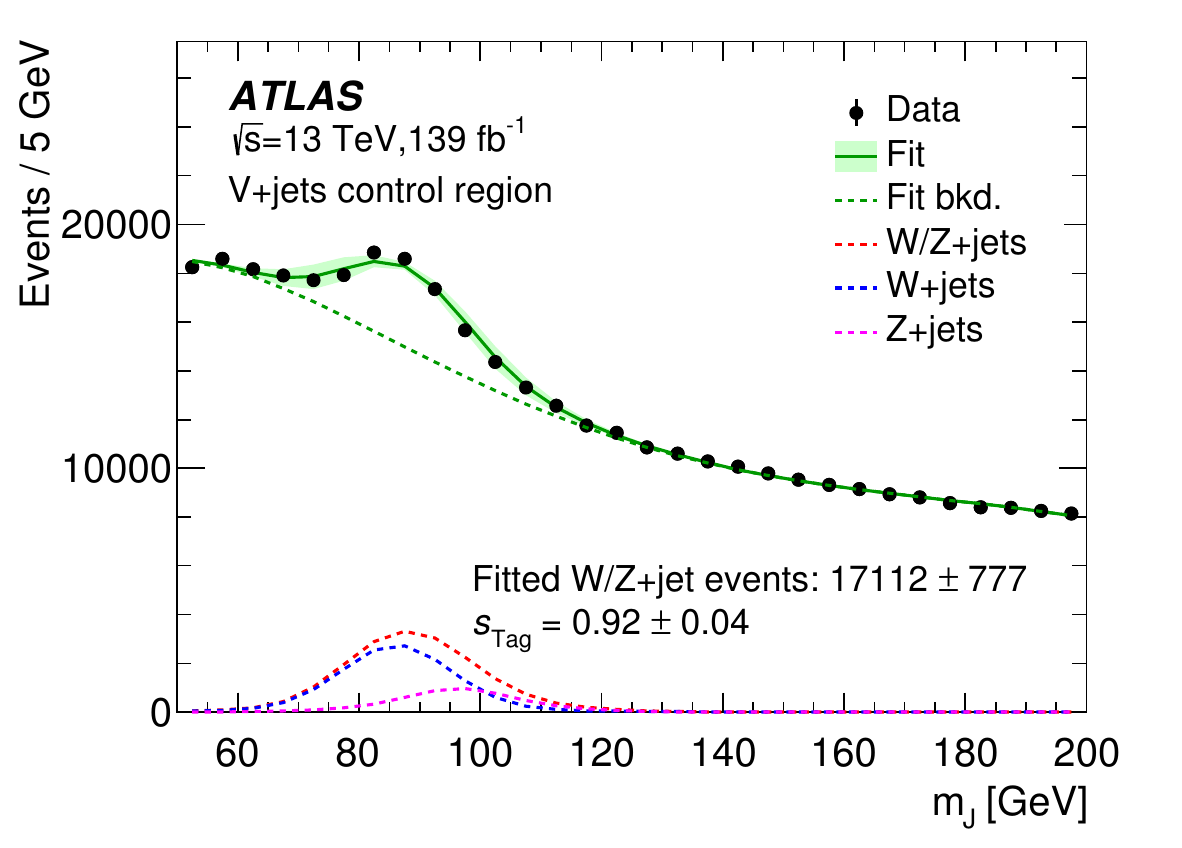}
 \label{fig:VV:JJ:ATLAS:CR}
}
\subfigure[ CMS, semi-leptonic $t\bar{t}$]{
 \includegraphics[width=0.42\textwidth]{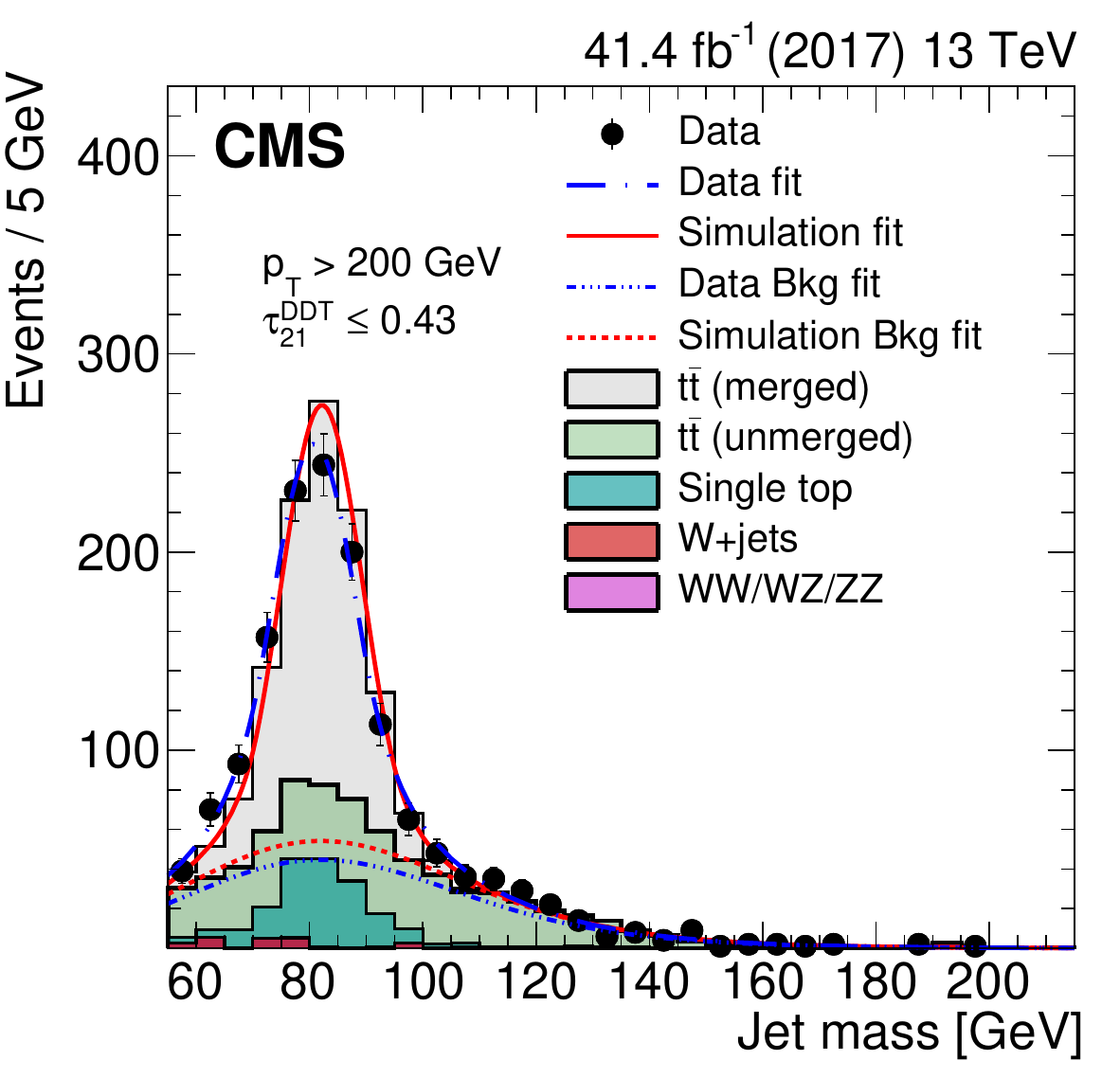}
 \label{fig:VV:JJ:CMS:CR}
}
\caption{(\textbf{a}) The $V$+jet control region used to extract the jet tagger scale factors for the ATLAS fully hadronic di-boson resonance search, using the full Run 2 dataset \cite{ATLAS:VV_JJ}. (\textbf{b}) The semi-leptonic $t\bar{t}$ control region used to extract the jet tagger scale factors for the CMS fully hadronic di-boson resonance search, using a partial Run 2 dataset \cite{CMS:VV_JJ}. \label{fig:VV:JJ:CR}}
\end{figure}

With the tagger performance under control, the remaining step is to define the background expectation in the signal region.
ATLAS does this by fitting the di-\largeR{}-jet invariant mass spectrum directly, using a functional form to describe the smoothly falling background shape.
The analysis further handles the fact that the $W$ and $Z$ taggers are not orthogonal by pre-combining the $WW$ and $WZ$ events into one signal region, and the $WW$ and $ZZ$ events into a second signal region, as these two combinations are useful in probing different possible signal interpretations.
This therefore directly handles the events that are identified as falling into both of the two tagger categories of interest; the resulting $WW+WZ$ and $WW+ZZ$ signal regions are shown in Figure \ref{fig:VV:JJ:ATLAS}a,b, respectively.
As no significant deviations are observed from the background expectation, the analysis proceeds to set limits on spin-0 Radion production using the $WW+ZZ$ region, spin-1 Heavy Vector Triplet (HVT) $W^\prime$ and $Z^\prime$ production using the $WW+WZ$ region, and spin-2 bulk RS graviton production using the $WW+ZZ$ region.
The $V^\prime$ production limits are shown in Figure \ref{fig:VV:JJ:ATLAS}c, while the graviton production limits are shown in Figure \ref{fig:VV:JJ:ATLAS}d.

Instead of fitting the di-\largeR{}-jet invariant mass spectrum alone, the CMS analysis simultaneously fits the invariant mass spectrum together with the individual jet mass spectra of both of the boson candidate jets.
By searching for peaks in this set of three distributions, it is possible that a resonance could be discovered in the di-\largeR{}-jet invariant mass spectrum that corresponds to boson masses other than those expected for $W$ and/or $Z$ bosons.
This possibility is further supported by the choice of the analysis to use a mass-decorrelated tagger, using methods discussed in Section \ref{sec:reco:largeR:tag} to make the tagger mass-independent; thus, the tagger used corresponds to a two-body-decay structure rather than strictly a $W$ or $Z$ boson decay.
The resulting di-\largeR{}-jet invariant mass spectra, with the background estimates taken from three-dimensional simultaneous fits, are shown for the HPHP category and the HPLP category in Figure \ref{fig:VV:JJ:CMS}a,b, respectively.
Two local excesses are observed, both at the level of 3.6 standard deviations, at 2.1\TeV{} and 2.9\TeV{}; these excesses both have a global significance of 2.3 standard deviations.
As no globally significant deviations are observed from the background expectation in either category, limits are set on spin-0 Radion models, spin-1 HVT $W^\prime$ and $Z^\prime$ models, and spin-2 bulk RS graviton models.
The resulting limits on $Z^\prime\to{}WW$ and $G_\mathrm{bulk}\to{}ZZ$ are shown in Figure \ref{fig:VV:JJ:CMS}c,d, respectively.

\begin{figure}[H]

\subfigure[]{
 \includegraphics[width=0.48\textwidth]{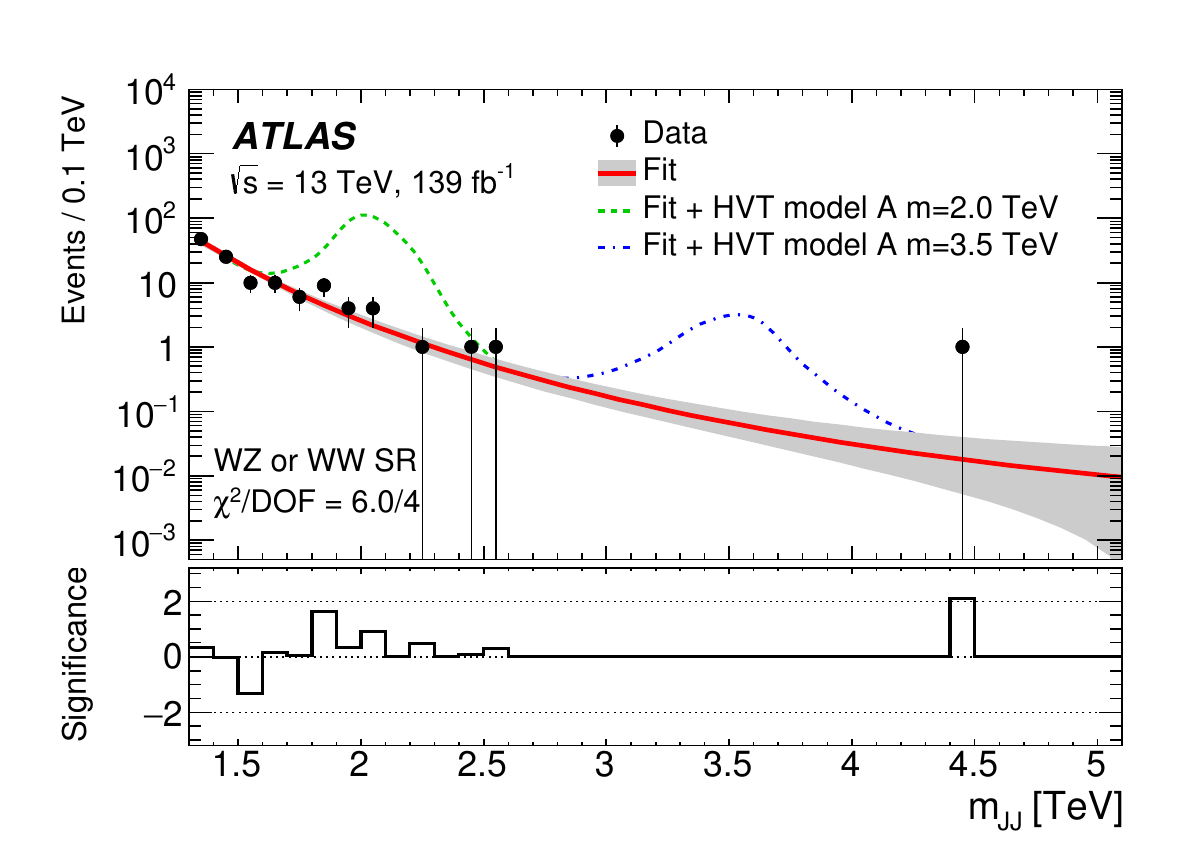}
 \label{fig:VV:JJ:ATLAS:SR1}
}
\subfigure[]{
 \includegraphics[width=0.48\textwidth]{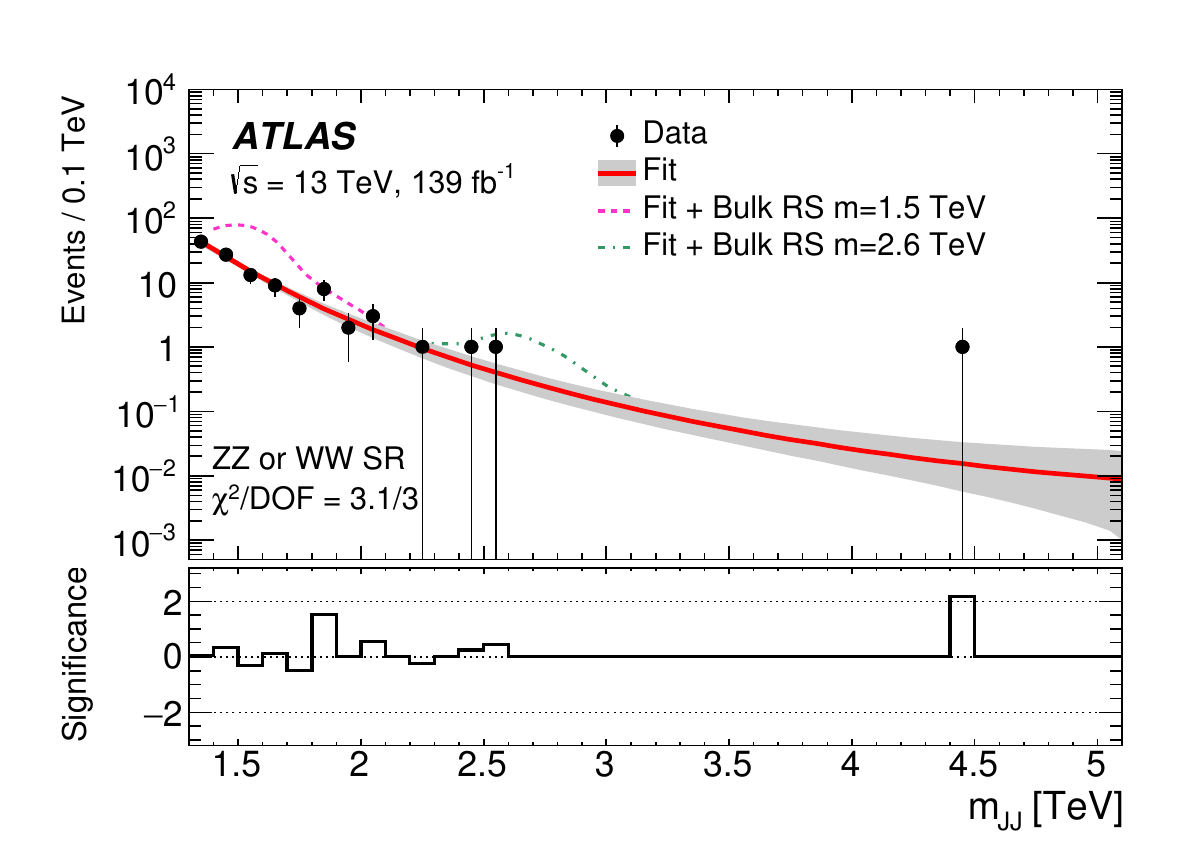}
 \label{fig:VV:JJ:ATLAS:SR2}
}\\
\subfigure[]{
 \includegraphics[width=0.48\textwidth]{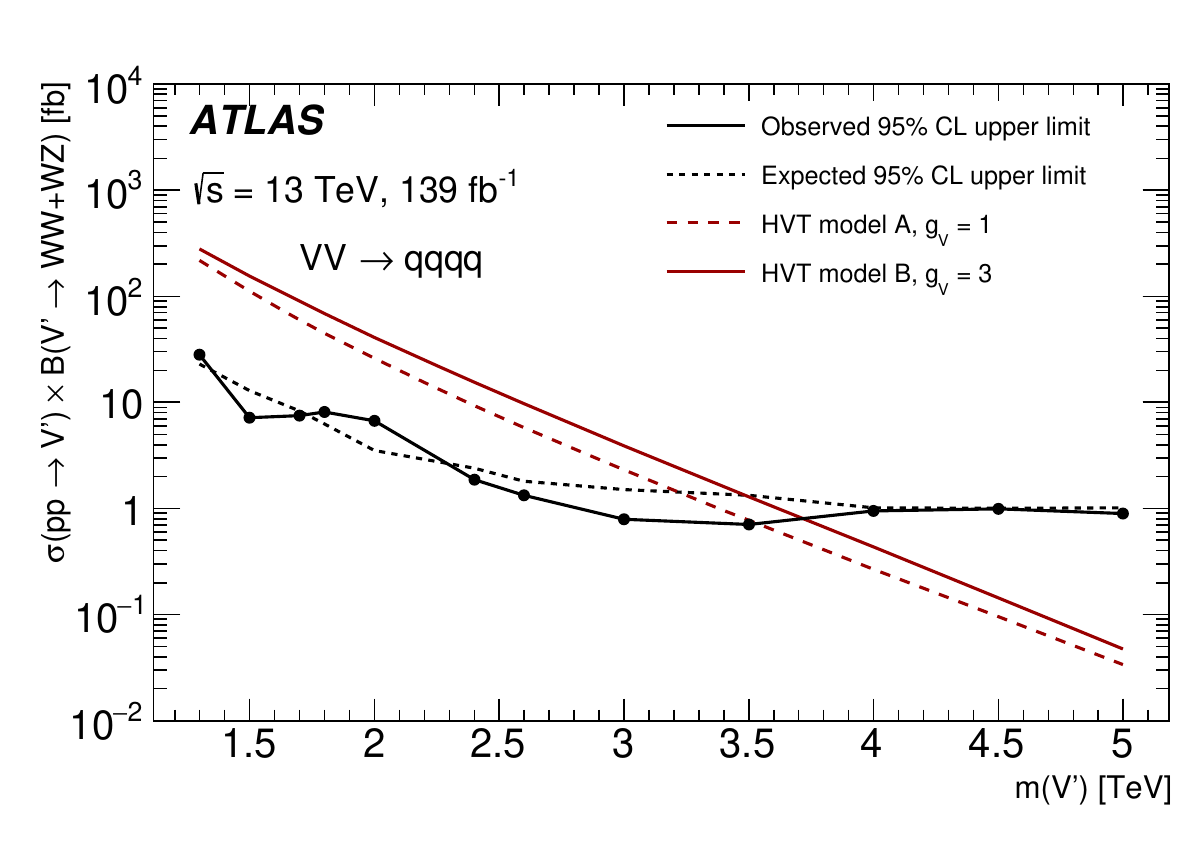}
 \label{fig:VV:JJ:ATLAS:L1}
}
\subfigure[]{
 \includegraphics[width=0.48\textwidth]{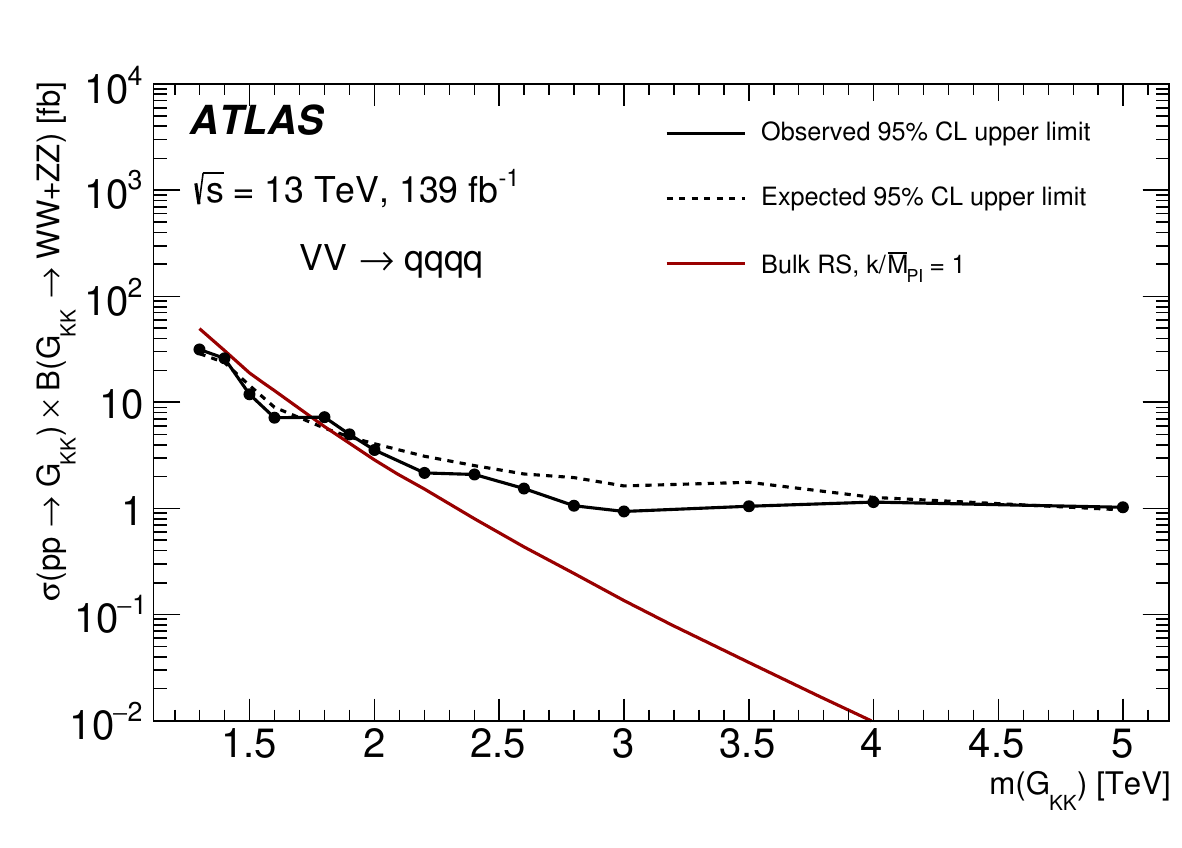}
 \label{fig:VV:JJ:ATLAS:L2}
}
\caption{The di-\largeR{}-jet invariant mass spectrum for the ATLAS fully hadronic di-boson resonance search, shown for the (\textbf{a}) $WW$ or $WZ$ region and the (\textbf{b}) $WW$ or $ZZ$ region, both using the full Run 2 dataset \cite{ATLAS:VV_JJ}. No significant deviations from the background expectation are observed, and thus limits are set on the production of (\textbf{c}) a new spin-1 $V^\prime$ decaying to $WW$ or $WZ$ boson pairs or (\textbf{d}) a new spin-2 $G_\mathrm{kk}$ decaying to $WW$ or $ZZ$ boson pairs. \label{fig:VV:JJ:ATLAS}}
\end{figure}

The need of fully hadronic di-boson searches to cut so tight on the signal in order to suppress the background, as shown earlier in Figure \ref{fig:VV:JJ:SigEff}, counteracts the aforementioned benefits of the larger branching fractions of $W$ and $Z$ bosons to hadronic final states.
Nonetheless, the hadronic final states are still complementary or competitive to other final states, even using the taggers that were available early in Run 2 \cite{ATLAS:VV_comb,CMS:VV_comb}.
The taggers used to reject the Standard Model backgrounds have improved quite a bit during Run 2; these improvements are likely to continue as advanced techniques are increasingly applied to this task.
Further tagger improvements would increase the fraction of signal events that can be retained, and thus have the potential to make the hadronic final state play an ever more important role in the search for new di-boson physics.

\subsection{Searches with Generic Bosons}
\label{sec:VV:generic}
% CWOLA

The previously described ATLAS and CMS searches for fully hadronic di-boson resonances were primarily optimized around the interpretation of Standard Model $W$ and $Z$ bosons.
However, a new physics particle $A$ may instead decay to other new particles $B$ and $C$, which then in turn decay back to pairs of quarks.
The CMS fully hadronic di-boson resonance search takes a first step in this direction in that the background estimation procedure fits the invariant mass distribution simultaneously with the individual jet mass distributions, but the analysis is not directly designed for new physics based around such alternative decays.

\vspace{-12pt}

\begin{figure}[H]

\subfigure[]{
 \includegraphics[width=0.48\textwidth]{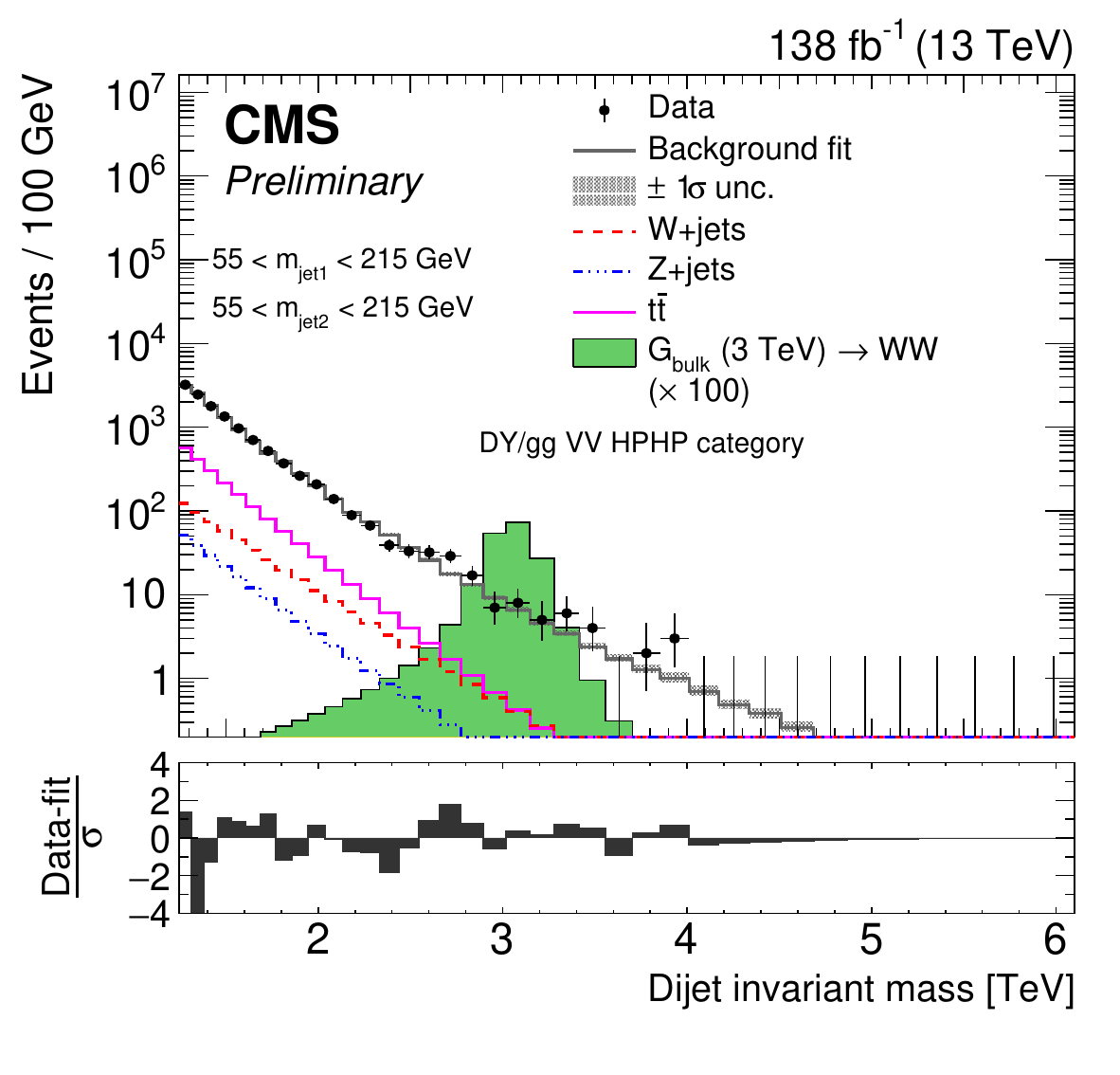}
 \label{fig:VV:JJ:CMS:SR1}
}
\subfigure[]{
 \includegraphics[width=0.48\textwidth]{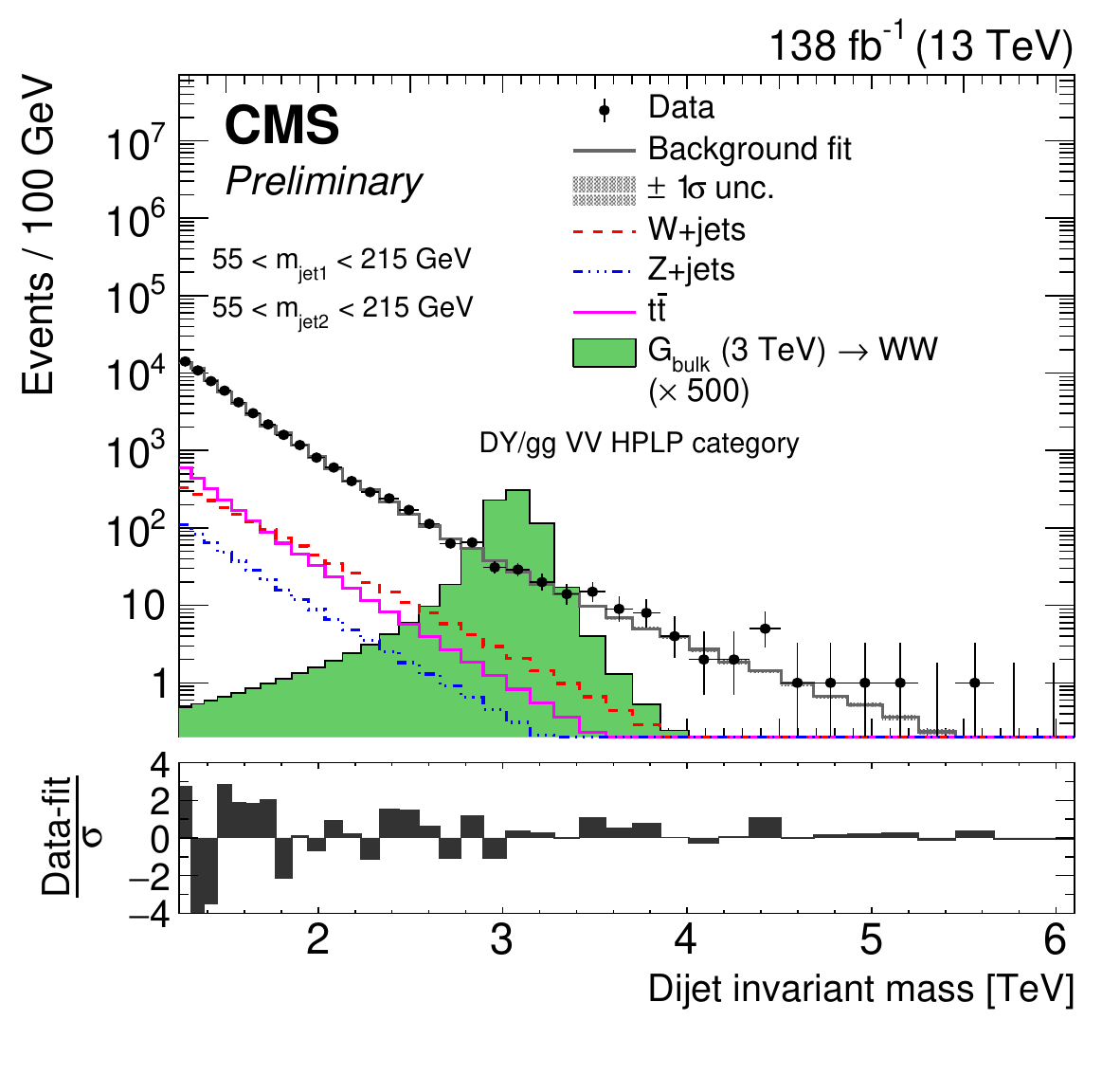}
 \label{fig:VV:JJ:CMS:SR2}
}\\
\subfigure[]{
 \includegraphics[width=0.48\textwidth]{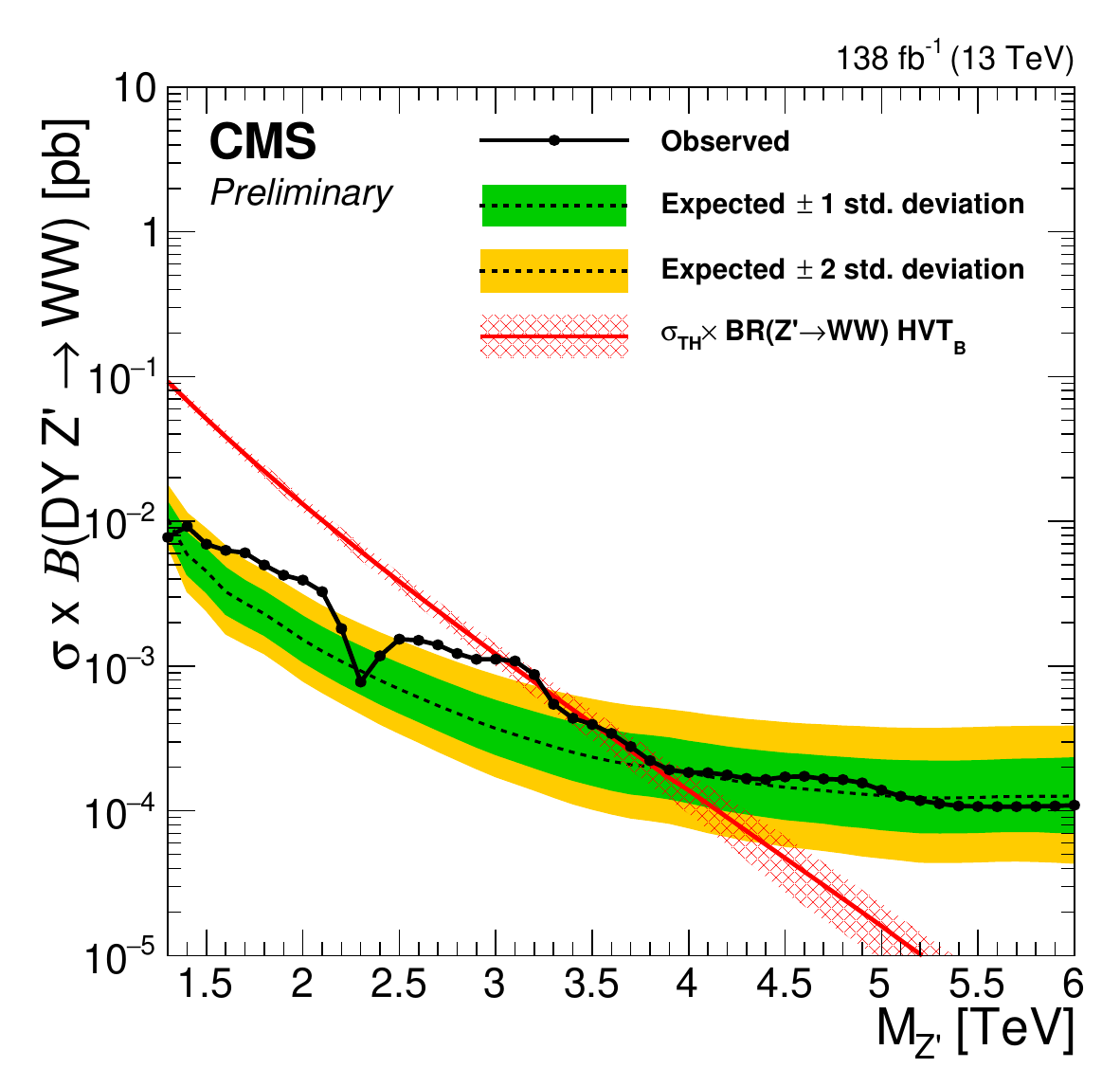}
 \label{fig:VV:JJ:CMS:L1}
}
\subfigure[]{
 \includegraphics[width=0.48\textwidth]{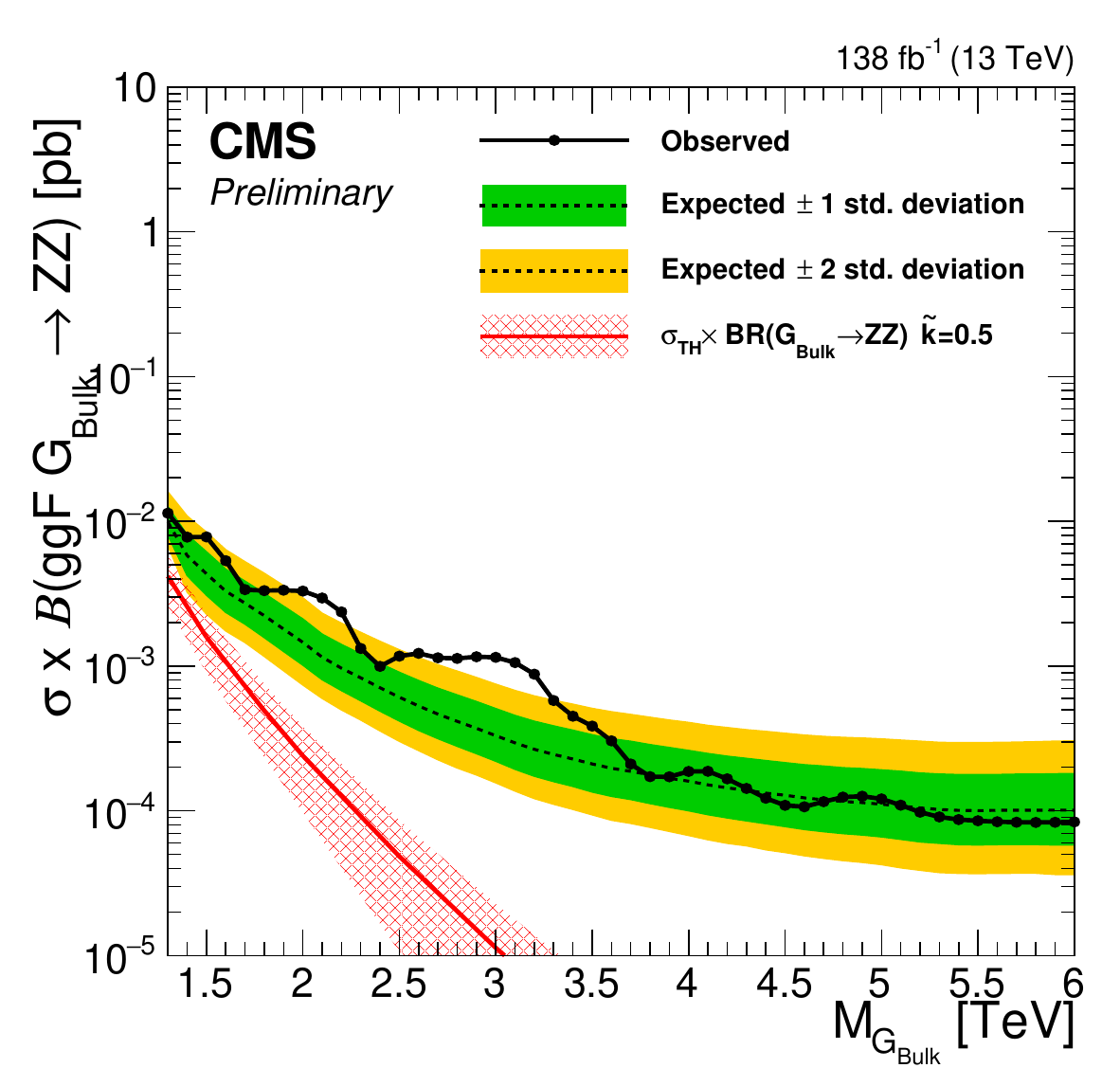}
 \label{fig:VV:JJ:CMS:L2}
}
\caption{The di-\largeR{}-jet invariant mass spectrum for the CMS fully hadronic di-boson resonance search, shown for the (\textbf{a}) $VV$ high-purity + high-purity (HPHP) region and the (\textbf{b}) $VV$ \mbox{high-purity + low-purity} (HPLP) region, both using the full Run 2 dataset \cite{CMS:VV_JJ_prelim}. No globally significant deviations from the background expectation are observed, and thus limits are set on the production of (\textbf{c}) a new spin-1 $Z^\prime$ decaying to $WW$ boson pairs or (\textbf{d}) a new spin-2 $G_\mathrm{bulk}$ decaying to $ZZ$ boson pairs. \label{fig:VV:JJ:CMS}}
\end{figure}

ATLAS has now conducted a search using the full Run 2 dataset, which has been optimised for the generic process $A\to{}BC$, where $B$ and $C$ both decay hadronically \cite{ATLAS:VV_CWOLA}.
This result is based on a very different analysis strategy than the aforementioned searches, built around the idea of Classification WithOut LAbels (CWOLA) \cite{Pheno:CWOLA}.
In this approach, the invariant mass spectrum is divided into eight regions.
The method scans over all but the extreme regions, considering one-by-one each region as a signal region, and the regions on either side as sidebands; this results in the study of six separate signal regions.
The analysis then trains a neural network to differentiate between the events in the signal window and the two sidebands.
Under the interpretation of there being new physics in the signal window more abundantly than in the sidebands, the network will learn a proxy for signal vs background discrimination.
The resulting network can then be applied to the full spectrum to enhance the contribution of signal events, and the spectrum can be fit to define the background expectation for a resonance search to be conducted within the \mbox{signal window.}

The network used to differentiate between signal-region-like and sideband-region-like events is trained using only the masses of the two \largeR{} jets.
It is furthermore applied using two different selections, one corresponding to keeping the 10\% most signal-region-like events ($\epsilon=10\%$), and another correspond to keeping the 1\% most signal-region-like events ($\epsilon=1\%$).
The resulting signal windows for $\epsilon=10\%$ are stitched together to form a single plot, shown in Figure \ref{fig:VV:CWOLA}a.
As the signal windows each apply a different neural network, there is no expectation that the resulting spectrum will be smooth at the stitching boundaries, as is quite clear at the 5.68\TeV{} boundary.

The approach described so far has no dependence on any simulated model, but if desired, a model can be injected into the neural network training process.
If this is done, the resulting network will become more sensitive to that specific model, at the cost of reduced sensitivity to other possible types of new physics.
This is useful to allow this generic search to be compared with other analyses searching for specific benchmark models, and thus an example of injecting 3\TeV{} and 5\TeV{} $W^\prime\to{}WZ$ signals is shown for $\epsilon=10\%$ in Figure \ref{fig:VV:CWOLA}b.
Comparing Figure \ref{fig:VV:CWOLA}a to Figure \ref{fig:VV:CWOLA}b, it is clear that the injection of a signal into the training process has resulted in a stronger classifier, and thus the number of events near 5\TeV{} is further suppressed with respect to the signal-model-independent training.

\vspace{-6pt}

\begin{figure}[H]

\subfigure[ SR, model-independent]{
 \includegraphics[width=0.48\textwidth]{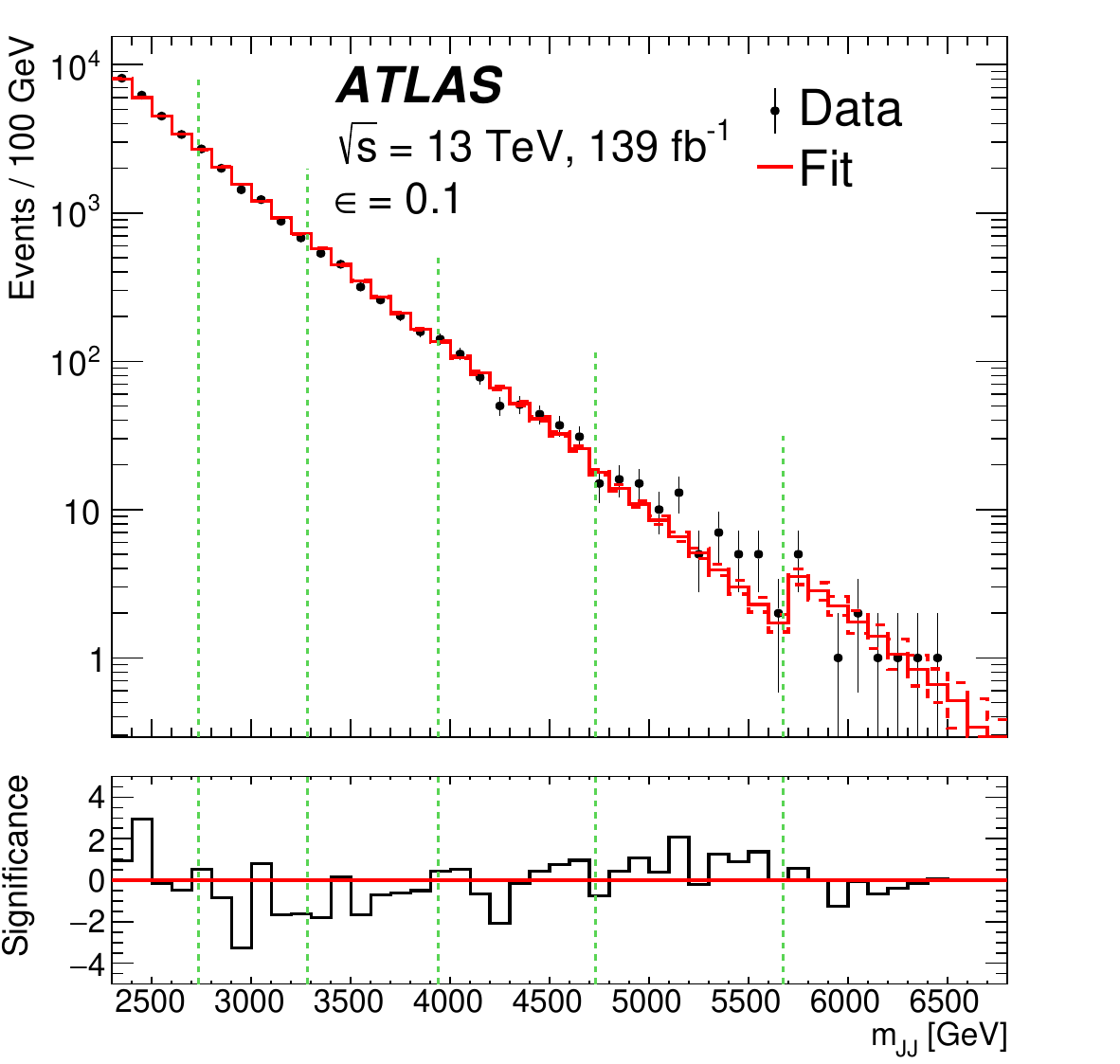}
 \label{fig:VV:CWOLA:ATLAS:SR1}
}
\subfigure[ SR, model-dependent]{
 \includegraphics[width=0.48\textwidth]{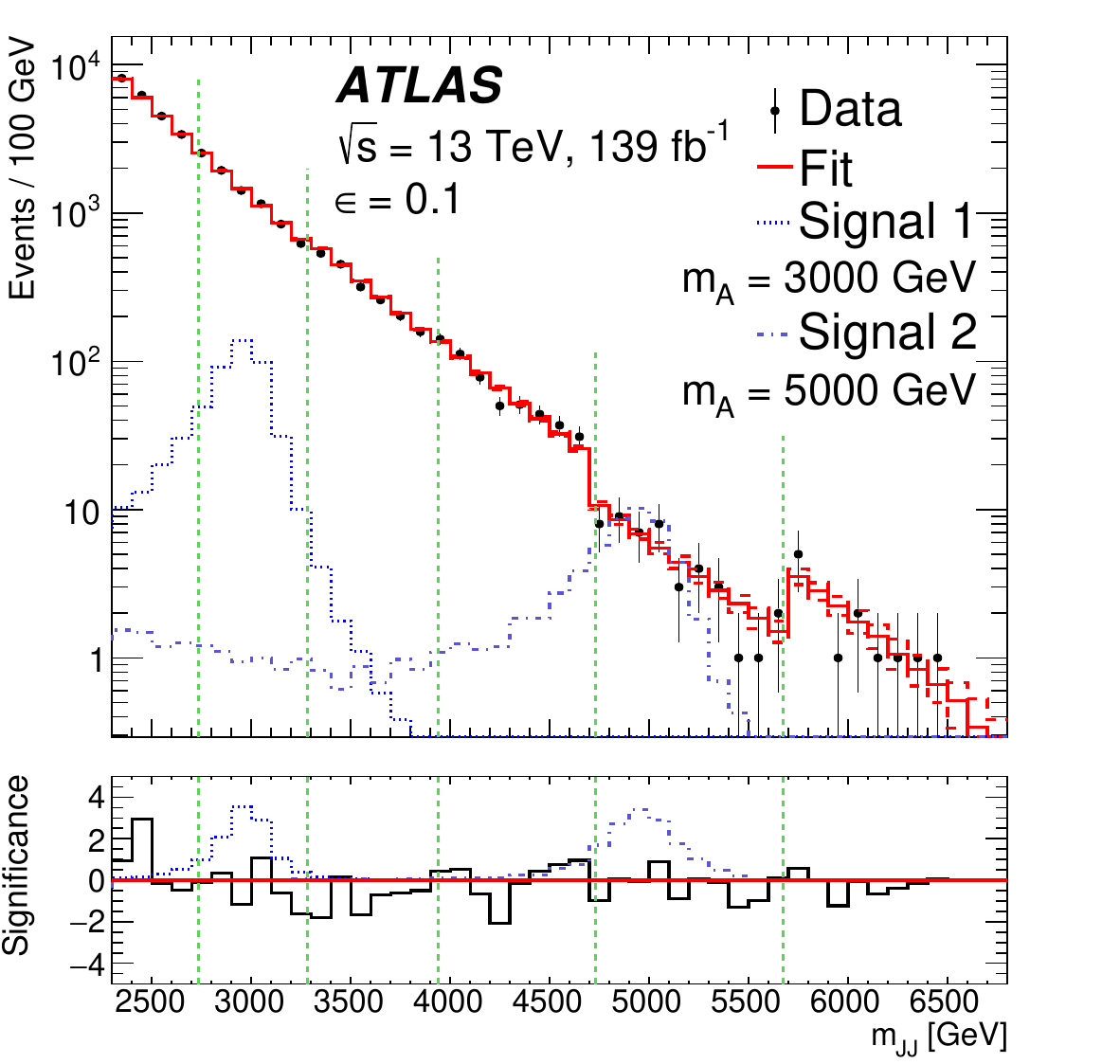}
 \label{fig:VV:CWOLA:ATLAS:SR2}
}\\
\subfigure[ Limits, 10\% selection]{
 \includegraphics[width=0.48\textwidth]{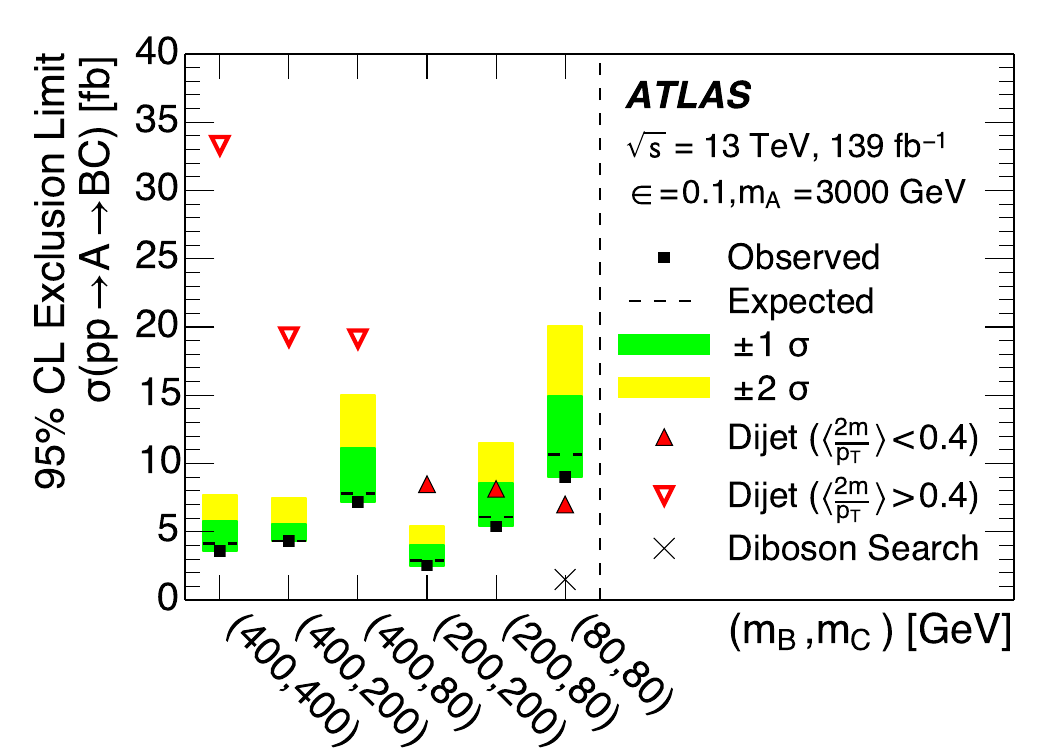}
 \label{fig:VV:CWOLA:ATLAS:L1}
}
\subfigure[ Limits, 1\% selection]{
 \includegraphics[width=0.48\textwidth]{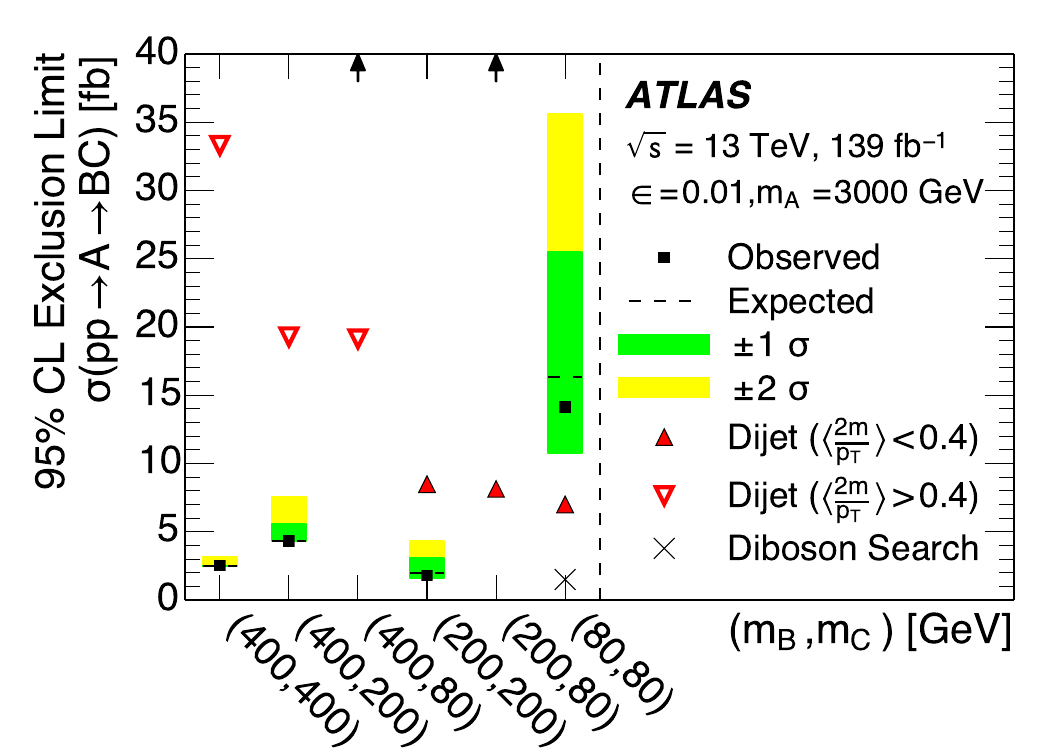}
 \label{fig:VV:CWOLA:ATLAS:L2}
}
\caption{The invariant mass distribution of the six signal regions considered in the di-\largeR{}-jet resonance search using weak supervision, showing (\textbf{a}) the signal regions for a cut of 10\% acceptance without any signal injected in the training process and (\textbf{b}) the signal regions for a cut of 10\% acceptance with signal injected at both 3\TeV{} and 5\TeV{} in the training process; both are done using the ATLAS full Run 2 dataset \cite{ATLAS:VV_CWOLA}. No significant deviations are observed from the background expectation in any of the signal regions, and thus limits are set on the production of a generic $A$ decaying to bosons $B$ and $C$, where different benchmark $B$ and $C$ masses are considered. Limits are shown for both (\textbf{c}) a 10\% selection and (\textbf{d}) a 1\% selection, and where limits are compared to traditional searches.\label{fig:VV:CWOLA}}
\end{figure}

No significant deviations are observed from the background expectation in any of the six regions, at either $\epsilon=10\%$ or $\epsilon=1\%$, and thus limits are set on benchmark $A\to{}BC$ model for a variety of different values of the masses of $A$, $B$, and $C$. The results for $A=3\TeV{}$ are shown for $\epsilon=10\%$ and $\epsilon=1\%$ in Figure \ref{fig:VV:CWOLA}c,d, respectively, where the limits from dedicated analyses are overlaid.
The dedicated fully hadronic di-boson resonance search from ATLAS is more sensitive than the generic search when $A$ and $B$ have the $W$ mass, but for other values of $A$ and $B$ the traditional analysis has no sensitivity, as the taggers used are heavily optimised around the Standard Model $W/Z$ boson interpretation.
Instead, the limits from the high-mass di-jet search discussed in Section \ref{sec:dijet:classic} are relevant, as the $A$ can instead decay back directly to pairs of quarks instead of decaying to $B$ and $C$.
The same figures show that the search for generic $A\to{}BC$ resonances is often much more sensitive than the di-jet search to this model, as the neural network is exploiting the structure of the final state to reject the Standard Model multi-jet background while retaining the signal candidates of interest.
This nicely demonstrates the utility of such an approach to searching for generic $A\to{}BC$ di-boson resonances: it cannot out-perform a dedicated search for a given combination of the masses of $B$ and $C$, but it can increase the sensitivity to other mass assumptions beyond what is possible from re-interpreting searches that do not consider the structure of the decay process.

% Complementarity
\section{Complementarity of Hadronic Physics Searches}
\label{sec:complementarity}

The di-jet searches presented in Section \ref{sec:dijet} are of great importance when considering complementarity between different methods of looking for new physics beyond the standard model.
Di-jet searches provide a generic means of constraining the presence of a whole class of new physics models: if a model assumes that the mediator can be produced through an s-channel process involving the annihilation of quark--antiquark pairs, then di-jet searches can probe that model, as the mediator can also decay back to quark--antiquark pairs (excluding mediators that are sufficiently long-lived for the decay to occur after leaving the detector volume).
This means that di-jet searches are directly complementary to the missing transverse momentum plus X searches discussed in Section \ref{sec:monoX} and the hadronic di-boson searches discussed in Section \ref{sec:VV}.

The complementarity of di-jet searches and missing transverse momentum plus X searches has been studied in detail during Run 2, with particular emphasis on the search for new axial-vector $Z^\prime$ bosons with couplings to dark matter.
ATLAS \cite{ATLAS:DMSummary} and CMS \cite{CMS:EXOSummary} have both created plots overlaying the sensitivity of their different types of searches to such a $Z^\prime$ model, as shown in Figure \ref{fig:comp:DM}.
The couplings of the postulated $Z^\prime$ boson to quarks ($g_q$), leptons ($g_\ell$), and dark matter ($g_\chi$) are not fixed parameters, and thus different coupling scenarios are considered, following the LPCC Dark Matter Working Group recommendations \cite{Theory:ZPrimeLHC1,Theory:ZPrimeLHC2,Theory:ZPrimeLHC3,Theory:ZPrimeLHC4}.
The different scenarios nicely demonstrate the complementary sensitivity of the different types of searches: if $g_q$ is large compared to $g_\ell$, di-jet searches lead the sensitivity across the full parameter space; if $g_q$ and $g_\ell$ are of similar size, di-lepton searches (not discussed in this review) take the lead; and if $g_q$ and $g_\ell$ are both small, missing transverse momentum plus X searches are of great importance.

Di-jet and di-boson searches are similarly complementary, but their joint sensitivity has not yet been compared to the same extent.
However, it is clear that similar behaviour would be observed, namely, di-jet searches would be of great importance when the coupling of the new mediator to quarks is large, and di-boson searches would provide leading sensitivity when the coupling of the new mediator to bosons is large.
%A full evaluation of the mutual coverage of di-jet and di-boson searches to models of new physics beyond the Standard Model would be of great interest, but such a publication remains to be seen.

% with couplings of $g_q = 0.25$, $q_\chi=1$, and $g_\ell=0$, as shown in Figures \ref{fig:comp:DM:ATLAS} and \ref{fig:comp:DM:CMS}, respectively.

\begin{figure}[H]

\subfigure[ ATLAS: $g_q=0.25$, $g_\ell=0$, $g_\chi=1$]{
 \includegraphics[width=0.47\textwidth]{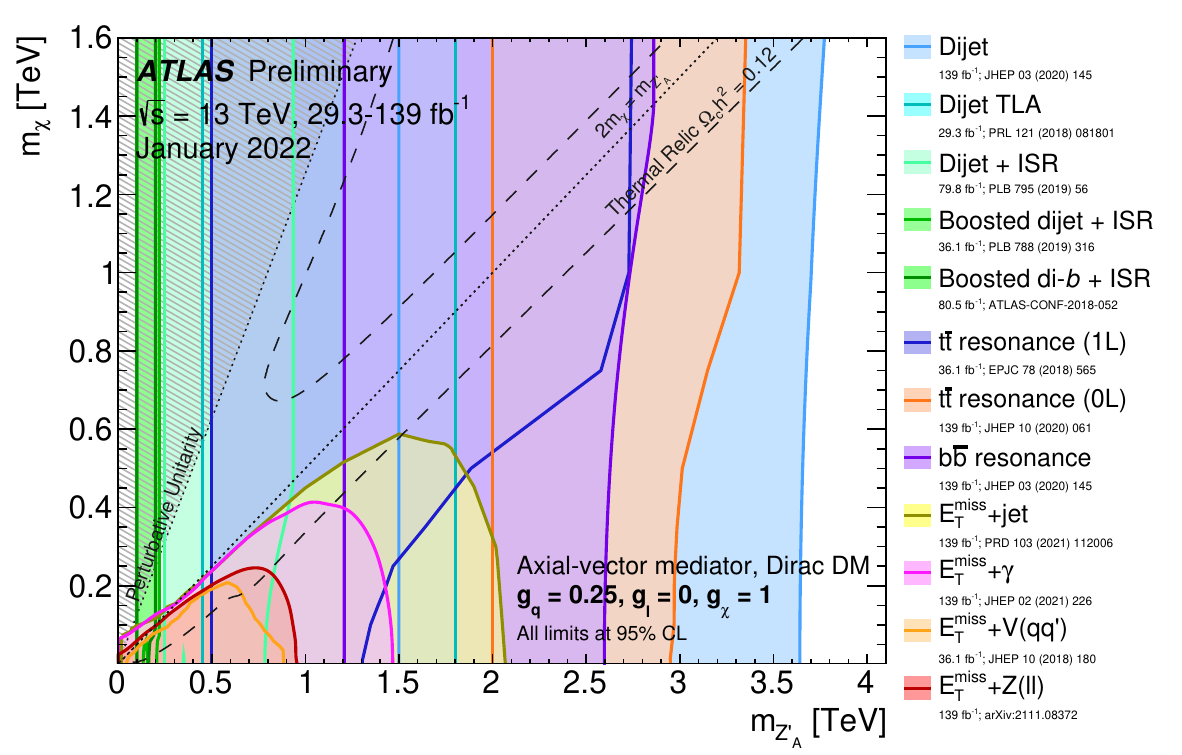}
 \label{fig:comp:DM:ATLAS}
}
\subfigure[ CMS: $g_q=0.25$, $g_\ell=0$, $g_\chi=1$]{
 \includegraphics[width=0.47\textwidth]{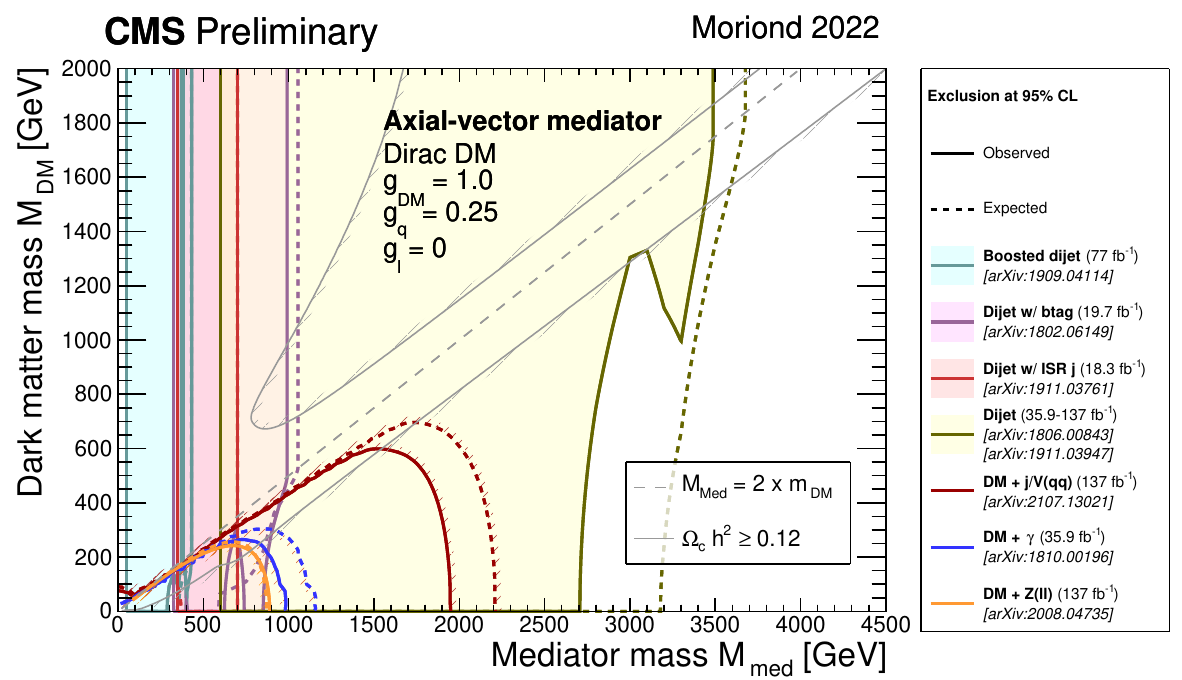}
 \label{fig:comp:DM:CMS}
}\\
\subfigure[$g_q=0.1$, $g_\ell=0.1$, $g_\chi=1$]{
 \includegraphics[width=0.47\textwidth]{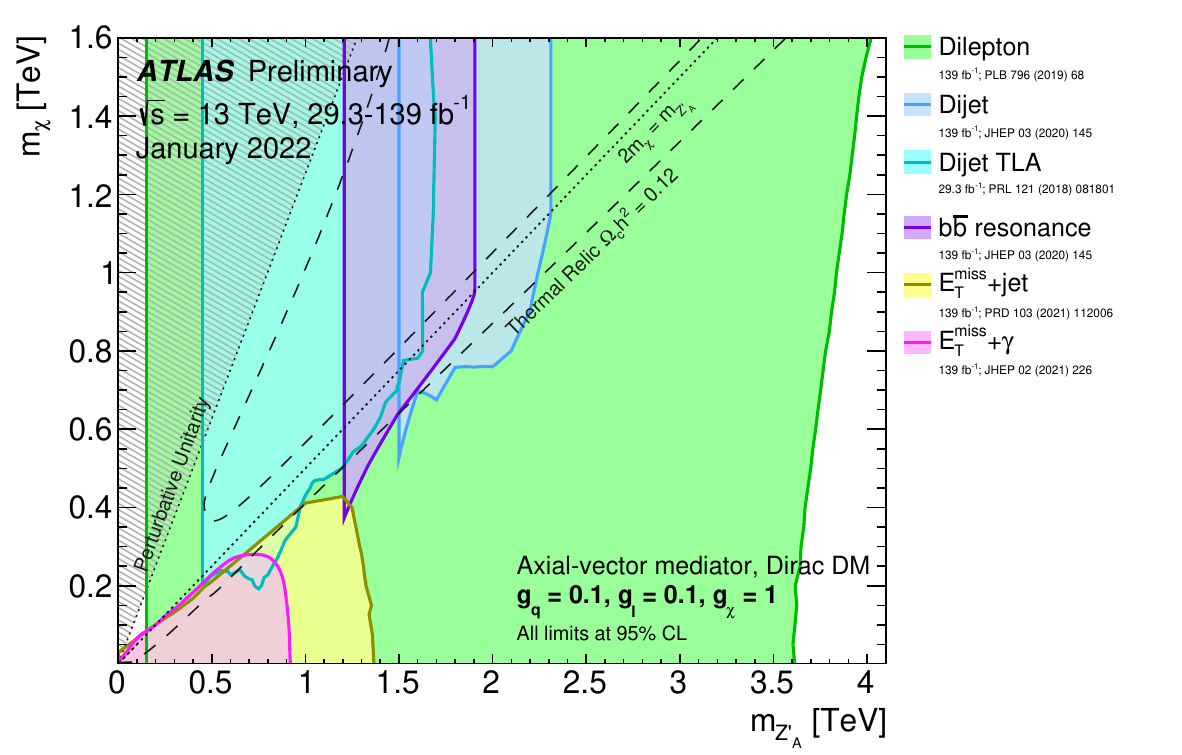}
 \label{fig:comp:DM:withLeptons}
}
\subfigure[$g_q=0.1$, $g_\ell=0.01$, $g_\chi=1$]{
 \includegraphics[width=0.47\textwidth]{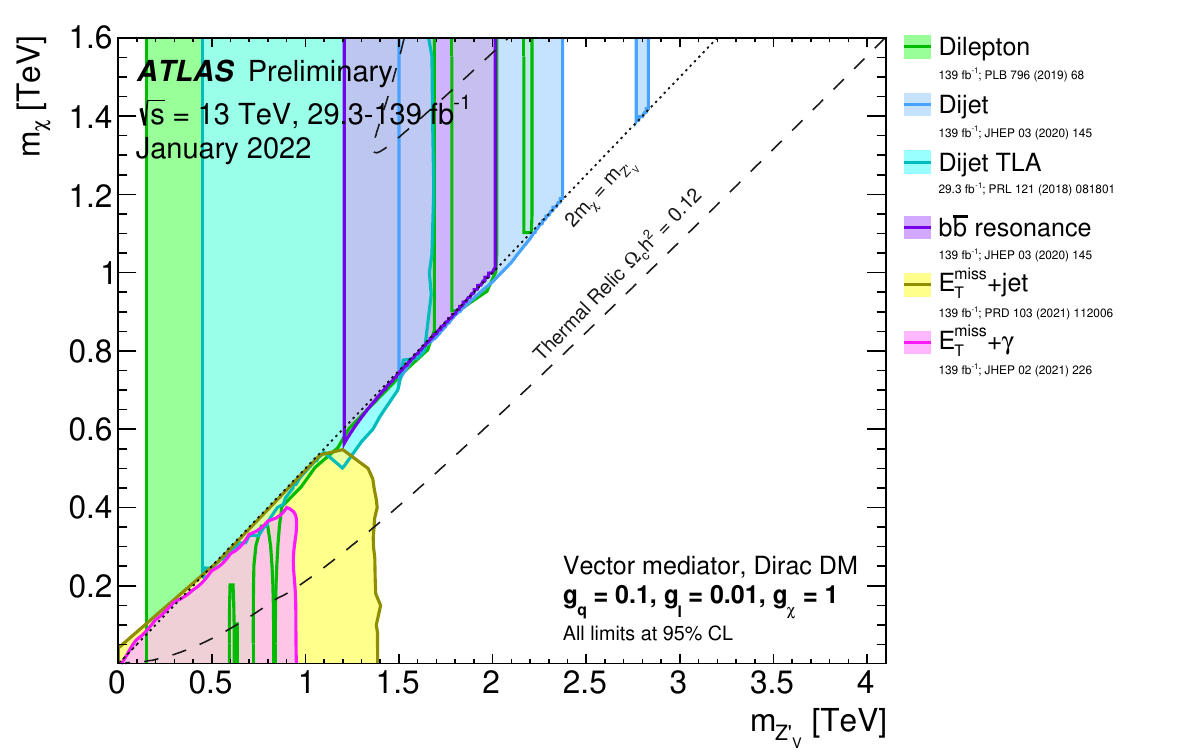}
 \label{fig:comp:DM:weakSM}
}
\caption{The parameter space coverage of axial-vector $Z^\prime$ models, with potential couplings to quarks ($g_q$), leptons ($g_\ell$), and dark matter particles ($g_\chi$,$g_\mathrm{DM}$). The sensitivity of a wide variety of relevant analyses is shown in an overlapping way, in order to present the total coverage of such models. The results are shown for (\textbf{a}) ATLAS \cite{ATLAS:DMSummary} and (\textbf{b}) CMS \cite{CMS:EXOSummary}, both with ($g_q=0.25$, $g_\ell=0$, $g_\chi=1$). Alternative assumptions for the $Z^\prime$ coupling values are shown for ATLAS results in (\textbf{c}) with ($g_q=0.1$, $g_\ell=0.1$, $g_\chi=1$) \cite{ATLAS:DMSummary}, and (\textbf{d}) with ($g_q=0.1$, $g_\ell=0.01$, $g_\chi=1$) \cite{ATLAS:DMSummary}. \label{fig:comp:DM}}
\end{figure}

% Outlook
\section{Summary and Outlook}
\label{sec:outlook}

This review provides an overview of ATLAS and CMS searches for new physics in hadronic final states, shortly before the start of Run 3.
Following a discussion on the motivations and challenges of physics involving hadronic final states at the LHC in Section \ref{sec:motivation}, the different jet reconstruction, calibration, and tagging strategies employed by ATLAS and CMS were presented in Section \ref{sec:reco}.
These provide the necessary background to understand how hadronic final states are observed and the precision that such hadronic observables have attained during Run 2 of the LHC.

With this baseline in place, the review shifted to searches for di-jet resonances in Section \ref{sec:dijet}.
These searches are of fundamental importance to the ATLAS and CMS physics programmes, as they are sensitive to a wide range of possible new physics models, due to their minimal set of assumptions: the new particle under study must be possible to create through quark--antiquark annihilation.
The di-jet search programme has increased in scope considerably during Run 2, including new strategies to circumvent the trigger barrier and to access the lower-mediator-mass regime with unprecedented precision.
New techniques to access this regime continue to be deployed, and several of the analyses presented have only been conducted by one of ATLAS or CMS so far; there is thus still scope for further improvements and possible discovery of new physics in these low-mass di-jet searches using only the Run 2 dataset, which will only be further improved during Run 3.

Many di-jet searches are conducted in order to probe the existence of new mediators between the Standard Model and dark matter.
Another means of probing the existence of such mediators is to directly study events including the production of dark matter particles, which must balance some initial-state-radiated object in order to be visible to the detector.
Searches for missing transverse momentum in association with jets, and briefly also with other objects, were presented in Section \ref{sec:monoX}.
While the flagship search of this type has already been published using the full Run 2 dataset by both ATLAS and CMS, there remain other signatures that are competitive for some models of mediators to the dark sector, and not all of those searches have yet been extended to the full dataset.
Moreover the flagship mono-jet analysis is generally systematically limited; thus, further refinements to the analysis strategy, the object reconstruction uncertainties, or theoretical uncertainties could provide sizeable improvements to the analysis sensitivity.

Another possibility is that the new mediator preferentially decays to pairs of bosons, and thus searches for resonances in fully hadronic decays of pairs of electroweak bosons were discussed in Section \ref{sec:VV}.
These searches depend crucially on modern developments in jet tagging, whereby jets consistent with originating from $W$ and/or $Z$ bosons are selected, while jets originating from quarks or gluons are rejected.
This is necessary to overcome the otherwise enormous Standard Model multi-jet background, and the ability to do so has improved substantially during Run 2.
These searches will continue to benefit from collecting more data, but ultimately further improvements to the tagger design or analysis strategy are likely to provide more significant gains in the coming years.
Additional searches for generic boson resonances, where the bosons do not necessarily have the $W$ or $Z$ boson mass, may also yield discoveries in regions that are not covered by dedicated analyses.

These three types of searches for new physics in hadronic final states are complementary, as discussed in Section \ref{sec:complementarity}.
The di-jet searches and missing transverse momentum plus associated object searches can both be directly interpreted under the context of the same model, which was shown for a new axial-vector mediator.
As there is no well-defined expectation for the couplings of such a new mediator to quarks, leptons, and dark matter particles, different signatures must all be studied in order to maximally cover the new physics parameter space.
Di-jet searches are also complementary to the fully hadronic di-boson searches for new mediators for similar reasons: di-jet searches would provide the strongest sensitivity when the quark coupling is large, while di-boson searches would provide better sensitivity when the coupling to bosons is large.
There is substantial scope for further comparisons and combinations of different search strategies, both hadronic or otherwise, which may indicate regions that are being missed by the current ATLAS and CMS search programmes.

The analyses presented in this review are only a subset of all possible searches for new physics in hadronic final states.
Hadronic final states are already of great interest at the LHC, yet the precision of hadronic physics observables has continued to improve, taggers to identify specific types of hadronic objects frequently provide large gains with respect to previous versions, and new hadronic analysis strategies are being developed and deployed.
The challenges of hadronic physics at the LHC are being slowly but surely mitigated, and with these advances, there will surely be many new opportunities for searches in hadronic final states in the years to come.

%%%%%%%%%%%%%%%%%%%%%%%%%%%%%%%%%%%%%%%%%%

\section*{Acknowledgements}
This review is part of a project that has received funding from the European Research Council (ERC) under the European Union’s Horizon 2020 research and innovation programme (Grant agreement No. 948254) and from the Swiss National Science Foundation (SNSF) under Eccellenza grant number PCEFP2\_194658.

\clearpage
%%%%%%%%%%%%%%%%%%%%%%%%%%%%%%%%%%%%%%%%%%

\addcontentsline{toc}{section}{Bibliography}
\printbibliography[title=References]

\end{document}